\documentclass[11pt,a4paper]{article}

\usepackage{jheppub}

\usepackage[english]{babel}

\usepackage{slashed}

\allowdisplaybreaks[2]

\newcommand{\half}{{\textstyle\frac{1}{2}}}

\newcommand{\ms}{\mskip 1.5mu}
\newcommand{\bs}{\mskip -1.5mu}

\newcommand{\lrD}{{D^{\hspace{-0.8em}%
      \raisebox{0.8ex}{$\scriptstyle\leftrightarrow$}}}{}}

\newcommand{\lrpartial}{{\partial^{\hspace{-0.65em}%
      \raisebox{0.8ex}{$\scriptstyle\leftrightarrow$}}}\hspace{-0.05em}{}}
\newcommand{\lpartial}{{\partial^{\hspace{-0.65em}%
      \raisebox{0.8ex}{$\scriptstyle\leftarrow$}}}\hspace{-0.05em}{}}
\newcommand{\rpartial}{{\partial^{\hspace{-0.65em}%
      \raisebox{0.8ex}{$\scriptstyle\rightarrow$}}}\hspace{-0.05em}{}}

\newcommand{\tvec}[1]{\boldsymbol{#1}}

\newcommand{\mvec}[1]{\smash{\vec{\mskip 0.5mu #1}}\mskip 1.5mu}

\newcommand{\tr}{\operatorname{tr}}

\newcommand{\one}{1\!\!1}

\newcommand{\mat}[1]{\big\langle\!\big\langle\ms #1
                     \ms\big\rangle\!\big\rangle}

\newcommand{\sing}[1]{{}^{1\!}#1}
\newcommand{\oct}[1]{{}^{8\!}#1}
\newcommand{\octA}[1]{{}^{A\!}#1}
\newcommand{\octS}[1]{{}^{S\!}#1}

\subheader{\hfill DESY 11-196}

\title{Elements of a theory for multiparton interactions in QCD}

\author[a]{Markus Diehl,}
\author[b]{Daniel Ostermeier}
\author[b]{and Andreas Sch\"afer}

\affiliation[a]{Deutsches Elektronen-Synchroton DESY, 22603 Hamburg,
  Germany}

\affiliation[b]{Institut f\"ur Theoretische Physik, Universit\"at
  Regensburg, 93040 Regensburg, Germany}

\abstract{We perform a detailed investigation of multiple hard
  interactions in hadron-hadron collisions.  We discuss the space-time,
  spin and color structure of multiple interactions, classify different
  contributions according to their power behavior and provide several
  elements required for establishing all-order factorization.  This also
  allows us to analyze the structure of Sudakov logarithms in double hard
  scattering.  We show how multiparton distributions can be constrained by
  connecting them with generalized parton distributions and by calculating
  their behavior at large transverse parton momenta.}

\begin{document}

\maketitle



\section{Introduction}
\label{sec:intro}

When two hadrons collide at high energies, more than one parton in one
hadron can have a hard interaction with a parton in the other hadron and
produce particles with large mass or transverse momentum.  The effects of
such multiparton interactions are suppressed or average out in
sufficiently inclusive observables, but they have important consequences
for the details of the hadronic final state.  The possible importance of
multiparton interactions has been realized long ago
\cite{Landshoff:1975eb,Landshoff:1978fq} and phenomenological estimates
have been given for many final states such as four jets (possibly
including $b$ quarks)
\cite{Humpert:1983pw,Humpert:1984ay,Ametller:1985tp,Mangano:1988sq,%
  DelFabbro:2002pw,Domdey:2009bg,Berger:2009cm}, jets associated with
photons or leptons \cite{Drees:1996rw}, four leptons produced by the
double Drell-Yan process \cite{Goebel:1979mi,Halzen:1986ue,Kom:2011nu} or
from two charmonium states
\cite{Kom:2011bd,Baranov:2011ch,Novoselov:2011ff}, as well as a number of
channels with electroweak gauge bosons \cite{Godbole:1989ti,Eboli:1997sv,%
  DelFabbro:1999tf,Cattaruzza:2005nu,Maina:2009vx,Maina:2009sj,Maina:2010vh,%
  Kulesza:1999zh,Gaunt:2010pi,Berger:2011ep}.  Experimental evidence for
multiple hard scattering has been found in the production of multijets
\cite{Akesson:1986iv,Alitti:1991rd,Abe:1993rv} and of a photon associated
with three jets \cite{Abe:1997bp,Abe:1997xk,Abazov:2009gc,Abazov:2011rd}.
A mini-review of the subject can be found in \cite{Sjostrand:2004pf} and
an overview of how multiparton interactions are modeled in current Monte
Carlo event generators is given in \cite{Buckley:2011ms}.

At LHC energies, the phase space for having several hard interactions in a
proton-proton collision is greatly increased compared with previous
experiments, and it is expected that the effects of multiple interactions
will be important in many processes
\cite{Alekhin:2005dx,Jung:2009eq,Bartalini:2010su,Bartalini:2011jp}.  This
poses a 
challenge in searches for new physics and at the same time offers the
possibility to study multiple interactions in much more detail than
before.  First experimental results on multiple hard scattering at the LHC
have already appeared \cite{Atlas:2011co} and more can be expected in the
near future \cite{Bartalini:2011xj}.

Understanding multiparton interactions is also important for heavy-ion
physics, where $pp$ or proton-nucleus collisions are used as a baseline
for collective effects in nucleus-nucleus collisions.  Compared with $pp$
collisions, multiparton interactions with nuclei have the additional
feature that the different scattering partons may come from the same
nucleon or from different nucleons in the nucleus.  Dedicated
investigations of multiple interactions in $pA$ collisions can be found in
\cite{Accardi:2001ih,%
  Strikman:2001gz,Cattaruzza:2004qb,Calucci:2010wg,Strikman:2010bg}.

Phenomenological estimates of multiparton interactions, as well as their
implementation in event generators, are based on a rather simple and
physically intuitive picture, whose basic ingredient is the probability to
find several partons inside a proton.  On the other hand, a systematic
description of multiparton interactions in QCD has not been achieved so far.
In the present work, we present a number of steps in this direction.  A brief
account of our main results has been given in \cite{Diehl:2011tt}.  We require
all parton-level scatters to have a hard scale, so that the concepts of
hard-scattering factorization and of parton distributions can be applied.
Since transverse momenta of final-state particles play a crucial role in the
characterization of multiple interactions, we fully keep track of this degree
of freedom and base our discussion on transverse-momentum dependent
multiparton distributions.

In section~\ref{sec:tree} we give a lowest-order analysis of multiple hard
scattering.  We find that the intuitive picture just mentioned emerges for
a subset of all relevant contributions to the cross section, but that
there are other contributions which may be of comparable size and hence
call for further investigation.  In section~\ref{sec:factorization} we
take first steps to extend existing factorization theorems for single-hard
scattering processes with measured transverse momentum
\cite{Collins:1981uk,Collins:2007ph,Collins:2011,Ji:2004wu} to the case of
multiple hard scattering.  While many ingredients for a full proof of
factorization are still missing (and the possibility that factorization is
broken cannot be ruled out), we obtain a number of encouraging results
that allow us in particular to analyze the structure of Sudakov
logarithms.  Section~\ref{sec:quark} gives more details about the
distribution of two quarks or antiquarks in the proton, in particular
about the effects of spin correlations and the possibility to learn more
about multiparton distributions by calculating their moments in lattice
QCD or by linking them to generalized parton distributions.  The
predictive power of perturbation theory is increased in kinematics where
all observed transverse momenta (as well as their vectors sums) are large
on a perturbative scale.  Complications and simplifications that arise in
this regime are discussed in section~\ref{sec:high-qt}, where we will also
encounter the conceptual problem of separating single from multiple
hard-scattering contributions in a systematic and consistent fashion.
Section~\ref{sec:conclude} contains our conclusions.

\section{Lowest order analysis}
\label{sec:tree}

\subsection{Momentum and position space structure}
\label{sec:momentum}

In this section we investigate the structure of multiparton interactions
in momentum and position space, restricting ourselves to graphs with the
lowest order in the strong coupling.  To avoid a clutter of indices we
consider scalar partons described by a hermitian field $\phi$, deferring
the inclusion of spin and color degrees of freedom to sections
\ref{sec:tree:spin} and \ref{sec:color}.  Our derivation of the cross
section formula for multiparton interactions uses standard methods.  For
cross sections integrated over transverse momenta in the final state,
similar derivations can be found in the literature
\cite{Paver:1982yp,Mekhfi:1983az}.  The extension to cross sections
differential in transverse momenta is new.
For ease of language we refer to the colliding hadrons as protons
throughout this work, bearing in mind that our results apply without
change to $p\bar{p}$ collisions or to any other hadron-hadron collision.


\subsubsection{Definition of multiparton distributions}
\label{sec:dist-def}

We begin by defining the multiparton distributions that appear in the
cross section formula we will derive shortly.  The following definitions
need to be completed by a prescription to renormalize ultraviolet
divergences and by Wilson lines that take into account collinear and soft
gluons as required to achieve factorization for the cross section.  These
issues will be discussed in section~\ref{sec:factorization}.

\begin{figure}
\begin{center}
\includegraphics[width=0.65\textwidth]{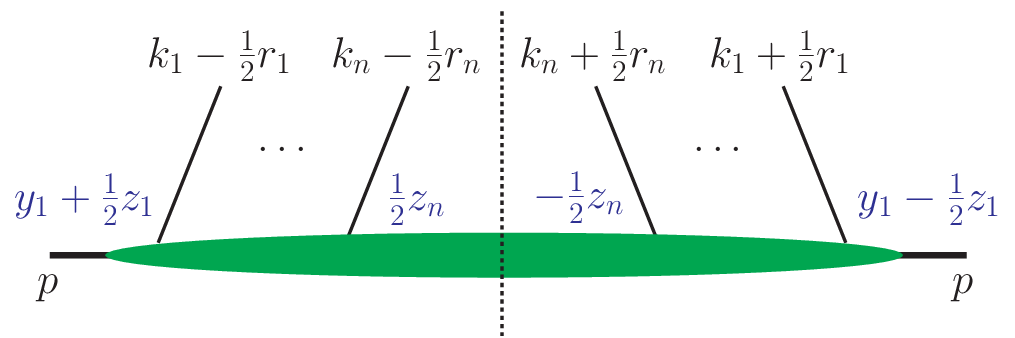}
\end{center}
\caption{\label{fig:labeling} Assignment of momentum and position
  arguments in the multiparton correlation functions and distributions.
  The dashed line denotes the final-state cut.}
\end{figure}

The building block from which multiparton distributions can be defined is
the $n$ parton correlation function
\begin{align}
  \label{Phi-def}
\Phi(l^{}_i, l'_i) &=
  \biggl[\, \prod_{i=1}^{n-1}
  \int \frac{d^4\xi^{}_i}{(2\pi)^4}\, \frac{d^4\xi'_i}{(2\pi)^4}\,
    e^{i \xi^{}_i l^{}_i - i \xi'_i \ms l'_i} \,\biggr]
\nonumber \\
& \quad \times
  \int \frac{d^4\xi^{}_n}{(2\pi)^4}\, e^{i \xi^{}_n l^{}_n}\,
\big\langle p \big|
    \bar{T} \biggl[\, \phi(0) \prod_{i=1}^{n-1} \phi(\xi'_{i}) \biggr]
    T \biggl[\, \prod_{i=1}^{n} \phi(\xi_i^{}) \biggr]
  \big| p \big\rangle \,,
\end{align}
where $T$ denotes time-ordering and $\bar{T}$ anti-time-ordering of the
fields.  This function describes the emission of $n$ partons in a
scattering amplitude and in its complex conjugate.  Throughout this work
we assume an unpolarized target: if the target carries spin then an
average over its polarization is implicit in \eqref{Phi-def} and all
subsequent expressions.  The parton four-momenta in the correlation
function are subject to the constraint
\begin{equation}
  \label{l-constraint}
\sum_{i=1}^n l^{}_i = \sum_{i=1}^n l'_i \,.
\end{equation}
In \eqref{Phi-def} we have chosen the position of the first field in the
matrix element to be $\xi'_{n}=0$.  Taking this position as arbitrary and
integrating over it with a factor $\exp(-i \xi'_n \ms l'_n)$ yields a
delta function for the constraint \eqref{l-constraint}.
The structure of the cross section will be more transparent if we use
symmetric variables
\begin{align}
  \label{symmetric-mom}
l^{}_i &= k_i - \half r_i \,, &
l'_i   &= k_i + \half r_i \,.
\end{align}
The constraint \eqref{l-constraint} then turns into
\begin{equation}
  \label{r-constraint}
\sum_{i=1}^n r_i = 0
\end{equation}
and we can rewrite the correlation function \eqref{Phi-def} as
\begin{align}
\Phi(k_i, r_i) &= \biggl[\, \prod_{i=1}^{n-1}
  \int \frac{d^4\xi^{}_i}{(2\pi)^4}\, \frac{d^4\xi'_i}{(2\pi)^4}\,
    e^{i (\xi^{}_i - \xi'_i )\ms k^{}_i
     - i (\xi^{}_i + \xi'_i)\ms  r^{}_i/2} 
  \biggr]\
  \int \frac{d^4\xi_n}{(2\pi)^4}\,
  e^{i \xi^{}_n k^{}_n + i \xi^{}_n \sum_{i=1}^{n-1} r^{}_i /2}
\nonumber \\
 & \quad \times
    \big\langle p \big|
    \bar{T} \biggl[\, \phi\bigl(-\half\xi^{}_n\bigr) \prod_{i=1}^{n-1} 
             \phi\bigl(\xi'_{i} - \half\xi^{}_n\bigr) \biggr]
    T \biggl[\, \phi\bigl(\half\xi^{}_n\bigr) \prod_{i=1}^{n-1} 
             \phi\bigl(\xi^{}_i - \half\xi^{}_n\bigr) \biggr]
    \big| p \big\rangle \,,
\end{align}
where we have replaced $r_n$ using \eqref{r-constraint}.  In addition we
have used translation invariance to shift position arguments in the matrix
element by $\xi_n/2$.  Substituting position variables according to
\begin{align}
y^{}_i + \half z^{}_i &= \xi^{}_i - \half \xi^{}_n , &
y^{}_i - \half z^{}_i &= \xi'_i - \half \xi^{}_n &
\text{for}~ i=1\ldots n-1
\end{align}
and $z_n = \xi_n$, we obtain
\begin{align}
  \label{Phi-symmetric}
\Phi(k_i, r_i) &=
  \biggl[\, \prod_{i=1}^{n}   \int \frac{d^4 z_i}{(2\pi)^4}\,
     e^{i z_i k_i} \biggr]
  \biggl[\, \prod_{i=1}^{n-1} \int \frac{d^4 y_i}{(2\pi)^4}\,
     e^{- i y_i r_i} \biggr]
\nonumber \\
 & \quad \times
    \big\langle p \big| 
    \bar{T} \biggl[\, \phi\bigl(-\half z_n\bigr) \prod_{i=1}^{n-1}
             \phi\bigl(y_i - \half z_i\bigr) \biggr]
    T \biggl[\, \phi\bigl(\half z_n\bigr) \prod_{i=1}^{n-1}
             \phi\bigl(y_i + \half z_i\bigr) \biggr]
    \big| p \big\rangle \,.
\end{align}
The assignment of momentum and position arguments is shown in
figure~\ref{fig:labeling}.

We now introduce light-cone coordinates $v^\pm = (v^0 \pm v^3)/\sqrt{2}$
and $\tvec{v} = (v^1, v^2)$ for any four-vector $v$.  In a frame where
$\tvec{p} = \tvec{0}$ we define multiparton distributions
\begin{align}
  \label{F-def}
F(x_i, \tvec{k}_i, \tvec{r}_i) &=
  \biggl[\, \prod_{i=1}^n k_i^+ \int dk_i^- \biggr]
  \biggl[\, \prod_{i=1}^{n-1}\,  (2\pi)^{3}\,
      2p^+ \!\! \int dr_i^- \biggr]\,
  \Phi(k_i, r_i) \,\bigg|_{k_i^+ = x_i^{} p^+,\, r_i^+ = 0} \,.
\end{align}
This can be written as
\begin{align}
  \label{momentum-F}
F(x_i, \tvec{k}_i, \tvec{r}_i) &= \biggl[\, \prod_{i=1}^n
       \int \frac{dz_i^-}{2\pi}\, e^{i x_i^{} z_i^- p^+} 
       \int \frac{d^2\tvec{z}_i}{(2\pi)^2}\, e^{-i \tvec{z}_i \tvec{k}_i}
     \biggr]\, \biggl[\, \prod_{i=1}^{n-1}\, 2p^+ \!\!
        \int dy_i^-\, d^2\tvec{y}_i\, e^{i \tvec{y}{}_i \tvec{r}_i} \biggr]
\nonumber \\
 & \quad \times
    \big\langle p \big|\, \mathcal{O}(0, z_n)
    \prod_{i=1}^{n-1} \mathcal{O}(y_i, z_i) \big| p \big\rangle \,,
\end{align}
where we have used the abbreviation
\begin{equation}
  \label{op-def}
\mathcal{O}(y_i, z_i) = \phi\bigl(y_i - \half z_i\bigr)
         \,i \lrpartial^+ \phi\bigl(y_i + \half z_i\bigr)
\Big|_{z_i^+ = y_i^+ = 0}
\end{equation}
for the bilinear parton operators and traded the factors $k_i^+$ for
derivatives $\lrpartial^+ = \half (\rpartial - \lpartial)^+$ acting on the
fields.  When going from \eqref{F-def} to \eqref{momentum-F} we have
replaced the time- or anti-time-ordered products appearing in
\eqref{Phi-symmetric} by usual products, which are understood to be
\emph{normal} ordered.  To justify this it is crucial that the arguments
of all fields in the operators \eqref{op-def} have a vanishing
plus-component.  For a generic configuration with all $\tvec{y}{}_i$ and
$\tvec{z}_i$ different from zero and from each other, all fields in
\eqref{op-def} have a spacelike separation, so that they commute because
of causality and can be written in any order.  The case where fields have
a lightlike separation requires special treatment, and different methods
for this case have been used in the literature for related matrix
elements, see \cite{Landshoff:1971xb,Diehl:1998sm} and
\cite{Jaffe:1983hp}.  As we shall see in section \ref{sec:high-qt},
lightlike field separations in \eqref{momentum-F} also lead to divergences
that need to be regulated.

We also introduce distributions that depend partially or entirely on
transverse positions ($\tvec{y}{}_i$ and $\tvec{z}_i$) instead of
transverse momenta ($\tvec{k}_i$ and $\tvec{r}_i$):
\begin{align}
  \label{mixed-F}
F(x_i, \tvec{k}_i, \tvec{y}{}_i) &=
  \biggl[\, \prod_{i=1}^{n-1} \int \frac{d^2\tvec{r}_i}{(2\pi)^2}\,
    e^{-i \tvec{y}{}_i \tvec{r}_i} \biggr]\,
  F(x_i, \tvec{k}_i, \tvec{r}_i)
\nonumber \\
 &= \biggl[\, \prod_{i=1}^n
       \int \frac{dz_i^-}{2\pi}\, e^{i x_i^{} z_i^- p^+} 
       \int \frac{d^2\tvec{z}_i}{(2\pi)^2}\, e^{-i \tvec{z}_i \tvec{k}_i}
    \biggr]\, \biggl[\, \prod_{i=1}^{n-1}\, 2p^+ \!\! \int dy_i^- \biggr]
\nonumber \\
 & \quad \times
    \big\langle p \big|\, \mathcal{O}(0, z_n)
    \prod_{i=1}^{n-1} \mathcal{O}(y_i, z_i) \big| p \big\rangle
\end{align}
and
\begin{align}
  \label{position-F}
F(x_i, \tvec{z}_i, \tvec{y}{}_i) &= \biggl[\, \prod_{i=1}^n
   \int d^2\tvec{k}_i\, e^{i \tvec{z}_i \tvec{k}_i} \biggr]
  F(x_i, \tvec{k}_i, \tvec{y}{}_i)
\nonumber \\
 &= \biggl[\, \prod_{i=1}^n
       \int \frac{dz_i^-}{2\pi}\, e^{i x_i^{} z_i^- p^+} 
    \biggr]\, \biggl[\, \prod_{i=1}^{n-1}\, 2p^+ \!\! \int dy_i^- \biggr]\,
    \big\langle p \big|\,\mathcal{O}(0, z_n)
    \prod_{i=1}^{n-1} \mathcal{O}(y_i, z_i) \big| p \big\rangle
\end{align}
In the arguments of \eqref{mixed-F} and \eqref{position-F} it is
understood that the average transverse position of the first two field
operators is $\tvec{y}_n = \tvec{0}$.  The three forms \eqref{momentum-F},
\eqref{mixed-F} and \eqref{position-F} can be used interchangeably, and
each of them has advantages in different situations.  As we shall see, the
momentum representation \eqref{momentum-F} naturally appears in Feynman
graph calculations, the mixed representation \eqref{mixed-F} has a rather
simple physical interpretation, and the position space representation
\eqref{position-F} is most convenient for the discussion of Sudakov
logarithms.

The factors of $2\pi$, $k_i^+$, and $2p^+$ in \eqref{F-def} to
\eqref{position-F} have been chosen such that the collinear (i.e.\
transverse-momentum integrated) distribution
\begin{align}
F(x_i, \tvec{y}{}_i) =
\biggl[\, \prod_{i=1}^n \int d^2\tvec{k}_i \biggr]\,
   F(x_i, \tvec{k}_i, \tvec{y}{}_i)
&= F(x_i, \tvec{z}_i=\tvec{0}, \tvec{y}{}_i)   
\label{position-F-coll}
\intertext{as well as the distribution}
\biggl[\, \prod_{i=1}^{n-1} \int d^2\tvec{y}{}_i \biggr]\,
   F(x_i, \tvec{k}_i, \tvec{y}{}_i)
&= F(x_i, \tvec{k}_i, \tvec{r}_i=\tvec{0})
\label{momentum-F-proba}
\end{align}
admit a probability interpretation.  $F(x_i, \tvec{y}{}_i)$ is the
probability to find $n$ partons with plus-momentum fractions $x_i$ and
transverse distances $\tvec{y}{}_i$ from parton number $n$, and $F(x_i,
\tvec{k}_i, \tvec{r}_i=\tvec{0})$ is the probability to find $n$ partons
with plus-momentum fractions $x_i$ and transverse momenta $\tvec{k}_i$.

By contrast, $F(x_i, \tvec{k}_i, \tvec{y}{}_i)$ is not a probability (due
to the uncertainty relation one cannot simultaneously fix transverse
momentum and transverse position) but rather has the structure of a Wigner
distribution \cite{Hillery:1984} in the transverse variables.  Its
integral over all $\tvec{k}_i$ gives the probability to find partons at
transverse positions $\tvec{y}{}_i$, and its integral over all
$\tvec{y}{}_i$ gives the probability to find partons with transverse
momenta~$\tvec{k}_i$.  A related interpretation for generalized parton
distributions can be found in \cite{Belitsky:2003nz}.  In
figure~\ref{fig:labeling} we can identify $\tvec{k}_i$ as the ``average''
transverse momenta of the partons and $\tvec{y}{}_i$ as their ``average''
transverse position, where the ``average'' is taken between the partons to
the left and to the right of the final-state cut in the figure.  In a
physical process, this corresponds to an average between partons in the
scattering amplitude and its complex conjugate.

The interpretation of multiparton distributions becomes more explicit if
one represents them in terms of the light-cone wave functions of the
target, see \cite{Blok:2010ge}.  Most conveniently derived in the
framework of light-cone quantization, this representation is analogous to
the wave function representation for single-parton densities
\cite{Brodsky:1989} and generalized parton distributions
\cite{Diehl:2000xz,Diehl:2002he}.  The distributions in
\eqref{position-F-coll} and \eqref{momentum-F-proba} can be written in
terms of \emph{squared} wave functions in impact parameter or
transverse-momentum space, which makes their probability interpretation
manifest.  The wave function representation also offers a way to model
multiparton distributions in the region of large momentum fractions, where
one can expect a small number of partonic Fock states to be dominant.  We
shall not pursue this avenue in the present work.

In later chapters we will also need collinear distributions that depend on
the momentum transfer variables $\tvec{r}_i$,
\begin{align}
  \label{momentum-F-coll}
F(x_i, \tvec{r}_i)
 = \biggl[\, \prod_{i=1}^n \int d^2\tvec{k}_i \biggr]\,
   F(x_i, \tvec{k}_i, \tvec{r}_i)
&= \biggl[\, \prod_{i=1}^{n-1} \int d^2\tvec{y}{}_i\,
     e^{i \tvec{y}{}_i \tvec{r}_i} \biggr]\,
   F(x_i, \tvec{y}{}_i) \,.
\end{align}
We will see in the following section that $F(x_i, \tvec{y}{}_i)$ or
equivalently $F(x_i, \tvec{r}_i)$ appear in multiple-scattering cross
sections.  This is not the case for the distributions
\begin{align}
  \label{F-x}
F(x_i) = F(x_i, \tvec{r}_i = \tvec{0})
  &= \biggl[\, \prod_{i=1}^n \int d^2\tvec{k}_i \biggr]\,
     \biggl[\, \prod_{i=1}^{n-1} \int d^2\tvec{y}{}_i \biggr]\,
   F(x_i, \tvec{k}_i, \tvec{y}{}_i) \,,
\end{align}
which give the probability to find $n$ partons with momentum fractions
$x_i$ and unspecified transverse positions or transverse momenta.  We note
that the integrals over $\tvec{k}_i$ in \eqref{position-F-coll},
\eqref{momentum-F-coll} and \eqref{F-x} are logarithmically divergent and
require appropriate regularization, which will be discussed in sections
\ref{sec:ladders-fact} and~\ref{sec:evolution}.

The definitions in this section are given for right-moving partons, with
$x_i$ being plus-momentum fractions.  Analogous definitions for
left-moving partons are obtained by exchanging the plus- and
minus-components of all position and momentum vectors.


\subsubsection{Cross section for \texorpdfstring{$n$}{n} hard scatters}
\label{sec:tree:cross-sect}

\begin{figure}
\begin{center}
\includegraphics[width=0.65\textwidth]{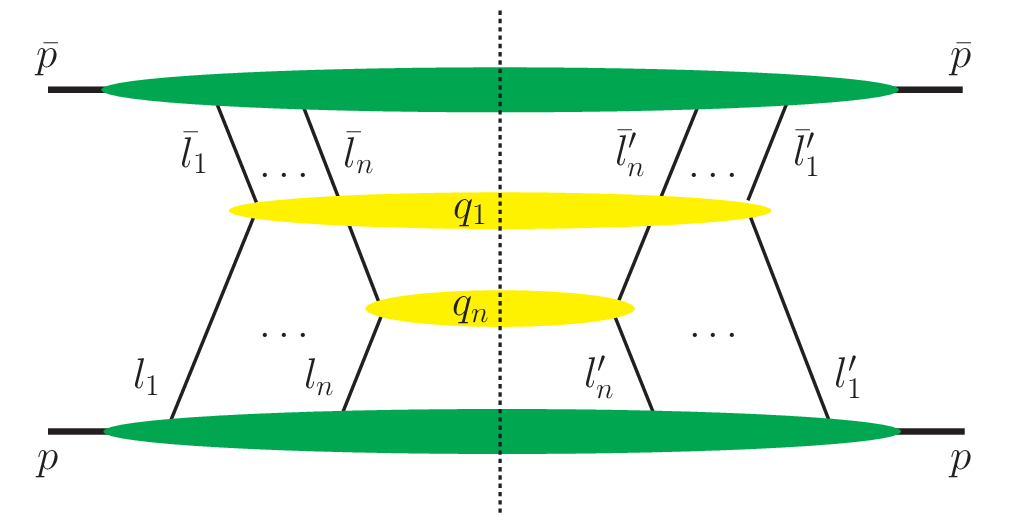}
\end{center}
\caption{\label{fig:n-scatters} Graph for the cross section of a collision
  with $n$ hard scatters at parton level.  The dashed line denotes the
  final-state cut.  Here and in the following, the lower blob is
  associated with the right-moving proton and the upper blob with the
  left-moving proton in the collision.}
\end{figure}

We now evaluate the cross section for a process with $n$ scatters at
parton level, as sketched in figure~\ref{fig:n-scatters}.  We work in a
reference frame with $\tvec{p} = \bar{\tvec{p}} = \tvec{0}$ and consider
kinematics where the squared c.m.\ energy $q_i^2$ of each scatter is large
and where each transverse momentum $|\tvec{q}_i|$ is much smaller than
$q_i^+$ and $q_i^-$.  Defining
\begin{align}
x_i^{} &= q_i^+/p^+ \,, &
\bar{x}_i^{} &= q_i^-/\bar{p}^- \,,
\end{align}
we can then approximate
\begin{equation}
  \label{q2-approx}
q_i^2 \approx 2 q_i^+ q_i^- \approx x_i^{} \bar{x}_i^{} s \,,
\end{equation}
where $s = (p+\bar{p})^2$ is the squared overall c.m.\ energy.  We neglect
the target mass throughout, so that $s \approx 2 p\bar{p} \approx 2 p^+
\bar{p}^-$ and the flux factor in the cross section is $1/(4p\bar{p})$.
One can trade the momentum fractions $x_i$ and $\bar{x}_i$ for $q_i^2$ and
the rapidities
\begin{equation}
Y_i = \frac{1}{2}\ms \log\frac{q_i^+}{q_i^-}
\end{equation}
with
\begin{equation}
  \label{phase-space-el}
dx_i\, d\bar{x}_i = \frac{1}{s}\, d(q_i^2)\, dY_i \,,
\end{equation}
where we have again used \eqref{q2-approx}.  We note that for the very
high $s$ achieved at the LHC, both $x_i$ and $\bar{x}_i$ are rather small,
except if $|Y_i|$ or $q_i^2$ is very large.

The cross section for $n$ hard scatters is given by
\begin{align}
  \label{X-section-start}
d\sigma &= \frac{1}{C}\, \frac{1}{4 p\bar{p}}\, 
  \biggl[\, \prod_{i=1}^{n} \frac{d^4q_i}{(2\pi)^4} \biggr]
  \sum_{X,\bar{X}} \,
    \biggl[\, \prod_{j=1}^{m} \int
            \frac{d^3p_{X,j}}{(2\pi)^3 2p_{X,j}^0} \biggr]
    \biggl[\, \prod_{j=1}^{\bar{m}} \int
            \frac{d^3p_{\bar{X},j}}{(2\pi)^3 2p_{\bar{X},j}^0} \biggr]
\nonumber \\
 & \quad \times \biggl[\, \prod_{i=1}^{n-1}
          \int \frac{d^4l_i}{(2\pi)^4} \,
               \frac{d^4\bar{l}_i}{(2\pi)^4} \,
          (2\pi)^4 \delta^{(4)}(q_i - l_i - \bar{l}_i)
          \int \frac{d^4l'_i}{(2\pi)^4} \,
               \frac{d^4\bar{l}'_i}{(2\pi)^4} \,
          (2\pi)^4 \delta^{(4)}(q_i - l'_i - \bar{l}'_i) \biggr]
\nonumber \\
 & \quad \times
   (2\pi)^{4} \delta^{(4)}\biggl( \sum_{i=1}^n q_i
      + \sum_{j=1}^{m} p_{X,j}
      + \sum_{j=1}^{\bar{m}} p_{\bar{X},j} - p - \bar{p} \biggr)
   \biggl[\, \prod_{i=1}^n  H_i(q_i, l_i,\bar{l}_i, l'_i,\bar{l}'_i)
   \biggr]
\nonumber \\
 & \quad \times
\big\langle p \big| \bar{T} \biggl[\, \phi(0)
  \prod_{i=1}^{n-1} d^4\xi'_i\, e^{-i \xi'_i \ms l'_i}\, 
  \phi(\xi'_i) \biggr] \big| X \big\rangle \,
\big\langle X \big| T \biggl[\, \phi(0)
  \prod_{i=1}^{n-1} d^4\xi_i\, e^{i \xi_i l_i}\, \phi(\xi_i)
  \biggr] \big| p \big\rangle
\nonumber \\
 & \quad \times
\big\langle \bar{p} \big| \bar{T} \biggl[\, \phi(0)
  \prod_{i=1}^{n-1} d^4\bar{\xi}'_i\, e^{-i \bar{\xi}'_i \ms \bar{l}'_i}\,
  \phi(\bar{\xi}'_i) \biggr] \big| \bar{X} \big\rangle \,
\big\langle \bar{X} \big| T \biggl[\, \phi(0)
  \prod_{i=1}^{n-1} d^4\bar{\xi}_i\, e^{i \bar{\xi}_i \bar{l}_i}\,
  \phi(\bar{\xi}_i) \biggr]
  \big| \bar{p} \big\rangle \,,
\end{align}
where the combinatorial factor $C$ contains a factor $k!$ for each set of $k$
identical hard-scattering final states.\footnote{An often used notation
  for two hard scatters is to write $m/2$ in the place of $1/C$, with
  $m=1$ if the hard-scattering final states are identical and $m=2$ if
  they are distinct.}
The remnant of proton $p$
($\bar{p}$) consists of $m$ ($\bar{m}$) spectators with momenta $p_{X,j}$
($p_{\bar{X},j}$).
$H_i$ denotes the squared matrix element for the $i$th hard scatter, with
truncated propagators of the incoming parton lines.  $H_i$ includes
integration over the internal phase space of the final state produced by
the hard scatter, with only the four-momentum $q_i$ kept fixed.  If such a
final state is the decay product of a single particle with mass $M$ and
width $\Gamma$ (e.g.\ a $W$ or a Higgs boson) then $H_i$ includes a factor
\begin{equation}
  \label{narrow-width}
\frac{1}{q_i^2 - M^2 + i\Gamma M}\, \frac{1}{q_i^2 - M^2 - i\Gamma M}
  \stackrel{\Gamma \ll M}{\approx}
  \frac{\pi}{\Gamma M}\, \delta(q_i^2 - M^2) \,,
\end{equation}
which in the limit of narrow width constrains $q_i$ to be on the mass
shell.  If the final state is a stable single particle with mass $M$, then
$H_i$ includes a delta function
\begin{align}
2\pi \delta( q_i^2 - M^2 )
\end{align}
so that together with the integration element $d^4 q_i /(2\pi)^4$ in
\eqref{X-section-start} one obtains the correct one-particle integration
measure $d^3 q_i^{}\big/ \big[ 2q_i^0\, (2\pi)^3 \ms\big]$.
We now rewrite the cross section in terms of the correlation functions
\eqref{Phi-def}.  To this end we use
\begin{align}
& \sum_{X}\, \biggl[\, \prod_{j=1}^{m} \int
             \frac{d^3p_{X,j}}{(2\pi)^3 2p_{X,j}^0} \biggr]
  (2\pi)^4 \delta^{(4)}\biggl(
           \sum_{i=1}^n l_i + \sum_{j=1}^{m} p_{X,j} - p \biggr)
\nonumber \\
&\qquad \times
\big\langle p \big| \bar{T} \biggl[\, \phi(0)
  \prod_{i=1}^{n-1} \int d^4\xi'_i\, e^{-i \xi'_i \ms l'_i}\, 
  \phi(\xi'_i) \biggr] \big| X \big\rangle \,
\big\langle X \big| T \biggl[\, \phi(0)
  \prod_{i=1}^{n-1} \int d^4\xi_i\, e^{i \xi_i l_i}\, \phi(\xi_i)
  \biggr] \big| p \big\rangle
\nonumber \\
&\quad =
\sum_{X}\, \biggl[\, \prod_{j=1}^{m} \int
           \frac{d^3p_{X,j}}{(2\pi)^3 2p_{X,j}^0} \biggr]
\int d^4\xi_n\,
   e^{- i \xi_n \left( p \ms - \ms \sum_{i=1}^n l_i
        \ms - \ms \sum_{j=1}^{m} p_{X,j} \right)} \;
   e^{i \xi_n \left( p \ms - \ms \sum_{j=1}^{m} p_{X,j} \right)}
\nonumber \\
&\qquad \times
\big\langle p \big| \bar{T} \biggl[\, \phi(0)
  \prod_{i=1}^{n-1} \int d^4\xi'_i\, e^{-i \xi'_i \ms l'_i}\, 
  \phi(\xi'_i) \biggr] \big| X \big\rangle \,
\big\langle X \big| T \biggl[\, \phi(\xi_n)
  \prod_{i=1}^{n-1} \int d^4\xi_i\, e^{i \xi_i l_i} \phi(\xi_i+\xi_n)
  \biggr] \big| p \big\rangle
\nonumber \\
&\quad =
\sum_{X}\, \biggl[\, \prod_{j=1}^{m} \int
           \frac{d^3p_{X,j}}{(2\pi)^3 2p_{X,j}^0} \biggr]
\nonumber \\
&\qquad \times
\big\langle p \big| \bar{T} \biggl[\, \phi(0)
  \prod_{i=1}^{n-1} \int d^4\xi'_i\, e^{-i \xi'_i \ms l'_i}\, 
  \phi(\xi'_i) \biggr] \big| X \big\rangle \,
\big\langle X \big| T \biggl[\,
  \prod_{i=1}^{n} \int d^4\xi_i\, e^{i \xi_i l_i}\, \phi(\xi_i)
  \biggr] \big| p \big\rangle
\nonumber \\
&\quad =
\biggl[\, 
  \prod_{i=1}^{n-1} \int d^4\xi'_i\, e^{-i \xi'_i \ms l'_i} \biggr]
\biggl[\,
  \prod_{i=1}^{n} \int d^4\xi_i\, e^{i \xi_i l_i} \biggr]
\big\langle p \big|
  \bar{T} \biggl[\, \phi(0) \prod_{i=1}^{n-1} \phi(\xi'_i) \biggr]
  T \biggl[\, \prod_{i=1}^{n} \phi(\xi_i) \biggr]
\big| p \big\rangle
\nonumber \\
&\quad = (2\pi)^{4(2n-1)}\, \Phi(l^{}_i, l'_i)  \,.
 \phantom{\biggl[\, \biggr]}
\end{align}
Using the analogous relation for the matrix element between $\bar{X}$ and
$\bar{p}$ and rewriting the momentum conservation constraint in
\eqref{X-section-start} as
\begin{align}
& (2\pi)^{4} \delta^{(4)}\biggl( \sum_{i=1}^n q_i
      + \sum_{j=1}^{m} p_{X,j}
      + \sum_{j=1}^{\bar{m}} p_{\bar{X},j} - p - \bar{p} \biggr)
\nonumber \\
&\quad = \int \frac{d^4l_n}{(2\pi)^4}\, \frac{d^4\bar{l}_n}{(2\pi)^4}\,
   (2\pi)^4 \delta^{(4)}\biggl( \sum_{i=1}^n q_i
            - \sum_{i=1}^n l_i - \sum_{i=1}^n \bar{l}_i \biggr)
\nonumber \\
&\qquad \times
(2\pi)^4 \delta^{(4)}\biggl(
   \sum_{i=1}^n l_i + \sum_{j=1}^{m} p_{X,j} - p \biggr) \,
(2\pi)^4 \delta^{(4)}\biggl( \sum_{i=1}^n \bar{l}_i
   + \sum_{j=1}^{\bar{m}} p_{\bar{X},j} - \bar{p} \biggr) \,,
\end{align}
we can express the cross section as
\begin{align}
  \label{X-section-corr}
d\sigma &= \frac{1}{C}\, \frac{1}{4 p\bar{p}}\, 
  \biggl[\, \prod_{i=1}^{n} \frac{d^4q_i}{(2\pi)^4} \biggr]
  \biggl[\, \prod_{i=1}^{n} \int d^4l_i\, d^4\bar{l}_i\;
          (2\pi)^4 \delta^{(4)}(q_i - l_i - \bar{l}_i) \biggr]
\nonumber \\
 & \quad \times
   \biggl[\, \prod_{i=1}^{n-1}
          \int d^4l'_i\, d^4\bar{l}'_i\;
          (2\pi)^4 \delta^{(4)}(q_i - l'_i - \bar{l}'_i) \biggr]
\nonumber \\
 & \quad \times
   \biggl[\, \prod_{i=1}^n  H_i(q_i, l_i,\bar{l}_i, l'_i,\bar{l}'_i)
   \biggr] \, \Phi(l^{}_i, l'_i)\, \bar{\Phi}(\bar{l}^{}_i, \bar{l}'_i)
\nonumber \\
 & =
\frac{1}{C}\, \frac{1}{4 p\bar{p}}\, 
  \biggl[\, \prod_{i=1}^{n} \frac{d^4q_i}{(2\pi)^4} \biggr]
  \biggl[\, \prod_{i=1}^{n} \int d^4k_i\, d^4\bar{k}_i\;
          (2\pi)^4 \delta^{(4)}(q_i - k_i - \bar{k}_i) \biggr]
\nonumber \\
 & \quad \times
   \biggl[\, \prod_{i=1}^{n-1}
          \int d^4r_i\, d^4\bar{r}_i\;
          (2\pi)^4 \delta^{(4)}(r_i + \bar{r}_i) \biggr]
\nonumber \\
 & \quad \times
   \biggl[\, \prod_{i=1}^n  H_i(q_i, k_i,\bar{k}_i, r_i,\bar{r}_i)
   \biggr] \, \Phi(k_i, r_i)\, \bar{\Phi}(\bar{k}_i, \bar{r}_i) \,,
\end{align}
where in the last step we have switched to the set of symmetric variables
\eqref{symmetric-mom}.  They have the important property that the
kinematic constraints on $r_i$ and $\bar{r}_i$ do not involve the
final-state momenta $q_i$, which will lead to a great simplification
below.


\paragraph{Hard-scattering approximation.}

The parton-level scattering processes involve a hard scale, which we
collectively denote by $Q^2 \sim q_i^2$ without assuming a particular
hierarchy among the individual squared momenta $q_i^2$.  The case where
one of them is much larger than the others is of particular relevance for
the description of the underlying event, but we shall not investigate the
consequences of such a hierarchy in the present work.  In the graph of
figure~\ref{fig:n-scatters} it is understood that partons emerging from
the shaded blobs have virtualities much smaller than $Q^2$.  The
components of the various four-momenta thus scale like
\begin{align}
k_i^+ \sim r_i^+ \sim p^+ \sim q_i^+ & \sim Q \,, &
\bar{k}_i^- \sim \bar{r}_i^- \sim \bar{p}^- \sim q_i^- & \sim Q \,,
\nonumber \\
k_i^- \sim r_i^- \sim p^- & \sim \Lambda^2/Q \,, &
\bar{k}_i^+ \sim \bar{r}_i^+ \sim \bar{p}^+ & \sim \Lambda^2/Q
\end{align}
and 
\begin{align}
|\tvec{k}_i| \sim |\tvec{r}_i| \sim
|\bar{\tvec{k}}_i| \sim |\bar{\tvec{r}}_i| \sim
|\tvec{q}_i| &\sim \Lambda \,,
\end{align}
where $\Lambda$ denotes the size of the transverse momenta $|\tvec{q}_i|$
or the scale of non-perturbative interactions, whichever is larger.
The momentum conservation constraint $\delta^{(4)}(r_i + \bar{r}_i)$
enforces that the components
\begin{equation}
r_i^+ \sim \bar{r}_i^- \sim \Lambda^2/Q
\end{equation}
are small, although by general scaling arguments they could be of order
$Q$.  The constraint $\delta^{(4)}(q_i - k_i - \bar{k}_i)$ leads to
\begin{align}
  \label{mom-frac-coll}
k_i^+ &\approx q_i^+ &
\bar{k}_i^- &\approx q_i^-
\end{align}
up to relative corrections of order $\Lambda^2/Q^2$.  We make these
approximations in the correlation functions $\Phi$ and $\bar{\Phi}$ and
see that the longitudinal momenta of the partons entering the hard
scattering are fixed by the final-state kinematics.  In the squared
hard-scattering matrix element $H_i(q_i, k_i,\bar{k}_i, r_i,\bar{r}_i)$ we
can neglect all transverse momenta and all components of order
$\Lambda^2/Q$.  With \eqref{mom-frac-coll} this only leaves a dependence
on the independent variables $q_i^+$ and $q_i^-$.  Since $H_i$ is
invariant under a boost along the $z$ axis, it can then only depend on $2
q_i^+ q_i^- \approx q_i^2$.  Altogether we then have
\begin{align}
  \label{pre-X-section}
& \biggl[\, \prod_{i=1}^{n}
\int dk_i^+ d\bar{k}_i^+\, 
  \delta(q_i^+ - k_i^+ - \bar{k}_i^+)\,
\int dk_i^- d\bar{k}_i^-\, 
  \delta(q_i^- - k_i^- - \bar{k}_i^-) \biggr]
\nonumber \\
& \qquad\times 
\biggl[\, \prod_{i=1}^{n-1}
 \int dr_i^+ d\bar{r}_i^+\, \delta(r_i^+ + \bar{r}_i^+)\,
 \int dr_i^- d\bar{r}_i^-\, \delta(r_i^- + \bar{r}_i^-)
\biggr]\,
\nonumber \\
& \qquad\times 
\biggl[\, \prod_{i=1}^n H_i(q_i, k_i,\bar{k}_i, r_i,\bar{r}_i)
\biggr]  \Phi(k_i, r_i)\, \bar{\Phi}(\bar{k}_i, \bar{r}_i)
\nonumber \\
& \quad = \biggl[\, \prod_{i=1}^{n} \int dk_i^+\, dk_i^- \biggr]
        \biggl[\, \prod_{i=1}^{n-1} \int dr_i^+\, dr_i^- \biggr]
\nonumber \\
& \qquad \times
\biggl[\, \prod_{i=1}^n H_i(q_i, k_i,\bar{k}_i, r_i,\bar{r}_i)
\biggr]\, \Phi(k_i, r_i)\, \bar{\Phi}(\bar{k}_i, \bar{r}_i)
\,\bigg|_{\substack{k_i^+ = q_i^+ - \bar{k}_i^+,\, r_i^+ = - \bar{r}_i^-
          \\ \bar{k}_i^- = q_i^- - k_i^-,\, \bar{r}_i^- = - r_i^+}}
\nonumber \\
& \quad\approx
\biggl[\, \prod_{i=1}^n H_i(q_i^2) \biggr]\,
    \biggl[\, \prod_{i=1}^{n} \int dk_i^- \biggr]
  \biggl[\, \prod_{i=1}^{n-1} \int dr_i^- \biggr]
  \Phi(k_i, r_i)\,\biggl|_{k_i^+ = q_i^+, r_i^+ = 0}
\nonumber \\
& \hspace{6.7em} \times
    \biggl[\, \prod_{i=1}^{n} \int d\bar{k}_i^+ \biggr]
  \biggl[\, \prod_{i=1}^{n-1} \int d\bar{r}_i^+ \biggr]
  \bar{\Phi}(\bar{k}_i, \bar{r}_i) 
  \,\biggl|_{\bar{k}_i^- = q_i^-, \bar{r}_i^- = 0} \,.
\end{align}
Inserting this into the cross section \eqref{X-section-corr} and using the
definition \eqref{F-def} of the multiparton distributions gives
\begin{align}
d\sigma &= \frac{1}{C}\, 
\frac{1}{4 p\bar{p}}\, \frac{1}{(4 p^+ \bar{p}^-)^{n-1}}
  \biggl[\, \prod_{i=1}^{n} d^4q_i\, \frac{1}{q_i^+ q_i^-} 
                            H_i(q_i^2) \biggr]\,
\biggl[\, \prod_{i=1}^{n} \int d^2\tvec{k}_i\, d^2\bar{\tvec{k}}_i\,
       \delta^{(2)}(\tvec{q}{}_i - \tvec{k}_i - \bar{\tvec{k}}_i) \biggr]
\nonumber \\
& \quad\times
\biggl[\, \prod_{i=1}^{n-1} \int \frac{d^2\tvec{r}_i}{(2\pi)^2} \biggr]
  F(x_i, \tvec{k}_i, \tvec{r}_i)\, 
  F(\bar{x}_i, \bar{\tvec{k}}_i, - \tvec{r}_i) \,.
\end{align}
Rewriting $d^4q_i = p^+ \bar{p}^- dx_i\, d\bar{x}_i\, d^2\tvec{q}{}_i$, we
obtain our final result for the cross section in momentum representation,
\begin{align}
  \label{X-sect-momentum}
\frac{d\sigma}{\prod_{i=1}^n dx_i\, d\bar{x}_i\, d^2\tvec{q}{}_i}
 &= \frac{1}{C}\, 
\biggl[\, \prod_{i=1}^{n} \,\hat{\sigma}_i(x_i \bar{x}_i s) \biggr]\,
\biggl[\, \prod_{i=1}^{n} 
     \int d^2\tvec{k}_i\, d^2\bar{\tvec{k}}_i\;
     \delta^{(2)}(\tvec{q}{}_i - \tvec{k}_i - \bar{\tvec{k}}_i) \biggr]
\nonumber \\
& \quad\times
\biggl[\, \prod_{i=1}^{n-1} \int \frac{d^2\tvec{r}_i}{(2\pi)^2} \biggr]
  F(x_i, \tvec{k}_i, \tvec{r}_i)\, 
  F(\bar{x}_i, \bar{\tvec{k}}_i, - \tvec{r}_i) \,,
\end{align}
where we have introduced the cross section
\begin{equation}
  \label{sigma-hat-def}
\hat\sigma_i(q_i^2) = \frac{1}{2 q_i^2}\, H_i(q_i^2)
\end{equation}
for the $i$th parton-level subprocess and used the approximation
\eqref{q2-approx}.
We have carried out the integrations over $\bar{\tvec{r}}_i$ using the
constraints $\delta^{(2)}(\tvec{r}_i + \bar{\tvec{r}}_i)$, so that the
distributions for the two protons are evaluated at opposite values of
their last arguments.  Fourier transforming these to position space, we
have
\begin{align}
  \label{X-sect-mixed}
\frac{d\sigma}{\prod_{i=1}^n dx_i\, d\bar{x}_i\, d^2\tvec{q}{}_i}
 &= \frac{1}{C}\, 
\biggl[\, \prod_{i=1}^{n} \,\hat{\sigma}_i(x_i \bar{x}_i s) \biggr]\,
\biggl[\, \prod_{i=1}^{n} 
     \int d^2\tvec{k}_i\, d^2\bar{\tvec{k}}_i\;
     \delta^{(2)}(\tvec{q}{}_i - \tvec{k}_i - \bar{\tvec{k}}_i) \biggr]
\nonumber \\
& \quad\times
\biggl[\, \prod_{i=1}^{n-1} \int d^2\tvec{y}{}_i \biggr]
  F(x_i, \tvec{k}_i, \tvec{y}{}_i)\, 
  F(\bar{x}_i, \bar{\tvec{k}}_i, \tvec{y}{}_i)
\end{align}
and the distributions are evaluated at equal values of $\tvec{y}{}_i$.
Transforming also the arguments $\tvec{k}_i$ and $\bar{\tvec{k}}_i$, we
have
\begin{align}
  \label{X-sect-position}
\frac{d\sigma}{\prod_{i=1}^n dx_i\, d\bar{x}_i\, d^2\tvec{q}{}_i}
 &= \frac{1}{C}\, 
\biggl[\, \prod_{i=1}^{n} \,\hat{\sigma}_i(x_i \bar{x}_i s) \biggr]\,
\biggl[\, \prod_{i=1}^{n} 
     \int \frac{d^2\tvec{z}_i}{(2\pi)^2}\,
          e^{-i \tvec{z}_i \tvec{q}{}_i} \biggr]
\nonumber \\
& \quad\times
 \biggl[\, \prod_{i=1}^{n-1} \int d^2\tvec{y}{}_i \biggr]
  F(x_i, \tvec{z}_i, \tvec{y}{}_i)\, 
  F(\bar{x}_i, \tvec{z}_i, \tvec{y}{}_i) \,,
\end{align}
where all position arguments in the two distributions coincide.  

The interpretation of the distributions $F(x_i, \tvec{k}_i, \tvec{y}{}_i)$
discussed in section~\ref{sec:dist-def} extends to the cross section
formula \eqref{X-sect-mixed}.  In each individual hard subprocess, two
partons with average transverse momenta $\tvec{k}_i$ and
$\bar{\tvec{k}}_i$ produce a final state with transverse momentum
$\tvec{q}_i$.  The $i$th scatter occurs at an average transverse distance
$\tvec{y}{}_i$ from the $n$th scatter.  The hard scatters are approximated
to be local in transverse space, so that their average distance is equal
to the average distance between the colliding partons in each proton.  We
thus find a rather intuitive interpretation of the variables in our cross
section formula, provided that we ``average'' the transverse momenta and
positions between the amplitude and its conjugate.  Let us however
emphasize that we have obtained \eqref{X-sect-mixed} from calculating
Feynman graphs using standard hard-scattering approximations, without any
appeal to classical or semi-classical arguments.

Integrating the cross section over all transverse momenta $\tvec{q}{}_i$
we obtain a simple result
\begin{align}
  \label{X-sect-integrated}
\frac{d\sigma}{\prod_{i=1}^n dx_i\, d\bar{x}_i}
 &= \frac{1}{C}\, 
\biggl[\, \prod_{i=1}^{n} \,\hat{\sigma}_i(x_i \bar{x}_i s) \biggr]
\biggl[\, \prod_{i=1}^{n-1} \int d^2\tvec{y}{}_i \biggr]
  F(x_i, \tvec{y}{}_i)\, F(\bar{x}_i, \tvec{y}{}_i)
\end{align}
in terms of collinear multiparton distributions.  This formula has long
been known and provides the basis of most phenomenological analyses of
multiple interactions in the literature.  It was derived in
\cite{Paver:1982yp} for scalar partons in a way very similar to the one we
have employed here.


\subsubsection{Single vs.\ multiple hard scattering}
\label{sec:sing-vs-mult}

The approximations we have made in the previous section give the leading
term of an expansion in powers of $\Lambda /Q$.  Let us investigate how
the resulting cross section \eqref{X-sect-mixed} scales with $Q$.  As can
readily be seen from its definition \eqref{mixed-F}, the mass dimension of
$F(x_i, \tvec{k}_i, \tvec{y}{}_i)$ is $-2$ and one has $F(x_i, \tvec{k}_i,
\tvec{y}{}_i) \sim \Lambda^{-2}$.  To obtain this power behavior, it is
essential that the distribution is invariant under a boost along the $z$
axis.  For instance, a hadronic matrix element that transforms like the
plus-component of a vector would be proportional to $p^+$ or another large
plus component and thus scale like $Q$ times the appropriate power of
$\Lambda$.  Note that the dependence of $F(x_i, \tvec{k}_i, \tvec{y}{}_i)$
on the large scale $Q$ via renormalization group or Sudakov logarithms
(see section \ref{sec:factorization}) is neglected at the level of power
counting.
The hard-scattering cross sections have a power behavior $\hat{\sigma}_i
\sim Q^{-2}$, and the integrations over transverse momenta count as
$d^2\tvec{k}_i\, d^2\bar{\tvec{k}}_i\; \delta^{(2)}(\tvec{q}{}_i -
\tvec{k}_i - \bar{\tvec{k}}_i) \sim \Lambda^2$.  Finally, the distances
$\tvec{y}{}_i$ in \eqref{X-sect-mixed} are generically of size $1/\Lambda$
so that $d^2\tvec{y}{}_i \sim \Lambda^{-2}$.  Putting all ingredients
together, one finds
\begin{align}
  \label{mult-hard-pow}
\frac{d\sigma}{\prod_{i=1}^n dx_i\, d\bar{x}_i\, d^2\tvec{q}{}_i}
  \,\bigg|_{\text{multiple}} 
&\sim \frac{1}{\Lambda^2\, Q^{2n}}
\end{align}
for the cross section of $n$ hard scatters.  One obtains of course the
same result if the power counting is done for the representations
\eqref{X-sect-momentum} in momentum space or \eqref{X-sect-position} in
position space, using $d^2 \tvec{r}_i \sim \Lambda^2$ or $d^2 \tvec{z}_i
\sim 1/\Lambda^2$.

Let us compare this with the cross section for producing the final states
with momenta $q_{i}$ in a single hard scattering.  With
\begin{align}
q       &= \sum_{i=1}^n q_i \,, &
x       &= \sum_{i=1}^n x_i = \frac{q^+}{p^+} \,, &
\bar{x} &= \sum_{i=1}^n \bar{x}_i = \frac{q^-}{\bar{p}^-}
\end{align}
the factorization formula for this case reads
\begin{align}
  \label{single-hard-int}
\frac{d\sigma}{dx\ms d\bar{x}\ms d\tvec{q}} \bigg|_{\text{single}}
 &= \hat{\sigma}(x\bar{x} s) \int d^2\tvec{k}\, d^2\bar{\tvec{k}}\;
    \delta^{(2)}(\tvec{q}{} - \tvec{k} - \bar{\tvec{k}})\,
    f(x,\tvec{k})\, f(\bar{x}, \bar{\tvec{k}}) \,,
\end{align}
where $\hat{\sigma}$ is the appropriate hard-scattering cross section and
$f(x,\tvec{k})$ and $f(\bar{x}, \bar{\tvec{k}})$ are trans\-verse-momentum
dependent single-parton densities.  The definition of $f(x,\tvec{k})$ can
be obtained from \eqref{mixed-F} by setting $n=1$, which gives a power
behavior $f(x,\tvec{k}) \sim \Lambda^{-2}$.  We now make
\eqref{single-hard-int} differential in the internal momentum variables of
the final state, which we choose as
\begin{align}
  \label{single-internal-vars}
u_i^{}       &= x_i^{} /x       = q_i^{\smash{+}} /q^+ \,, &
\bar{u}_i^{} &= \bar{x}_i^{} /x = q_i^{\smash{-}} /q^-
\end{align}
and $\tvec{q}_{i}$ with $i=1, \ldots, n-1$.  We then have
\begin{align}
  \label{single-hard-diff}
\frac{d\sigma}{\prod_{i=1}^n dx_i\, d\bar{x}_i\, d^2\tvec{q}{}_i}
  \,\bigg|_{\text{single}}
 &= \frac{d\hat{\sigma}}{\prod_{i=1}^{n-1}
                         du_i\, d\bar{u}_i\, d^2\tvec{q}{}_i}
    \int d^2\tvec{k}\, d^2\bar{\tvec{k}}\;
    \delta^{(2)}(\tvec{q}{} - \tvec{k} - \bar{\tvec{k}})\,
    \frac{f(x,\tvec{k})}{x^{n-1}}\,
    \frac{f(\bar{x}, \bar{\tvec{k}})}{\bar{x}^{n-1}} \,.
\end{align}
The differential hard-scattering cross section on the r.h.s.\ behaves as
$Q^{-2n}$, so that we have
\begin{align}
  \label{single-hard-pow}
\frac{d\sigma}{\prod_{i=1}^n dx_i\, d\bar{x}_i\, d^2\tvec{q}{}_i}
  \,\bigg|_{\text{single}}
&\sim \frac{1}{\Lambda^2\, Q^{2n}} \,.
\end{align}
We obtain the important result that if one leaves the cross section
differential in the transverse momenta $\tvec{q}_i$, the contributions
from single and from multiple hard scattering have the same power behavior
in the large scale $Q$, so that multiple hard scattering is \emph{not}
power suppressed.  It is easy to see that the power behavior in
\eqref{mult-hard-pow} and \eqref{single-hard-pow} holds for any
combination of single and multiple hard scatters, e.g.\ when producing the
final states with momenta $q_1$ and $q_2$ in a single hard scatter and
each final state with momentum $q_3$, $q_4$, etc.\ in a hard scatter of
its own.

Let us now see what happens if we integrate over the $\tvec{q}_i$.  In the
multiple-scattering mechanism, each transverse momentum $\tvec{q}_i$ is
the sum $\tvec{k}_i + \bar{\tvec{k}}_i$ of two parton momenta and thus
limited to be of size $\Lambda$, so that the phase space volume is
$\prod_{i=1}^n d^2\tvec{q}_i \sim \Lambda^{2n}$.  With a single hard
scattering, however, the individual momenta $\tvec{q}_i$ can be as large
as the hard scale $Q$, and only their sum $\tvec{q}$ is limited to be of
order $\Lambda$ by the constraint $\tvec{q} = \tvec{k} + \bar{\tvec{k}}$
in \eqref{single-hard-diff}.  The phase space volume in this case is
therefore $\prod_{i=1}^n d^2\tvec{q}_i = d^2\tvec{q}\ms \prod_{i=1}^{n-1}
d^2\tvec{q}_i \sim \Lambda^2\ms Q^{2n-2}$, and we have
\begin{align}
\frac{d\sigma}{\prod_{i=1}^n dx_i\, d\bar{x}_i}
  \,\bigg|_{\text{multiple}} &\sim \frac{\Lambda^{2n-2}}{Q^{2n}} \, ,  &
\frac{d\sigma}{\prod_{i=1}^n dx_i\, d\bar{x}_i}
  \,\bigg|_{\text{single}} &\sim \frac{1}{Q^{2}}
\end{align}
for the cross sections integrated over all transverse momenta.
Multiple-scattering contributions are now suppressed by at least one power
of $\Lambda^2 /Q^2$ and are hence power corrections to the contribution
from a single hard scattering, as has been known for a long time
\cite{Politzer:1980me}.  This is indeed necessary for the validity of the
familiar collinear factorization theorems, which only take into account
single hard scatters.

Power counting in the hard scale $Q$ provides an essential criterion for
determining which contributions to the cross section are important.  There
are, however, other important factors to keep in mind.  We already
mentioned Sudakov logarithms in $\tvec{q}_i^2 /Q^2$, which appear in the
cross section differential in $\tvec{q}_i$ and are different for single
and multiple hard scattering.  They will be discussed in
section~\ref{sec:cs-general}.  Another aspect in which single and multiple
scattering contributions differ is the dependence on the momentum
fractions $x_i$ and $\bar{x}_i$, which can be rather small as we remarked
after \eqref{phase-space-el}.  We will return to this point in
section~\ref{sec:power-counting}.


\subsubsection{Impact parameter representation}
\label{sec:impact}

The cross section in \eqref{X-sect-position} involves distributions
$F(x_i, \tvec{z}_i, \tvec{y}{}_i)$ that depend on the transverse positions
of the scattering partons but still refer to proton states with definite
(zero) transverse momenta.  In this section we give a formulation
completely in transverse position space, closely following the
construction of impact-parameter dependent parton distributions in
\cite{Soper:1976jc,Burkardt:2002hr,Diehl:2002he}.

To begin with, we define a non-forward correlation function $\Phi(l^{}_i,
l'_i; p, p')$ exactly as in \eqref{Phi-def} but with a state $\langle p'
|$ having a different momentum than the state $|p \rangle$.  Using the
same arguments as in section~\ref{sec:dist-def} we can derive a
representation of the form \eqref{Phi-symmetric} for $\Phi(l^{}_i, l'_i;
p, p')$, with $\langle p \ms|$ replaced by $\langle p' |$.  The
constraints on the parton momenta read
\begin{align}
  \label{prime-constraint}
p - \sum_{i=1}^n l^{}_i &= p' - \sum_{i=1}^n l'_i \,, &
\sum_{i=1}^n r_i &= p' - p \,.
\end{align}
in this case.  In the same manner we define multiparton distributions
$F(x_i, \tvec{k}_i, \tvec{r}_i; \tvec{p}, \tvec{p}')$, $F(x_i, \tvec{k}_i,
\tvec{y}{}_i; \tvec{p}, \tvec{p}')$ and $F(x_i, \tvec{z}_i, \tvec{y}{}_i;
\tvec{p}, \tvec{p}')$ as in \eqref{F-def} to \eqref{position-F}, but taken
between states $\langle p^+, \tvec{p}'|$ and $| p^+, \tvec{p} \rangle$.
Note that we take the same plus-momentum in the bra and ket state, even if
their transverse momenta are different.

We now consider a transverse boost, i.e.\ a Lorentz transformation that
changes the transverse components of a four-vector $v$ as
\begin{equation}
  \label{transv-boost}
\tvec{v}\to \tvec{v} - v^+\, \frac{\tvec{p} + \tvec{p}'}{2p^+}
\end{equation}
and leaves plus-components unchanged.  Invariance under this
transformation implies
\begin{equation}
  \label{transv-boost-invar}
F(x_i, \tvec{k}_i, \tvec{r}_i; \tvec{p}, \tvec{p}')
 = F\bigl(x_i, \tvec{k}_i - x_i \tvec{P}, \tvec{r}_i;
     - \half\tvec{\Delta}, \half\tvec{\Delta} \big)
\end{equation}
with
\begin{align}
\tvec{P} &= \half (\tvec{p} + \tvec{p}') \,, &
\tvec{\Delta} &= \tvec{p}' - \tvec{p} \,.
\end{align}
In impact parameter space we then have
\begin{align}
F(x_i, \tvec{z}_i, \tvec{y}{}_i; \tvec{p}, \tvec{p}')
 &= \biggl[\, \prod_{i=1}^n
   \int d^2\tvec{k}_i\, e^{i \tvec{z}_i \tvec{k}_i} \biggr]\,
\biggl[\, \prod_{i=1}^{n-1} \int \frac{d^2\tvec{r}_i}{(2\pi)^2}\,
    e^{-i \tvec{y}{}_i \tvec{r}_i} \biggr]\,
F(x_i, \tvec{k}_i, \tvec{r}_i; \tvec{p}, \tvec{p}')
\nonumber \\
 &= \biggl[\, \prod_{i=1}^n e^{i \tvec{z}_i x_i \tvec{P}} \biggr]
    \biggl[\, \prod_{i=1}^n
      \int d^2\tvec{k}_i\, e^{i \tvec{z}_i (\tvec{k}_i - x_i 
           \tvec{P})} \biggr]\,
\biggl[\, \prod_{i=1}^{n-1} \int \frac{d^2\tvec{r}_i}{(2\pi)^2}\,
    e^{-i \tvec{r}_i \tvec{y}{}_i} \biggr]\,
\nonumber \\
 &\qquad \times
  F\bigl(x_i, \tvec{k}_i - x_i \tvec{P}, \tvec{r}_i;
     - \half\tvec{\Delta}, \half\tvec{\Delta} \big)
  \phantom{\biggl[ \biggr]}
\nonumber \\
 &= e^{i \tvec{P} \sum_{i=1}^n x_i \tvec{z}_i}\,
  F\bigl(x_i, \tvec{z}, \tvec{y}{}_i;
     - \half\tvec{\Delta}, \half\tvec{\Delta} \big) \,.
\end{align}
We now introduce proton states with definite impact parameter:
\begin{equation}
  \label{impact-states}
|p^+, \tvec{b} \rangle = \int \frac{d^2\tvec{p}}{(2\pi)^2}\,
   e^{-i \tvec{b} \tvec{p}}\, |p^+, \tvec{p} \rangle \,.
\end{equation}
One readily obtains their normalization
\begin{equation}
  \label{impact-norm}
\langle p'^+, \tvec{b}' | p^+, \tvec{b} \rangle
  = 4\pi p^+ \delta(p'^+ - p^+)\,
    \delta^{(2)}(\tvec{b}' - \tvec{b})
\end{equation}
from the usual relativistic normalization $\langle p'^+, \tvec{p}' | p^+,
\tvec{p} \rangle = (2\pi)^3\, 2 p^+\, \delta(p'^+ - p^+)\,
\delta^{(2)}(\tvec{p}' - \tvec{p})$ of momentum eigenstates (recall that
at fixed $\tvec{p}$ one has $dp^0 /p^0 = dp^+ /p^+$ in the invariant
integration element).  For later use we also give the projector on
one-particle states,
\begin{equation}
  \label{impact-projector}
\one = \int \frac{dp^+\ms d^2\tvec{b}}{4\pi p^+} \;
  |p^+, \tvec{b} \rangle \langle p^+, \tvec{b} | \;,
\end{equation}
which is readily checked by taking the matrix element between the
one-particle states in \eqref{impact-states} and using
\eqref{impact-norm}.  We finally define the center of momentum of $m$
particles with plus-momenta $p_i^+$ and transverse positions
$\tvec{b}_i^{}$ as
\begin{align}
\tvec{b} &= \sum_{i=1}^m p_i^+ \ms \tvec{b}_i^{\phantom{+}}
   \bigg/ \sum_{i=1}^m p_i^+ \,.
\end{align}
By virtue of Lorentz invariance, this is a conserved quantity.  Note the
analogy between \eqref{transv-boost} and non-relativistic boosts if
$\tvec{v}$ is a momentum and if one replaces plus-momenta by masses.  The
center of momentum is thus the analog of the center of mass in the
non-relativistic case, which is of course conserved.

Let us consider the matrix element of the same operator as in
\eqref{position-F}, but taken between impact parameter instead of
transverse-momentum eigenstates.  We have
\begin{align}
  \label{impact-matr-el}
 & \biggl[\, \prod_{i=1}^n
       \int \frac{dz_i^-}{2\pi}\, e^{i x_i^{} z_i^- p^+} 
  \biggr]\, \biggl[\, \prod_{i=1}^{n-1}\, 2p^+ \!\! \int dy_i^- \biggr]
\big\langle p^+, - \tvec{b} - \half\tvec{d} \,\big|
  \mathcal{O}(0, z_n) \prod_{i=1}^{n-1} \mathcal{O}(y_i, z_i)
  \big| p^+, - \tvec{b} + \half\tvec{d} \,\big\rangle
\nonumber \\
 &\quad =
  \int \frac{d^2\tvec{p}'}{(2\pi)^2}\, \frac{d^2\tvec{p}}{(2\pi)^2}\;
  e^{- i (\tvec{p}' - \tvec{p})\, \tvec{b}
     - i (\tvec{p}' + \tvec{p})\, \tvec{d} /2}\,
F(x_i, \tvec{z}_i, \tvec{y}{}_i; \tvec{p}, \tvec{p}')
\nonumber \\
 &\quad =
  \int \frac{d^2\tvec{\Delta}}{(2\pi)^2}\,
  \frac{d^2\tvec{P}}{(2\pi)^2}\;
  e^{- i \tvec{b} \tvec{\Delta} - i \tvec{P}\ms \tvec{d}
     + i \tvec{P} \sum_{i=1}^n x_i \tvec{z}_i}\,
  F\bigl(x_i, \tvec{z}_i, \tvec{y}{}_i;
     - \half\tvec{\Delta}, \half\tvec{\Delta} \big)
\nonumber \\
 &\quad = 
    \delta^{(2)}\Bigl( \tvec{d} - \sum_{i=1}^n x_i \tvec{z}_i \Bigr)\;
 \int \frac{d^2\tvec{\Delta}}{(2\pi)^2}\; e^{-i \tvec{b} \tvec{\Delta}}
    F\bigl(x_i, \tvec{z}_i, \tvec{y}{}_i;
     - \half\tvec{\Delta}, \half\tvec{\Delta} \big) \,.
\end{align}
The delta function in the last line reflects the conservation of the
center of momentum, which equals
\begin{align}
- \tvec{b} - \half\tvec{d} &=
  \sum_{i=1}^{n-1} x_i \bigl( \tvec{y}{}_i - \half\tvec{z}_i \bigr)
  - \half x_n \tvec{z}_n  + x_s \tvec{z}_s \,,
\nonumber \\
- \tvec{b} + \half\tvec{d} &=
  \sum_{i=1}^{n-1} x_i \bigl( \tvec{y}{}_i + \half\tvec{z}_i \bigr)
  + \half x_n \tvec{z}_n  + x_s \tvec{z}_s
\end{align}
for the bra and ket states in the matrix element, respectively.  Here $x_s
= 1 - \smash{\sum_{i=1}^{n} x_i}$ and $\tvec{z}_s$ is the center of
momentum of the spectator partons.
We define impact-parameter dependent multiparton distributions by
\begin{equation}
  \label{impact-dist-def}
F(x_i, \tvec{z}_i, \tvec{y}{}_i; \tvec{b})
= \int \frac{d^2\tvec{\Delta}}{(2\pi)^2}\; e^{-i \tvec{b} \tvec{\Delta}}
    F\bigl(x_i, \tvec{z}_i, \tvec{y}{}_i;
     - \half\tvec{\Delta}, \half\tvec{\Delta} \big) \,.
\end{equation}
If we set $\tvec{z}_i = \tvec{0}$ then the matrix element in
\eqref{impact-matr-el} is taken at $\tvec{d} = \tvec{0}$ and hence becomes
diagonal.  We can interpret $F(x_i, \tvec{z}_i = \tvec{0}, \tvec{y}{}_i;
\tvec{b})$ as the probability to find $n$ partons with plus-momentum
fractions $x_i$ in a target that is localized in impact parameter space,
with parton number $n$ at a transverse distance $\tvec{b}$ from the center
of the target and partons $1$ to $n-1$ at relative transverse distances
$\tvec{y}{}_i$ from parton $n$.

\begin{figure}
\begin{center}
\includegraphics[width=0.55\textwidth]{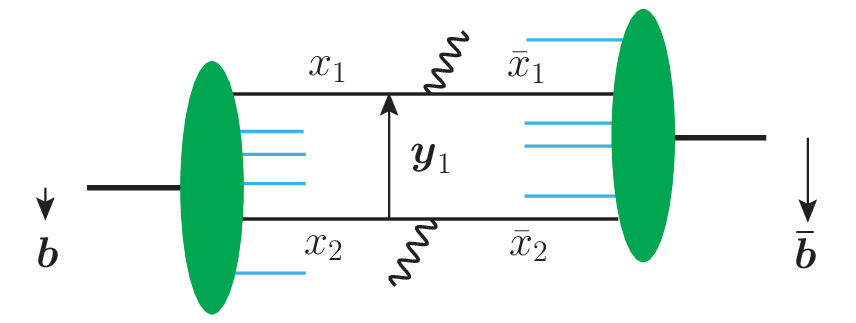}
\end{center}
\caption{\label{fig:impact-scatter} Visualization of the cross section
  formula \protect\eqref{X-sect-impact} for $n=2$ when $\tvec{q}_1$ and
  $\tvec{q}_2$ are integrated over.  Each hard scatter produces a heavy
  gauge boson in this example.}
\end{figure}

Inverting \eqref{impact-dist-def} and setting $\tvec{\Delta} = \tvec{0}$
we get
\begin{equation}
  \label{impact-dist-int}
F\bigl(x_i, \tvec{z}_i, \tvec{y}{}_i) =
  \int d^2\tvec{b}\, F(x_i, \tvec{z}_i, \tvec{y}{}_i; \tvec{b})
\end{equation}
and can therefore represent the multiple-scattering cross section
\eqref{X-sect-position} as
\begin{align}
  \label{X-sect-impact}
\frac{d\sigma}{\prod_{i=1}^n dx_i\, d\bar{x}_i\, d^2\tvec{q}{}_i}
 &= \frac{1}{C}\, 
\biggl[\, \prod_{i=1}^{n} \,\hat{\sigma}_i(x_i \bar{x}_i s) \biggr]\,
\biggl[\, \prod_{i=1}^{n} 
     \int \frac{d^2\tvec{z}_i}{(2\pi)^2}\,
          e^{-i \tvec{z}_i \tvec{q}{}_i} \biggr]
\nonumber \\
& \quad\times
 \biggl[\, \prod_{i=1}^{n-1} \int d^2\tvec{y}{}_i \biggr]
 \int d^2\tvec{b}\, d^2\bar{\tvec{b}}\;
  F(x_i, \tvec{z}_i, \tvec{y}{}_i; \tvec{b})\,
  F(\bar{x}_i, \tvec{z}_i, \tvec{y}{}_i; \bar{\tvec{b}}) \,.
\end{align}
Integration over $\tvec{q}_i$ leads to $\tvec{z}_i = \tvec{0}$ as in
\eqref{X-sect-integrated}.  The resulting cross section formula was
already derived in \cite{Calucci:2009ea}, and it has a very intuitive
geometric interpretation shown in figure~\ref{fig:impact-scatter}.  As
already noted after \eqref{X-sect-position}, the approximations we have
made for the hard-scattering subprocesses imply that each pair of
colliding partons in the hadrons $p$ and $\bar{p}$ must be at the same
position in impact parameter space.  The relative distances $\tvec{y}{}_i$
between the partons are hence the same in both hadrons, but the distance
of the partons from the center of their parent hadron is in general
different in $p$ and $\bar{p}$.  The relative transverse distance
$\tvec{b} - \bar{\tvec{b}}$ between the hadrons is integrated over in the
cross section.

Our result \eqref{X-sect-impact} shows that the representation of the
cross section in terms of impact-parameter dependent distributions remains
simple even if the transverse momenta $\tvec{q}_i$ are kept fixed.  In the
geometric interpretation just described, we then have to replace
``distances'' by ``average distances'', with the average taken between
the amplitude and its conjugate.  What is lost in this case is a
probability interpretation of the multiparton distributions.  The two
fields associated with a parton in the target are now taken at a relative
transverse distance $\tvec{z}_i$, whose typical size is $|\tvec{z}_i| \sim
1 /|\tvec{q}{}_i|$.


\subsubsection{Reduction to single-parton distributions}
\label{sec:reduct}

In order to build a phenomenology of multiple interactions, one needs a
simple ansatz for multiparton distributions as a starting point.  It is
natural to approximate those distributions that have a probability
interpretation by the product of single-parton densities.  In this section
we show how one can formally implement this approximation and generalize
it to the distributions $F(x_i, \tvec{z}_i, \tvec{y}{}_i)$ or $F(x_i,
\tvec{k}_i, \tvec{y}{}_i)$, which do not represent probabilities.

To this end we insert complete sets of intermediate hadron states in
the operator product appearing in the multiparton distributions:
\begin{align}
  \label{state-insertion}
& \mathcal{O}(0, z_n)
    \prod_{i=1}^{n-1} \,\mathcal{O}(y_i, z_i)
= \mathcal{O}(0, z_n) \biggl[\, \prod_{i=1}^{n-1}\,
    \sum_{X_i} \big| X_i \big\rangle \big\langle X_i \big|\,
    \mathcal{O}(y_i, z_i) \biggr]
\nonumber \\
&\quad =
  \sum_{X_{n-1}, \ldots, X_1}
     \mathcal{O}(0, z_n) \big| X_{n-1} \big\rangle
\biggl[\, \prod_{i=2}^{n-1} \big\langle X_i \big|\, \mathcal{O}(y_i, z_i)
          \big| X_{i-1} \big\rangle \biggr]
\big\langle X_1 \big|\, \mathcal{O}(y_1, z_1) \,.
\end{align}
Note that the two parton fields in each operator $\mathcal{O}(y_i, z_i)$
are associated with the same plus-momentum fraction $x_i$ in the
multiparton distributions.

The approximation that gives a product of single-parton distributions is
to \emph{assume} that among all intermediate states $|X_i \rangle$ the
dominant ones are single-proton states.  This reduces the complete sets of
intermediate states to the projection operators \eqref{impact-projector},
and one obtains
\begin{align}
& \delta^{(2)}\Bigl( \tvec{d} - \sum_{i=1}^n x_i \tvec{z}_i \Bigr)\;
  F(x_i, \tvec{z}_i, \tvec{y}{}_i; \tvec{b})
\approx \biggl[\, \prod_{i=1}^n
       \int \frac{dz_i^-}{2\pi}\, e^{i x_i^{} z_i^- p^+} 
  \biggr]\, \biggl[\, \prod_{i=1}^{n-1}\, 2p^+ \!\! \int dy_i^- \biggr]
\nonumber \\
 &\quad \times
\biggl[\, \prod_{i=1}^{n-1} 
   \int \frac{dp_i^+\ms d^2\tvec{b}_i^{}}{4\pi p_i^+} \biggr]
\big\langle p^+, - \tvec{b} - \half\tvec{d} \,\big|\, \mathcal{O}(0,z_n)
  \big| p_{n-1}^+, \tvec{b}_{n-1}^{} \big\rangle
\nonumber \\
 &\quad \times
\biggl[\, \prod_{i=2}^{n-1} 
\big\langle p_i^+, \tvec{b}_{i}^{} \,\big|\, \mathcal{O}(y_i, z_i)
   \big| p_{i-1}^+, \tvec{b}_{i-1}^{} \big\rangle
\biggr]
\big\langle p_1^+, \tvec{b}_{1}^{} \,\big|\, \mathcal{O}(y_1,z_1)
\big| p^+, - \tvec{b} + \half\tvec{d} \,\big\rangle \,.
\end{align}
Translation invariance and the definition \eqref{impact-states} of
impact-parameter states imply
\begin{equation}
\big\langle p_i^+, \tvec{b}_{i}^{} \,\big|\, \mathcal{O}(y_i, z_i)
   \big| p_{i-1}^+, \tvec{b}_{i-1}^{} \big\rangle
= e^{i y_i^- (p_i^{+} - p_{i-1}^{+})}\,
  \big\langle p_i^+, \tvec{b}_{i}^{} - \tvec{y}{}_i \,\big|
  \mathcal{O}(0, z_i) \big| p_{i-1}^+, \tvec{b}_{i-1}^{} - \tvec{y}_{i-1}
  \big\rangle
\end{equation}
and hence
\begin{align}
  \label{impact-product-1}
& \delta^{(2)}\Bigl( \tvec{d} - \sum_{i=1}^n x_i \tvec{z}_i \Bigr)\;
  F(x_i, \tvec{z}_i, \tvec{y}{}_i; \tvec{b})
\nonumber \\
&\quad \approx
\biggl[\, \prod_{i=1}^{n-1} \int d^2\tvec{b}_i \biggr]
  \int \frac{dz_n^-}{2\pi}\, e^{i x_n^{} z_n^- p^+}\,
    \big\langle p^+, - \tvec{b} - \half\tvec{d} \,\big|\,
    \mathcal{O}(0,z_n) \big| p^+, \tvec{b}_{n-1} \big\rangle
\nonumber \\
&\qquad \times
\biggl[\, \prod_{i=2}^{n-1} 
\int \frac{dz_i^-}{2\pi}\, e^{i x_i^{} z_i^- p^+}\,
  \big\langle p^+, \tvec{b}_{i} - \tvec{y}{}_i \,\big|
  \mathcal{O}(0, z_i) \big| p^+, \tvec{b}_{i-1}^{} - \tvec{y}_{i-1}
  \big\rangle
\biggr]
\nonumber \\
&\qquad\qquad \times
\int \frac{dz_1^-}{2\pi}\, e^{i x_1^{} z_1^- p^+}\,
  \big\langle p^+, \tvec{b}_{1} - \tvec{y}_1 \,\big|
  \mathcal{O}(0, z_1) 
  \big| p^+, - \tvec{b} - \tvec{y}_1 + \half\tvec{d} \,\big\rangle \,.
\end{align}
Using \eqref{impact-matr-el} for $n=1$, we have
\begin{equation}
  \label{impact-dist-0}
\int \frac{dz^-}{2\pi}\, e^{i x z^- p^+}\,
  \big\langle p^+, - \tvec{b} - \half\tvec{d} \,\big|
  \mathcal{O}(0, z) \big| p^+, - \tvec{b} + \half\tvec{d} \,\big\rangle
= \delta^{(2)}\bigl( \tvec{d} - x \tvec{z} \bigr)\,
  f(x, \tvec{z}, \tvec{b}) \,,
\end{equation}
where $f(x, \tvec{z}; \tvec{b})$ can be written as
\begin{align}
  \label{impact-dist-1}
f(x, \tvec{z}; \tvec{b}) &=
\int \frac{d^2\tvec{\Delta}}{(2\pi)^2}\; e^{-i \tvec{b} \tvec{\Delta}}
    f(x, \tvec{z}; \tvec{\Delta})
\intertext{with}
  \label{impact-dist-2}
f(x, \tvec{z}; \tvec{\Delta})
 &= \int \frac{dz^-}{2\pi}\, e^{i x z^- p^+}\,
  \big\langle p^+, \half\tvec{\Delta} \big|
  \mathcal{O}(0, z) \big| p^+, - \half\tvec{\Delta} \big\rangle \,.
\end{align}
A reader familiar with generalized parton distributions will recognize
that 
\begin{align}
  \label{mom-dist}
f(x, \tvec{k}; \tvec{\Delta})
&= \int \frac{d^2\tvec{z}}{(2\pi)^2}\;
  e^{- i \tvec{z} \tvec{k}}\, f(x, \tvec{z}; \tvec{\Delta})
\end{align}
is a transverse-momentum dependent generalized parton distribution at zero
skewness.  We will shortly need the collinear distributions
\begin{align}
f(x; \tvec{b}) &= f(x, \tvec{z} = \tvec{0}; \tvec{b}) \,,
&
f(x; \tvec{\Delta}) &= f(x, \tvec{z} = \tvec{0}; \tvec{\Delta})
\end{align}
as well.  Introduced long ago in \cite{Soper:1976jc,Burkardt:2002hr}, the
impact parameter density $f(x; \tvec{b})$ gives the probability to find a
parton with momentum fraction $x$ at a transverse distance $\tvec{b}$ from
the center of the proton.

\begin{figure}
\begin{center}
\includegraphics[width=0.99\textwidth]{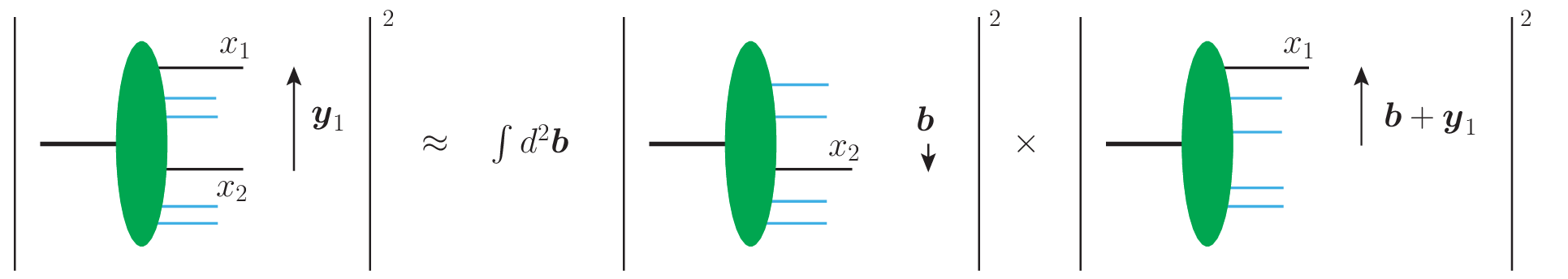}
\end{center}
\caption{\label{fig:impact-reduct} Illustration of the approximate
  relation \protect\eqref{impact-product-coll} for $n=2$.}
\end{figure}

The delta function on the r.h.s.\ of \eqref{impact-dist-0} implies that
\begin{align}
\tvec{b}_{n-1} &= - \tvec{b} - \half\sum_{i=1}^{n-1} x_i \tvec{z}_i
                  + \half x_n \tvec{z}_n &
\tvec{b}_{i-1} &= \tvec{b}_i + x_i \tvec{z}_i ~~\text{for}~ 1 < i < n-1
\end{align}
in \eqref{impact-product-1}, so that we obtain the desired approximation
\begin{multline}
  \label{impact-product-2}
F(x_i, \tvec{z}_i, \tvec{y}{}_i; \tvec{b}) \approx
f\bigl( x_n, \tvec{z}_n; \tvec{b}
   + \half (x_1 \tvec{z}_1 + \ldots + x_{n-1} \tvec{z}_{n-1} ) \bigl)
\\
  \times \biggl[\, \prod_{i=2}^{n-1}
f\bigl(x_{i}, \tvec{z}_{i}; \tvec{b} + \tvec{y}_{i}
   + \half ( x_1 \tvec{z}_1 + \ldots + x_{i-1} \tvec{z}_{i-1} )
   - \half ( x_{i+1} \tvec{z}_{i+1} - \ldots x_n \tvec{z}_n ) \bigr)
\biggr]
\\
  \times f\bigl(x_{1}, \tvec{z}_{1}; \tvec{b} + \tvec{y}_{1}
   - \half ( x_{2}\ms \tvec{z}_{2} + \ldots x_n \tvec{z}_n ) \bigr)
\end{multline}
of a multiparton distribution.  Setting $\tvec{z}_i = \tvec{0}$ and
integrating over $\tvec{b}$, we obtain in particular the collinear
multiparton distribution $F(x_i, \tvec{y}{}_i)$ in terms of impact-parameter
dependent single-parton densities,
\begin{align}
  \label{impact-product-coll}
F(x_i, \tvec{y}{}_i) &\approx \int d^2\tvec{b}\;
  f(x_n; \tvec{b})\, \prod_{i=1}^{n-1} f(x_i; \tvec{b} + \tvec{y}{}_i) \,.
\end{align}
This relation is illustrated in figure~\ref{fig:impact-reduct}, which uses
the representation of parton distributions as squared light-cone wave
functions we mentioned briefly before~\protect\eqref{momentum-F-coll}.

Let us now insert \eqref{impact-product-2} into the cross section
\eqref{X-sect-impact}.  For measured transverse momenta $\tvec{q}_i$, the
different single-parton distributions are entangled by their $\tvec{z}_i$
dependence.  By contrast, the $\tvec{q}_i$ integrated cross section
simplifies to
\begin{align}
  \label{X-sect-impact-product}
\frac{d\sigma}{\prod_{i=1}^n dx_i\, d\bar{x}_i}
 &\approx \frac{1}{C}\, 
\biggl[\, \prod_{i=1}^{n} \,\hat{\sigma}_i(x_i \bar{x}_i s) \biggr]\,
 \int d^2\tvec{b}\, d^2\bar{\tvec{b}}\;
    f(x_n; \tvec{b})\, f(\bar{x}_n; \bar{\tvec{b}})
\nonumber \\
&\quad \times
  \biggl[\, \prod_{i=1}^{n-1} \int d^2\tvec{y}{}_i\;
    f(x_i; \tvec{b} + \tvec{y}{}_i)\,
    f(\bar{x}_i; \bar{\tvec{b}} + \tvec{y}{}_i)
 \biggr]
\nonumber \\
&=  \frac{1}{C}\, \int d^2\tvec{\rho}\;
\biggl[\, \prod_{i=1}^{n} \,\hat{\sigma}_i(x_i \bar{x}_i s)
  \int d^2\tvec{y}{}_i\;
    f(x_i; \tvec{y}{}_i - \tvec{\rho})\,
    f(\bar{x}_i; \tvec{y}{}_i) \,\biggr] \,,
\end{align}
where the only integration variable linking the different factors is the
relative distance $\tvec{\rho} = \tvec{b} - \bar{\tvec{b}}$, and where we
have renamed the integration variable $\bar{\tvec{b}}$ to $\tvec{y}_n$ in
the second step.
In different forms, this relation (or more precisely its analog for quarks
and gluons instead of scalar partons) has long been used as a starting
point of phenomenological studies, see e.g.\
\cite{Sjostrand:1986ep,Sjostrand:1987su,%
  Durand:1987yv,Durand:1988ax,Ametller:1987ru} and
\cite{Frankfurt:2003td,Domdey:2009bg}.\footnote{%
  We note that in \protect\cite{Frankfurt:2003td,Domdey:2009bg} the impact
  parameter arguments of $f$ are $\tvec{\rho} - \tvec{y}{}_i$ and
  $\tvec{y}{}_i$ instead of $\tvec{y}{}_i - \tvec{\rho}$ and
  $\tvec{y}{}_i$ (if we translate to our notation).  This is equivalent in
  the spin independent sector, where the single-parton distributions are
  independent of the direction of the impact parameter.}

As observed in \cite{Blok:2010ge} for the case of collinear distributions,
the reduction of multiparton to single-parton distributions also takes a
simple form in the transverse-momentum representation.  This remains true
if one keeps the transverse parton momenta unintegrated.  To see this, we
integrate \eqref{impact-product-2} over $\tvec{b}$ and Fourier transform
w.r.t.\ $\tvec{y}{}_i$ and $\tvec{z}_i$ as specified by \eqref{mixed-F}
and \eqref{position-F}.  Changing integration variables from $\tvec{b}$
and $\tvec{y}{}_i$ to the impact parameter arguments of the distributions
on the r.h.s.\ of \eqref{impact-product-2}, we obtain
\begin{align}
  \label{mom-product-1}
F(x_i, \tvec{k}_i, \tvec{r}_i) & \approx
f\bigl( x_n, \tvec{k}_{n}
    - \half\ms x_n \bigl( \tvec{r}_1 + \ldots
         + \tvec{r}_{n-1} ); \tvec{r}_n \bigl)
\nonumber \\
 &\quad \times \biggl[\, \prod_{i=2}^{n-1}
f\bigl(x_{i}, \tvec{k}_{i}
    - \half\ms x_i\ms ( \tvec{r}_1 + \ldots + \tvec{r}_{i-1} )
    + \half\ms x_i\ms ( \tvec{r}_{i+1} + \ldots \tvec{r}_n );
    \tvec{r}_i \bigr)
\biggr]
\nonumber \\
 &\qquad \times f\bigl(x_{1}, \tvec{k}_{1}
    + \half\ms x_1 ( \tvec{r}_{2} + \ldots + \tvec{r}_n );
    \tvec{r}_1 \bigr) \,,
\end{align}
where we recall that $\tvec{r}_n = - \sum\limits_{i=1}^{n-1} \tvec{r}_i$.
Integrated over the momenta $\tvec{k}_i$ this simply reads
\begin{align}
  \label{mom-product-2}
  F(x_i, \tvec{r}_i) \,\approx\, \prod_{i=1}^{n} f(x_i; \tvec{r}_i) \,,
\end{align}
so that the cross section \eqref{X-sect-impact-product} becomes
\begin{align}
  \label{X-sect-mom-product}
\frac{d\sigma}{\prod_{i=1}^n dx_i\, d\bar{x}_i}
& \approx \frac{1}{C}\, 
\biggl[\, \prod_{i=1}^{n-1} \int \frac{d^2\tvec{r}_i}{(2\pi)^2}
\biggr]\,
\biggl[\, \prod_{i=1}^{n} \,\hat{\sigma}_i(x_i \bar{x}_i s)\,
          f(x_i; \tvec{r}_i)\, f(\bar{x}_i; - \tvec{r}_i) \biggr] \,.
\end{align}
The arguments $\tvec{r}_i$ in \eqref{mom-product-2} can easily be
anticipated from figure~\ref{fig:labeling}.

We emphasize that the relations \eqref{impact-product-2} to
\eqref{mom-product-2} have been obtained by restricting a sum over all
intermediate states to a single proton.  We do not have a motivation for
this restriction other than observing that it results in neglecting
correlations between different partons in the proton.  It seems plausible
to assume that this is a reasonable first approximation, at least in a
certain region of variables, but one should not expect it to be very
precise.  Possible deviations from this approximation and their
phenomenological consequences have recently been discussed in
\cite{Calucci:1997ii,Calucci:1999yz,DelFabbro:2000ds,%
  Rogers:2009ke,Domdey:2009bg,Flensburg:2011kj}.


\subsection{Parton spin}
\label{sec:tree:spin}

Let us now see how the scattering formulae \eqref{X-sect-momentum} to
\eqref{X-sect-integrated} are modified in QCD, where partons have nonzero
spin.  In \eqref{X-section-start} to \eqref{pre-X-section} the squared
amplitude $H_i$ of the $i$th hard scattering and the hadronic matrix
elements of parton field operators acquire spinor indices in the case of
quarks and Lorentz indices in the case of gluons.  These indices can be
treated as in the case of a single hard scattering.  For the time being we
still omit color degrees of freedom, which will be discussed in
section~\ref{sec:color}.

\subsubsection{Quarks}
\label{sec:spin:quarks}

The correlation function for $n$ quarks entering the hard scattering is
\begin{align}
  \label{Phi-quarks}
& \Phi_{\alpha_1 \beta_1 \ldots\ms \alpha_n \beta_n}(k_i, r_i)
 = \biggl[\, \prod_{i=1}^{n}   \int \frac{d^4 z_i}{(2\pi)^4}\,
       e^{i z_i k_i} \biggr]
   \biggl[\, \prod_{i=1}^{n-1} \int \frac{d^4 y_i}{(2\pi)^4}\,
       e^{- i y_i r_i} \biggr]
\nonumber \\
& \quad \times
    \big\langle p \big| 
    \bar{T} \Bigl[\, \bar{q}_{\beta_1}\bigl(y_1 - \half z_1\bigr)
         \cdots \bar{q}_{\beta_n} \bigl(- \half z_n\bigr) \Bigr]\,
    T \Bigl[\, q_{\alpha_n} \bigl(\half z_n\bigr)
         \cdots q_{\alpha_1}\bigl(y_1 + \half z_1\bigr) \Bigr]
    \big| p \big\rangle \,.
\end{align}
When this is integrated over the parton minus-momenta, the anti-time and
time ordering can be omitted and one can reorder the fields as
\begin{align}
  \label{reordered-fields}
\biggl[\, \prod_{i=1}^{n-1}
 \bar{q}_{\beta_i} \bigl(y_i - \half z_i\bigr)\,
   q_{\alpha_i}\bigl(y_i + \half z_i\bigr) \,\biggr]
 \bar{q}_{\beta_n} \bigl(- \half z_n\bigr)\,
   q_{\alpha_n}\bigl(\half z_n\bigr)
\end{align}
by an even permutation.  For antiquarks entering the scattering, one has
an operator product $q_{\alpha_i}(y_i - \half z_i)\, \bar{q}_{\beta_i}(y_i
+ \half z_i)$ instead of $\bar{q}_{\beta_i}(y_i - \half z_i)\,
q_{\alpha_i}(y_i + \half z_i)$.

Consider the case of a quark entering the hard scattering.  We wish to
rearrange the spinor indices in the product $\Phi_{\alpha\beta} H_{i,
  \beta\alpha}$, where for brevity we write $\alpha,\beta$ instead of
$\alpha_i, \beta_i$ and leave out all other indices on which $H_i$ and
$\Phi$ depend.  The rearrangement is achieved by the Fierz transform
\begin{align}
  \label{Fierz}
H_{i,\beta\alpha} &=
  \half\ms \delta_{\beta\alpha} \tr\bigl( \half H_i \bigr)
+ \half (\gamma_5)_{\beta\alpha} \tr\bigl( \half \gamma_5\ms H_i \bigr)
+ \half (\gamma^\mu)_{\beta\alpha} \tr\bigl( \half \gamma_\mu H_i \bigr)
\nonumber \\[0.2em]
&\quad
+ \half (\gamma^\mu \gamma_5)_{\beta\alpha}
        \tr\bigl( \half\gamma_5 \gamma_\mu H_i \bigr)
+ \half i (\sigma^{\mu\nu} \gamma_5)_{\beta\alpha}
        \tr\bigl( \tfrac{1}{4} i \sigma_{\nu\mu} \gamma_5\ms H_i \bigr) \,.
\end{align}
The Dirac matrices with open indices on the r.h.s.\ multiply fields
$\bar{q}_\beta(y_i - \half z_i)\, q_\alpha(y_i + \half z_i)$ in the
correlation function $\Phi_{\alpha\beta}$ for the right-moving proton.
The dominant terms in the cross section are those where that matrix is
$\Gamma = \half\gamma^+, \half\gamma^+ \gamma_5$ or $\half i \sigma^{+j}
\gamma_5$ with $j=1,2$, because $\Gamma_{\beta\alpha} \Phi_{\alpha\beta}$
is then proportional to the large momentum component $p^+ \sim Q$ by
virtue of Lorentz invariance.  The traces over the hard scattering matrix
$H_i$ on the r.h.s.\ of \eqref{Fierz} have both large plus and minus
components since $H_i$ depends on the boson momentum $q_i$.  One thus has
\begin{align}
  \label{Fierz-dominant}
\Phi_{\alpha\beta}\ms H_{i, \beta\alpha} &=
  \tr\bigr( \half \gamma^+ \Phi \bigr) \tr\bigl( \half \gamma^- H_i \bigr)
+ \tr\bigr( \half \gamma^+ \gamma_5\ms \Phi \bigr)
     \tr\bigl( \half\gamma_5 \gamma^- H_i \bigr)
\nonumber \\[0.2em]
&\quad
 + \tr\bigr( \half i \sigma^{j +} \gamma_5\ms \Phi \bigr)
     \tr\bigl( \half i \sigma^{j -} \gamma_5\ms H_i \bigr)
 + \{ \text{power suppressed terms} \} \,,
\end{align}
where a sum over the transverse index $j=1,2$ is understood.

When defining distributions for scalar partons in \eqref{F-def}, we
included a factor $k_i^+$ for each parton $i=1,\ldots,n$.  For quarks we
do not do this, but instead include this factor $k_i^+$ in the definition
of the parton-level cross section $\sigma_i$ from the squared matrix
element $H_i$.  Writing $k_{i,c}^{}$ for the collinear approximation of
$k_i^{}$ (i.e.\ $k_{i,c}^+ = k_i^+$, $k_{i,c}^- = 0$ and
$\tvec{k}_{i,c}^{} = \tvec{0}$) we recognize in
\begin{align}
  \label{quark-spin-av}
k_i^+ \tr\bigl( \half \gamma^-\ms H_i \bigr) &=
   \half \tr\bigl( \slashed{k}{}_{i,c}\ms H_i \bigr)
 = \half \sum_{s} \bar{u}_s(k_{i,c})\ms H_i\, u_s(k_{i,c})
\end{align}
the spin averaged squared amplitude for an incoming on-shell quark.  The
corresponding terms with $\gamma^- \gamma_5$ and $i\sigma^{-j} \gamma_5 =
\gamma_5\ms \gamma^j \gamma^-$ are respectively associated with scattering
on a longitudinally and transversely polarized quark.

Integrating \eqref{Phi-quarks} over the minus components of the parton
momenta, one obtains multi-parton distributions as in \eqref{momentum-F}
with the scalar field operators \eqref{op-def} replaced by quark bilinears
\begin{align}
  \label{quark-bilinears}
\mathcal{O}_{a}(y_i, z_i) &= \bar{q}(y_i - \half z_i)\,
  \Gamma_a \ms q(y_i + \half z_i) \Big|_{z_i^+ = y_i^+ = 0} \,,
\end{align}
where $a = q, \Delta q, \delta q$ labels the polarization and
\begin{align}
  \label{Gamma-twist-2}
\Gamma_{q}          &= \half \gamma^+ \, , &
\Gamma_{\Delta q}   &= \half \gamma^+ \gamma_5 \, , &
\Gamma_{\delta q}^j &= \half i \sigma^{j +} \gamma_5 \,.
\end{align}
We recognize the operators that appear in the definition of single-parton
densities for unpolarized, longitudinally polarized and transversely
polarized quarks, see e.g.\ \cite{Ralston:1979ys,Tangerman:1994eh}.
For antiquarks entering the hard scattering one proceeds in an analogous
way.  The corresponding operators are
\begin{align}
  \label{antiquark-bilinears}
\mathcal{O}_{\bar{a}}(y_i, z_i) &=
- \bar{q}(y_i + \half z_i)\,
  \Gamma_{\bar a}\ms q(y_i - \half z_i) \Big|_{z_i^+ = y_i^+ = 0}
\end{align}
with
\begin{align}
\Gamma_{\bar{q}} &= \Gamma_{q} \,,
&
\Gamma_{\Delta \bar{q}} &= -\Gamma_{\Delta q} \,,
&
\Gamma_{\delta \bar{q}}^j &= \Gamma_{\delta q}^j \,.
\end{align}
The overall minus sign in \eqref{antiquark-bilinears} reflects a change in
the order of field operators from $q_\alpha(y_i - \half z_i)\,
\bar{q}_\beta(y_i + \half z_i)$ to $\bar{q}_\beta(y_i + \half z_i)\,
q_\alpha(y_i - \half z_i)$, cf.\ our remark after
\eqref{reordered-fields}.  In the case of $\mathcal{O}_{\Delta\bar{q}}$ a
further minus sign is included in $\Gamma_{\Delta \bar{q}}$, so that the
operator corresponds to the difference of antiquarks with positive and
negative helicity.

\begin{figure}
\begin{center}
\includegraphics[width=0.77\textwidth]{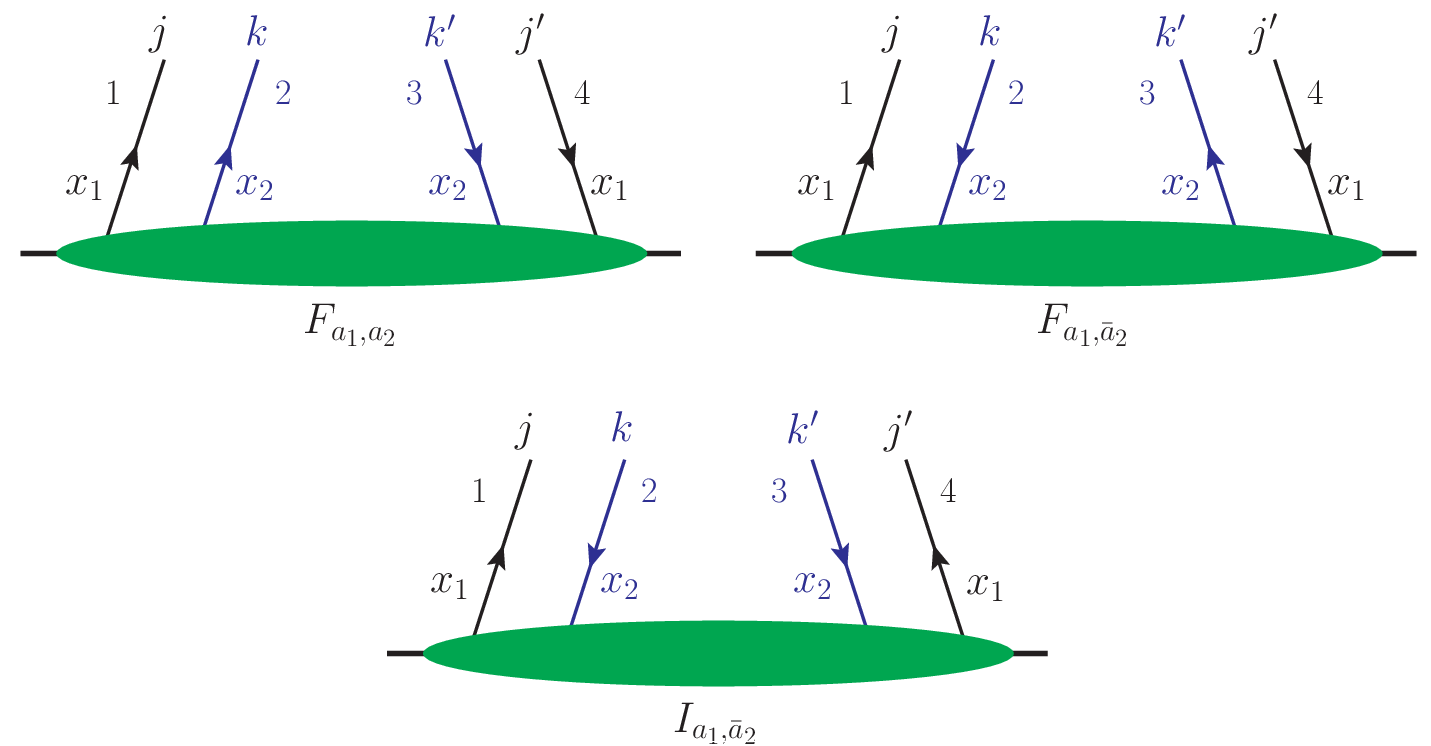}
\end{center}
\caption{\label{fig:distrib} The numbering of fields in two-parton
  distributions specified in \protect\eqref{index-key}.  The color indices
  $j,j',k$ and $k'$ will be discussed in
  section~\protect\ref{sec:color:quarks}.}
\end{figure}

From now on we concentrate on two-parton distributions.  The formalism can
be extended without conceptual difficulties to higher multiple
interactions, but the resulting expressions become rather unwieldy.  As
one encounters nontrivial features already for double hard scattering, it
is natural to elaborate this case first.  To simplify the discussion, we
introduce a compact notation
\begin{align}
  \label{shorthand}
& \mat{\varphi_4\, \varphi_3\, \varphi_2\, \varphi_1}
 = \biggl[\, \prod_{i=1}^2
       \int \frac{dz_i^-\, d^2\tvec{z}_i^{}}{(2\pi)^3}\,
       e^{i x_i^{} z_i^- p^+ -i \tvec{z}_i^{} \tvec{k}_i^{}}
    \biggr]\,
\nonumber \\
 &\qquad \times
  2p^+ \!\! \int dy^-
    \big\langle p \big|\,
       \varphi(y - \half z_1)\, \varphi(- \half z_2)\,
       \varphi(\half z_2)\, \varphi(y + \half z_1)
    \big| p \big\rangle \Big|_{z_1^+ = z_2^+ = y^+_{\phantom{1}} = 0}
\end{align}
for the Fourier transformed matrix element of a product of field operators
$\varphi$.  Their indices are assigned according to
\begin{alignat}{3}
  \label{index-key}
1 &\;\leftrightarrow\; y + \half z_1 
 &&\;\leftrightarrow\; \text{momentum fraction $x_1$ in amplitude}
\nonumber \\[0.1em]
2 &\;\leftrightarrow\; \hspace{1.8em} \half z_2
 &&\;\leftrightarrow\; \text{momentum fraction $x_2$ in amplitude}
\nonumber \\[0.1em]
3 &\;\leftrightarrow\; \phantom{y} - \half z_2
 &&\;\leftrightarrow\; \text{momentum fraction $x_2$ in conjugate
                             amplitude}
\nonumber \\[0.1em]
4 &\;\leftrightarrow\; y - \half z_1
 &&\;\leftrightarrow\; \text{momentum fraction $x_1$ in conjugate
                             amplitude}
\end{alignat}
as shown in figure~\ref{fig:distrib}.
Throughout this paper we consider unpolarized incident hadrons, so that an
average over the proton spin is understood in \eqref{shorthand}.
A two-quark distribution is then given by
\begin{align}
  \label{quark-mixed-F}
F_{a_1, a_2}(x_i, \tvec{k}_i, \tvec{y}) &=
  \mat{(\bar{q}_3\ms \Gamma_{a_2}\ms q_2)\,
       (\bar{q}_4\ms \Gamma_{a_1}\ms q_1)} \,,
\end{align}
and if the parton with momentum fraction $x_2$ is an antiquark one has
instead
\begin{align}
  \label{qqbar-mixed-F}
F_{a_1, \bar{a}_2}(x_i, \tvec{k}_i, \tvec{y}) &=
  \mat{(\bar{q}_2\ms \Gamma_{\bar{a}_2}\ms q_3)\,
       (\bar{q}_4\ms \Gamma_{a_1}\ms q_1)} \,.
\end{align}
In straightforward extension of the case of single-parton distributions
\cite{Tangerman:1994eh}, the matrix elements defining distributions for
quarks and antiquarks are thus connected as
\begin{align}
  \label{quark-antiquark}
F_{a_1, \bar{a}_2}(x_1,x_2, \tvec{k}_1,\tvec{k}_2, \tvec{y})
  = \sigma_{a_2} F_{a_1, a_2}(x_1,-x_2, \tvec{k}_1,-\tvec{k}_2, \tvec{y})
\end{align}
with sign factors $\sigma_{q} = \sigma_{\delta q} = +1$ and
$\sigma_{\Delta q} = -1$.  Definitions and relations analogous to
\eqref{quark-mixed-F}, \eqref{qqbar-mixed-F} and \eqref{quark-antiquark}
hold for the case where the parton with momentum fraction $x_1$ is an
antiquark.

The previous arguments can be repeated for the partons in the left-moving
proton, with the roles of plus and minus components interchanged.  We
define the hard-scattering cross section for a right-moving quark and a
left-moving antiquark as
\begin{align}
  \label{quark-X-sect}
\hat{\sigma}_{i, a \bar{a}} &= \frac{1}{2 q_i^2}\,
  \bigl[ P_{a}(k_{i}) \bigr]{}_{\alpha\beta}\,
  \bigl[ P_{\bar{a}}(\bar{k}_{i}) \bigr]{}_{\bar{\beta}\bar{\alpha}}\;
  H_{i, \beta\alpha\ms \bar{\alpha}\bar{\beta}}
\end{align}
with spin projectors
\begin{alignat}{3}
  \label{quark-projectors}
P_{q}(k) &= P_{\bar{q}}(k_c) &&= \half \slashed{k}{}_c \, , &
\hspace{3em}
P_{\Delta q}(k) = - P_{\Delta\bar{q}}(k_c) =
  \half \gamma_5\ms \slashed{k}{}_c \,,
\nonumber \\
P^j_{\delta q}(k) &= P^j_{\delta\bar{q}}(k_c) &&=
  \half \gamma_5\ms  \slashed{k}{}_c\ms \gamma^j_{}
\end{alignat}
constructed from the collinear momenta $k_{i,c}$ introduced before
\eqref{quark-spin-av}, i.e.\ $k_{i,c}^+ = k_i^+$ for right-moving partons
and $k_{i,c}^- = k_i^-$ for left-moving ones, with all other components
equal to zero.  The spin projectors match the Fierz decomposition
\eqref{Fierz-dominant} and the operators in \eqref{quark-bilinears} and
\eqref{antiquark-bilinears}, and they can be expressed in terms of quark
or antiquark spinors as in \eqref{quark-spin-av}.  It is understood that
for each label $\delta q$ or $\delta\bar{q}$ the cross section
\eqref{quark-X-sect} depends on a transverse Lorentz index, which has not
been explicitly displayed.
In most reactions the partonic subprocess involves only chirality
conserving interactions.  Since incoming quarks and antiquarks are
approximated as massless in the hard scattering, only the combinations
$\hat{\sigma}_{q, \bar{q}}$, $\hat{\sigma}_{\Delta q, \Delta\bar{q}}$,
$\hat{\sigma}_{q, \Delta\bar{q}}$, $\hat{\sigma}_{\Delta q, \bar{q}}$ and
$\hat{\sigma}_{\delta q, \delta\bar{q}}$ are then nonzero.  For parity
conserving processes such as the production of a virtual photon, one is
left with only $\hat{\sigma}_{q, \bar{q}}$, $\hat{\sigma}_{\Delta q,
  \Delta\bar{q}}$ and $\hat{\sigma}_{\delta q, \delta\bar{q}}$.
Hard-scattering cross sections $\hat{\sigma}_{i, \bar{a} a}$ for
right-moving antiquarks and left-moving quarks are defined as in
\eqref{quark-X-sect} with an appropriate change of spinor indices.

\begin{figure}
\begin{center}
\includegraphics[width=0.88\textwidth]{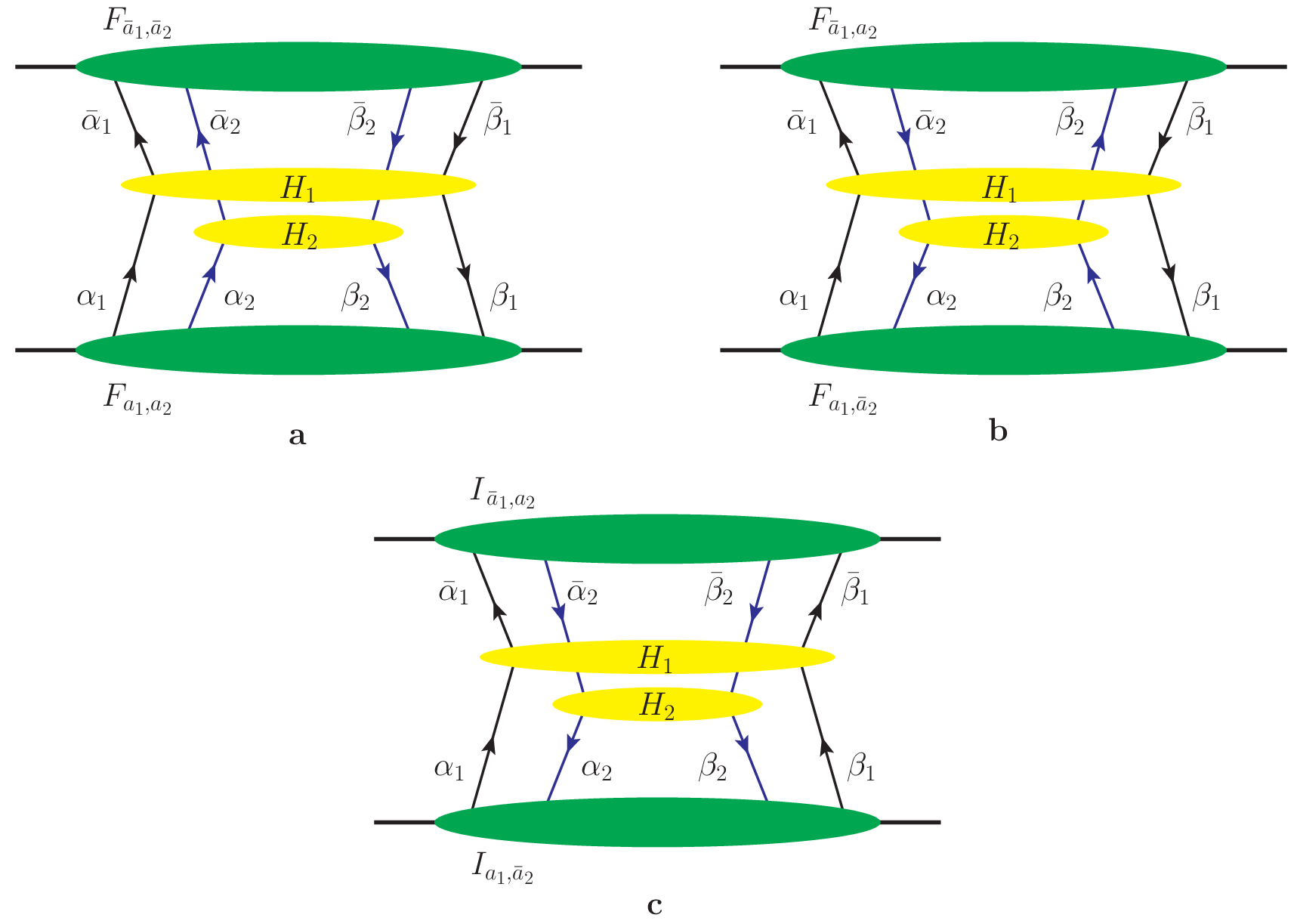}
\end{center}
\caption{\label{fig:double-scatt} Graphs for double hard scattering
  initiated by different combinations of quarks and antiquarks in the
  amplitude and the conjugate amplitude.}
\end{figure}

We now have everything at hand to write down the expression for the
double-scattering graphs of figure~\ref{fig:double-scatt}a and b.  For a
single quark flavor, one has
\begin{align}
  \label{X-sect-qq}
& \frac{d\sigma}{\prod_{i=1}^2 dx_i\, d\bar{x}_i\, d^2\tvec{q}{}_i}
  \Bigg|_{\text{fig.~\protect\ref{fig:double-scatt}a,b}}
= \frac{1}{C}\, 
\sum_{\genfrac{}{}{0pt}{1}{a_1, a_2 = q, \Delta q, \delta q}{%
      \bar{a}_1, \bar{a}_2 = \bar{q}, \Delta\bar{q}, \delta\bar{q}}}\,
\biggl[\, \prod_{i=1}^{2} 
     \int d^2\tvec{k}_i\, d^2\bar{\tvec{k}}_i\;
     \delta^{(2)}(\tvec{q}{}_i - \tvec{k}_i - \bar{\tvec{k}}_i) \biggr]
\nonumber \\[0.2em]
& \quad \times
\int d^2\tvec{y} \,
\biggl[
  \hat{\sigma}_{1, a_1 \bar{a}_1}(x_1 \bar{x}_1 s)\,
  \hat{\sigma}_{2, a_2 \bar{a}_2}(x_2\ms \bar{x}_2 s)\,
  F_{a_1, a_2}(x_i, \tvec{k}_i, \tvec{y})\, 
  F_{\bar{a}_1, \bar{a}_2}(\bar{x}_i, \bar{\tvec{k}}_i, \tvec{y})
\nonumber \\
& \hspace{4.4em} +
  \hat{\sigma}_{1, a_1 \bar{a}_1}(x_1 \bar{x}_1 s)\,
  \hat{\sigma}_{2, \bar{a}_2 a_2}(x_2\ms \bar{x}_2 s)\,
  F_{a_1, \bar{a}_2}(x_i, \tvec{k}_i, \tvec{y})\, 
  F_{\bar{a}_1, a_2}(\bar{x}_i, \bar{\tvec{k}}_i, \tvec{y})
\biggr] \,,
\end{align}
where $S=2$ if the final states of the two hard scatters are identical and
$S=1$ otherwise.
It is straightforward to Fourier transform the previous expressions either
from the interparton distance $\tvec{y}$ to the relative transverse
momentum $\tvec{r}$, or from average transverse momenta $\tvec{k}_i,
\bar{\tvec{k}}_i$ to transverse positions $\tvec{z}_i$, as we did in
\eqref{momentum-F}, \eqref{position-F} and \eqref{X-sect-momentum},
\eqref{X-sect-position} for scalar partons.

Notice that \eqref{X-sect-qq} involves a polarization dependence in the
multiparton distributions and hard-scattering cross section.  This is
because, even for unpolarized hadron beams, the polarization of the two
partons with momentum fractions $x_1$ and $x_2$ can be correlated among
themselves.  We will discuss this in more detail in
section~\ref{sec:spin-decomp}.

The two-quark and quark-antiquark distributions considered so far have the
form $\mat{\mathcal{O}_1\, \mathcal{O}_2}$, where the $\mathcal{O}_i$ are
bilinear operators from \eqref{quark-bilinears} or
\eqref{antiquark-bilinears}.  As we discussed after
\eqref{momentum-F-proba}, these distributions can be interpreted as
probabilities or pseudo-probabilities in the sense of Wigner distributions
for two partons in the proton that carry momentum fractions $x_1$ and
$x_2$, respectively.

There are further double-scattering graphs that contribute to the cross
section and involve distributions which represent interference terms
rather than probabilities.  In figure~\ref{fig:double-scatt}c we show the
case where the parton with momentum fraction $x_1$ is a quark in the
scattering amplitude and an antiquark in the conjugate scattering
amplitude.  Such interference terms in fermion number have no equivalent
in single hard-scattering processes, where they are forbidden by fermion
number conservation.  For their description we introduce interference
distributions
\begin{align}
  \label{fermion-interference}
I_{a_1, \bar{a}_2}(x_i, \tvec{k}_i, \tvec{y}) &=
\mat{(\bar{q}_2\ms \Gamma_{\bar{a}_2}\ms q_4)\,
     (\bar{q}_3\ms \Gamma_{a_1}\ms q_1)} \,,
\nonumber \\
I_{\bar{a}_1, a_2}(x_i, \tvec{k}_i, \tvec{y}) &=
\mat{(\bar{q}_4\ms \Gamma_{a_2}\ms q_2)\,
     (\bar{q}_1\ms \Gamma_{\bar{a}_1}\ms q_3)} \,.
\end{align}
In the absence of a probability interpretation, the choice of quark vs.\
antiquark labels in the Dirac matrices is pure convention.  We assign
labels such that $a$ indicates a quark and $\bar{a}$ an antiquark in the
amplitude, i.e.\ for the parton indices 1 and 2 in
figure~\ref{fig:distrib}.
The graph in figure~\ref{fig:double-scatt}c contributes to the cross
section as
\begin{align}
  \label{X-sect-interf}
& \frac{d\sigma}{\prod_{i=1}^2 dx_i\, d\bar{x}_i\, d^2\tvec{q}{}_i}
  \Bigg|_{\text{fig.~\protect\ref{fig:double-scatt}c}}
= \frac{1}{C}\, 
\sum_{\genfrac{}{}{0pt}{1}{a_1, a_2 = q, \Delta q, \delta q}{%
      \bar{a}_1, \bar{a}_2 = \bar{q}, \Delta\bar{q}, \delta\bar{q}}}
  H_{1,\, \alpha_1 \beta_1
          \bar{\alpha}_1 \bar{\beta}_1}(k_{1}, \bar{k}_{1}) \,
  \bigl[ P_{a_1}(k_{1}) \bigr]{}_{\alpha_1 \beta_2}\,
  \bigl[ P_{\bar{a}_2}(k_{2}) \bigr]{}_{\beta_1 \alpha_2}\,
\nonumber \\[0.2em]
& \qquad \times
  H_{2,\, \alpha_2 \beta_2
          \bar{\alpha}_2 \bar{\beta}_2}(k_{2}, \bar{k}_{2}) \,
  \bigl[ P_{\bar{a}_1}(\bar{k}_{1})
         \bigr]{}_{\bar{\beta}_2 \bar{\alpha}_1}\,
  \bigl[ P_{a_2}(\bar{k}_{2})
         \bigr]{}_{\bar{\alpha}_2 \bar{\beta}_1}
\phantom{\int}
\nonumber \\[0.2em]
& \quad \times
\biggl[\, \prod_{i=1}^{2} 
     \int d^2\tvec{k}_i\, d^2\bar{\tvec{k}}_i\;
     \delta^{(2)}(\tvec{q}{}_i - \tvec{k}_i - \bar{\tvec{k}}_i) \biggr]
\int d^2\tvec{y} \,
  I_{a_1, \bar{a}_2}(x_i, \tvec{k}_i, \tvec{y})\, 
  I_{\bar{a}_1, a_2}(\bar{x}_i, \bar{\tvec{k}}_i, \tvec{y}) \,.
\phantom{\int}
\end{align}
We see that the contraction of Dirac indices ties together the two
hard-scattering kernels and the spin projectors $P$, so that one cannot
define separate partonic cross sections $\hat{\sigma}_1$ and
$\hat{\sigma}_2$.  The power behavior of this contribution is the same as
in \eqref{X-sect-qq}.

\begin{figure}
\begin{center}
\includegraphics[width=0.58\textwidth]{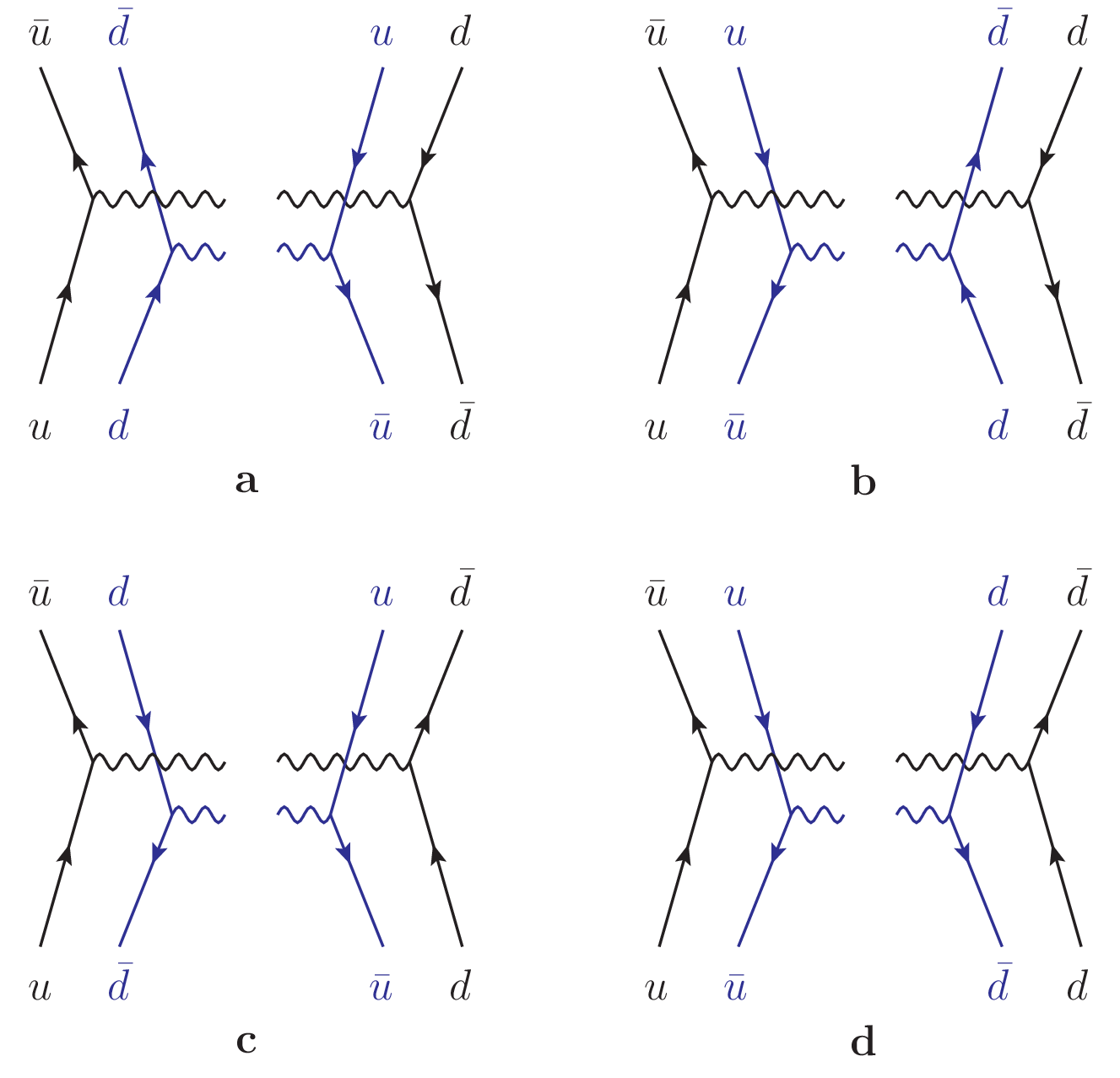}
\end{center}
\caption{\label{fig:interference} Graphs for double hard scattering with
  interference in quark flavor.  For simplicity, the blobs indicating the
  hadronic matrix elements are not shown.}
\end{figure}

Taking different quark flavors into account, we obtain further
interference terms.  The contributions in figure~\ref{fig:interference}a
and b involve the interference of different quark flavors, and those in
figure~\ref{fig:interference}c and d the combined interference in fermion
number and flavor.  The relevant matrix elements are easily written down,
reading e.g.\ $\mat{(\bar{u}_3\ms \Gamma_{a_2}\ms d_2)\, (\bar{d}_4\ms
  \Gamma_{a_1}\ms u_1)}$ for the lower part of
figure~\ref{fig:interference}a.  

Which interference distributions are of appreciable size is interesting
from the point of view of nucleon structure and important for
phenomenology.  One may for instance imagine that diquark-like
correlations in the nucleon wave function play an important role in this
context.  In section~\ref{sec:small-x} we will argue that for small values
of $x_i$ both fermion number and quark flavor interference distributions
should become relatively small.


\subsubsection{Gluons}
\label{sec:spin:gluons}

If gluons enter a hard-scattering subprocess, special attention needs to
be paid to their polarization.  In covariant gauges such as Feynman gauge,
an unlimited number of right-moving gluons with polarization in the plus
direction can be attached to a hard scattering graph without any power
suppression.  The effect of these gluons is resummed into Wilson lines,
which we will discuss in detail in section~\ref{sec:coll}.  Alternatively,
one may work in light-cone gauge $A^+ = 0$, where the corresponding gluon
polarization is absent.  One then has to be careful about subtle effects
from Wilson lines at infinity, see our remark at the end of
section~\ref{sec:coll}.

Once the right-moving gluons with plus polarization (and the left-moving
gluons with minus polarization) are taken into account, the leading
contribution to the cross section comes from gluons with transverse
polarization, corresponding to field operators $A^j$ with $j=1,2$.  It is
for these gluons that one introduces parton distributions similar to those
of quarks.  To decompose the product of two gluon potentials with
transverse polarization, we use the relations
\begin{align}
  \label{transverse-tensors}
\epsilon^{jj'} \epsilon^{kk'} &=
   \delta^{jk} \delta^{j'k'} - \delta^{jk'} \delta^{j'k} \,,
\nonumber \\
\tau^{jj',kk'} &= \half \bigl(
   \delta^{jk} \delta^{j'k'} + \delta^{jk'} \delta^{j'k}
   - \delta^{jj'} \delta^{kk'} \bigr) \,,
\end{align}
where the indices $j,j',k,k' = 1,2$ are restricted to be transverse and
where $\epsilon^{jj'}$ is the two-dimensional antisymmetric tensor with
$\epsilon^{12}=1$.  The tensor $\tau^{jj',kk'}$ is symmetric and traceless
in each of the index pairs $(jj')$ and $(kk')$.  As an analog of the
decomposition \eqref{Fierz} for fermions, we can thus write
\begin{align}
  \label{gluon-Fierz}
H_i^{jj'} &= \bigl( \half \delta^{jj'} \delta^{kk'}
  + \half \epsilon^{jj'} \epsilon^{kk'} + \tau^{jj',kk'} \bigr)\,
  H_i^{kk'}
\nonumber \\[0.2em]
&= \delta^{jj'} \bigl( \half \delta^{kk'} H_i^{kk'} \bigr)
 - i\epsilon^{jj'} \bigl( \half i\epsilon^{kk'} H_i^{kk'} \bigr)
 + \tau^{jj',l\ms l'} \bigl( \tau^{l\ms l', kk'} H_i^{kk'} \bigr)
\end{align}
for the squared hard-scattering matrix element, where in the last step we
have used the relation $\tau^{jj',l\ms l'} \tau^{l\ms l', kk'} =
\tau^{jj',k' k}$.

The tensors depending on $j,j'$ in \eqref{gluon-Fierz} are to be
contracted with a product $A^{j'} A^{j}$ of gluon potentials in the
multigluon correlation function
\begin{align}
  \label{Phi-gluons}
& \Phi_{}^{j_1^{} j'_1 \ldots\ms j_n^{} j'_n}(k_i, r_i) =
  \biggl[\, \prod_{i=1}^{n}   \int \frac{d^4 z_i}{(2\pi)^4}\,
             e^{i z_i k_i} \biggr]
  \biggl[\, \prod_{i=1}^{n-1} \int \frac{d^4 y_i}{(2\pi)^4}\,
             e^{- i y_i r_i} \biggr]
\nonumber \\
 & \qquad \times
    \big\langle p \big| 
    \bar{T} \biggl[\, A^{j'_n} \bigl(-\half z_n\bigr) \prod_{i=1}^{n-1}
             A^{j'_i}\bigl(y_i - \half z_i\bigr) \biggr] \,
    T \biggl[\, A^{j_n^{}} \bigl(\half z_n\bigr) \prod_{i=1}^{n-1}
                A^{j_i^{}} \bigl(y_i + \half z_i\bigr) \biggr]
    \big| p \big\rangle \,.
\end{align}
In analogy to the definition \eqref{F-def} for scalar partons (and in
contrast to the one for quarks) we include a factor $k_i^+$ for each gluon
$i$ when defining multi-gluon distributions $F$ from $\Phi$.  One then
obtains
\begin{align}
  \label{gluon-mixed-F}
F_{a_1, \ldots, a_n}(x_i, \tvec{k}_i, \tvec{y}{}_i)
 &= \biggl[\, \prod_{i=1}^n\,
       \frac{1}{x_i\ms p^+}
       \int \frac{dz_i^-}{2\pi}\, e^{i x_i^{} z_i^- p^+} 
       \int \frac{d^2\tvec{z}_i}{(2\pi)^2}\, e^{-i \tvec{z}_i \tvec{k}_i}
    \biggr]\, \biggl[\, \prod_{i=1}^{n-1}\, 2p^+ \!\! \int dy_i^- \biggr]
\nonumber \\
 & \quad \times
    \big\langle p \big|\, \mathcal{O}_{a_n}(0, z_n)
    \prod_{i=1}^{n-1} \mathcal{O}_{a_i}(y_i, z_i) \big| p \big\rangle \,,
\end{align}
where
\begin{align}
  \label{gluon-bilinears}
\mathcal{O}_{a}^{}(y_i, z_i) &=
\Pi_{a}^{jj'}
  G^{+ j'}(y_i - \half z_i)\, G^{+ j}(y_i + \half z_i)
\end{align}
with polarization labels $a = g, \Delta g, \delta g$ and
\begin{align}
\Pi_{g}^{jj'} &= \delta^{jj'} \, , &
\Pi_{\Delta g}^{jj'} &= i\epsilon^{jj'} \, , &
\bigl[ \Pi_{\delta g}^{\ms l\ms l'} \bigr]{}^{jj'} &= \tau^{jj',l\ms l'} \,.
\end{align}
The operators $\mathcal{O}_{g}$ and $\mathcal{O}_{\Delta g}$ appear in the
usual densities for unpolarized and longitudinally polarized gluons.  By
contrast, $\mathcal{O}_{\delta g}^{\ms l\ms l'}$ describes the
interference of two gluons whose helicities differ by two units, or
equivalently the difference between linear gluon polarization in two
orthogonal directions.  Such distributions have previously been discussed
in different contexts, see \cite{Jaffe:1989xy,Belitsky:2000jk} and
\cite{Mulders:2000sh,Nadolsky:2007ba,Catani:2010pd,%
  Boer:2010zf,Qiu:2011ai,Boer:2011kf}.

In going from \eqref{Phi-gluons} to \eqref{gluon-mixed-F} we have traded
gluon potentials for field strengths using the relation $G^{+j} =
\partial^+ A^j$ valid in the light-cone gauge $A^+=0$.  Under the Fourier
transform this turns $k_i^+ A^{j'} A^{j}$ into $(k_i^+)^{-1}\, G^{+ j'}
G^{+ j}$ and explains the factor $1/(x_i\ms p^+)$ for each parton in
\eqref{gluon-mixed-F}.  It is plausible that gluon field strengths rather
than potentials should appear in the definition of gluon distributions,
since $G^{\mu\nu}$ has a simple behavior under gauge transformations and
can be used to construct gauge invariant operators.  How a definition with
$G^{+ j}$ emerges in Feynman gauge is rather involved and has been shown
explicitly for the case of a single hard scattering in
\cite{Collins:2008sg}.

Parton-level cross sections for gluons are defined as in
\eqref{quark-X-sect} with spin projectors
\begin{align}
  \label{gluon-projectors}
P_{g}^{kk'}        &= \half \delta^{kk'} \, , &
P_{\Delta g}^{kk'} &= - \half i\epsilon^{kk'} \, , &
\bigl[ P_{\delta g}^{l\ms l'} \bigr]{}^{kk'} &= \tau^{l\ms l',kk'}
\end{align}
following from \eqref{gluon-Fierz}.  In $P_{g}$ one readily recognizes the
average over the two transverse gluon polarization.  The expressions
\eqref{gluon-bilinears} to \eqref{gluon-projectors} are for right-moving
gluons.  For left-moving gluons, one has to change $+$ into $-$
coordinates in \eqref{gluon-bilinears} and reverse the sign of the
$\epsilon$ tensor in $\Pi_{\Delta g}$ and $P_{\Delta g}$.  This is because
in a covariant decomposition of the matrix elements the two-dimensional
$\epsilon$ tensor arises from the four-dimensional one as $\epsilon^{jj'}
= \epsilon^{+-jj'}$ and thus changes sign when $+$ and $-$ coordinates are
interchanged.

Using our shorthand notation \eqref{shorthand} we can write two-gluon
distributions as
\begin{align}
F_{a_1, a_2}(x_i, \tvec{k}_i, \tvec{y})
&= (x_1\ms p^+)^{-1} (x_2\ms p^+)^{-1}\,
  \mat{(\Pi_{a_2}^{kk'} G^{+k'}_3\ms G^{+k}_2)\,
       (\Pi_{a_1}^{jj'} G^{+j'}_4\ms G^{+j}_1)} \,.
\end{align}
Of course, there are also multiparton distributions involving both quarks
and gluons.  When discussing the mixing of two-quark and two-gluon
distributions in section~\ref{sec:ladders-color} we shall need quark-gluon
distributions of the type
\begin{align}
  \label{qg-mixed-F}
F_{a_1, a_2}(x_i, \tvec{k}_i, \tvec{y})
&= (x_1\ms p^+)^{-1}\,
  \mat{(\bar{q}_3\ms \Gamma_{a_2}\ms q_2)\,
       (\Pi_{a_1}^{jj'} G^{+j'}_4\ms G^{+j}_1)}
\end{align}
with $a_1 = g, \Delta g$ and $a_2 = q, \Delta q$.


\subsection{Color}
\label{sec:color}

In contrast to single-parton densities, where two parton fields are always
coupled to a color singlet, multiparton distributions have a nontrivial
color structure.  We limit ourselves to two-parton distributions here,
i.e.\ to correlation functions with four parton fields.  In this section
we give general decompositions of their color structure.  Dynamical
aspects where color plays an essential role will be encountered throughout
section~\ref{sec:factorization}, as well as in
sections~\ref{sec:ladders-color} and \ref{sec:splitting-dist}.

\subsubsection{Quarks}
\label{sec:color:quarks}

For two-quark distributions we write
\begin{align}
  \label{quark-color-decomp}
F_{jj', kk'} &=
  \mat{( \bar{q}_{3,k'}\ms \Gamma_{a_2}\ms q_{2,k} )\,
       ( \bar{q}_{4,j'}\ms \Gamma_{a_1}\ms q_{1,j} )}
 = \frac{1}{N^2} \biggl[ \sing{F}\, \delta_{jj'} \delta_{kk'}
     + \frac{2N}{\sqrt{N^2-1}}\, \oct{F}\, t^a_{jj'} t^a_{kk'}
   \biggr] \,,
\end{align}
where $j,j'$ and $k,k'$ are color indices and $N$ is the number of colors.
The indices $1$ and $2$ on the quark fields are shorthand for the position
space arguments associated with momentum fractions $x_1$ and $x_2$, as
given in \eqref{quark-mixed-F}.  For ease of writing we do not display the
polarization labels $a_1, a_2$ of $F$ when not necessary.
The functions $\sing{F}$ and $\oct{F}$ can be projected out as
\begin{alignat}{3}
  \label{quark-color-project}
\sing{F} &= \delta_{j'j}\, \delta_{k'k}\, F_{jj', kk'}
 &&= \mat{(\bar{q}_3\ms \Gamma_{a_2}\ms q_2)\,
          (\bar{q}_4\ms \Gamma_{a_1}\ms q_1)} \,,
\nonumber \\
\oct{F} &= \frac{2N}{\sqrt{N^2-1}}\,
             t^a_{j'j}\, t^a_{k'k}\,  F_{jj', kk'}^{}
 &&= \frac{2N}{\sqrt{N^2-1}}\,
     \mat{(\bar{q}_3\ms \Gamma_{a_2}\ms t^a q_2)\,
          (\bar{q}_4\ms \Gamma_{a_1} t^a q_1)} \,.
\end{alignat}
We see that for $N=3$ the quark lines carrying the same longitudinal
momentum are coupled to color singlets and color octets in $\sing{F}$ and
$\oct{F}$, respectively.\footnote{For convenience we use the notation
  $\oct{F}$ for general $N$.}
Obviously, only $\sing{F}$ admits an interpretation as the joint density
of quarks with momentum fractions $x_1$ and $x_2$, summed over their
respective colors.  The prefactor of $\oct{F}$ in
\eqref{quark-color-decomp} has been chosen such that it also appears in
the projection \eqref{quark-color-project}.  For this choice the color
singlet and color octet distributions enter with equal weight
\begin{align}
\bigl( \sing{F}\ms \sing{F}
     + \oct{F}\ms  \oct{F} \,\bigr) \big/ N^2
\end{align}
in the cross section of processes where hard scatters produce
color-singlet systems.  In this sense, the size of $\oct{F}$ relative to
$\sing{F}$ directly indicates its relevance to phenomenology.

For parameterizing the color structure of $F_{jj', kk'}$ one can
alternatively use $\sing{F}$ and the matrix element
\begin{align}
  \label{skewed-color}
\delta_{j'k}\, \delta_{k'j}\, F_{jj', kk'} &=
  \mat{(\bar{q}_{3,j}\ms \Gamma_{a_2}\ms q_{2,k})\,
       (\bar{q}_{4,k}\ms \Gamma_{a_1}\ms q_{1,j})}
 = \frac{\sqrt{N^2-1}}{N}\, \oct{F} + \frac{1}{N}\, \sing{F} \,,
\end{align}
in which quark lines carrying different longitudinal momentum couple to
color singlets.  We note that this combination becomes equal to $\oct{F}$
in the limit of large $N$.  It can be rewritten in terms of matrix
elements
\begin{align}
  \label{skewed-mom}
\sing{\tilde{F}} &=
\mat{(\bar{q}_{4}\ms \Gamma_{a_2}\ms q_{2})\,
     (\bar{q}_{3}\ms \Gamma_{a_1}\ms q_{1})}
\end{align}
that involve bilinear quark operators with no uncontracted color or spinor
indices.  This is achieved by a Fierz transform of $\Gamma_{a_2}\,
\Gamma_{a_1}$ w.r.t.\ the spinor indices of $\bar{q}_{3,j}$ and $q_{1,j}$,
followed by a Fierz transform w.r.t.\ the other two indices.  Writing
\begin{align}
\mathcal{O}_{a_1, a_2} &=
  (\bar{q}_{3,j}\ms \Gamma_{a_2}\ms q_{2,k})\,
  (\bar{q}_{4,k}\ms \Gamma_{a_1}\ms q_{1,j}) \,,
&
\tilde{\mathcal{O}}_{a_1, a_2} &=
  (\bar{q}_{4,k}\ms \Gamma_{a_2}\ms q_{2,k})\,
  (\bar{q}_{3,j}\ms \Gamma_{a_1}\ms q_{1,j})
\end{align}
one has
\begin{align}
  \label{spin-rearrange-1}
\renewcommand{\arraystretch}{1.1}
\begin{pmatrix}
  \tilde{\mathcal{O}}_{q,q} + \tilde{\mathcal{O}}_{\Delta q,\Delta q} \\
  \tilde{\mathcal{O}}_{q,q} - \tilde{\mathcal{O}}_{\Delta q,\Delta q} \\
  \tilde{\mathcal{O}}_{q,\Delta q} + \tilde{\mathcal{O}}_{\Delta q,q} \\
  \tilde{\mathcal{O}}_{q,\Delta q} - \tilde{\mathcal{O}}_{\Delta q,q} \\
  \tilde{\mathcal{O}}_{\delta q,\delta q}^{jj'}
\end{pmatrix}
 &= {}-
\renewcommand{\arraystretch}{1.1}
\begin{pmatrix}
 1 & 0 & 0 & 0 & 0 \\
 0 & 0 & 0 & 0 & \delta^{kk'} \\
 0 & 0 & 1 & 0 & 0 \\
 0 & 0 & 0 & 0 & i\epsilon^{kk'} \\
 0 & \half \delta^{jj'} & 0 & -\half i\epsilon^{jj'} & \tau^{jj',kk'} \\
\end{pmatrix}
\begin{pmatrix}
  \mathcal{O}_{q,q} + \mathcal{O}_{\Delta q,\Delta q} \\
  \mathcal{O}_{q,q} - \mathcal{O}_{\Delta q,\Delta q} \\
  \mathcal{O}_{q,\Delta q} + \mathcal{O}_{\Delta q,q} \\
  \mathcal{O}_{q,\Delta q} - \mathcal{O}_{\Delta q,q} \\
  \mathcal{O}_{\delta q,\delta q}^{kk'}
\end{pmatrix}
\intertext{and}
  \label{spin-rearrange-2}
\renewcommand{\arraystretch}{1.1}
\begin{pmatrix}
  \tilde{\mathcal{O}}_{q,\delta q}^j
  + \tilde{\mathcal{O}}_{\delta q,q}^j \\
  \tilde{\mathcal{O}}_{q,\delta q}^j
  - \tilde{\mathcal{O}}_{\delta q,q}^j \\
  \tilde{\mathcal{O}}_{\Delta q,\delta q}^j
  + \tilde{\mathcal{O}}_{\delta q,\Delta q}^j \\
  \tilde{\mathcal{O}}_{\Delta q,\delta q}^j
  - \tilde{\mathcal{O}}_{\delta q,\Delta q}^j
\end{pmatrix}
 &= {}-
\renewcommand{\arraystretch}{1.1}
\begin{pmatrix}
  -\delta^{jk} & 0 & 0 & 0 \\
  0 & 0 & 0 & i\epsilon^{jk} \\
  0 & 0 & -\delta^{jk} & 0 \\
  0 & i\epsilon^{jk} & 0 & 0
\end{pmatrix}
\begin{pmatrix}
  \mathcal{O}_{q,\delta q}^k + \mathcal{O}_{\delta q,q}^k \\
  \mathcal{O}_{q,\delta q}^k - \mathcal{O}_{\delta q,q}^k \\
  \mathcal{O}_{\Delta q,\delta q}^k + \mathcal{O}_{\delta q,\Delta q}^k \\
  \mathcal{O}_{\Delta q,\delta q}^k - \mathcal{O}_{\delta q,\Delta q}^k
\end{pmatrix} \,,
\end{align}
where $\tau$ is defined in \eqref{transverse-tensors} and the transverse
indices $k,k'$ of the tensor operators on the r.h.s.\ are summed over as
appropriate.  The global minus sign in both equations comes from the
reordering of fermion fields between $\mathcal{O}$ and
$\tilde{\mathcal{O}}$.  The inverse transformation goes with the same
matrices.  Of course, the distributions $\sing{\tilde{F}}$ do not have a
probability interpretation since the quark fields coupled to color
singlets carry different momentum fractions.

To illustrate that the color octet combination $\oct{F}$ need not be small
let us consider a three-quark system, as is done in constituent quark
models.  Irrespective of the details in the model, the color part of the
three-quark wave function is $\epsilon_{jkl}$.  The color structure of a
two-quark distribution is thus given by
\begin{align}
F_{jj',kk'} &\;\propto\; 
   \epsilon_{jkl}\, \epsilon_{j'k'l}
 = \delta_{jj'}\, \delta_{kk'} - \delta_{jk'}\, \delta_{kj'} \,,
\end{align}
where $l$ is the color index of the spectator quark and therefore summed
over.  With \eqref{quark-color-project} one readily finds $\oct{F} = -
\sqrt{2}\; (\sing{F})$.  The combination in \eqref{skewed-color}, where
the quark lines $\{ 13 \}$ and $\{ 24 \}$ are coupled to color singlets is
then $\frac{1}{3} \bigl( \sqrt{8}\; \oct{F} + \sing{F} \bigr) = - \ms
(\sing{F})$ and thus as large as $\sing{F}$ itself.

The preceding expressions can easily be adapted for the quark-antiquark
distributions $F_{a_1, \bar{a}_2}$ defined in \eqref{qqbar-mixed-F}.  With
color indices labeled as in figure \ref{fig:distrib}, the corresponding
matrix element reads $\mat{( \bar{q}_{2,k}\ms \Gamma_{\bar{a}_2}\ms
  q_{3,k'} )\, ( \bar{q}_{4,j'}\ms \Gamma_{a_1}\ms q_{1,j} )}$ and is
decomposed as on the r.h.s.\ of \eqref{quark-color-decomp} with
interchanged indices $k$ and $k'$.  An extra minus sign appears in the
transformation laws \eqref{spin-rearrange-1} and \eqref{spin-rearrange-2}
whenever a label $\Delta q$ is changed to $\Delta\bar{q}$, because
$\Gamma_{\Delta \bar{q}} = -\Gamma_{\Delta q}$.

An analogous color decomposition can finally be made for the interference
distributions $I_{a_1, \bar{a}_2}$ defined in
\eqref{fermion-interference},
\begin{align}
I_{jj', kk'} &=
  \label{interf-color-decomp}
  \mat{( \bar{q}_{2,k}\ms \Gamma_{\bar{a}_2}\ms q_{4,j'} )\,
       ( \bar{q}_{3,k'}\ms \Gamma_{a_1}\ms q_{1,j} )}
 = \frac{1}{N^2} \Bigl[ \sing{I}\, \delta_{jk'} \delta_{j'k}
     + \frac{2N}{\sqrt{N^2-1}}\ms \oct{I}\, t^a_{jk'} t^a_{j'k}
   \Bigr]
\end{align}
with
\begin{alignat}{3}
\sing{I} &= \delta_{k'j}\, \delta_{kj'}\, I_{jj', kk'}
 &&= \mat{(\bar{q}_2\ms \Gamma_{\bar{a}_2}\ms q_4)\,
          (\bar{q}_3\ms \Gamma_{a_1}\ms q_1)} \,,
\nonumber \\
\oct{I} &= \frac{2N}{\sqrt{N^2-1}}\,
           t^a_{k'j}\, t^a_{kj'}\, I_{jj', kk'}^{}
 &&= \frac{2N}{\sqrt{N^2-1}}\,
     \mat{(\bar{q}_2\ms \Gamma_{\bar{a}_2}\ms t^a q_4)\,
          (\bar{q}_3\ms \Gamma_{a_1} t^a q_1)} \,.
\end{alignat}
In analogy to \eqref{skewed-color} one can alternatively use $\sing{I}$
together with
\begin{align}
\delta_{k'j'}\, \delta_{kj}\, I_{jj', kk'} &=
  \mat{(\bar{q}_{2,j}\ms \Gamma_{\bar{a}_2}\ms q_{4,j'})\,
       (\bar{q}_{3,j'}\ms \Gamma_{a_1}\ms q_{1,j})}
 = \frac{\sqrt{N^2-1}}{N}\, \oct{I} + \frac{1}{N}\ms \sing{I} \,.
\end{align}
By the same transformation as in \eqref{spin-rearrange-1} and
\eqref{spin-rearrange-2}, with appropriate sign changes for the antiquark
matrices $\Gamma_{\bar{a}_2}$, one can rewrite this as a linear
combination of matrix elements
\begin{align}
\sing{\tilde{I}}
&= \mat{(\bar{q}_3\ms \Gamma_{\bar{a}_2}\ms q_4)\,
        (\bar{q}_2\ms \Gamma_{a_1}\ms q_1)} \,,
\end{align}
where the quark bilinears have no uncontracted spin or color indices.
Using the relation $t^a_{jk'} t^a_{j'k} = \half \delta_{jk}\ms
\delta_{k'j'} - \frac{1}{2 N}\ms \delta_{jk'}\ms \delta_{j'k}$ one can
also rewrite \eqref{interf-color-decomp} as
\begin{align}
  \label{interf-color-decomp-alt}
I_{jj', kk'} &=
\frac{1}{N^2} \Biggl[ 
  \sqrt{\frac{N}{2 (N-1)}}\, {}^{\bar{3}\!}I
    \bigl( \delta_{jk'}\ms \delta_{j'k}
         - \delta_{jk}\ms \delta_{j'k'} \bigr)
+ \sqrt{\frac{N}{2 (N+1)}}\, {}^{6\!}I
    \bigl( \delta_{jk'}\ms \delta_{j'k}
         + \delta_{jk}\ms \delta_{j'k'} \bigr)
\Biggr]
\end{align}
with
\begin{align}
  \label{sextet-def}
{}^{\bar{3}\!}I &= 
  \sqrt{\frac{N-1}{2N}}\, \sing{I} - \sqrt{\frac{N+1}{2N}}\, \oct{I} \,,
&
{}^{6\!}I &= 
  \sqrt{\frac{N+1}{2N}}\, \sing{I} + \sqrt{\frac{N-1}{2N}}\, \oct{I} \,.
\end{align}
The transformation between $\sing{I}, \oct{I}$ and ${}^{\bar{3}\!}I,
{}^{6\!}I$ is orthogonal.  For $N=3$ we can rewrite $\delta_{jk'}\ms
\delta_{j'k} - \delta_{jk}\ms \delta_{j'k'} = \epsilon_{jj'l}\,
\epsilon_{k'kl}$ and recognize that ${}^{\bar{3}\!}I$ describes the case
where the quarks with momentum fraction $x_1$ are coupled to a color
antitriplet, whereas ${}^{6\!}I$ describes the case where they form a
sextet.


\subsubsection{Gluons}
\label{sec:color:gluon}

The color structure for multi-gluon distributions requires the coupling of
several color octets and is hence more involved than for quarks.  For a
two-gluon distribution we proceed by first coupling each of the gluon
pairs $\{ 14 \}$ and $\{ 23 \}$ to irreducible representations and then
coupling these pairs to an overall color singlet.  For the color
structures that can mix with quarks we write
\begin{align}
  \label{gluon-color-decomp}
F^{aa',bb'} &= 
(x_1\ms p^+)^{-1} (x_2\ms p^+)^{-1}\,
  \mat{(G^{b'}_3\ms \Pi_{a_2}\ms G^{b}_2)\,
       (G^{a'}_4\ms \Pi_{a_1}\ms G^{a}_1)}
\phantom{\frac{1}{1}}
\nonumber \\
&= \frac{1}{(N^2-1)^2}\, \biggl[ \sing{F}\, \delta^{aa'} \delta^{bb'}
   - \frac{\sqrt{N^2-1}}{N}\, \octA{F}\, f^{aa'\!c}\ms f^{bb'\!c}
  + \frac{N \sqrt{N^2-1}}{N^2-4}\,
     \octS{F}\, d^{\ms aa'\!c}\ms d^{\ms bb'\!c}
   + \cdots \biggr]
\nonumber \\
\end{align}
with a shorthand notation $G^{a'} \Pi_{a_i} G^{a} = \Pi_{a_i}^{jj'} G^{a',
  +j'}\ms G^{a, +j}$ for the contractions of gluon polarization indices.
As is readily seen from
\begin{align}
(x_1\ms p^+) (x_2\ms p^+)\,
\sing{F} &= \mat{(G^{b}_3\ms \Pi_{a_2}\ms G^{b}_2)\,
                 (G^{a}_4\ms \Pi_{a_1}\ms G^{a}_1)} \,,
            \phantom{\frac{\sqrt{1}}{1}}
\nonumber \\
(x_1\ms p^+) (x_2\ms p^+)\,
\octA{F} &= {}- \frac{\sqrt{N^2-1}}{N}\
            \mat{(f^{cbb'} G^{b'}_3\ms \Pi_{a_2}\ms G^{b}_2)\,
                 (f^{caa'} G^{a'}_4\ms \Pi_{a_1}\ms G^{a}_1)} \,,
\nonumber \\
(x_1\ms p^+) (x_2\ms p^+)\,
\octS{F} &= \frac{N \sqrt{N^2-1}}{N^2-4}\, 
            \mat{(d^{cbb'} G^{b'}_3\ms \Pi_{a_2}\ms G^{b}_2)\,
                 (d^{caa'} G^{a'}_4\ms \Pi_{a_1}\ms G^{a}_1)} \,,
\end{align}
each of the pairs $\{ 14 \}$ and $\{ 23 \}$ in $\sing{F}$, $\octA{F}$ and
$\octS{F}$ is respectively coupled to a singlet, an antisymmetric and a
symmetric octet.  For hard-scattering processes producing color singlet
states, these distributions enter the cross section as
\begin{align}
  \label{gluon-color-X}
\bigl[\, \sing{F}\, \sing{F}
  + \octA{F}\,  \octA{F}
  + \octS{F}\,  \octS{F} + \cdots \,\bigr] \big/ (N^2-1)^2 \,.
\end{align}
The ellipsis in \eqref{gluon-color-decomp} and \eqref{gluon-color-X}
stands for terms where the gluon pairs are in higher color
representations.  For $SU(3)$ these are $10$, $\overline{10}$, $27$, and
the full decomposition reads
\begin{align}
  \label{gluon-decomp-8}
F^{aa',bb'} & \underset{N=3}{=}\;
   \frac{1}{64}\, \Bigl[ \sing{F}\, \delta^{aa'} \delta^{bb'}
   - \frac{\sqrt{8}}{3}\, \octA{F}\, f^{aa'\!c}\ms f^{bb'\!c}
   + \frac{3 \sqrt{8}}{5}\; \octS{F}\, d^{\ms aa'\!c}\ms d^{\ms bb'\!c}
\nonumber \\
& \qquad\qquad
   + \frac{2}{\sqrt{10}}\, {}^{10\!}F\; t_{10}^{aa',bb'}
   + \frac{2}{\sqrt{10}}\, {}^{\overline{10}\!}F\;
            \bigl( t_{10}^{aa',bb'} \bigr)^*
   + \frac{4}{\sqrt{27}}\, {}^{27\!}F\; t_{27}^{aa',bb'}
\Bigr]
\end{align}
with tensors \cite{Mekhfi:1985dv}
\begin{align}
  \label{gluon-higher-reps}
t_{10}^{aa',bb'} &= \delta^{ab} \delta^{a'b'} - \delta^{ab'} \delta^{a'b}
  - \tfrac{2}{3}\ms f^{aa'\!c} f^{bb'\!c}
  - i ( d^{abc} f^{a'b'c} + f^{abc} d^{a'b'c} ) \,,
\nonumber \\
t_{27}^{aa',bb'} &= \delta^{ab} \delta^{a'b'} + \delta^{ab'} \delta^{a'b}
  - \tfrac{1}{4}\ms \delta^{aa'} \delta^{bb'}
  - \tfrac{6}{5}\ms d^{\ms aa'\!c} d^{\ms bb'\!c} \,.
\end{align}
In ${}^{10\!}F$ the indices $(aa')$ are coupled to $10$ and $(bb')$ to
$\overline{10}$, whereas in ${}^{\overline{10}\!}F$ the opposite is the
case.  The normalization factors in \eqref{gluon-decomp-8} are such that
the production of color singlet particles involves the combination $\bigl[
\sing{F}\, \sing{F} + \octA{F}\, \octA{F} + \octS{F}\, \octS{F} +
{}^{10\!}F\, {}^{\overline{10}\!}F + {}^{\overline{10}\!}F\, {}^{10\!}F +
{}^{27\!}F\, {}^{27\!}F \bigl] \big/ 64$.  Useful relations between the
$f$ and $d$ tensors can be found in \cite{MacFarlane:1968vc}.

We conclude this section with the color decomposition of the quark-gluon
distributions introduced in \eqref{qg-mixed-F}.  The quark lines can only
couple to a color singlet or octet, which has to be matched by the gluon
lines in order to obtain an overall singlet.  A complete decomposition is
thus given by
\begin{align}
  \label{qg-color-decomp}
F_{jj'}^{aa'} &= (x_1\ms p^+)^{-1}\,
  \mat{(\bar{q}_{3,j'}\ms \Gamma_{a_2}\ms q_{2,j})\,
       (G^{a'}_4\ms \Pi_{a_1}\ms G^{a}_1)}
\nonumber \\
&= \frac{1}{N (N^2-1)}\, \Bigl[ \sing{F}\, \delta^{aa'}\, \delta_{jj'}
   - \octA{F}\, \sqrt{2}\, i f^{\ms aa'\!c}\, t^{c}_{jj'}
   + \sqrt{\frac{2N^2}{N^2-4}}\, \octS{F}\, d^{\ms aa'\!c}\, t^{c}_{jj'}
  \Bigr]
\end{align}
with
\begin{align}
(x_1\ms p^+)\, \sing{F} &=
   \mat{(\bar{q}_{3}\ms \Gamma_{a_2}\ms q_{2})\,
        (G^{a}_4\ms \Pi_{a_1}\ms G^{a}_1)} \,,
   \phantom{\frac{\sqrt{1}}{1}}
\nonumber \\
(x_1\ms p^+)\, \octA{F} &= \sqrt{2}\;
   \mat{(\bar{q}_{3}\ms \Gamma_{a_2}\ms t^c q_{2})\,
        (i f^{caa'} G^{a'}_4\ms \Pi_{a_1}\ms G^{a}_1)} \,,
   \phantom{\frac{\sqrt{1}}{1}}
\nonumber \\
(x_1\ms p^+)\, \octS{F} &= \sqrt{\frac{2N^2}{N^2-4}}\;
   \mat{(\bar{q}_{3}\ms \Gamma_{a_2}\ms t^c q_{2})\,
        (d^{caa'} G^{a'}_4\ms \Pi_{a_1}\ms G^{a}_1)} \,.
\end{align}
The factor $i$ in \eqref{qg-color-decomp} has been chosen so that
$\octA{F}$ is real valued (since $i f^{\ms aa'\!c}$ is Hermitian w.r.t.\
the indices $a$ and $a'$).  The normalization factors multiplying
$\octA{F}$ and $\octS{F}$ are the geometric means of their counterparts in
\eqref{quark-color-decomp} and \eqref{gluon-color-decomp}.


\subsection{Power counting and dominant graphs}
\label{sec:power-counting}

In section~\ref{sec:sing-vs-mult} we have already compared the power
behavior in $\Lambda /Q$ of single and multiple hard scattering cross
sections.  We now take a closer look at this issue and extend our analysis
to the interference of single and multiple scattering.

\begin{figure}
\begin{center}
\begin{tabular}{c@{\qquad}cc@{\quad}c}
 & & $\displaystyle\frac{s\ms d\sigma}{\prod_{i=1}^2 dx_i\,
                                  d\bar{x}_i\, d^2\tvec{q}{}_i}$ &
     $\displaystyle\frac{s\ms d\sigma}{\prod_{i=1}^2 dx_i\, d\bar{x}_i}$
\\[1.6em]
\textbf{a} & \parbox[c]{0.37\textwidth}{%
  \includegraphics[width=0.37\textwidth]{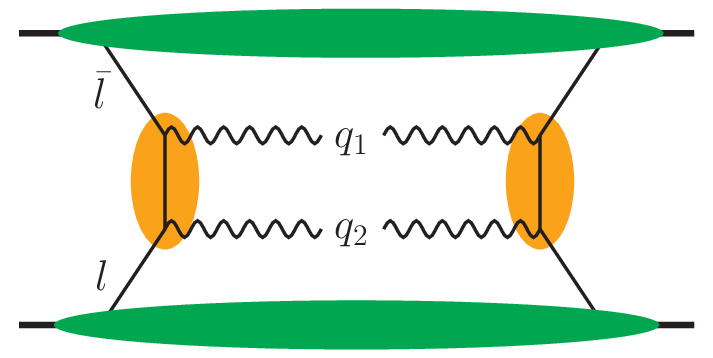}} &
  $\displaystyle\frac{1}{\Lambda^2\ms Q^2}$ &
  $1$ \\[5.0em]
\textbf{b} & \parbox[c]{0.37\textwidth}{%
  \includegraphics[width=0.37\textwidth]{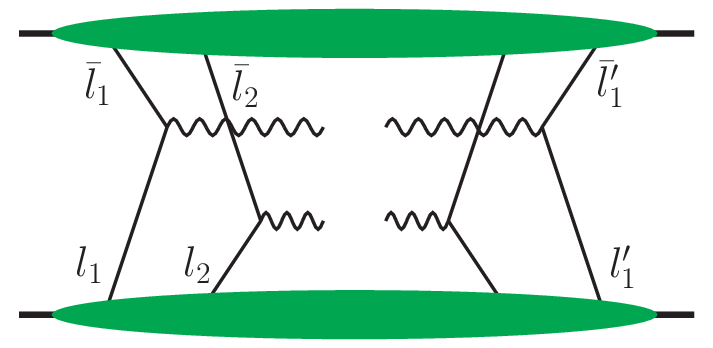}} &
  $\displaystyle\frac{1}{\Lambda^2\ms Q^2}$ &
  $\displaystyle\frac{\Lambda^2}{Q^2}$ \\[3.5em]
\textbf{c} & \parbox[c]{0.37\textwidth}{%
  \includegraphics[width=0.37\textwidth]{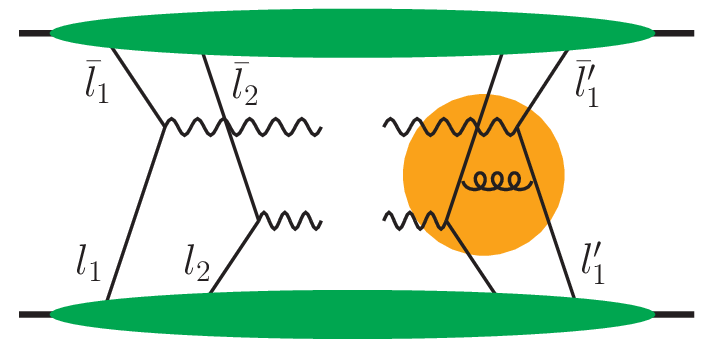}} &
  $\displaystyle\frac{1}{Q^4}$ &
  $\displaystyle\frac{\Lambda^4}{Q^4}$ \\[3.5em]
\textbf{d} & \parbox[c]{0.37\textwidth}{%
  \includegraphics[width=0.37\textwidth]{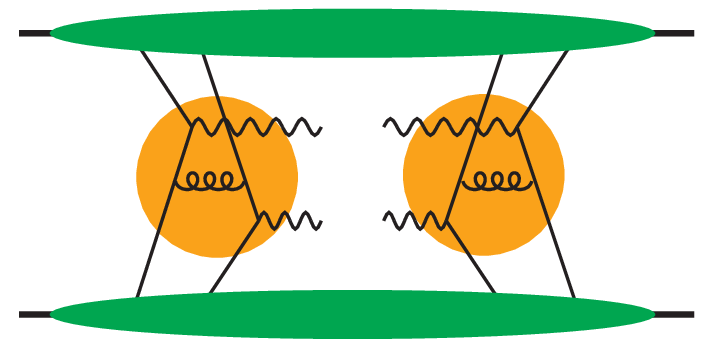}} &
  $\displaystyle\frac{\Lambda^2}{Q^6}$ &
  $\displaystyle\frac{\Lambda^4}{Q^4}$
\end{tabular}
\end{center}
\caption{\label{fig:power-beh} Example graphs and power behavior for
  different combinations of single and double hard scattering
  contributions to gauge boson pair production.  It is understood that
  internal lines of the hard-scattering subgraphs are off shell by order
  $Q^2$, whereas partons emerging from the proton matrix elements are
  off shell by order $\Lambda^2$.}
\end{figure}

As building blocks for establishing the power behavior of the cross
section we take correlation functions $\Phi_{n}$ involving $n$ parton
fields and amplitudes $T_{k\to m}$ for hard-scattering processes with $k$
incoming partons and $m$ final-state particles.  The relevant parton
correlation functions are obtained by replacing the scalar parton fields
in \eqref{Phi-def} by quark fields $\bar{q}$ and $q$, or by the transverse
components $A^j$ of the gluon potential as in \eqref{Phi-gluons}.  To
treat quarks and gluons on a common footing, it is convenient to use
modified correlation functions $\Phi'_{n}$ that are divided by
$\sqrt{l^+}$ for each quark or antiquark line with momentum $l$, and
modified hard-scattering amplitudes $T'_{k\to m}$ that are multiplied with
the corresponding factor $\sqrt{l^+}$.  Furthermore, pairs $\bar{q}_\beta$
and $q_\alpha$ of quark fields in $\Phi'_{n}$ are contracted with one of
the Dirac matrices $\Gamma_{a}$ in \eqref{Gamma-twist-2} that give the
dominant contributions to the cross section.\footnote{%
  For the purpose of power counting, it is not important which of the
  matrices $\Gamma_a$ is taken and which pairs of quark fields are
  contracted together if there are more than two of them.  We will not
  specify these details in the present section and use $\Phi'_n$ in a
  generic sense.  Likewise, color indices are not relevant for power
  counting and will be omitted.}
The products $T'_{k\to m}\, T^{\prime *}_{k'\to m}$ of modified
hard-scattering amplitudes with their complex conjugates are to be
contracted with the corresponding Dirac matrices specified in
\eqref{Fierz-dominant}.  Since
\begin{align}
& \frac{1}{\sqrt{l^+ l'^+}}\, \bar{q} \gamma^+ q \,, &
& \frac{1}{\sqrt{l^+ l'^+}}\, \bar{q} \gamma^+\gamma_5\ms q \,, &
& \frac{1}{\sqrt{l^+ l'^+}}\, \bar{q} i\sigma^{+j}\gamma_5\ms q \,, &
& A^{j} A^{k}
\end{align}
have the same mass dimension and are invariant under boosts along the $z$
axis, the power behavior of the modified correlation functions is
\begin{align}
  \label{Phi-scaling}
\Phi'_{n} \sim \Lambda^{2 - 3n}
\end{align}
regardless of the parton species.  The power on the r.h.s.\ is just the
mass dimension of $\Phi'_{n}$.  By definition, all internal lines of the
hard-scattering subgraphs are off shell by order $Q^2$, so that the power
behavior of the amplitudes $T'_{k\to m}$ (where the propagators of
external particles are truncated) is also determined by their mass
dimension.  For the processes considered in the following, one has
\begin{align}
  \label{T-scaling}
T'_{k\to m} &\sim Q^{4-k-m} \,,
\end{align}
as can readily be checked for the example graphs in
figure~\ref{fig:power-beh}.

For definiteness we consider the production of two particles with large
masses and respective four-momenta $q_1, q_2$.  Examples are the weak
gauge bosons $W$, $Z$ or a Higgs boson.  The power behavior of the cross
section is the same if we replace one or both of the heavy particles by a
set of light particles such as a lepton pair or a pair of jets, provided
that we integrate over the internal phase space of the final-state
particles while keeping $q_i$ fixed.  Replacing for instance a particle
with momentum $q_i$ and mass $M_i$ by two massless particles with momenta
$p_1$ and $p_2$, we have to change
\begin{align}
  \label{one-particle}
d^4 q_i\; 2\pi \delta(q_i^2 - M_i^2) \,
    T'_{k\to 1}\, T^{\prime *}_{k'\to 1}
 &= dx_i\, d\bar{x}_i\, d^2\tvec{q}_i\, \pi\ms
    \delta\Bigl( x_i \bar{x}_i - \frac{M_i^2 + \tvec{q}_i^2}{s} \Bigr)\,
    T'_{k\to 1}\, T^{\prime *}_{k'\to 1}
\end{align}
into
\begin{align}
  \label{two-particles}
\frac{d^4 q_i}{(2\pi)^2} \int
    \frac{d^3 p_1}{2 p_1^0}\, \frac{d^3 p_2}{2 p_2^0}\,
    \delta^{(4)}( q_i - p_1 - p_2)\,
    T'_{k\to 2}\, T^{\prime *}_{k'\to 2}
 &= dx_i\, d\bar{x}_i\, d^2\tvec{q}_i\;
    s \int \frac{d\Omega_1}{(8\pi)^2}\;
    T'_{k\to 2}\, T^{\prime *}_{k'\to 2} \,,
\end{align}
where $d\Omega_1$ is the solid angle of $p_1$ in the rest frame of $q_i$.
We recall that $x_i^{} = q_i^{\smash{+}} /p^+$ and $\bar{x}_i^{} =
q_i^{\smash{-}} / \bar{p}^-$ are defined in terms of final-state momenta
and thus directly observable.  According to \eqref{T-scaling} the scaling
behavior of the squared hard-scattering amplitudes changes by $1/Q^2$ when
going from \eqref{one-particle} to \eqref{two-particles}, which is
compensated by the phase space volume $s\ms d\Omega_1 \sim Q^2$.  One may
put restrictions on the phase space integration, such as a minimum
transverse momentum of $p_1$, as long as $d\Omega_1$ remains of order $1$.
For each further final-state particle, the squared amplitude acquires an
extra $1/Q^2$, which is compensated by an extra phase space integration
with volume of order $Q^2$.

After these preliminaries we can establish the power behavior of the
conventional mechanism with a single hard scattering, shown in
figure~\ref{fig:power-beh}a.  The cross section formula can be obtained in
exactly the same way as in section~\ref{sec:tree:cross-sect}.  Omitting
all factors that are not relevant for power counting (including the
$\delta$ functions constraining $x_i\ms \bar{x}_i$ in
\eqref{one-particle}) we have
\begin{align}
  \label{pow-fig-a}
 \frac{s\ms d\sigma}{\prod_{i=1}^2 dx_i\, d\bar{x}_i\, d^2\tvec{q}{}_i}
  \bigg|_{\text{fig.~\protect\ref{fig:power-beh}a}}
& \sim\;  \int d^4 l\, d^4\bar{l}\;
  \delta^{(4)}(q_1 + q_2 - l - \bar{l})\;
  \big| T'_{2\to 2} \big|^2 \; \Phi'_{2}\; \bar{\Phi}'_{2}
\nonumber \\[0.2em]
& \approx\;
  \big| T'_{2\to 2} \big|^2 \;
  \int d^2\tvec{l}\, d^2\bar{\tvec{l}}\;
       \delta^{(2)}(\tvec{q}_1 + \tvec{q}_2 - \tvec{l} - \bar{\tvec{l}})
\nonumber \\[0.2em]
& \quad \; \times
  \int dl^-\, \Phi'_{2}   \;\Big|_{l^+ \,=\, (x_1 + x_2)\, p^+}
  \int d\bar{l}^+\, \bar{\Phi}'_{2}
  \;\Big|_{\bar{l}^- \,=\, (\bar{x}_1 + \bar{x}_2)\, \bar{p}^-} \;,
\end{align}
where for simplicity we have not displayed the momentum arguments of $T'$,
$\Phi'$ and $\bar{\Phi}'$ (which can readily be inferred from
figure~\ref{fig:power-beh}a).  It is understood that in the second step we
have made the collinear approximation and neglected $\tvec{l}$ and
$\bar{\tvec{l}}$ in the hard scattering, as well as $l^-$ compared with
$\bar{l}^-$, and $\bar{l}^+$ compared with $l^+$.  The power behavior of
the integration regions is $d^2\tvec{l} \sim \Lambda^2$ and $dl^- \sim
d\bar{l}^+ \sim \Lambda^2 /Q$, so that together with \eqref{Phi-scaling}
and \eqref{T-scaling} we obtain
\begin{align}
\frac{s\ms d\sigma}{\prod_{i=1}^2 dx_i\, d\bar{x}_i\, d^2\tvec{q}{}_i}
\bigg|_{\text{fig.~\protect\ref{fig:power-beh}a}}
 &\;\sim\;  Q^{0} \times \Lambda^2 \times
            \Bigl(\ms \frac{\Lambda^2}{Q} \times \Lambda^{-4} \Bigr)^2
  \;=\; \frac{1}{\Lambda^2\ms Q^2} \,.
\end{align}
For the double hard-scattering contribution in figure~\ref{fig:power-beh}b
one has
\begin{align}
  \label{pow-fig-b}
& \frac{s\ms d\sigma}{\prod_{i=1}^2 dx_i\, d\bar{x}_i\, d^2\tvec{q}{}_i}
  \bigg|_{\text{fig.~\protect\ref{fig:power-beh}b}}
  \;\sim\;
  \biggl[\, \prod_{i=1}^2
     \int d^4 l_i^{}\, d^4\bar{l}_i^{}\;
     \delta^{(4)}(q_i^{} - l_i^{} - \bar{l}_i^{}) \,\biggr]\,
\nonumber \\[0.2em]
& \qquad \; \times
  \int d^4 l'_1\, d^4\bar{l}'_1\; \delta^{(4)}(q_1 - l'_1 - \bar{l}'_1)\,
  T'_{2\to 1}\, T^{\prime *}_{2\to 1}\;
  T'_{2\to 1}\, T^{\prime *}_{2\to 1}\;
  \Phi'_{4}\; \bar{\Phi}'_{4}
\nonumber \\
& \quad \approx\; 
  \big| T'_{1\to 2} \big|^2\, \big| T'_{1\to 2} \big|^2\;
  \biggl[\, \prod_{i=1}^2 \int d^2\tvec{l}_i\, d^2\bar{\tvec{l}}_i\,
    \delta^{(2)}(\tvec{q}_i - \tvec{l}_i - \bar{\tvec{l}}_i) \,\biggr]\,
  \int d^2\tvec{l}'_1\, d^2\bar{\tvec{l}}'_1\,
    \delta^{(2)}(\tvec{q}_1 - \tvec{l}'_1 - \bar{\tvec{l}}'_1)\,
\nonumber \\
& \qquad \; \times
  \int dl_1^-\, dl_2^-\, dl_1^{\prime -}\, 
    \Phi'_{4} \;\Big|_{l^+_i \,=\, l_i^{\prime +} \,=\, x_i^{}\, p^+}
  \int d\bar{l}_1^+\, d\bar{l}_2^+\, d\bar{l}_1^{\prime +}\,
    \bar{\Phi}'_{4} \;\Big|_{\bar{l}^-_i \,=\,
      \bar{l}_i^{\prime -} \,=\, \bar{x}_i^{}\, \bar{p}^-} \,.
\end{align}
Note that we have used the constraint $\delta(q_1^+ - l_1^{\prime +} -
\bar{l}_1^{\prime +})$ to fix the large component $l_1^{\prime +}$ at its
value $q_1^+$ in the collinear approximation, thus leaving the integral
over the small component $\bar{l}_1^{\prime +}$.  If instead one uses the
constraint to fix $\bar{l}_1^{\prime +} = q_1^+ - l_1^{\prime +}$ one
would have to count the integration element $d\bar{l}_1^{\prime +}$ as
order $\Lambda^2 /Q$ since $\bar{l}_1^{\prime +}$ can only vary by that
amount.  An analogous remark applies to the constraint $\delta(q_1^- -
l_1^{\prime -} - \bar{l}_1^{\prime -})$.  The power behavior of
\eqref{pow-fig-b} is
\begin{align}
\frac{s\ms d\sigma}{\prod_{i=1}^2 dx_i\, d\bar{x}_i\, d^2\tvec{q}{}_i}
  \bigg|_{\text{fig.~\protect\ref{fig:power-beh}b}}
\;\sim\; Q^4 \times \Lambda^6 \times
         \Bigl(\ms \frac{\Lambda^{6}}{Q^3} \times \Lambda^{-10} \Bigr)^2
  \;=\; \frac{1}{\Lambda^2\ms Q^2}
\end{align}
and hence the same as for single hard scattering, in agreement with the
result we obtained for scalar partons in section~\ref{sec:sing-vs-mult}.

Let us now see how the power behavior changes if on one side of the
final-state cut the two quark-antiquark annihilation graphs are connected
by a hard gluon.  We then have an interference between double hard
scattering and a single hard-scattering process as shown in
figure~\ref{fig:power-beh}c,
\begin{align}
  \label{pow-fig-c}
& \frac{s\ms d\sigma}{\prod_{i=1}^2 dx_i\, d\bar{x}_i\, d^2\tvec{q}{}_i}
  \bigg|_{\text{fig.~\protect\ref{fig:power-beh}c}}
\nonumber \\[0.2em]
& \quad \sim\;
  \biggl[\, \prod_{i=1}^2
     \int d^4 l_i^{}\, d^4\bar{l}_i^{}\;
     \delta^{(4)}(q_i^{} - l_i^{} - \bar{l}_i^{}) \,\biggr]\,
  \int d^4 l'_1\, d^4\bar{l}'_1\;
  T'_{2\to 1}\, T'_{2\to 1}\, T^{\prime *}_{4\to 2}\;
  \Phi'_{4}\; \bar{\Phi}'_{4}
\nonumber \\
& \quad \approx\; 
  T'_{2\to 1}\, T'_{2\to 1}\, 
  \int dl_1^{\prime +}\, d\bar{l}_1^{\prime -}\, T^{\prime *}_{4\to 2}\;
  \biggl[\, \prod_{i=1}^2 \int d^2\tvec{l}_i\, d^2\bar{\tvec{l}}_i\,
    \delta^{(2)}(\tvec{q}{}_i - \tvec{l}_i - \bar{\tvec{l}}_i) \,\biggr]\,
\nonumber \\
& \qquad \; \times
  \int dl_1^-\, dl_2^-\, dl_1^{\prime -}\, d^2\tvec{l}_1'\, \Phi'_{4}
     \;\Big|_{l^+_i \,=\, x_i^{}\, p^+}
 \int d\bar{l}_1^+\, d\bar{l}_2^+\,
       d\bar{l}_1^{\prime +}\, d^2\bar{\tvec{l}}_1'\, \bar{\Phi}'_{4}
     \;\Big|_{\bar{l}^-_i \,=\, \bar{x}_i^{}\, \bar{p}^-}
\nonumber \\
& \quad \sim\; Q^2 \times \Lambda^4 \times
        \Bigl(\ms \frac{\Lambda^{6}}{Q^3}
              \times \Lambda^2 \times \Lambda^{-10} \Bigr)^2
  \;=\; \frac{1}{Q^4} \,.
\end{align}
This is power suppressed compared with the contributions in
figures~\ref{fig:power-beh}a and b and may therefore be neglected.  It is
instructive to see why the power counting changes between
\eqref{pow-fig-b} and \eqref{pow-fig-c}.  Compared with $T^{\prime
  *}_{2\to 1}\, T^{\prime *}_{2\to 1}$, the hard-scattering amplitude
$T^{\prime *}_{4\to 2}$ is down by a factor of $1/Q^4$, which in the
example of figure~\ref{fig:power-beh}c is due to two additional quark
propagators and one additional gluon propagator relative to
figure~\ref{fig:power-beh}b.  The additional loop integrations over the
large components $l_1^{\prime +}$ and $\bar{l}_1^{\prime -}$ in
\eqref{pow-fig-c} each scale like $Q$, but for the transverse momentum
integrations one now has $d^2\tvec{l}_1'\, d^2\bar{\tvec{l}}_1' \sim
\Lambda^4$ compared with $d^2\tvec{l}'_1\, d^2\bar{\tvec{l}}'_1\,
\delta^{(2)}(\tvec{q}_1 - \tvec{l}'_1 - \bar{\tvec{l}}'_1) \sim \Lambda^2$
before.  Altogether one has thus lost a factor of $\Lambda^2 /Q^2$.

By an analogous argument one finds that the differential cross section for
the pure single hard-scattering mechanism in figure~\ref{fig:power-beh}d
is power suppressed by a factor of $\Lambda^2 /Q^2$ compared with the one
in figure~\ref{fig:power-beh}c.

\begin{figure}
\begin{center}
\begin{tabular}{c@{\qquad}cc@{\quad}c}
 & & $\displaystyle\frac{s\ms d\sigma}{\prod_{i=1}^2 dx_i\,
                                  d\bar{x}_i\, d^2\tvec{q}{}_i}$ &
     $\displaystyle\frac{s\ms d\sigma}{\prod_{i=1}^2 dx_i\, d\bar{x}_i}$
\\[1.6em]
\textbf{a} & \parbox[c]{0.37\textwidth}{%
  \includegraphics[width=0.37\textwidth]{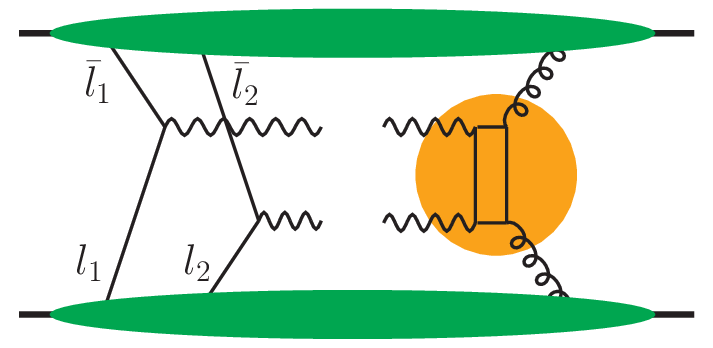}} &
  $\displaystyle\frac{1}{\Lambda^2\ms Q^2}$ &
  $\displaystyle\frac{\Lambda^2}{Q^2}$ \\[3.5em]
\textbf{b} & \parbox[c]{0.37\textwidth}{%
  \includegraphics[width=0.37\textwidth]{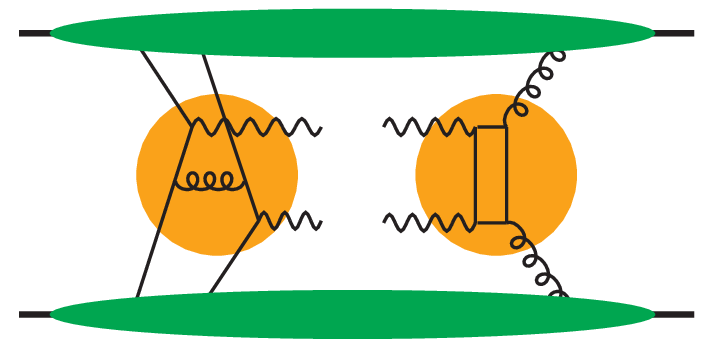}} &
  $\displaystyle\frac{1}{Q^4}$ &
  $\displaystyle\frac{\Lambda^2}{Q^2}$ \\[3.5em]
\textbf{c} & \parbox[c]{0.37\textwidth}{%
  \includegraphics[width=0.37\textwidth]{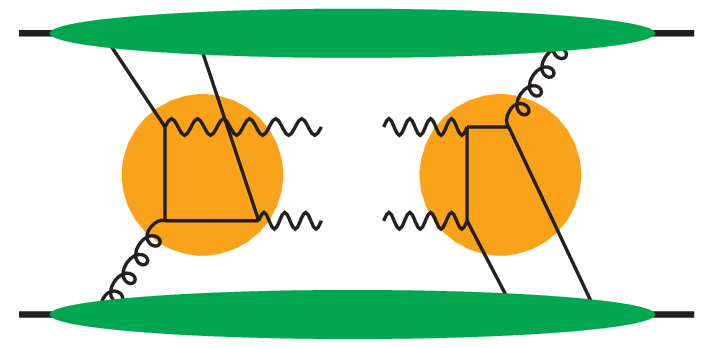}} &
  $\displaystyle\frac{1}{Q^4}$ &
  $\displaystyle\frac{\Lambda^2}{Q^2}$ \\[3.5em]
\textbf{d} & \parbox[c]{0.37\textwidth}{%
  \includegraphics[width=0.37\textwidth]{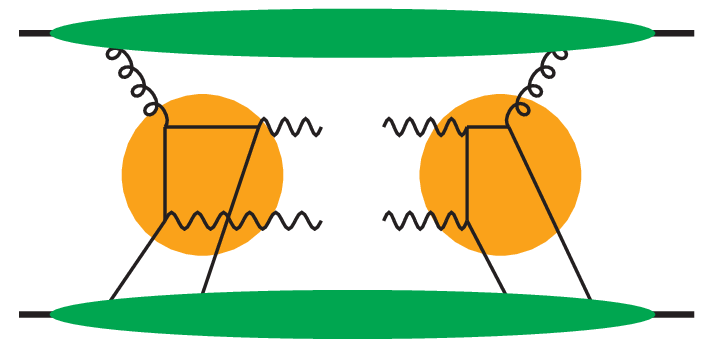}} &
  $\displaystyle\frac{1}{Q^4}$ &
  $\displaystyle\frac{\Lambda^2}{Q^2}$
\end{tabular}
\end{center}
\caption{\label{fig:power-beh-2} As figure~\protect\ref{fig:power-beh},
  but with parton correlation functions that involve gluons.}
\end{figure}

The graphs in figure~\ref{fig:power-beh}c and d involve single hard
scatters with four incoming partons.  There is, however, also an
interference between double hard scattering and single hard scattering
with \emph{two} incoming partons.  This involves correlation functions for
three partons, of which at least one must be a gluon due to fermion number
conservation.  An example is shown in figure~\ref{fig:power-beh-2}a, which
gives
\begin{align}
  \label{pow-2-fig-a}
& \frac{s\ms d\sigma}{\prod_{i=1}^2 dx_i\, d\bar{x}_i\, d^2\tvec{q}{}_i}
  \bigg|_{\text{fig.~\protect\ref{fig:power-beh-2}a}}
\;\sim\;
  \biggl[\, \prod_{i=1}^2
     \int d^4 l_i^{}\, d^4\bar{l}_i^{}\;
     \delta^{(4)}(q_i^{} - l_i^{} - \bar{l}_i^{}) \,\biggr]\,
  T'_{2\to 1}\, T'_{2\to 1}\, T^{\prime *}_{2\to 2}\;
  \Phi'_{3}\; \bar{\Phi}'_{3}
\nonumber \\
&\quad \approx\; 
  T'_{2\to 1}\, T'_{2\to 1}\, T^{\prime *}_{2\to 2}\;
  \biggl[\, \prod_{i=1}^2 \int d^2\tvec{l}_i\, d^2\bar{\tvec{l}}_i\,
    \delta^{(2)}(\tvec{q}_i - \tvec{l}_i - \bar{\tvec{l}}_i) \,\biggr]\,
\nonumber \\
&\qquad \; \times
  \int dl_1^-\, dl_2^-\, \Phi'_{3}
    \;\Big|_{l^+_i \,= x_i\, p^+}
  \int d\bar{l}_1^+\, d\bar{l}_2^+\, \bar{\Phi}'_{3}
    \;\Big|_{\bar{l}^-_i =\, \bar{x}_i\, \bar{p}^-}
\nonumber \\
&\quad \sim\; Q^2 \times \Lambda^4 \times
       \Bigl(\ms \frac{\Lambda^{4}}{Q^2} \times \Lambda^{-7} \Bigr)^2
  \;=\; \frac{1}{\Lambda^2\ms Q^2} \,.
\end{align}
This is the same power behavior as the squared single and double hard
scattering contributions in figures~\ref{fig:power-beh}a and b, so that
interference terms of this type are \emph{not} power suppressed.  The
example graph at hand has a suppression by $\alpha_s$ since two-gluon
fusion into gauge bosons only starts at one-loop level, but for other
final states like jets there is no such suppression.  We will encounter
these interference terms again in section~\ref{sec:splitting-power} (see
figure~\ref{fig:2-3-X}).

Adding a hard gluon between the two single scatters on the left of
figure~\ref{fig:power-beh-2}a leads to the interference between different
single hard-scattering processes in figure~\ref{fig:power-beh-2}b.  In the
same way as above one finds that it is power suppressed by $\Lambda^2
/Q^2$ compared with the leading contributions to the cross section.

The contributions discussed so far have hard-scattering subprocesses with
the same number of incoming partons from one and the other proton.  One
can, however, also have a parton in one proton scatter on two partons in
the other proton.  Examples for this are shown in
figures~\ref{fig:power-beh-2}c and d, and one finds that their power
behavior is the same as for figure~\ref{fig:power-beh-2}b.

The pattern emerging from the preceding examples is clear: leading-power
contributions are obtained as long as all hard-scattering processes
involve only two incoming partons.  This includes contributions from
single scattering, double scattering and their interference.  For each
hard scattering initiated by four partons one has a suppression by
$\Lambda^2 /Q^2$, and each hard scattering initiated by three partons
comes with a suppression factor~$\Lambda /Q$.


\subsubsection{Cross section integrated over transverse momenta}
\label{sec:power-counting:integrated}

So far we have considered the cross section differential in $\tvec{q}_1$
and $\tvec{q}_2$.  We now discuss how the power counting is changed when
the cross section is integrated over these transverse momenta.  As we
already observed in section~\ref{sec:sing-vs-mult}, the integration
measure $d^2\tvec{q}_1\, d^2\tvec{q}_2$ counts differently depending on
whether $\tvec{q}_1$ and $\tvec{q}_2$ are both restricted to be of order
$\Lambda$, or whether they can individually be of order $Q$ and only their
sum $\tvec{q}_1 + \tvec{q}_2$ is restricted to size $\Lambda$.  The latter
requires a single hard-scattering process in both the amplitude and its
conjugate.  For our examples we thus find an integration volume
$d^2\tvec{q}_1\, d^2\tvec{q}_2$ of order $\Lambda^2\ms Q^2$ for
graphs~\ref{fig:power-beh}a, d and \ref{fig:power-beh-2}b, c, d, whereas
in the other cases the integration volume is of order $\Lambda^4$.  The
resulting power behavior of the cross section is given in the figures.

We see that the pattern of power suppression is different from the one we
found for the cross section differential in $\tvec{q}_1$ and $\tvec{q}_2$.
In particular, the leading-power contribution now comes only from the
standard single hard-scattering in graph \ref{fig:power-beh}a.

The power behavior of the other contributions can be made more transparent
by taking a closer look at the correlation functions they involve.  As is
evident from \eqref{pow-fig-a}, the single-hard-scattering contribution of
graph \ref{fig:power-beh}a goes with the transverse-momentum integrated
correlation function $\int dl^-\, d^2\tvec{l}\, \Phi'_2$ and its
counterpart for $\bar{\Phi}'_2$, which are proportional to the usual
collinear quark or antiquark densities.
By contrast, integration of \eqref{pow-fig-c} over $\tvec{q}_1$ and
$\tvec{q}_2$ gives a four-parton correlation function
\begin{align}
  \label{twist-four-corr}
& \int dl_1^-\, dl_2^-\, dl_1^{\prime -}\,
  d^2\tvec{l}_1^{}\, d^2\tvec{l}_2^{}\, d^2\tvec{l}_1'\, \Phi'_{4}
\;\propto\;
  \int d\xi_1^-\, d\xi_2^-\, d\xi_1^{\prime -}\,
  e^{i \xi_1^- l_1^+ + i \xi_2^- l_2^+ - i\xi_1^{\prime -} l_1^{\prime +}}
\nonumber \\
& \qquad\qquad \times
  \bigl\langle p \big|\ms \bar{q}(0)\ms \Gamma_{a_2}\ms q(\xi_2^{})\;\,
     \bar{q}(\xi'_1)\ms \Gamma_{a_1}\ms q(\xi_1^{})
  \ms\big| p \bigr\rangle
  \Big|_{\xi_1^+ = \xi_2^+ = \xi_1^{\prime +} = 0, \,
         \tvec{\xi}_1^{} = \tvec{\xi}_2^{} = \tvec{\xi}'_1 = \tvec{0}}
\end{align}
and its counterpart for $\bar{\Phi}'_{4}$.  In these correlation functions
\emph{all} independent transverse parton momenta are integrated over, and
correspondingly all field operators have the same transverse position.  In
physical terms, the single hard scattering in the conjugate amplitude has
forced all hard scatters in figure~\ref{fig:power-beh}c to take place at
the same transverse position.\footnote{%
  As is well known, integrals over transverse parton momenta in the
  correlation functions are logarithmically divergent.  If these
  divergences are avoided by a transverse-momentum cutoff of order $Q$
  (which is the largest scale in the process) then the relative transverse
  positions of the partons are of order $1/Q$.}
By contrast, the double hard scattering contribution in
figure~\ref{fig:power-beh}b has two pairs of fields with a relative
transverse distance $\tvec{y}$ as we have seen in
section~\ref{sec:tree:cross-sect}, corresponding to two hard scatters
taking place at positions that can be separated by a typical hadron size.
This difference has recently been pointed out in \cite{Calucci:2009ea}.
One obtains the same twist-four correlators \eqref{twist-four-corr} when
integrating the contribution of graphs \ref{fig:power-beh}d and
\ref{fig:power-beh-2}d over $\tvec{q}_1$ and $\tvec{q}_2$.

Similarly, one finds that the transverse-momentum integrated cross
sections from graphs~\ref{fig:power-beh-2}a, b and c involve correlation
functions
\begin{align}
  \label{twist-three-corr}
& \int dl_1^-\, dl_2^-\,
  d^2\tvec{l}_1^{}\, d^2\tvec{l}_2^{}\, \bar{\Phi}'_{3}
\nonumber \\
& \quad \;\propto\;
  \int d\xi_1^-\, d\xi_2^-\,
  e^{i \xi_1^- l_1^+ + i \xi_2^- l_2^+}\,
  \bigl\langle p \big|\ms A^j(0)\;
     \bar{q}(\xi_2)\ms \Gamma_{a}\ms q(\xi_1)
  \ms\big| p \bigr\rangle \Big|_{\xi_1^+ = \xi_2^+ = 0, \,
         \tvec{\xi}_1^{} = \tvec{\xi}_2^{} = \tvec{0}} \,,
\end{align}
which are proportional to collinear twist-three distributions.  Again, a
single hard scattering in the amplitude or its conjugate is enough to
put all fields at the same transverse position.

The power behavior like $\Lambda^4 /Q^4$ of the integrated cross sections
for graphs \ref{fig:power-beh}c and d is now readily understood, as it
involves a collinear twist-four distribution for both colliding protons,
each of which is responsible for a power suppression by $\Lambda^2 /Q^2$.
Likewise, graphs~\ref{fig:power-beh-2}a, b and c involve the product of
two collinear twist-three distributions, and graph~\ref{fig:power-beh-2}d
the product of a twist-two with a twist-four distribution.  In both cases
the integrated cross section is therefore suppressed by $\Lambda^2 /Q^2$
(which happens to be the same suppression factor as for the
double-hard-scattering contribution of graph \ref{fig:power-beh}b).

In the transverse-momentum integrated cross section, graphs
\ref{fig:power-beh}c and \ref{fig:power-beh-2}a with a double hard
scattering in the amplitude play no particular role compared with their
respective counterparts, graphs \ref{fig:power-beh}d and
\ref{fig:power-beh-2}b, which involve the same correlation functions and
have the same power behavior.  Indeed, one may regard graphs
\ref{fig:power-beh}c and \ref{fig:power-beh-2}a simply as higher-twist
contributions with disconnected hard-scattering graphs on one side of the
final-state cut, rather than associating them with multiple hard
scattering.  This was recently advocated in \cite{Calucci:2009ea}.

We emphasize that such a view is appropriate \emph{only} if the cross
section is integrated over transverse momenta.  For observed transverse
momenta $\tvec{q}_1$ and $\tvec{q}_2$ we have a \emph{different} power
behavior for graphs \ref{fig:power-beh}c and d, as well as for graphs
\ref{fig:power-beh-2}a and b.  In particular, the interference
contribution from graph~\ref{fig:power-beh-2}a then has the same
leading-power behavior as graphs \ref{fig:power-beh}a and b.  Let us also
note that for graph~\ref{fig:power-beh-2}a the quark and antiquark in each
proton are not at the same transverse position for fixed $\tvec{q}_1$ and
$\tvec{q}_2$.  If we express the correlation functions $\Phi'_3$ and
$\bar{\Phi}'_{3}$ through matrix elements $\langle p |\ms A^j(0)\;
\bar{q}(\xi_2)\ms \Gamma_{a}\ms q(\xi_1) \ms| p \rangle$ and $\langle p
|\ms A^{k}(0)\; \bar{q}(\bar{\xi}_1)\ms \Gamma_{\bar{a}}\ms q(\bar{\xi}_2)
\ms| p \rangle$ then the transverse-momentum integrations in
\eqref{pow-2-fig-a} can be carried out and give
\begin{align}
\int d^2\tvec{l}_i\, d^2\bar{\tvec{l}}_i\;
    \delta^{(2)}(\tvec{q}{}_i - \tvec{l}_i - \bar{\tvec{l}}_i) \,
 e^{-i \tvec{\xi}{}_i \tvec{l}_i
   - i \bar{\tvec{\xi}}{}_i \bar{\tvec{l}}_i}
 &= (2\pi)^2 \delta^{(2)}(\tvec{\xi}_i - \bar{\tvec{\xi}}_i)\;
    e^{-i \tvec{\xi}_i \tvec{q}_i}
\end{align}
for $i=1,2$.  With $|\tvec{q}_i| \sim \Lambda$ we thus have a typical
quark-antiquark distance $|\tvec{\xi}_1 - \tvec{\xi}_2| \sim 1/\Lambda$.


\subsection{Effects at small \texorpdfstring{$x$}{x}}
\label{sec:small-x}

Typical values of $x_i$ and $\bar{x}_i$ at the LHC can be quite small, as
we already noted after \eqref{phase-space-el}.  At $\sqrt{s} = 7
\operatorname{TeV}$ and $q_i^2 = m_Z^2$ one has for instance $\sqrt{x_i\ms
  \bar{x}_i \rule{0pt}{1.7ex}} = 1.3 \times 10^{-2}$.  Although phenomena
at small $x$ are not the main focus of this work, we wish to make a few
comments on them in the present section.

We begin by recalling that the usual densities for quarks, antiquarks and
gluons rise with small $x$.  This rise can be approximately described by
power laws $q(x) \sim \bar{q}(x) \sim x^{-1-\lambda_q}$ and $g(x) \sim
x^{-1-\lambda_g}$, with exponents $\lambda_q$ and $\lambda_g$ between $0$
and $1$ that depend on the factorization scale $\mu$.  The abundance of
small-$x$ partons can be understood as a consequence of repeated
radiation, which is essentially described by ladder graphs.  Such graphs
are in particular resummed by the DGLAP evolution equations, which make
the rise at small $x$ steeper when $\mu$ is increased.

In the simple approximation where correlations between partons are
neglected, multiparton distributions are the product of single-parton
densities as discussed in section~\ref{sec:reduct}.  The distribution of
$n$ quarks or antiquarks then approximately behaves like $F(x_i,
\tvec{k}_i, \tvec{y}{}_i) \sim (x_1 x_2 \cdots x_n)^{-1-\lambda_q}$ if all
momentum fractions are sufficiently small.  If all momentum fractions are
of similar size, $x_i \sim \bar{x}_i \sim x$, this gives a factor $x^{- 2n
  (1+ \lambda_q)}$ in the cross section \eqref{X-sect-mixed}.  If the same
final state is produced by a single quark-antiquark annihilation, the
corresponding factor is only $x^{-2(n + \lambda_q)}$ according to
\eqref{single-hard-diff}.\footnote{For this comparison it is important
  that the hard-scattering cross section on the r.h.s.\ of
  \protect\eqref{single-hard-diff} depends only on the momenta $q_i$ and
  not on $p$ or $\bar{p}$.  It is hence proportional to $Q^{-2n}$ (without
  any further factors of $x_i$ or $\bar{x}_i$) and thus of the same order
  as the product $\hat{\sigma}_1 \hat{\sigma}_2 \cdots \hat{\sigma}_n$ in
  \protect\eqref{X-sect-mixed}.}
The multiple scattering mechanism is thus enhanced for small momentum
fractions, both for observed and integrated transverse momenta
$\tvec{q}_i$.  In terms of graphs, this enhancement can be traced back to
multiple ladders, one for each pair of partons with the same momentum
fraction $x_i$ in figure~\ref{fig:n-scatters}.  We expect that such an
enhancement exists, although the above estimate based on completely
uncorrelated partons is likely too simplistic.

Note that a strong rise at small $x$ is only observed for parton densities
that mix with gluons under evolution, but not for combinations like $q(x)
- \bar{q}(x)$ or $u(x) - d(x)$, which rise more slowly than $x^{-1}$.  A
corresponding pattern is expected for multi-parton distributions.  Since
they cannot mix with multigluon distributions, the interference
distributions in fermion number or quark flavor discussed at the end of
section~\ref{sec:spin:quarks} are not enhanced at small $x$.  We hence
expect them to play a minor role in small-$x$ kinematics.

The preceding arguments apply to both quark and gluon distributions in the
framework of hard-scattering factorization, and based on the experience
with single-parton densities one expects them to be relevant for momentum
fractions $x_i \sim \bar{x}_i$ of order $10^{-2}$ or smaller.
At very small $x$ the gluon is by far the dominant parton species in the
proton, and one may use high-energy factorization and the BFKL approach to
describe the dynamics of gluon ladders.  The primary expansion variable of
this approach is $\log \frac{1}{x}$, rather than the ratio $Q/\Lambda$
used in the power counting arguments on which hard-scattering
factorization is based.  Basic quantities in high-energy factorization are
Green functions depending on transverse gluon momenta, which bear close
resemblance with the transverse-momentum dependent gluon distributions
discussed in this work and naturally allow one to keep track of transverse
momenta $\tvec{q}_i$ in the final state.

Investigations of multiparton scattering in the BFKL approach can be found
in \cite{Ragazzon:1995cb,Braun:2000ua,Bartels:2005wa,%
  Levin:2008jf,Bartels:2011qi}.  In agreement with the arguments given
above, one finds that the two-gluon distribution receives a contribution
from two independent BFKL ladders, with a small-$x$ exponent twice as
large as for a single BFKL ladder \cite{Braun:2000ua}.  More complicated
graphs with four gluons in the $t$ channel have been analyzed in
\cite{Braun:2000ua,Levin:2008jf,Bartels:2011qi}.  As to the high-energy
behavior of three $t$ channel gluons, all solutions found so far for the
corresponding evolution equations have a \emph{weaker} small-$x$ growth
than a single BFKL pomeron \cite{Bartels:1999yt,Janik:1998xj}.

Let us see how small-$x$ dynamics affects the different graphs
investigated in section~\ref{sec:power-counting}.  In line with our above
discussion, we assume that correlation functions for four partons give a
faster growth of the cross section with energy than correlation functions
for two partons.  We have no definite expectation for the corresponding
behavior of three-parton correlation functions; if the results just
mentioned for gluons in the small-$x$ limit are a guide, then
contributions with three $t$ channel partons are not favored for small
momentum fractions.  For the cross section differential in $\tvec{q}_i$ we
then find that the multiple-scattering mechanism of figure
\ref{fig:power-beh}b is actually favored over the single-scattering graph
\ref{fig:power-beh}a, which has the same power behavior in $\Lambda/Q$ but
a weaker rise at small $x$.  With the caveat just mentioned, the
interference contribution in figure \ref{fig:power-beh-2}a is expected to
be less important.

The cross section integrated over $\tvec{q}_i$ is dominated by the
conventional single-scattering mechanism in figure~\ref{fig:power-beh}a by
power counting in $\Lambda/Q$.  Among the contributions that are
suppressed by $\Lambda^2 /Q^2$ the double-scattering
graph~\ref{fig:power-beh}b is enhanced at small $x$.  To a lesser extent
the same is true for graph~\ref{fig:power-beh-2}d, which involves four $t$
channel partons in only one of the two protons.  There may be situations
where the small-$x$ enhancement overcompensates the power suppression by
$\Lambda^2 /Q^2$, for instance in the high-energy production of minijets,
where the hard scale $Q$ is not too large.  In such cases the BFKL
approach may be more adequate than the one using hard-scattering
factorization.


\subsection{The ``effective cross section''}
\label{sec:sigma-eff}

The cross section for double hard scattering is often written as
$\sigma_{\text{ds}} = \sigma_1\, \sigma_2 /(C\ms \sigma_{\text{eff}})$,
where $\sigma_1$ and $\sigma_2$ are single hard scattering cross sections,
$C$ is the combinatorial factor introduced below \eqref{X-section-start} and
$\sigma_{\text{eff}}$ is an ``effective cross section'' characterizing the
strength of multiple interactions.  Let us see to which extent such a
formula holds true in the light of the results we have derived so far.

Under the assumption that there are no correlations between different
partons in the target hadron we derived the factorized form
\eqref{impact-product-coll} for multiparton distributions in a model
theory with scalar partons.  This derivation carries over to the color
singlet distributions of two unpolarized quarks, antiquarks or gluons,
i.e.\ to $\sing{F}_{q_1, q_2}$, $\sing{F}_{q_1,\bar{q}_2}$,
$\sing{F}_{\bar{q}_1,q_2}$, $\sing{F}_{\bar{q}_1,\bar{q}_2}$ and
$\sing{F}_{g,g}$, where the two quark flavors $q_1$ and $q_2$ may be
different.  If one further assumes that the impact-parameter dependent
distributions of a single quark, antiquark and gluon have the form
$f_{c}(x; \tvec{b}) \approx F(\tvec{b})\, f_{c}(x)$ with a common impact
parameter profile $F(\tvec{b})$ for all parton species $c$, then the cross
section \eqref{X-sect-impact-product} for double hard scattering takes the
form
\begin{align}
  \label{X-sect-global-prod}
\frac{d\sigma_{\text{ds}}}{\prod_{i=1}^2 dx_i\, d\bar{x}_i}
 &\approx \frac{1}{C\ms \sigma_{\text{eff}}}\;
  \frac{d\sigma_1}{dx_1\, d\bar{x}_1}\,
  \frac{d\sigma_2}{dx_2\, d\bar{x}_2}  
\end{align}
with
\begin{align}
\frac{d\sigma_i}{d x_i\, d\bar{x}_i}
 &= \sum_{c=q,\bar{q},g\phantom{d}} \sum_{d=q,\bar{q},g}
    \hat{\sigma}^{}_{i, c d}(x_i \bar{x}_i s)\ms
          f_{c}(x_i) f_{d}(\bar{x}_i)
\end{align}
and
\begin{align}
  \label{sigma-eff}
\frac{1}{\sigma_{\text{eff}}}
 &= \int d^2\bs \tvec{\rho}\; \biggl[\,
    \int d^2\tvec{y}\; F(\tvec{y} - \tvec{\rho}) F( \tvec{y})
    \,\biggr]^2
  = \int \frac{d^2\tvec{r}}{(2\pi)^2}\; \bigl[ F(\tvec{r}) \bigr]^4 \,.
\end{align}
The second form of \eqref{sigma-eff} has recently been given in
\cite{Blok:2011bu} and uses that the Fourier transform $F(\tvec{r})$ of
$F(\tvec{b})$ depends only on $\tvec{r}^2$ because of rotation invariance.

It is natural to ask whether \eqref{X-sect-global-prod} extends to the
cross section differential in $\tvec{q}_i$.  If one has $f_{c}(x,
\tvec{z}; \tvec{b}) \approx F(\tvec{b})\, f_{c}(x, \tvec{z})$ for all
parton types $c$ then the factorized approximation in
\eqref{impact-product-2} gives $F_{c_1,c_2}(x_i, \tvec{z}_i, \tvec{y};
\tvec{b}) \approx F(\tvec{b} + \half x_1 \tvec{z}_1)\, F(\tvec{b} +
\tvec{y} - \half x_2 \tvec{z}_2)\, f_{c_2}(x_2, \tvec{z}_2)\, f_{c_1}(x_1,
\tvec{z}_1)$.  Inserting this into the cross section formula
\eqref{X-sect-position} does not lead to a factorized form, because $x_i
\tvec{z}_i$ appears in the arguments of the impact parameter profile $F$.
A simplification occurs however if the measured transverse momenta
$\tvec{q}_i$ are large compared to a hadronic scale $\Lambda$ (while being
much smaller than $Q$).\footnote{This kinematic region is examined in
  detail in section~\protect\ref{sec:high-qt}.  Notice the change of
  notation compared with the previous sections, where $\Lambda$ denotes a
  hadronic scale \emph{or} the size of $|\tvec{q}_i|$, whichever is
  larger.}
The typical values of $\tvec{z}_i$ in the cross section are then small
compared with $1/\Lambda$, whereas typical values of $\tvec{b}$ and
$\tvec{b} + \tvec{y}$ are of hadronic size.  We can thus approximate
$F(\tvec{b} + \half x_1 \tvec{y}_1)\, F(\tvec{b} + \tvec{y} - \half x_2
\tvec{z}_2) \approx F(\tvec{b})\, F(\tvec{b} + \tvec{y})$ and obtain
\begin{align}
  \label{X-sect-qt-global-prod}
\frac{d\sigma_{\text{ds}}}{\prod_{i=1}^2
                           dx_i\, d\bar{x}_i\, d^2\tvec{q}{}_{i}}
 &\approx \frac{1}{C\ms \sigma_{\text{eff}}}\;
  \frac{d\sigma_1}{dx_1\, d\bar{x}_1\, d^2\tvec{q}{}_{1}}\,
  \frac{d\sigma_2}{dx_2\, d\bar{x}_2\, d^2\tvec{q}{}_{2}}
\qquad \text{for $|\tvec{q}{}_i| \gg \Lambda$}
\end{align}
with
\begin{align}
\frac{d\sigma_i}{dx_i\, d\bar{x}_i \, d^2\tvec{q}{}_{i}}
 &= \int \frac{d^2\tvec{z}_i}{(2\pi)^2}\;
          e^{-i \tvec{z}_i \tvec{q}{}_i}
   \sum_{c=q,\bar{q},g\phantom{d}} \sum_{d=q,\bar{q},g}
          \hat{\sigma}^{}_{i, c d}(x_i \bar{x}_i s)\ms
          f_{c}(x_i, \tvec{z}_i) f_{d}(\bar{x}_i, \tvec{z}_i)
\end{align}
and $\sigma_{\text{eff}}$ as in \eqref{sigma-eff}.  Both
\eqref{X-sect-global-prod} and \eqref{X-sect-qt-global-prod} can be made
differential in further variables describing the sets of particles
produced by the two hard scatters.  If one integrates these relations over
kinematic variables in the presence of cuts, they only retain their
validity if each cut refers to particles in one of the two sets but not in
both.

The assumptions that allow one to relate the cross sections for double and
single hard scattering by a single process independent constant
$\sigma_{\text{eff}}$ are quite strong, and a number of effects can
invalidate \eqref{X-sect-global-prod} and \eqref{X-sect-qt-global-prod}:
\begin{itemize}
\item an impact parameter profile $F_{c\,}(\tvec{b})$ that is not the same
  for different parton distributions.  The effect of this was estimated
  for a specific model in \cite{DelFabbro:2000ds}.
\item a correlation between the $x$ and $\tvec{b}$ dependence in the
  single-parton distributions $f_c(x; \tvec{b})$ or $f_c(x, \tvec{z};
  \tvec{b})$.  
  Evidence that such a correlation is appreciable for $x$ above $0.1$
  comes from the calculation of the Mellin moments $\int dx\, x^{n-1}
  \bigl[ f_q(x; \tvec{b}) + (-1)^{n} f_{\bar{q}}(x; \tvec{b}) \bigr]$ with
  $n=1,2,3$ in lattice QCD, see \cite[section 4.4.5]{Hagler:2009ni} and
  references therein.  The interpretation of HERA measurements
  \cite{Aktas:2005xu,Chekanov:2002xi} for $\gamma p \to J\!/\Psi\, p$ in
  terms of generalized parton distributions shows that the average squared
  impact parameter $\langle \tvec{b}^2 \rangle$ of small-$x$ gluons in the
  proton has a weak logarithmic dependence on $x$
  \cite{Frankfurt:2003td,Diehl:2007zu,Frankfurt:2010ea}.  An estimate of
  how a correlation between $x$ and $\tvec{b}$ in $f_c(x; \tvec{b})$
  affects multiparton interactions has been made in \cite{Corke:2011yy}.
\item correlations between different partons in the proton, which
  invalidate the relations \eqref{impact-product-2} and
  \eqref{impact-product-coll} between two-parton and single-parton
  distributions.  In \cite{Frankfurt:2004kn} it was argued that such
  correlations are significant.
\item an appreciable size of multiparton distributions that describe spin
  correlations between two partons (section~\ref{sec:tree:spin}), of
  distributions where partons with the same momentum fraction $x_i$ are
  not coupled to color singlets (section~\ref{sec:color}), or of
  interference distributions in fermion number or quark flavor
  (section~\ref{sec:spin:quarks}).
\end{itemize}
Finally, the assumption that the observed cross section is given by
contributions from either single or double hard scattering is invalid if
their interference (see figure~\ref{fig:power-beh-2}a) is important.  All
in all, we feel that \eqref{X-sect-global-prod} or
\eqref{X-sect-qt-global-prod} may be useful for order-of-magnitude
estimates but should be used with great caution.  Of course, one may
\emph{define} $\sigma_{\text{eff}}$ as the ratio
$(d\sigma_1/d\Gamma_{\!1})\, (d\sigma_2/d\Gamma_2) \big/ (S\ms
d\sigma_{\text{ds}}/d\Gamma_{\!1}\, d\Gamma_2)$ of differential cross
sections for single and double scattering.  Since this ratio can depend on
the process and on all kinematic variables, $\sigma_{\text{eff}}$ is then
\emph{not} a universal constant.

\section{Beyond lowest order: factorization and Sudakov logarithms}
\label{sec:factorization}

So far we have analyzed the lowest-order graphs that contribute to
multiple scattering processes.  For a systematic treatment in QCD we need
to go beyond this approximation and in particular take into account graphs
where additional gluons are exchanged.  A complete analysis should
eventually establish whether an all-order factorization formula can be
written down for a given observable.  We will not attempt to do this here,
but provide some building blocks for such an analysis.  We use the
framework of hard-scattering factorization, which essentially organizes
the dynamics according to virtualities (as opposed to high-energy or
small-$x$ factorization, where the organizing principle is based on
rapidity).  We focus on the cross section differential in small transverse
momenta and in particular investigate the structure of Sudakov logarithms.
In section~\ref{sec:coll-fact} we will make some remarks on
transverse-momentum integrated cross sections, described by collinear
factorization.  For reasons given in section~\ref{sec:coll-soft} we will
concentrate on the double Drell-Yan process, i.e.\ on the production of
two electroweak gauge bosons, which for definiteness we take to be virtual
photons.  Likewise, we will use the single Drell-Yan process as an example
when we recall the ingredients for factorization with a single hard
scattering.


\subsection{Dominant graphs}
\label{sec:libby-sterman}

One of the first tasks when establishing factorization for a given process
is to identify the dominant graphs in the kinematic limit one is
interested in.  The appropriate tool for hard-scattering factorization is
the method of Libby and Sterman \cite{Sterman:1978bi,Libby:1978bx}, which
we briefly recapitulate.  The first step is to trade the limit of large
kinematic invariants (which we collectively denoted by $Q$ earlier) for
the limit of vanishing masses of all partons.  In doing so, one uses that
up to an overall normalization the quantities of interest depend on the
ratio of $Q$ and the masses.  If we keep small transverse momenta in the
differential cross section, those must be sent to zero as
well.\footnote{This was not stated in the original work by Libby and
  Sterman, who considered transverse-momentum integrated quantities.}
One is thus led to examine which graphs and which phase space regions give
rise to mass divergences.  Such divergences come from the poles of Feynman
propagators, but only if for a suitable loop integration variable there
are poles on both sides of the real axis, which ``pinch'' the contour of
the loop integration.  If there is no pinch, the poles can be avoided by
deforming the integration contour.  One finds that lines that give pinch
singularities are either soft (i.e.\ all their momentum components are
close to zero) or collinear to one of the incoming or outgoing particles
of the process.  All other lines are far off-shell (possibly after complex
contour deformation).  The leading contribution in the large $Q$ limit of
a given graph comes from regions of phase space in the vicinity of the
pinch singular configurations just described.  To obtain a factorization
formula, one has to express subgraphs with collinear or soft lines in
terms of matrix elements that make sense beyond perturbation theory.
Parton densities and related quantities are constructed from these matrix
elements.  Off-shell lines are organized into hard subgraphs, which can be
calculated perturbatively.

A physically intuitive interpretation of the previous construction is
provided by the Coleman-Norton theorem \cite{Coleman:1965xm}.  The pinch
singular configurations of a graph correspond to a scattering process
where the lines with collinear momenta correspond to classical
trajectories in space-time.  The trajectory associated with each line is
proportional to its four-momentum, so that it shrinks to a point for soft
lines.  In the ``reduced graph'' that represents the corresponding
classical process, off-shell lines in the original graph are likewise
contracted to points.

The preceding analysis is based on the denominators of Feynman propagators
and gives only a necessary condition for the occurrence of mass
singularities.  A power counting analysis taking into account the
numerators of Feynman graphs (similar to the one we gave in
section~\ref{sec:power-counting}) provides further restrictions on the
contributions that actually dominate a given observable.  At this level,
the polarization of gluon lines is found to play a crucial role.

\begin{figure}
\begin{center}
\includegraphics[height=0.32\textwidth]{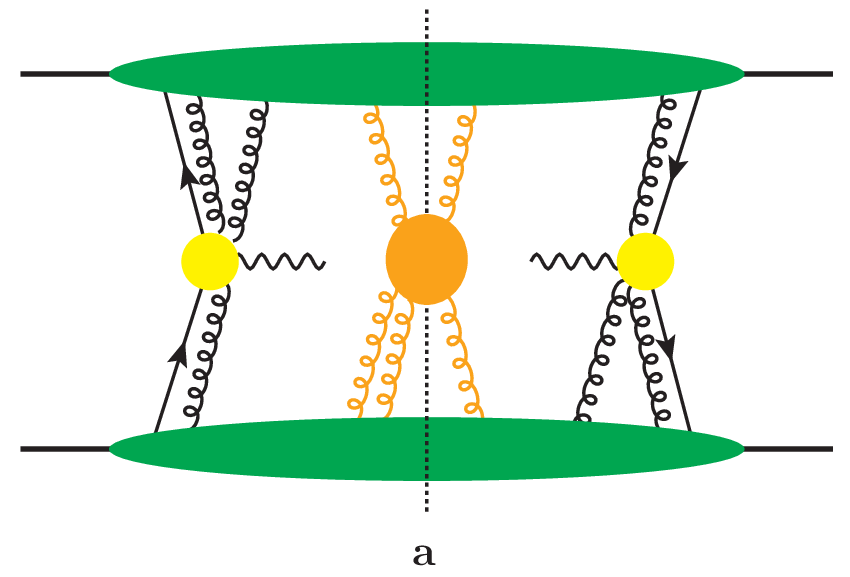}
\hfill
\includegraphics[height=0.32\textwidth]{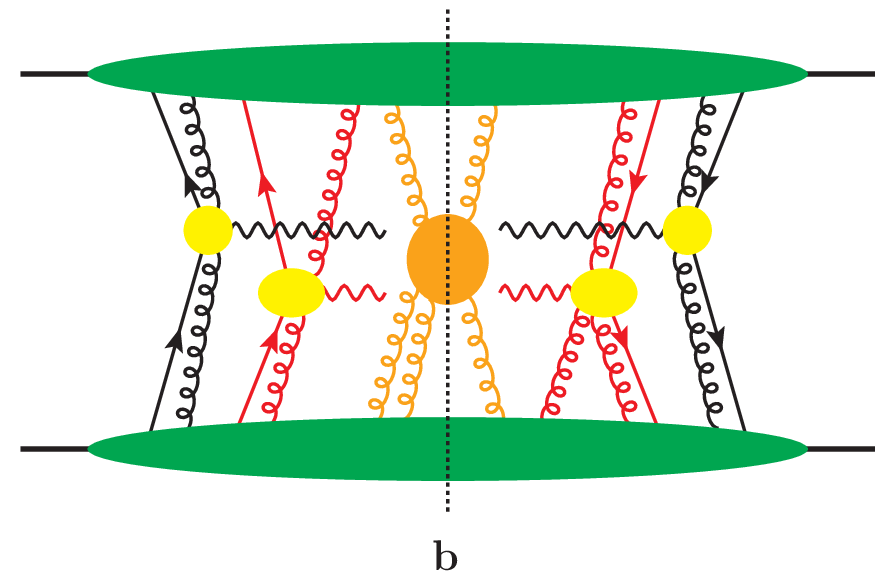}
\end{center}
\caption{\label{fig:hard-coll-soft} Leading graphs for the single (a) and
  double (b) Drell-Yan process at measured $q_T \ll Q$.  The upper and
  lower blobs denote collinear subgraphs, the blob crossing the
  final-state cut (dashed line) denotes a soft subgraph, and the blobs
  with a final-state gauge boson denote hard subgraphs.}  
\end{figure}

For single Drell-Yan production at fixed small transverse photon momentum,
one finds that the dominant graphs have the structure shown in
figure~\ref{fig:hard-coll-soft}a.  For each of the colliding protons there
is a collinear subgraph.  On either side of the final-state cut there is
one hard subgraph producing the final state boson, connected with each
collinear subgraph by exactly one fermion line and an arbitrary number of
gluon lines, which must have polarization in the plus direction for
right-moving and in the minus direction for left-moving collinear gluons.
Finally, there is a soft subgraph with soft gluons attaching to either of
the collinear subgraphs.  There are no soft gluons coupling to the hard
subgraphs.

The dominant graphs for double Drell-Yan production are easily identified
and just have an additional hard subgraph for the second produced gauge
boson on either side of the final-state cut.  As we have already seen in
section~\ref{sec:power-counting}, hard subgraphs that are connected to
each collinear graph by a single parton line have leading power behavior.
The power counting for the soft graph is not affected by having one or two
hard subprocesses.  Finally, the absence of soft gluons coupling to a hard
subgraph has the same reason as in the single Drell-Yan case, namely that
such soft gluons increase the number of hard propagator denominators in
the hard subgraph, without providing a compensating large numerator factor
or phase space volume.

\begin{figure}
\begin{center}
\includegraphics[width=0.8\textwidth]{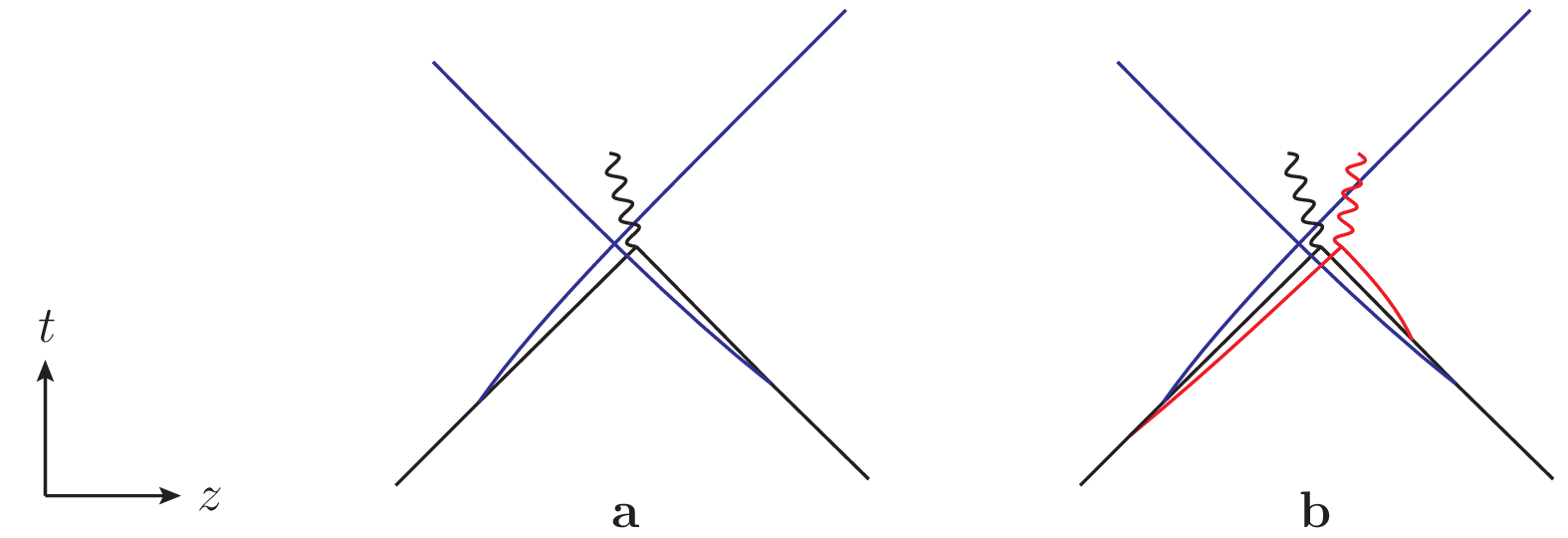}
\end{center}
\caption{\label{fig:space-time} Space-time representation of the leading
  graphs for the single (a) and double (b) Drell-Yan process at measured
  $q_T \ll Q$.  The parton lines move along light-like paths and have been
  drawn with a slight curvature only for clarity.  Likewise, the two
  bosons in figure (b) are meant to be produced at the same point in $t$
  and $z$.}
\end{figure}

The space-time representation in the sense of the Coleman-Norton theorem
is shown in figure~\ref{fig:space-time} for the graphs in
figure~\ref{fig:hard-coll-soft}.  Parton lines from one and the other
proton meet at one point in the $t$-$z$ plane and annihilate into a gauge
boson.  For double Drell-Yan production, the two bosons are produced at
the same point in $t$ and $z$.  The transverse momenta of partons and the
produced bosons are neglected in this interpretation (see above), so that
the classical scattering process takes place at fixed transverse
coordinates ($x$ and $y$).


\subsubsection{``Rescattering'' contributions}
\label{sec:rescatter}

\begin{figure}
\begin{center}
\includegraphics[width=0.71\textwidth]{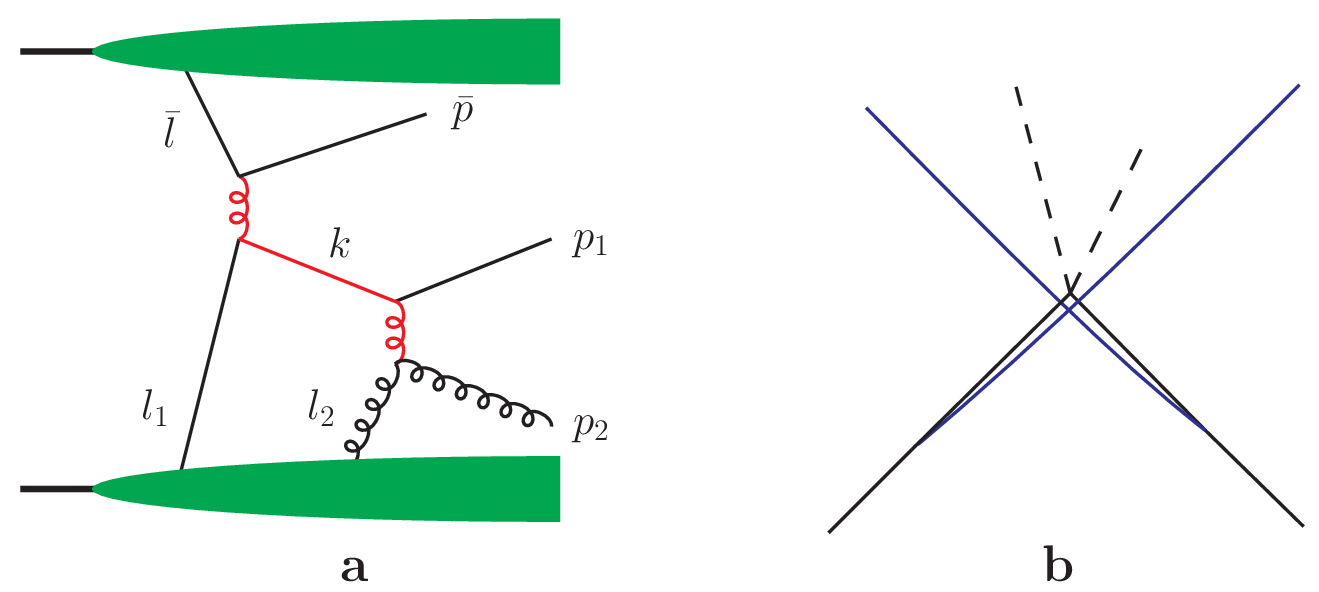}
\end{center}
\caption{\label{fig:rescatter} (a) A graph associated with
  ``rescattering'' in the literature.  (b) Space-time representation for
  the production of two high-$p_T$ partons in a single hard $2\to 2$
  scattering process.  The dashed lines have momentum components in the
  transverse $x$-$y$ plane; all other lines move on the light-cone $z =
  \pm t$.}
\end{figure}

Before discussing in detail the leading graphs in
figure~\ref{fig:hard-coll-soft}, we wish to comment on graphs of the type
shown in figure~\ref{fig:rescatter}a.  They have been associated with
``rescattering'' in the literature \cite{Paver:1984ux,Corke:2009tk} and
were calculated in terms of two hard $2\to 2$ QCD processes, where the
parton with momentum $k$ in the figure is treated as an outgoing parton in
the first scattering and as an incoming parton in the second one.  It is
understood that the transverse momenta $\tvec{p}_1$, $\tvec{p}_2$ and
$\bar{\tvec{p}}$ are all large (we denote their order by $p_T$ below).

We argue here that this is not a correct way to calculate the graph, at
least not within the usual hard-scattering factorization framework used in
\cite{Paver:1984ux,Corke:2009tk}.  According to our discussion in the
previous section, the lines that enter or exit a hard-scattering subgraph
must correspond to pinched singularities and thus admit a classical
space-time interpretation in the sense of the Coleman-Norton theorem.
This is not possible for the line with momentum $k$ in
figure~\ref{fig:rescatter}a.  As illustrated in
figure~\ref{fig:rescatter}b, the two partons emerging from a hard $2\to 2$
process have large transverse momenta and, being on shell, thus have
finite rapidities.  In other words, their velocity in the $z$ direction is
smaller than the speed of light.  As soon as such a parton has propagated
over a finite distance, it can no longer scatter on another parton from
one of the two initial protons, since those partons do move with the speed
of light along $z$.  The proper treatment of the parton with momentum $k$
is thus to regard it as an internal line in a single hard-scattering
process with three incoming partons ($l_1, l_2, \bar{l}$) and three
outgoing ones ($p_1, p_2, \bar{p}$).  As we saw in
section~\ref{sec:power-counting} such a contribution is power suppressed
(if $\tvec{p}_1 + \tvec{p}_2 + \bar{\tvec{p}}$ is integrated over, it
involves a parton distribution of higher twist) and can hence be
neglected.

Put differently, the parton with momentum $k$ is generically far off-shell
in the leading region of the graph in figure~\ref{fig:rescatter}a.  A
kinematical analysis readily shows that the final-state momenta $p_1, p_2,
\bar{p}$ fix the sum $l_1^+ + l_2^+$ to a large value of order $p_T$, up
to small corrections of order $1/p_T$.  The value of $l_1^+$ is however
integrated over a large interval of order $p_T$.  For a particular value
of $l_1^+$ in this interval, the propagator of $k$ does have a pole, but
this pole is not a pinch singularity (the gluons adjacent to $k$ are far
off-shell when $k^2=0$ and their propagator poles are a distance of order
$p_T$ away in the complex $l_1^+$ plane).  One can thus deform the
integration contour of $l_1^+$ such that $k^2$ is always of order $p_T^2$
and thus large.\footnote{When actually calculating the hard scattering,
  one can nevertheless integrate $l_1^+$ along the real axis; the pole of
  $1/(k^2+i\epsilon)$ then provides an absorptive part to the
  hard-scattering amplitude.  The possibility to deform the integration
  contour of $l_1^+$ justifies the perturbative treatment of the
  propagator for $k$.}


\subsection{Collinear and soft gluons}
\label{sec:coll-soft}

We now return to the graphs in figure~\ref{fig:hard-coll-soft}.  They
contain an arbitrary number of collinear and soft gluons, and further
simplifications are required to obtain a useful factorization formula that
involves a limited number of nonperturbative quantities.

In existing factorization theorems, the effects of collinear and of soft
gluons are described by Wilson line operators, to all orders in the strong
coupling.  The possibility to obtain such a simple structure is crucial
for establishing factorization.  Detailed analyses of this issue can be
found in \cite{Collins:1981uk,Collins:2007ph,Collins:2011,Ji:2004wu} for
single Drell-Yan production or for its crossed-channel analogs, the
production of back-to-back hadrons in $e^+e^-$ annihilation or
semi-inclusive deep inelastic scattering (SIDIS).  By contrast, for
hadron-hadron collisions producing back-to-back jets or hadrons with
measured transverse momenta, serious obstacles to establishing
factorization have been identified in \cite{Rogers:2010dm} and in previous
work cited therein.  A systematic treatment of transverse-momentum
dependent factorization for jet or hadron production in multiple hard
scattering will probably need to wait until a suitable formulation for
single hard scattering has been found.

We therefore limit our considerations in this section to the double
Drell-Yan process.  Extending our arguments to the production of other
colorless particles is trivial if the hard scattering is initiated by
quarks or antiquarks and should be possible if it is initiated by gluons.
We shall not attempt to give a full proof of factorization even for double
Drell-Yan production.  Instead, we will analyze the lowest-order graphs
with an additional exchanged collinear or soft gluon.  To a large part
this will be a recapitulation of the corresponding analysis for the single
Drell-Yan process.  We nevertheless give the necessary steps in some
detail, in order to see how the arguments generalize to double hard
scattering.  We will pay particular attention to the color indices for
quarks and antiquarks, since the color structure of two-parton
distributions is nontrivial compared with the single-parton case.
Finally, we will point out which further issues need to be settled to
obtain a full proof of factorization.


\subsubsection{From collinear gluons to Wilson lines in parton
  distributions}
\label{sec:coll}

\begin{figure}
\begin{center}
\includegraphics[width=0.5\textwidth]{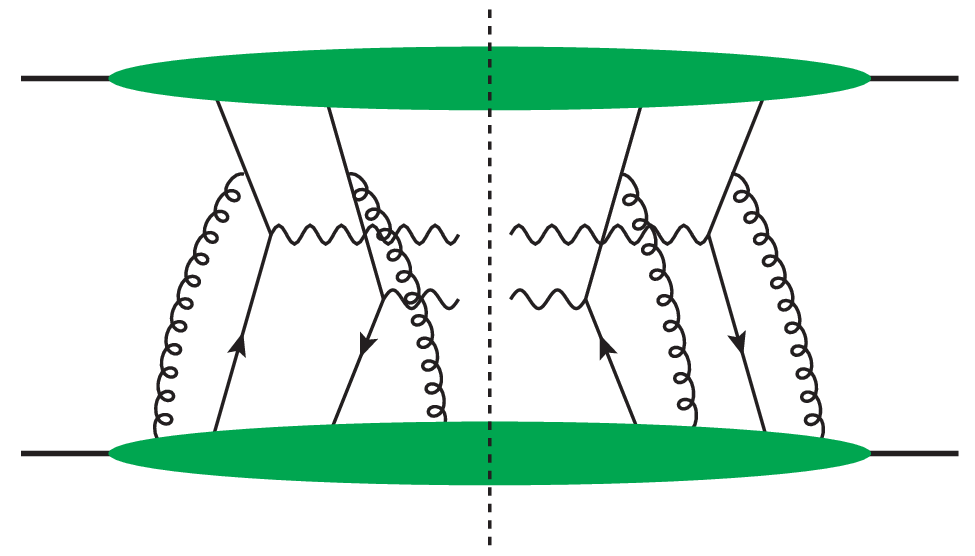}
\end{center}
\caption{\label{fig:many-coll} Example graph for the double Drell-Yan
  process with collinear gluons coupling to left-moving quarks or
  antiquarks before those undergo a hard scattering.}
\end{figure}

Figure~\ref{fig:many-coll} shows an example where several gluons collinear
to the right-moving proton $p$ couple to a left-moving quark or antiquark.
The quark or antiquark is thus taken far off shell, so that its propagator
and its coupling to the gluon belong to one of the hard-scattering
subprocesses.

\begin{figure}
\begin{center}
\includegraphics[width=0.63\textwidth]{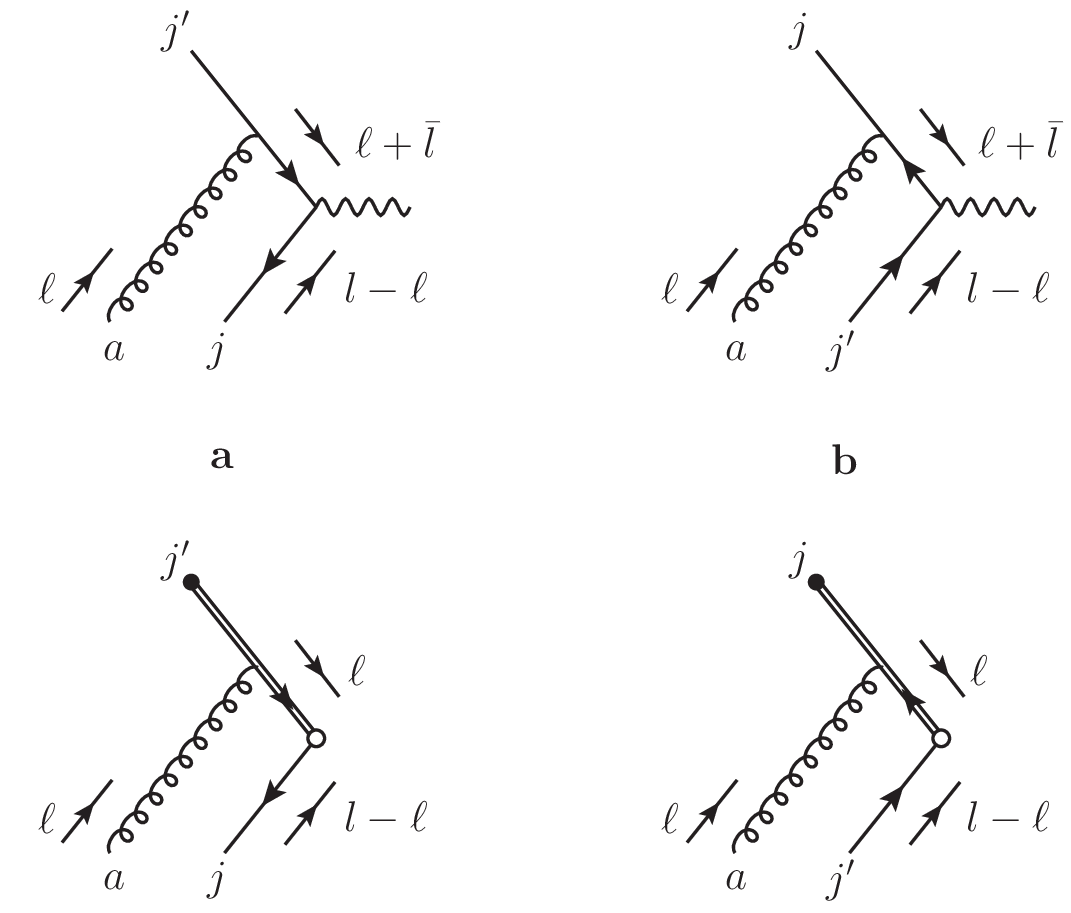}
\end{center}
\caption{\label{fig:dy-coll} Collinear gluons in the Drell-Yan process.
  Top row: subgraphs with a right-moving gluon coupling to a left-moving
  quark or antiquark before it annihilates.  Bottom row: corresponding
  graphs after the off-shell propagators have been replaced by eikonal
  lines.}
\end{figure}

We now recapitulate the analysis of one such coupling, which is well-known
from single Drell-Yan production, taking particular care of color indices
and of the distinction between quarks and antiquarks.  The relevant part
of the graph in figure~\ref{fig:dy-coll}a can be written as
\begin{align}
  \label{dy-coll-a}
T_{a} &=
\langle \ldots \bar{q}_j\, A^{\alpha,a} \ldots \rangle\,
  \frac{i}{\gamma\ms (\ell + \bar{l}_c) \rule{0pt}{2.2ex}}\,
    (-ig)\, t^{a}_{jj'} \gamma_\alpha\, u(\bar{l}_c)\,
  \langle \ldots q_{j'} \ldots \rangle \,,
\end{align}
where in a shorthand notation we write $\langle \ldots \bar{q}_j\,
A^{\alpha,a} \ldots \rangle$ and $\langle \ldots q_{j'} \ldots \rangle$
for the hadronic matrix elements of the right and left moving proton,
respectively.  The subscript $c$ on $\bar{l}$ indicates the collinear
approximation specified after \eqref{quark-projectors}, i.e.\ $\bar{l}_c^-
= \bar{l}^-$, $\bar{l}_c^+ = 0$ and $\bar{\tvec{l}}_c = \tvec{0}$.
Instead of the spinor $u(\bar{l}_c)$ for the incoming quark we could also
use the projection operator $P(\bar{l})$, see the discussion after
\eqref{quark-spin-av} and \eqref{quark-projectors}.  The vertex with the
produced photon and the spinor for the incoming right-moving quark are not
needed for our argument and have been omitted.  Our sign convention for
the strong coupling $g$ is such that the covariant derivative reads $D^\mu
= \partial^\mu + ig A^\mu$.

The expression in \eqref{dy-coll-a} has the structure $R^\alpha H_\alpha$,
where $R$ is the matrix element of the right-moving proton and $H$ a
hard-scattering amplitude.  One therefore has $|R^+| \gg |R^-|,
|\tvec{R}|$, whereas all components of $H^\alpha$ are generically of the
same size.  To leading-power accuracy we therefore have $R^\alpha H_\alpha
\approx R^+ H^-$.  We now introduce an auxiliary spacelike vector $v$ with
\begin{align}
  \label{v-left}
v^- &> 0 \,, & v^+ &< 0 \,, & \tvec{v} = \tvec{0}
\end{align}
and either $|v^+| \sim v^-$ or $|v^+| \ll v^-$.  We can then write
\begin{align}
  \label{grammer-yennie-1}
R H \approx R^+ H^- 
  &= R^+ v^-  \frac{1}{\ell^+ v^- + i\epsilon}\, \ell^+ H^-
   \approx R v\,  \frac{1}{\ell v + i\epsilon}\, \ell H \,,
\end{align}
where in the last step we have used the conditions on the components of
$R^\alpha$, $H^\alpha$ and $v^\alpha$ just stated, as well as $|\ell^+|
\gg |\ell^-|, |\tvec{\ell}|$ for the momentum $\ell$ of the right-moving
gluon.  For reasons given in the next section, we have provided an
$i\epsilon$ prescription to the factor $1/\ell v$ in
\eqref{grammer-yennie-1} such that the pole in $\ell^+$ is on the same
side of the real axis as in the propagator of the off-shell quark that
couples to the photon in figure~\ref{fig:dy-coll}a:
\begin{align}
  \label{coll-denominator-1}
\ell v + i\epsilon &= \ell^+ v^- - \ell^- |v^+| + i\epsilon \,,
&
(\ell + \bar{l}_c)^2 + i\epsilon &= 2 \ell^+ \bar{l}^- + i\epsilon \,,
\end{align}
where it is important that $\bar{l}^- > 0$.  With \eqref{grammer-yennie-1}
we can rewrite \eqref{dy-coll-a} as
\begin{align}
  \label{dy-coll-a-inter}
T_a &=
\langle \ldots \bar{q}_j\, A^{\alpha,a} \ldots \rangle
  (-ig t^{a}_{jj'}\ms v_\alpha)\, \frac{i}{\ell v + i\epsilon}
 \biggl[ \frac{1}{\gamma (\ell + \bar{l}_c)}\,
   (\gamma\ell)\, u(\bar{l}_c) \biggr]
  \langle \ldots q_{j'} \ldots \rangle \,.
\end{align}
With $\gamma\ell = \gamma\ms (\ell + \bar{l}_c) -
\gamma\bar{l}_c$ and $(\gamma \bar{l}_c)\, u(\bar{l}_c) = 0$ we
finally obtain
\begin{align}
  \label{dy-coll-a-fin}
T_a &=
\langle \ldots \bar{q}_j\, A^{\alpha,a} \ldots \rangle
  (-ig t^{a}_{jj'}\ms v_\alpha)\, \frac{i}{\ell v + i\epsilon}\,
  u(\bar{l}_c)\, \langle \ldots q_{j'} \ldots \rangle \,.
\end{align}
In the hard-scattering amplitude we have thus traded the coupling $-ig
t^{a} \gamma_\alpha$ of the gluon to the quark and the adjacent quark
propagator $i\big/ \gamma (\ell + \bar{l}_c)$ for the coupling $-ig t^a
v_\alpha$ of the gluon to a so-called eikonal line and the eikonal
propagator $i /(\ell v + i\epsilon)$.

Repeating the same steps for the graph in figure~\ref{fig:dy-coll}b
gives
\begin{align}
  \label{dy-coll-b}
T_{b} &=
\langle \ldots \bar{q}_j \ldots \rangle\,
  \bar{v}(\bar{l}_c)\, \gamma_\alpha\,
    \frac{-i}{\gamma\ms (\ell + \bar{l}_c) \rule{0pt}{2.2ex}}\,
    (-ig)\, t^{a}_{jj'}\,
  \langle \ldots A^{\alpha,a}\, q_{j'} \ldots \rangle
\nonumber \\
&= \langle \ldots \bar{q}_j \ldots \rangle\,
  \bar{v}(\bar{l}_c)\, \frac{-i}{\ell v + i\epsilon}\,
    (-ig t^{a}_{jj'}\ms v_\alpha)\,
  \langle \ldots A^{\alpha,a}\, q_{j'} \ldots \rangle \,.
\end{align}
The change from an incoming quark to an incoming antiquark in the hard
scattering has changed the overall sign of the propagator $i\big/ \gamma
(\ell + \bar{l}_c)$, which is reflected in an overall sign change of the
eikonal propagator $i /(\ell v + i\epsilon)$.  On the other hand, the
momentum flow in the graph and the resulting $i\epsilon$ prescriptions
have remained the same.

\begin{figure}
\begin{center}
\includegraphics[width=0.65\textwidth]{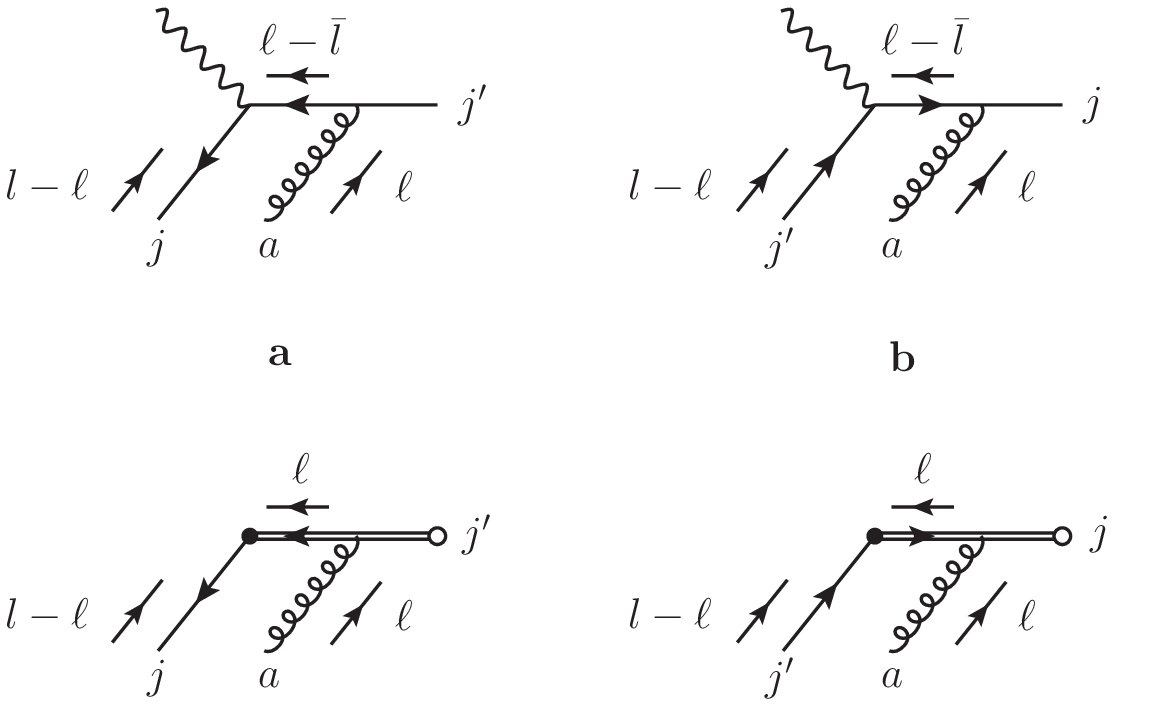}
\end{center}
\caption{\label{fig:sidis-coll} As figure~\protect\ref{fig:dy-coll}, but
  for semi-inclusive deep inelastic scattering.}
\end{figure}

It is instructive in this context to compare Drell-Yan production with
SIDIS, where one has an outgoing quark or antiquark in the hard
scattering.  The corresponding graphs are shown in figure
~\ref{fig:sidis-coll}a and b.  Taking the same vector $v$ as before, we
have
\begin{align}
  \label{sidis-coll}
T'_{a} &=
\langle \ldots \bar{q}_j\, A^{\alpha,a} \ldots \rangle\,
  \frac{i}{\gamma\ms (\ell - \bar{l}_c) \rule{0pt}{2.2ex}}\,
    (-ig)\, t^{a}_{jj'} \gamma_\alpha\, v(\bar{l}_c)\,
  \langle \ldots q_{j'} \ldots \rangle \,,
\nonumber \\
 &= \langle \ldots \bar{q}_j\, A^{\alpha,a} \ldots \rangle
  (-ig t^{a}_{jj'} v_\alpha)\, \frac{i}{\ell v - i\epsilon}\,
  v(\bar{l}_c)\, \langle \ldots q_{j'} \ldots \rangle \,.
\nonumber \\
T'_{b} &=
\langle \ldots \bar{q}_j \ldots \rangle\,
  \bar{u}(\bar{l}_c)\, \gamma_\alpha\,
    \frac{-i}{\gamma\ms (\ell - \bar{l}_c) \rule{0pt}{2.2ex}}\,
    (-ig)\, t^{a}_{jj'}\,
  \langle \ldots A^{\alpha,a}\, q_{j'} \ldots \rangle
\nonumber \\
&= \langle \ldots \bar{q}_j \ldots \rangle\,
  \bar{u}(\bar{l}_c)\, \frac{-i}{\ell v - i\epsilon}\,
    (-ig t^{a}_{jj'} v_\alpha)\,
  \langle \ldots A^{\alpha,a}\, q_{j'} \ldots \rangle \,.
\end{align}
Compared with graphs \ref{fig:dy-coll}a and b, the relative flow of the
momenta $\ell$ and $\bar{l}_c$ in the off-shell quark or antiquark has
changed.  Hence the corresponding propagator has a denominator
\begin{align}
(\ell - \bar{l}_c)^2 + i\epsilon &= - 2 \ell^+ \bar{l}^- + i\epsilon
  \label{coll-denominator-2}
\end{align}
instead of the one in the second equation of \eqref{coll-denominator-1}.
As a result, the sign of $i\epsilon$ in the eikonal propagator is now
reversed.

\begin{figure}
\begin{center}
\includegraphics[width=0.6\textwidth]{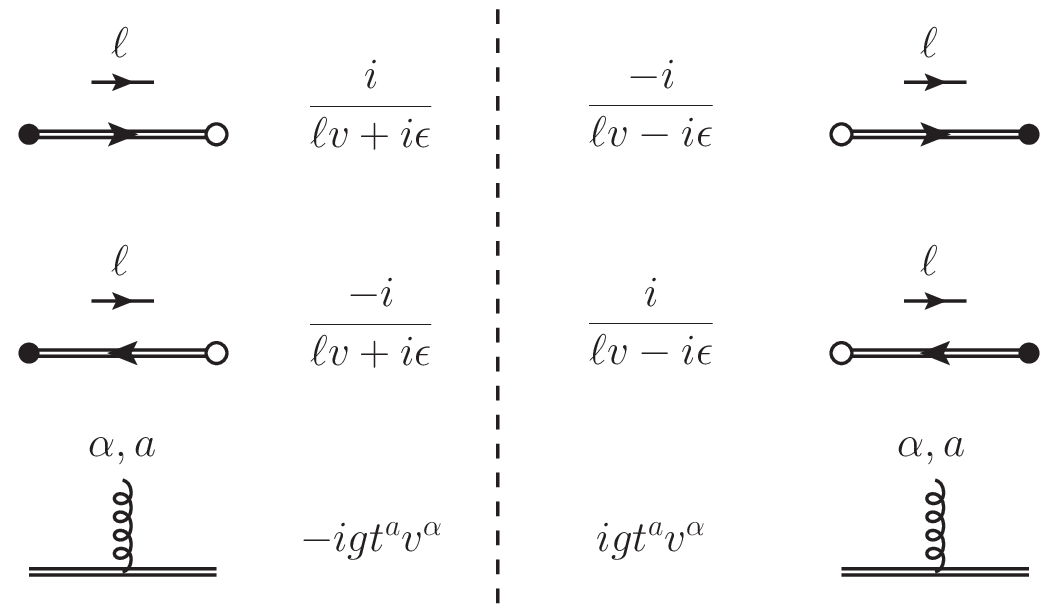}
\end{center}
\caption{\label{fig:eikonal-rules} Feynman rules for eikonal lines
  representing quarks or antiquarks.  The rules for lines to the left of
  the final-state cut (denoted by the dashed line) directly follow from
  \protect\eqref{dy-coll-a-fin} to \protect\eqref{sidis-coll}, and those
  for lines to the right of the final-state cut are obtained as usual by
  complex conjugation.}
\end{figure}

A graphical notation for eikonal lines needs to specify the flow of the
momentum $\ell$ relative to
\begin{enumerate}
\item the color flow (and hence the fermion number flow in the quark line
  which is represented by the eikonal line).  This determines the overall
  sign of the eikonal propagator.  We denote the color flow by an arrow on
  the eikonal line, which points in the same direction as the arrow on the
  original fermion line.
\item the flow of the large momentum $\bar{l}_c$ in the original fermion
  line, which is either an incoming or an outgoing line in the
  hard-scattering subprocess.  This determines the sign of $i\epsilon$ in
  the eikonal propagator.  We indicate this graphically by a full or an
  empty circle at the end of the eikonal line, such that the large
  momentum flows from the full to the empty circle.  Since incoming and
  outgoing partons in the hard scattering can be associated with a
  classical path in space-time according to
  section~\ref{sec:libby-sterman}, the full circle represents the past and
  the empty circle the future time direction.
\end{enumerate}
The corresponding Feynman rules are given in
figure~\ref{fig:eikonal-rules}, and the graphs corresponding to the
eikonal representation in \eqref{dy-coll-a-fin}, \eqref{dy-coll-b} and
\eqref{sidis-coll} are shown in the bottom rows of
figures~\ref{fig:dy-coll} and \ref{fig:sidis-coll}.\footnote{Our graphical
  notation differs from that in the literature.  In
  \protect\cite{Collins:1981uk} for instance, an arrow on the eikonal line
  was associated with the flow of the large momentum, and the overall sign
  due to the color flow was indicated by explicit color indices and taken
  into account in the vertex between a gluon and an eikonal line, rather
  than in the eikonal propagator.}

We now briefly review how eikonal lines are generated by Wilson line
operators in the hadronic matrix elements that appear in a factorization
formula.  The relevant part of the expression \eqref{dy-coll-a-fin},
together with the relevant integrations over momentum and position
variables reads
\begin{align}
X_{j'} &= \int d^4\ell\,
  e^{i \xi (l - \ell)}\, \bar{q}_{j}(\xi)
  \int \frac{d^4 \zeta}{(2\pi)^4}\, e^{i \zeta \ell}\, 
  v A^{a}(\zeta)\, (-ig t^a_{jj'})\, \frac{i}{\ell v + i\epsilon} \,.
\end{align}
Using the representation
\begin{align}
  \label{eikonal-fourier}
\frac{i}{\ell v + i\epsilon} &= \int_0^{\infty} d\lambda\,
  e^{i \lambda (\ell v + i\epsilon)}
\end{align}
we can rewrite this as
\begin{align}
  \label{dy-wilson-a}
X_{j'} &= e^{i \xi l}\ms \bar{q}_{j}(\xi) \int d^4\ell\,
  \int \frac{d^4 \zeta}{(2\pi)^4}\,
  \int_0^{\infty} d\lambda\; e^{i (\lambda v + \zeta - \xi) \ell}\,
  v A^{a}(\zeta)\, (-ig t^a_{jj'})
\nonumber \\
 &= e^{i \xi l}\ms \bar{q}_{j}(\xi)
    \biggl[ -ig \int_0^{\infty} d\lambda\,
    v A^a(\xi - \lambda v)\, t^a_{jj'} \biggr] \,.
\end{align}
Introducing the Wilson line
\begin{align}
  \label{wilson-dy}
W(\xi; v) &= \operatorname{P} \exp\biggl[ ig \int_0^{\infty}
             d\lambda\, v A^a(\xi - \lambda v)\, t^a \biggr] \,,
\end{align}
where $\operatorname{P}$ denotes path ordering, we recognize the term in
square brackets in \eqref{dy-wilson-a} as the term of order $g$ in the
expansion of $W^\dagger(\xi; v)$.  In a full factorization proof, one has
to show that the coupling of two or more collinear gluons to the incoming
quark line in figure~\ref{fig:dy-coll}a exponentiates, so that their
combined effect is the replacement
\begin{align}
\bar{q}_{j'}(\xi) &\to \bar{q}_{j}(\xi)\,
  \bigl[ W^\dagger(\xi; v) \bigr]{}_{jj'}
\end{align}
in the operator defining the parton distribution.  Likewise, the
expression in \eqref{dy-coll-b} corresponds to the one-gluon term in the
replacement
\begin{align}
q_{j}(\xi) &\to \bigl[ W(\xi; v) \bigr]{}_{jj'}\, q_{j'}(\xi) \,.
\end{align}
The conditions we imposed on $v$ after \eqref{v-left} hold in a frame
where $p$ moves fast to the right.  One readily finds that in the rest
frame of $p$ one has $v^0 > 0$, so that the Wilson line \eqref{wilson-dy}
relevant for Drell-Yan production has a path pointing into the past.  By
contrast, the reversed sign of $i\epsilon$ in the eikonal propagators for
SIDIS corresponds to
\begin{align}
\bar{q}_{j'}(\xi) &\underset{\text{SIDIS}}{\to} \bar{q}_{j}(\xi)\,
  \bigl[ W'^{\ms\dagger}(\xi; v) \bigr]{}_{jj'} \,,
&
q_{j}(\xi) &\underset{\text{SIDIS}}{\to}
  \bigl[ W'(\xi; v) \bigr]{}_{jj'}\, q_{j'}(\xi)
\end{align}
with a future-pointing Wilson line
\begin{align}
  \label{wilson-sidis}
W'(\xi; v) &= \operatorname{P} \exp\biggl[ - ig \int_0^{\infty}
             d\lambda\, v A^a(\xi + \lambda v)\, t^a \biggr] \,.
\end{align}
The preceding discussion was for right-moving collinear gluons and
generalizes trivially to left-moving collinear gluons in the proton with
momentum $\bar{p}$.  The corresponding Wilson lines are to be defined with
an auxiliary vector $w$ that satisfies
\begin{align}
  \label{w-right}
w^+ &> 0 \,, & w^- &< 0 \,, & \tvec{w} = \tvec{0}
\end{align}
and either $|w^-| \sim w^+$ or $|w^-| \ll w^+$ in a frame where $\bar{p}$
moves fast to the left.  In the rest frame of $\bar{p}$ one then has $w^0
> 0$.

The manipulations in the preceding arguments are all concerned with a
single hard-scattering subprocess at a time, so that they readily apply to
double Drell-Yan graphs such as in figure~\ref{fig:many-coll}, where they
give the order $g$ part of a Wilson line for each quark or antiquark
operator in the multiparton distributions.  The full operator for a
two-quark distribution then reads for instance
\begin{align}
  \label{full-wilson-op}
& \bigl[\ms \bar{q}(- \half z_2)\ms 
     W^\dagger(- \half z_2; v) \bigr]{}_{k'}\, \Gamma_{a_2} \,
  \bigl[ W(\half z_2;v)\, q(\half z_2) \bigr]{}_{k}\,
\nonumber \\
& \quad \times
\bigl[\ms \bar{q}(y - \half z_1)\ms
     W^\dagger(y - \half z_1; v) \bigr]{}_{j'}\, \Gamma_{a_1} \,
  \bigl[ W(y + \half z_1;v)\, q(y + \half z_1) \bigr]{}_{j}\,
\Big|_{z_2^+ = z_1^+ = y^+ = 0} \,.
\end{align}
The open color indices $j, j', k, k'$, which were carried by quark fields
in the lowest-order formula, are now carried by the ``ends'' of the four
past-pointing Wilson lines.  The projection on color singlet and color
octet distributions is done as in \eqref{quark-color-decomp}.

Let us now mention how the previous arguments need to be generalized to
obtain a complete factorization proof for double Drell-Yan production.
\begin{itemize}
\item The step from \eqref{dy-coll-a-inter} to \eqref{dy-coll-a-fin},
  which eliminates an internal fermion propagator in the hard-scattering
  graph, is elementary when applied to the lowest-order hard scattering.
  For more complicated graphs (with loop corrections or further external
  gluons) one needs a Ward identity to achieve this simplification.  In a
  model theory with Abelian gluons, this is quite simple to establish, see
  e.g.\ \cite[chapter 10.8]{Collins:2011}.  The formulation for QCD is
  more complicated and involves external ghost lines in addition to
  external gluons in the hard scattering (see \cite[chapters 11.3 and
  11.9]{Collins:2011}).
\item We have considered only one gluon coupling to each hard-scattering
  subgraph.  One needs to show that the coupling of an arbitrary number of
  gluons exponentiates and gives a full Wilson line $W(y;v)$ or its
  complex conjugate.  Again, this is simple to show for Abelian gluons
  (see \cite[chapter 10.8]{Collins:2011}).  To the best of our knowledge,
  an explicit proof for transverse-momentum dependent distributions in QCD
  has not yet been given.

  We note that the present and the previous point only concern one
  hard-scattering subprocess at a time.  It should therefore be
  straightforward to extend arguments valid for the single Drell-Yan
  process to the case of double Drell-Yan production.
\item The two Wilson lines $W(\half z_2;v)$ and $W(y + \half z_1;v)$ in
  \eqref{full-wilson-op} correspond to gluons in the scattering amplitude,
  where all gluons fields should be time ordered.  With $v^2<0$ the gluon
  operators in one Wilson line have a spacelike separation, so that they
  commute and can readily be brought into the order required by path
  ordering.  Two gluon operators in different Wilson lines do not
  necessarily have this property, and the possibility to reorder the
  fields needs to be investigated.  A similar statement holds for the two
  Wilson lines $ W^\dagger(- \half z_2; v)$ and $W^\dagger(y - \half z_1;
  v)$ that correspond to gluons in the conjugate scattering amplitude.
\item The operator in \eqref{full-wilson-op} is not explicitly gauge
  invariant, because the Wilson lines end at different positions at
  infinity, namely at $a_i - \infty\ms v$ with finite spacelike $a_i$ for
  $i=1,2,3,4$.  The same issue already arises for single-parton
  distributions and has been discussed in
  \cite{Belitsky:2002sm,Boer:2003cm} for lightlike Wilson lines, i.e.\ for
  $v^2=0$.  In a gauge where the gluon potential (and any product of gluon
  potentials) has zero expectation value at $a - \infty\ms v$, one can
  trivially complement the operator \eqref{full-wilson-op} with Wilson
  lines that go in the transverse direction and connect the lightlike
  Wilson lines to a common reference point, e.g.\ to $- \infty\ms v$.
  After projecting the open color indices at this reference point onto
  color-singlet or color-octet combinations, the resulting operator is
  explicitly invariant under local gauge transformations.  The extra
  Wilson lines in the transverse direction are essential in the gauge $v A
  = 0$, where the Wilson lines in \eqref{full-wilson-op} reduce to unity,
  see the discussion in \cite{Belitsky:2002sm}.

  As we will see in section~\ref{sec:full-fact}, the choice $v^2 = 0$ is
  not suitable for transverse-momentum dependent factorization.  To obtain
  a gauge invariant definition of the relevant parton distributions, one
  needs to extend the procedure just described to the case where $v^2 <
  0$.  This holds both for single and multiple hard scattering.
\end{itemize}


\subsubsection{Soft gluons and the soft factor}
\label{sec:soft}

We now turn to the exchange of soft gluons between right- and left-moving
partons, i.e.\ to the soft subgraph in figure~\ref{fig:hard-coll-soft},
and show how it can be described in terms of a soft factor that is defined
as the vacuum expectation value of Wilson lines.

\begin{figure}
\begin{center}
\includegraphics[width=0.67\textwidth]{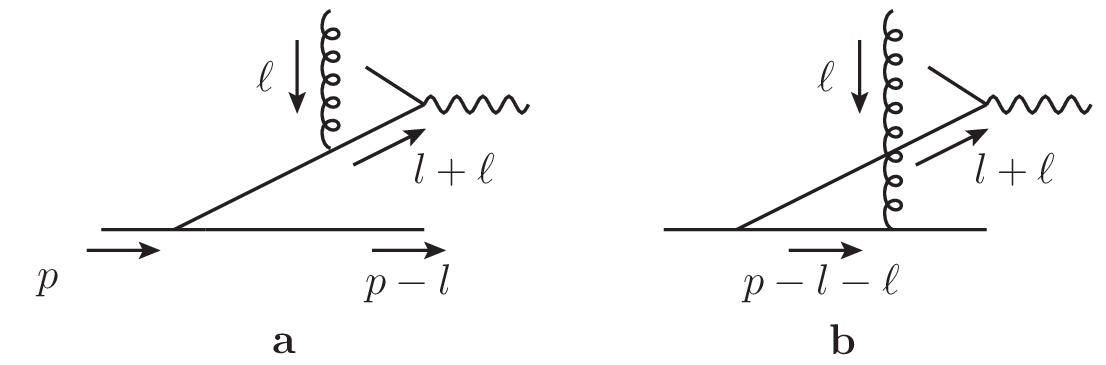}
\end{center}
\caption{\label{fig:soft-gluon} Coupling of a soft gluon to a collinear
  parton that (a) enters the hard scattering or (b) is a spectator.}
\end{figure}

For definiteness we consider one soft gluon with momentum $\ell$,
exchanged between the soft subgraph $S^\alpha$ and the collinear subgraph
$R^\alpha$ of the right-moving partons.  Here $\alpha$ is the polarization
index of the gluon considered, and the indices for any other external
gluons are omitted for simplicity.  We assume that the components of
$\ell$ are of comparable size, $|\ell^+| \sim |\ell^-| \sim
|\tvec{\ell}|$, as well as the momentum components of all other soft
gluons attached to~$S$.  The components of $S^\alpha$ are then also
comparable to each other.  Since $|R^+| \gg |R^-|, |\tvec{R}|$ we then
have
\begin{align}
  \label{soft-approx-0}
\ell R & \approx \ell^- R^+ \,, &
S R & \approx S^{-} R^+ \,.
\end{align}
Introducing an auxiliary spacelike vector $w$ as in \eqref{w-right} with
$|w^-| \ll w^+$, we furthermore have $S w \approx S^- w^+$, so that we can
write
\begin{align}
  \label{soft-approx-1}
S_{\alpha} R^\alpha &\approx 
   S^- \frac{w^+}{\ell^- w^+ + i\epsilon}\, \ell^- R^+
\approx S_{\alpha}\, \frac{w^\alpha}{\ell w + i\epsilon}\, \ell R
\end{align}
with a factor $w^\alpha /(\ell w + i\epsilon)$ that will eventually turn
into a Wilson line.  The $i\epsilon$ prescription for the pole at $\ell w
= 0$ is adequate for $\ell$ flowing from $S$ into $R$ in the scattering
amplitude, i.e.\ on the left of the final-state cut in
figure~\ref{fig:hard-coll-soft}.  We note that this prescription
corresponds to the one for collinear gluons in the previous section, cf.\
figures~\ref{fig:dy-coll} and~\ref{fig:soft-gluon}a.

The approximations in \eqref{soft-approx-0} and \eqref{soft-approx-1}
break down in the so-called Glauber region, i.e.\ for soft momenta
dominated by their transverse components, $|\tvec{\ell}| \gg |\ell^+|,
|\ell^-|$.  A major part of a factorization proof for hadron-hadron
collisions is to establish that this momentum region does not contribute
to the final factorization formula.  With the $i\epsilon$ prescription we
have chosen, the pole of $1/(\ell w + i\epsilon)$ is on the same side of
the real $\ell^-$ axis as the propagator pole of the quark with momentum
$l + \ell$ in figure~\ref{fig:soft-gluon}a, which is readily seen by
adapting \eqref{coll-denominator-1}.  In the graph of
figure~\ref{fig:soft-gluon}a one can avoid the Glauber region by a contour
deformation to complex $\ell^-$.  With the same contour deformation one
can however not avoid propagator poles in graphs where the gluon couples
to a spectator parton (rather than to the parton entering the hard
subprocess).  An example is shown in figure~\ref{fig:soft-gluon}b, where
the pole in $\ell^-$ of the propagator for the line $p-l-\ell$ is on the
opposite side of the real axis than the pole of the propagator for the
line $l+\ell$ in figure~\ref{fig:soft-gluon}a.  To apply the Ward
identities discussed below, one has to make the same contour deformation
for $\ell^-$ in both graphs of figure~\ref{fig:soft-gluon} and will thus
pick up a residue contribution from the propagator of the spectator
parton.  For single Drell-Yan production one can show that the sum over
all such residue contributions cancels due to unitarity, see the
discussion in \cite[chapters 14.3 and 14.4]{Collins:2011} and in the
original literature cited therein.  We do not know whether and how such
arguments can be extended to the case of double hard scattering and leave
this issue as an important task for further investigation.  We will
proceed under the assumption that such an extension can be made.

\begin{figure}
\begin{center}
\includegraphics[width=0.99\textwidth,bb=71 477 975 720]{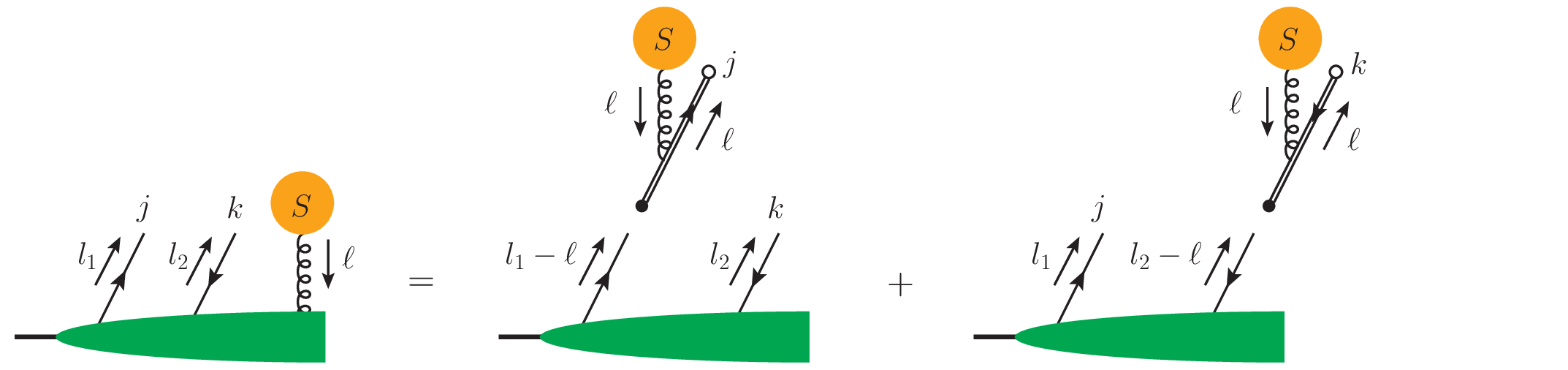}
\end{center}
\caption{\label{fig:ward-0} Graphical illustration of the Ward identity
  \protect\eqref{soft-Ward}.  Shown is only the part of the
  quark-antiquark distribution to the left of the final-state cut.  The
  Feynman rules for eikonal lines are given in
  figure~\protect\ref{fig:eikonal-rules}.}
\end{figure}

Following the procedure for single Drell-Yan production, the next step in
our argument is to use a Ward identity to relate the collinear subgraph
with a gluon attachment to the same subgraph without a gluon.  For the
correlation function describing quark-antiquark emission and an additional
gluon in the amplitude, this identity reads
\begin{align}
  \label{soft-Ward}
\frac{S w}{\ell w + i\epsilon}\;  \ell_\alpha\ms
  \Phi^{\alpha, a}_{jj', kk'}(\ell; l_1^{}, l_2^{}, l_1', l_2')
 &= S w\, (-ig t^{a}_{jm})\, \frac{i}{\ell w + i\epsilon}\,
    \Phi_{mj', kk'}(l_1^{} - \ell, l_2^{}, l_1', l_2')
\nonumber \\
 &\quad
  + S w\, (-ig t^{a}_{mk})\, \frac{-i}{\ell w + i\epsilon}\,
    \Phi_{jj',mk'}(l_1^{}, l_2^{} - \ell, l_1', l_2')
\end{align}
and is depicted in figure \ref{fig:ward-0}.  Analogous identities can be
written down for the emission of two quarks or two antiquarks, with a
factor $i/(\ell w + i\epsilon)$ for each quark line and $-i/(\ell w +
i\epsilon)$ for each antiquark line in the amplitude.  We leave it to
future work to give a general proof of these identities, but verify them
here for two simple examples.

\begin{figure}
\begin{center}
\includegraphics[width=0.6\textwidth]{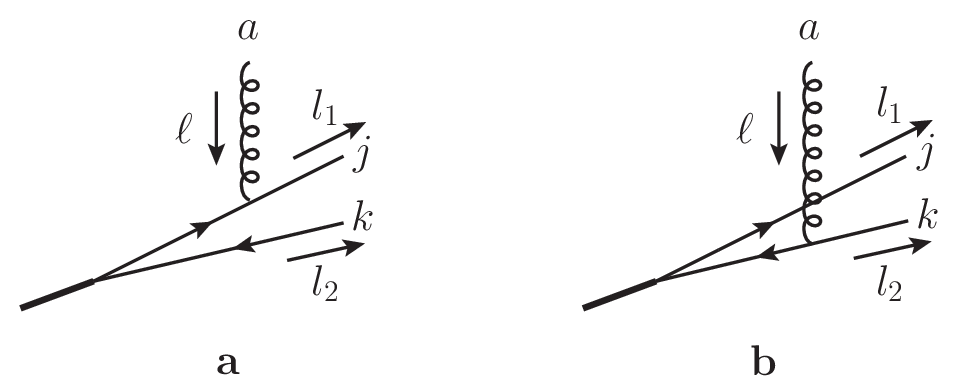}
\end{center}
\caption{\label{fig:ward-1} Graphs for a gluon coupling to a
  quark-antiquark system that originates from a colorless target via a
  pointlike vertex.  The indices $j$, $k$ and $a$ refer to color.}
\end{figure}

Our first example is a quark-antiquark pair with a pointlike coupling to a
target.  The corresponding two-parton distribution is then proportional to
$\delta_{jk}\, i\ms (\gamma\ms l_1)^{-1} \otimes (-i) (\gamma\ms
l_2)^{-1}$, where $l_1$ and $l_2$ are the respective momenta of the quark
and antiquark, and $j$ and $k$ are their respective color indices.  The
tensor product $\otimes$ refers to the spinor indices, whose coupling at
the vertex with the target we need not specify for our argument.
Attaching a soft gluon in the amplitude, we have to add the graphs in
figure \ref{fig:ward-1}a and b.  Contracting the gluon polarization index
$\alpha$ with $\ell^\alpha$ and using the same trick as in the step from
\eqref{dy-coll-a-inter} to \eqref{dy-coll-a-fin}, we obtain
\begin{align}
& (-ig)\, t^a_{jk}\, \frac{i}{\gamma\ms l_1}\,
    (\gamma \ell)\, \frac{i}{\gamma\ms (l_1 - \ell)}
    \otimes \frac{-i}{\gamma\ms l_2}
+ (-ig)\, t^a_{jk}\, \frac{i}{\gamma\ms l_1} \otimes
    \frac{-i}{\gamma\ms (l_2 - \ell)}\, (\gamma \ell)\,
    \frac{-i}{\gamma\ms l_2}
\nonumber \\
&\quad = g\ms t^a_{jk}\,
   \biggl\{ \frac{i}{\gamma\ms (l_1 - \ell)}
          - \frac{i}{\gamma\ms l_1} \biggr\} \otimes
   \frac{-i}{\gamma\ms l_2}
 - g\ms t^a_{jk}\, \frac{i}{\gamma\ms l_1} \otimes
   \biggl\{ \frac{-i}{\gamma\ms (l_2 - \ell)}
          - \frac{-i}{\gamma\ms l_2} \biggr\}
\nonumber \\
&\quad = + i\, (-ig t^{a}_{jm})\, \delta_{mk}^{}\,
   \frac{i}{\gamma\ms (l_1 - \ell)} \otimes \frac{-i}{\gamma\ms l_2}
 - i\, (-ig t^{a}_{mk})\, \delta_{jm}^{}\,
   \frac{i}{\gamma\ms l_1} \otimes \frac{-i}{\gamma\ms (l_2 - \ell)} \,.
\end{align}
Multiplication with $S w/(\ell w + i\epsilon)$ gives
\eqref{soft-Ward} for this particular case.

\begin{figure}
\begin{center}
\includegraphics[width=0.95\textwidth]{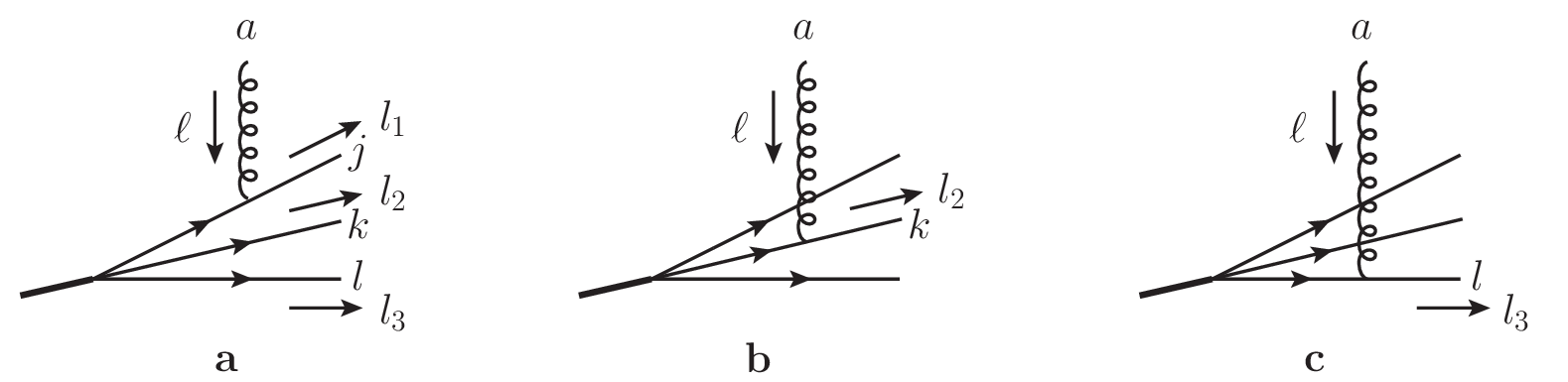}
\end{center}
\caption{\label{fig:ward-2} Graphs for a gluon coupling to a three-quark
  system that originates from a colorless fermion via a pointlike vertex.
  $j$, $k$, $l$ and $a$ are color indices.}
\end{figure}

As a second example we take a colorless fermion target coupled to three
quarks by a pointlike vertex.  The two-quark distribution is then
proportional to
\begin{align}
\epsilon_{jkl} \,\frac{i}{\gamma\ms l_1} \otimes
   \frac{i}{\gamma\ms l_2} \otimes \bar{u}(l_3) \,,
\end{align}
where $l_3$ is the momentum of the spectator quark.  Coupling a gluon to
this system, we get the three graphs shown in figure~\ref{fig:ward-2},
which after contraction with $\ell^\alpha$ give
\begin{align}
& (-ig)\, t^a_{jm} \epsilon^{}_{mkl}\, \frac{i}{\gamma\ms l_1}\,
    (\gamma \ell)\, \frac{i}{\gamma\ms (l_1 - \ell)} \otimes
    \frac{i}{\gamma\ms l_2} \otimes \bar{u}(l_3)
\nonumber \\
&\quad + (-ig)\, t^a_{km} \epsilon^{}_{jml}\, \frac{i}{\gamma\ms l_1}
    \otimes \frac{i}{\gamma\ms l_2}\, (\gamma \ell)\,
    \frac{i}{\gamma\ms (l_2 - \ell)} \otimes \bar{u}(l_3)
\nonumber \\
&\quad + (-ig)\, t^a_{lm} \epsilon^{}_{jkm}\, \frac{i}{\gamma\ms l_1}
    \otimes \frac{i}{\gamma\ms l_2} \otimes
    \bar{u}(l_3)\, (\gamma \ell)\, \frac{i}{\gamma\ms (l_3 - \ell)}
\nonumber \\
&= g\ms t^a_{jm} \epsilon^{}_{mkl}\,
   \biggl\{ \frac{i}{\gamma\ms (l_1 - \ell)}
            - \frac{i}{\gamma\ms l_1} \biggr\} \otimes
    \frac{i}{\gamma\ms l_2} \otimes \bar{u}(l_3)
\nonumber \\
&\quad + g\ms t^a_{km} \epsilon^{}_{jml}\, \frac{i}{\gamma\ms l_1}
    \otimes \biggl\{ \frac{i}{\gamma\ms (l_2 - \ell)}
                     - \frac{i}{\gamma\ms l_2} \biggr\}
    \otimes \bar{u}(l_3)
  - g\ms t^a_{lm} \epsilon^{}_{jkm}\, \frac{i}{\gamma\ms l_1} \otimes
    \frac{i}{\gamma\ms l_2} \otimes \bar{u}(l_3)
\nonumber \\
&= + i\, (-ig)\, t^{a}_{jm}\,
   \epsilon^{}_{mkl} \,\frac{i}{\gamma\ms (l_1 - \ell)} \otimes
   \frac{i}{\gamma\ms l_2} \otimes \bar{u}(l_3)
 + i\, (-ig)\, t^{a}_{km}\,
   \epsilon^{}_{jml} \,\frac{i}{\gamma\ms l_1} \otimes
   \frac{i}{\gamma\ms (l_2 - \ell)} \otimes \bar{u}(l_3)
\nonumber \\
&\quad - g\ms V^a_{ijk}\,\frac{i}{\gamma\ms l_1}
   \otimes \frac{i}{\gamma\ms l_2} \otimes \bar{u}(l_3)
\end{align}
with the tensor $V^a_{ijk} = t^a_{jm} \epsilon^{}_{mkl} + t^a_{km}
\epsilon^{}_{jml} + t^a_{lm} \epsilon^{}_{jkm}$.  This tensor is zero,
because it is completely antisymmetric and hence proportional to
$V^a_{ijk}\, \epsilon^{}_{ijk} = 2 (t^a_{jj} + t^{a}_{kk} + t^{a}_{ll}) =
0$.  Multiplication with $S w/(\ell w + i\epsilon^{})$ finally gives the
equivalent of \eqref{soft-Ward} for a two-quark distribution.

The preceding arguments can readily be adapted for a soft gluon attached
to left-moving collinear partons by exchanging $+$ and $-$ components of
the relevant vectors.  The auxiliary vector $w$ is then replaced by $v$ as
in \eqref{v-left}, with $|v^+| \ll v^-$.  Likewise, one can repeat all
arguments for soft gluons in the conjugate amplitude, i.e.\ to the right
of the final-state cut in figure~\ref{fig:hard-coll-soft}.  In the
corresponding Ward identities one then has to use the Feynman rules on the
r.h.s.\ of figure~\ref{fig:eikonal-rules}.

\begin{figure}
\begin{center}
\includegraphics[width=0.5\textwidth]{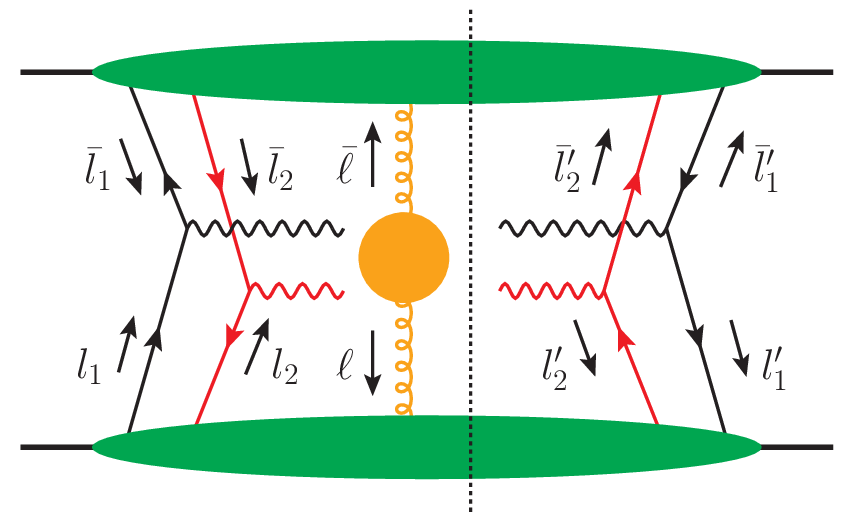}
\end{center}
\caption{\label{fig:soft-proc} Diagram with one soft gluon exchanged
  between the left- and right-moving partons to the left of the
  final-state cut.}
\end{figure}

Consider now the diagram in figure~\ref{fig:soft-proc}, where in the
amplitude one gluon is exchanged between the left- and right-moving
partons.  Its contribution to the cross section is proportional to
\begin{align}
  \label{soft-0}
& \int \frac{d^4\ell}{(2\pi)^4}\, \frac{d^4\bar{\ell}}{(2\pi)^4}\,
  (2\pi)^4 \delta^{(4)}(\ell + \bar{\ell})\,
  S_{\alpha\beta}^{ab}(\ell, \bar{\ell})\,
\nonumber \\
&\quad \times \biggl[\, \prod_{i=1}^2 \int d^4 l_i\, d^4\bar{l}_i\,
          (2\pi)^4 \delta^{(4)}(q_i - l_i - \bar{l}_i) \,\biggr]
 \int d^4 l'_1\, d^4\bar{l}'_1\,
          (2\pi)^4 \delta^{(4)}(q_i - l'_1 - \bar{l}'_1)\,
\nonumber \\
&\quad \times 
 \bigl[\Phi_{a_1,\bar{a}_2}\bigr]{}^{\alpha, a}_{jj', kk'}(\ell;
        l_1^{}, l_2^{}, l_1', l_2')\,
 \bigl[\Phi_{\bar{a}_1,a_2}\bigr]{}^{\beta, b}_{jj', kk'}(\bar{\ell};
        \bar{l}_1^{}, \bar{l}_2^{}, \bar{l}_1', \bar{l}_2')
 \phantom{\int}
\nonumber \\
&\approx
\int \frac{d^4\ell}{(2\pi)^4}\, \frac{d^4\bar{\ell}}{(2\pi)^4}\,
  (2\pi)^4 \delta^{(4)}(\ell + \bar{\ell})\,
  \frac{i w^\alpha}{\ell w + i\epsilon}\,
  S_{\alpha\beta}^{ab}(\ell, \bar{\ell})\,
  \frac{i v^\beta}{\bar{\ell} v + i\epsilon}
\nonumber \\
&\quad \times
\biggl[\, \prod_{i=1}^2 \int d^4 l_i\, d^4\bar{l}_i\,
          (2\pi)^4 \delta^{(4)}(q_i - l_i - \bar{l}_i) \,\biggr]
  \int d^4 l'_1\, d^4\bar{l}'_1\,
          (2\pi)^4 \delta^{(4)}(q_1 - l'_1 - \bar{l}'_1)\,
\nonumber \\
&\quad \times
\biggl[ (-ig t^{a}_{jm})\ms
        \bigl[\Phi_{a_1,\bar{a}_2}\bigr]{}_{mj', kk'}(
              l_1^{} - \ell, l_2^{}, l_1', l_2')
      - (-ig t^{a}_{mk})\ms
        \bigl[\Phi_{a_1,\bar{a}_2}\bigr]{}_{jj', mk'}(
              l_1^{}, l_2^{} - \ell, l_1', l_2') 
\biggr]\,
\nonumber \\
&\quad \times
\biggl[ (-ig t^{b}_{kn})\ms
  \bigl[\Phi_{\bar{a}_1,a_2}\bigr]{}_{jj', nk'}(
        \bar{l}_1^{}, \bar{l}_2^{} - \bar{\ell}, \bar{l}_1', \bar{l}_2')
      - (-ig t^{b}_{nj})\ms
  \bigl[\Phi_{\bar{a}_1,a_2}\bigr]{}_{nj', kk'}(
        \bar{l}_1^{} - \bar{\ell}, \bar{l}_2^{}, \bar{l}_1', \bar{l}_2')
\biggr]
\nonumber \\
&=
\int \frac{d^4\ell}{(2\pi)^4}\, \frac{d^4\bar{\ell}}{(2\pi)^4}\,
  (2\pi)^4 \delta^{(4)}(\ell + \bar{\ell})\,
  (-ig t^{a}_{jm})\, \frac{i w^\alpha}{\ell w + i\epsilon}\,
  S_{\alpha\beta}^{ab}(\ell, \bar{\ell})\,
  (-ig t^{b}_{kn})\, \frac{i v^\beta}{\bar{\ell} v + i\epsilon}
\nonumber \\
&\quad \times
  \int d^4 l_1\, d^4\bar{l}_1\,
          (2\pi)^4 \delta^{(4)}(q_1 - l_1 - \bar{l}_1 - \ell)\,
  \int d^4 l_2\, d^4\bar{l}_2\,
          (2\pi)^4 \delta^{(4)}(q_2 - l_2 - \bar{l}_2 - \bar{\ell})\,
\nonumber \\
&\quad \times
  \int d^4 l'_1\, d^4\bar{l}'_1\,
          (2\pi)^4 \delta^{(4)}(q_1 - l'_1 - \bar{l}'_1)\,
\nonumber \\
&\quad \times
  \bigl[\Phi_{a_1,\bar{a}_2}\bigr]{}_{mj', kk'}(
        l_1^{}, l_2^{}, l_1', l_2')\,
  \bigl[\Phi_{\bar{a}_1,a_2}\bigr]{}_{jj', nk'}(
        \bar{l}_1^{}, \bar{l}_2^{}, \bar{l}_1', \bar{l}_2')
 + \{ \text{three more terms} \} \,,
 \phantom{\int}
\end{align}
where in the last step we have shifted the integration variables $l_1$ and
$\bar{l}_2$.  For simplicity we have omitted a global factor, as well as
the expressions for $q\bar{q}\to \gamma^*$, which in the hard-scattering
approximation only depend on the external momenta $q_1$ and $q_2$ and thus
do not appear under the loop integrals (see
section~\ref{sec:tree:cross-sect}).

To provide a representation beyond perturbation theory, we represent the
soft subgraph (which for two external gluons is just the gluon propagator)
as a matrix element,
\begin{align}
  \label{soft-graph-mat}
(2\pi)^4 \delta^{(4)}(\ell + \bar{\ell})\,
  S_{\alpha\beta}^{ab}(\ell, \bar{\ell})
&= \int d^4\xi\, d^4\bar{\xi}\;
  e^{i \xi \ell + i \bar{\xi} \bar{\ell}}\,
  \langle 0 | A_{\alpha}^a(\xi)\, A_{\beta}^b(\bar{\xi}) |\ms 0 \rangle \,.
\end{align}
Here we have omitted the time ordering between the fields, which requires
justification when $\xi$ and $\bar{\xi}$ do not have a spacelike
separation.  We gloss over this point here (see also our discussion at the
end of section~\ref{sec:coll}) but return to it briefly at the end of
section~\ref{sec:soft-basic}.
Using \eqref{soft-graph-mat} and \eqref{eikonal-fourier} we then have for
the first term in \eqref{soft-0}
\begin{align}
  \label{soft-1}
& \int \frac{d^4\bar{\ell}\, d^4\bar{\xi}}{(2\pi)^4}
  \int \frac{d^4\ell\, d^4\xi}{(2\pi)^4}\;
    e^{i \xi \ell + i \bar{\xi} \bar{\ell}}\,
  \int_0^{\infty}\!\!\! d\lambda \int_0^{\infty}\!\!\! d\bar{\lambda}\;
    e^{i \lambda \ell w + i \bar{\lambda} \bar{\ell} v}\,
 (-ig t^{a}_{jm})\, (-ig t^{b}_{kn})
    \langle 0 | w A^a(\xi)\, v A^b(\bar{\xi}) | 0 \rangle
\nonumber \\
&\quad \times
  \int d^4 l_1\, d^4\bar{l}_1\,
  (2\pi)^2\, \delta(q_1^+ - l_1^+)\, \delta(q_1^- - \bar{l}_1^-)
  \int d^2\tvec{\xi}_1\,
  e^{-i \tvec{\xi}_1 (\tvec{q}_1 - \tvec{l}_1^{}
                      - \tvec{\bar{l}}_1^{} - \tvec{\ell})}
\nonumber \\
&\quad \times
  \int d^4 l_2\, d^4\bar{l}_2\,
  (2\pi)^2\, \delta(q_2^+ - l_2^+)\, \delta(q_2^- - \bar{l}_2^-)
 \int d^2\tvec{\xi}_2\,
  e^{-i \tvec{\xi}_2 (\tvec{q}_2 - \tvec{l}_2^{}
                      - \tvec{\bar{l}}_2^{} - \bar{\tvec{\ell}})}
\nonumber \\
&\quad \times
  \int d^4 l_1'\, d^4\bar{l}_1'\,
  (2\pi)^2\, \delta(q_1^+ - l_1^{\prime +})\,
             \delta(q_1^- - \bar{l}_1^{\,\prime -})
  \int d^2\tvec{\xi}_1'\, e^{-i \tvec{\xi}_1'
  (\tvec{q}_1 - \tvec{l}_1' - \tvec{\bar{l}}_1^{\,\prime})}
\nonumber \\
&\quad \times
  \bigl[\Phi_{a_1,\bar{a}_2}\bigr]{}_{mj', kk'}(
        l_1^{}, l_2^{}, l_1', l_2')\,
  \bigl[\Phi_{\bar{a}_1,a_2}\bigr]{}_{jj', nk'}(
        \bar{l}_1^{}, \bar{l}_2^{}, \bar{l}_1', \bar{l}_2')
  \phantom{\int}
\nonumber \\
&= \int d^2\tvec{\xi}_1\, d^2\tvec{\xi}_1\, d^2\tvec{\xi}_1'\;
   e^{-i \tvec{\xi}_1 \tvec{q}_1 -i \tvec{\xi}_2 \tvec{q}_2
      -i \tvec{\xi}_1' \tvec{q}_1}
\nonumber \\
&\quad \times
  \int d^4 l_1^{}\, d^4 l_2^{}\, d^4 l_1'\;
  e^{i \tvec{\xi}_1 \tvec{l}_1^{} + i \tvec{\xi}_2 \tvec{l}_2^{}
   + i \tvec{\xi}_1' \tvec{l}_1^{\,\prime}}\,
  (2\pi)^3\, \delta(q_1^+ - l_1^+)\, \delta(q_2^+ - l_2^+)\,
     \delta(q_1^+ - l_1^{\prime +})\,
\nonumber \\
&\qquad \times
  \bigl[\Phi_{a_1,\bar{a}_2}\bigr]{}_{mj', kk'}(
        l_1^{}, l_2^{}, l_1', l_2')
  \phantom{\int}
\nonumber \\
&\quad \times
  \int d^4 \bar{l}_1^{}\, d^4 \bar{l}_2^{}\, d^4 \bar{l}_1'\;
  e^{i \tvec{\xi}_1 \bar{\tvec{l}}_1^{}
   + i \tvec{\xi}_2 \bar{\tvec{l}}_2^{}
   + i \tvec{\xi}_1' \bar{\tvec{l}}_1^{\,\prime}}\,
  (2\pi)^3\, \delta(q_1^- - \bar{l}_1^-)\, \delta(q_2^- - \bar{l}_2^-)\,
     \delta(q_1^- - \bar{l}_1^{\,\prime -})\,
\nonumber \\
&\qquad \times
  \bigl[\Phi_{\bar{a}_1,a_2}\bigr]{}_{jj', nk'}(
        \bar{l}_1^{}, \bar{l}_2^{}, \bar{l}_1', \bar{l}_2')
  \phantom{\int}
\nonumber \\
&\quad \times
  \bigl\langle 0 \big|
     \biggl[ -ig \int_0^{\infty}\!\!\! d\lambda\,
     w A^a(\xi_{1T} - \lambda w)\, t^{a}_{jm} \biggr]
     \biggl[ -ig \int_0^{\infty}\!\!\! d\bar{\lambda}\,
     v A^b(\xi_{2T} - \bar{\lambda} v)\, t^{b}_{kn} \biggr]
  \big| 0 \big\rangle \,,
\end{align}
where $\xi_{iT}^{}$ denotes the four-vector with $\xi_{iT}^+ = \xi_{iT}^-
= 0$ and transverse components $\tvec{\xi}_{i}$.  The corresponding
expression for the diagram without soft gluon exchange is obtained by
replacing the last line in \eqref{soft-1} by $\delta_{jm} \delta_{kn}$.
Using \eqref{wilson-dy}, we recognize the factors in square brackets in
that line as the order $g$ terms in conjugate Wilson lines
$W^\dagger(\xi_{1T}; w)$ and $W^\dagger(\xi_{2T}; v)$.  In the transverse
plane, the paths of these Wilson lines are at the positions that are
Fourier conjugate to the transverse quark momenta $\tvec{l}_1$ and
$\bar{\tvec{l}}_2$ in \eqref{soft-1}.  The three other terms in
\eqref{soft-0} give analogous contributions, with Wilson lines
$W(\xi_{2T}; w)$ and $W(\xi_{1T}, v)$ at the positions that are Fourier
conjugate to the transverse antiquark momenta $\tvec{l}_2$ and
$\bar{\tvec{l}}_1$, respectively.

After a change to symmetric momentum and position variables as specified
between \eqref{Phi-def} and \eqref{Phi-symmetric}, and after restoration
of global kinematic factors, the second to fifth lines on the r.h.s.\ of
\eqref{soft-1} turn into the product $F_{a_1,\bar{a}_2}(x_i, \tvec{z}_i,
\tvec{y}) F_{\bar{a}_1,a_2}(\bar{x}_i, \tvec{z}_i, \tvec{y})$ of
two-parton distributions in transverse position space, and the Wilson
lines are to be evaluated at the appropriate transverse positions of the
quark or antiquark fields in the definition of these distributions.

It is straightforward to repeat the preceding derivation for a soft gluon
exchanged to the right of the final-state cut, as well as for the case
where the gluon crosses this cut.  For an model theory with Abelian
gluons, it is not difficult to see how soft subgraphs with an arbitrary
number of external gluons add up to full Wilson lines, in close analogy to
the case of single Drell-Yan production.  We do not attempt here to give a
corresponding proof for the nonabelian theory, given that even for the
single Drell-Yan process this is quite involved.  The structure suggested
by our analysis of one-gluon exchange is however clear: the effect of all
soft subgraphs is to multiply the Born-level cross section
\eqref{X-sect-position} in position space representation by a soft factor.
This factor is the vacuum expectation value of a product of Wilson lines,
with one Wilson line for each external quark or antiquark in the
multiparton distributions.  We thus have
\begin{align}
  \label{X-sect-soft-1}
& \frac{d\sigma}{\prod_{i=1}^2 dx_i\, d\bar{x}_i\, d^2\tvec{q}{}_i}
 = \frac{1}{C}\, 
\biggl[\, \prod_{i=1}^{2} \,\hat{\sigma}_i(x_i \bar{x}_i s) \biggr]\,
\biggl[\, \prod_{i=1}^{2} 
     \int \frac{d^2\tvec{z}_i}{(2\pi)^2}\,
          e^{-i \tvec{z}_i \tvec{q}{}_i} \biggr]\, \int d^2\tvec{y}
\nonumber \\
& \qquad\times 
  \bigl[F_{\bar{a}_1,a_2}\bigr]{}_{mm',nn'}(
        \bar{x}_i, \tvec{z}_i, \tvec{y})\,
  \bigl[S_{q\bar{q}} \bigr]{}_{mm',nn';jj',kk'}(\tvec{z}_i, \tvec{y})
  \bigl[F_{a_1,\bar{a}_2}\bigr]{}_{jj',kk'}(
        x_i, \tvec{z}_i, \tvec{y})\, 
\phantom{\frac{1}{1}}
\nonumber \\
& \quad + \{ \text{further terms} \} \,,
\end{align}
where the ``further terms'' describe the remaining combinations of quarks
or antiquarks in the two-parton distributions, as discussed in
section~\ref{sec:spin:quarks}.  The soft factor reads
\begin{align}
  \label{double-soft-fact}
& \bigl[S_{q\bar{q}} \bigr]{}_{mm',nn';jj',kk'}(\tvec{z}_i, \tvec{y})
\nonumber \\
&\quad = \bigl\langle\ms 0 \,\big| 
   \bigl[ W(y_T + \half z_{1T}; v)\ms
          W^\dagger(y_T + \half z_{1T}; w) \bigr]{}_{mj}\,
   \bigl[ W(y_T - \half z_{1T}; w)\ms
          W^\dagger(y_T - \half z_{1T}; v) \bigr]{}_{j'm'}\,
\phantom{\int}
\nonumber \\
& \qquad \times \bigl[ W(\half z_{2T}; w)\ms
          W^\dagger(\half z_{2T}; v) \bigr]{}_{kn}\,
   \bigl[ W(- \half z_{2T}; v)\ms
          W^\dagger(- \half z_{2T}; w) \bigr]{}_{n'k'}
   \big|\, 0 \ms\bigr\rangle \,.
\end{align}
We notice that Wilson lines $W(\xi;v)\, W^\dagger(\xi;w)$ and $W(\xi;w)\,
W^\dagger(\xi;v)$ are contracted pairwise in their color indices.  {}From
our derivation we see that this color contraction follows from the fact
that the hard scatters produce color-singlet particles, so that the color
indices of annihilating quarks and antiquarks are directly contracted with
each other.  The ``further terms'' in \eqref{X-sect-soft-1} have a soft
factor $S_{qq}$ multiplying $F_{\bar{a}_1, \bar{a}_2}\ms F_{a_1, a_2}$ and
a soft factor $S_I$ multiplying the product of interference distributions
$I_{\bar{a}_1, a_2}\ms I_{a_1, \bar{a}_2}$.  These factors are defined in
analogy to \eqref{double-soft-fact} with an appropriate interchange of
arguments and indices for $W$ and $W^\dagger$.

In analogy to two-parton distributions, we can represent $S_{q\bar{q}}$ in
a singlet-octet basis for index pairs $jj'$, $kk'$, etc.
\begin{align}
  \label{soft-color-decomp}
& \bigl[S_{q\bar{q}} \bigr]{}_{mm',nn';jj',kk'} 
 = \frac{1}{N^2} \biggl[ {}^{11\!}S_{q\bar{q}} \; \delta_{mm'}^{}\ms
      \delta_{n'n}^{}\ms \delta_{j'j}^{}\ms \delta_{kk'}^{}
+ \frac{2N}{\sqrt{N^2-1}}\, {}^{18\!}S_{q\bar{q}} \;
    \delta_{mm'}^{}\ms \delta_{n'n}^{}\ms t^a_{j'j}\ms t^a_{kk'}
\nonumber \\
&\quad
+ \frac{2N}{\sqrt{N^2-1}}\, {}^{81\!}S_{q\bar{q}} \;
    t^b_{mm'}\ms t^b_{n'n}\ms \delta_{j'j}^{}\ms \delta_{kk'}^{}
+ \frac{4N^2}{N^2-1}\,  {}^{88\!}S_{q\bar{q}} \;
    t^b_{mm'}\ms t^b_{n'n}\ms t^a_{j'j}\ms t^a_{kk'} \biggr] \,.
\end{align}
Defining the matrix
\begin{align}
  \label{soft-matrix}
\mathbf{S}_{q\bar{q}} &= 
\begin{pmatrix}
    {}^{11\!}S_{q\bar{q}} \ & {}^{18\!}S_{q\bar{q}} \\
    {}^{81\!}S_{q\bar{q}} \ & {}^{88\!}S_{q\bar{q}}
\end{pmatrix}
\end{align}
we then have
\begin{align}
  \label{X-sect-soft-2}
& \frac{d\sigma}{\prod_{i=1}^2 dx_i\, d\bar{x}_i\, d^2\tvec{q}{}_i}
 = \frac{1}{C}\, 
\biggl[\, \prod_{i=1}^{2} \,\hat{\sigma}_i(x_i \bar{x}_i s) \biggr]\,
\biggl[\, \prod_{i=1}^{2} 
     \int \frac{d^2\tvec{z}_i}{(2\pi)^2}\,
          e^{-i \tvec{z}_i \tvec{q}{}_i} \biggr]\, \int d^2\tvec{y}
\nonumber \\[0.2em]
& \quad\times
  \begin{pmatrix}
    \sing{F}_{\bar{a_1},a_2}(\bar{x}_i, \tvec{z}_i, \tvec{y}) \\
    \oct{F}_{\bar{a}_1,a_2}(\bar{x}_i, \tvec{z}_i, \tvec{y})
  \end{pmatrix}{\rule{0pt}{3.5ex}}^{\!\! \rm T} \;
\mathbf{S}_{q\bar{q}}(\tvec{z}_i, \tvec{y})
 \begin{pmatrix}
    \sing{F}_{a_1,\bar{a}_2}(x_i, \tvec{z}_i, \tvec{y}) \\
    \oct{F}_{a_1,\bar{a}_2}(x_i, \tvec{z}_i, \tvec{y})
  \end{pmatrix}
+ \{ \text{further terms} \} \,.
\end{align}
One can of course rewrite the cross section in terms of distributions
$F(x_i, \tvec{k}_i, \tvec{y})$ or $F(x_i, \tvec{k}_i, \tvec{r})$ depending
on transverse momenta.  The result involves a Fourier transformed soft
factor and is a convolution in transverse-momentum variables.

The soft factor \eqref{double-soft-fact} for double Drell-Yan production
generalizes the corresponding factor appearing in the single Drell-Yan
process, which reads
\begin{align}
  \label{single-soft-fact}
S_{q}(\tvec{z})
&= \frac{1}{N}\, \bigl\langle\ms 0 \,\big| 
   \bigl[ W(\half z_{T}; v)\ms
          W^\dagger(\half z_{T}; w) \bigr]{}_{mj}\,
   \bigl[ W(- \half z_{T}; w)\ms
          W^\dagger(- \half z_{T}; v) \bigr]{}_{jm}\,
   \big|\, 0 \ms\bigr\rangle
\end{align}
for the annihilation of a right-moving quark with a left-moving antiquark.
The color indices are now contracted to an overall singlet, as they are in
${}^{11\!}S_{q\bar{q}}$.  The analog of \eqref{X-sect-soft-2} is
\begin{align}
  \label{X-sect-soft}
& \frac{d\sigma}{dx\, d\bar{x}\, d^2\tvec{q}}
 = \hat{\sigma}_i(x \bar{x} s)\,
   \int \frac{d^2\tvec{z}}{(2\pi)^2}\,
          e^{-i \tvec{z} \tvec{q}}\,
   f_{\bar{q}}(\bar{x}, \tvec{z})\, S_{q}(\tvec{z})\, f_q(x, \tvec{z})
+ \{ \text{further term} \} \,,
\end{align}
where the ``further term'' corresponds to a right-moving antiquark and a
left-moving quark.  For the discussion in subsequent sections we note that
at $\tvec{z} = \tvec{0}$ the product of Wilson lines in
\eqref{single-soft-fact} reduces to the trace of the unit matrix, so that
$S_{q}(\tvec{0}) = 1$.  Similarly, one finds from \eqref{double-soft-fact}
and \eqref{soft-color-decomp} that
\begin{align}
  \label{soft-matrix-unity}
\mathbf{S}_{q\bar{q}}(\tvec{z}_i = \tvec{0}, \tvec{y} = \tvec{0})
= \begin{pmatrix} 1 \ & 0 \\ 0 \ & 1
  \end{pmatrix} \,.
\end{align}

To close this section let us collect the issues in the soft-gluon sector
that need to be worked out for a full factorization proof.  Some of them
we have already mentioned.
\begin{itemize}
\item One needs to show that the exchange of gluons in the Glauber region
  cancels in the cross section.  Such a cancellation requires a specific
  choice of $i\epsilon$ prescription in the eikonal propagators.  For the
  prescription in \eqref{soft-approx-1}, which corresponds to
  past-pointing Wilson lines, one can show that Glauber gluons do cancel
  in single Drell-Yan production.  It is natural to expect that the same
  prescription is appropriate for the double Drell-Yan process, if there
  is \emph{any} choice for which Glauber gluons decouple in that case.
\item The Ward identity \eqref{soft-Ward} for attaching one gluon to a
  collinear subgraph needs to be proven, and it needs to be extended to
  the case where additional gluons are attached to the subgraph.  One then
  needs to show that the attachment of an arbitrary number of gluons
  exponentiates to the Wilson lines in the soft factor
  \eqref{double-soft-fact}.

  In a model theory with Abelian gluons, a corresponding proof should be a
  rather simple extension of the corresponding arguments for single-parton
  distributions, which can be found in \cite[chapter~10.8]{Collins:2011}.
  An explicit proof for transverse-momentum dependent factorization in QCD
  is still lacking even for single hard scattering, as far as we know.
\item It must be shown that explicit time ordering of the gluon operators
  in the soft factor \eqref{double-soft-fact} can be omitted.  It must
  also be established that one can complement the Wilson lines along $v$
  and $w$ in the soft factor in such a way that one has an explicitly
  gauge invariant definition.  We expect that for both issues it should be
  possible to extend a proof for single Drell-Yan production to the double
  Drell-Yan process, but we are not aware of an explicit proof for the
  single Drell-Yan case.
\end{itemize}
The second and third bullet items are closely connected with the
corresponding points for collinear gluons, which we discussed at the end
of the previous section.


\subsubsection{Towards a factorization formula}
\label{sec:full-fact}

In section~\ref{sec:coll} we have seen how collinear gluons give rise to
the Wilson line operators \eqref{full-wilson-op} in the matrix elements
defining multiparton distributions.  However, these Wilson line operators
contain not only collinear but also soft gluons, which are already taken
into account in the soft factor \eqref{double-soft-fact}.  At the level of
graphs, this is reflected in the fact that the gluon momentum $\ell$ in
figures \ref{fig:dy-coll} and \ref{fig:soft-gluon}a can be either
collinear or soft.  To prevent double counting of soft gluon
contributions, the factorization formula for the cross section requires
appropriate subtractions.

Let us briefly recapitulate how this problem can be solved for single
Drell-Yan production.  The necessary subtractions can be performed by
dividing out vacuum matrix elements of the form \eqref{single-soft-fact}.
There is a certain freedom of whether to absorb these matrix elements into
the soft factor or into the parton distributions that appear in the final
factorization formula .  The former choice was made in the original work
\cite{Collins:1981uk} of Collins and Soper,\footnote{This may not be quite
  obvious in \protect\cite{Collins:1981uk} but has been clearly pointed
  out in section~X.A of \protect\cite{Collins:2007ph}.}
whereas both \cite{Collins:2007ph} and \cite{Ji:2004wu} have made the
latter choice.  Finally, in recent work by Collins \cite{Collins:2011}
(see \cite{Aybat:2011zv} for a brief summary) all matrix elements of the
form \eqref{single-soft-fact} have been absorbed into the parton
distributions, which gives a factorization formula without an explicit
soft factor.  Whichever choice is made, a consistent formulation requires
one to take matching $i\epsilon$ prescriptions in eikonal lines when
treating collinear or soft gluon attachments, as we did in
\eqref{grammer-yennie-1} and \eqref{soft-approx-1}.

Another detail that admits several choices is the direction of the path in
Wilson lines.  In section~\ref{sec:coll} we have seen that the
approximations needed for right-moving collinear gluons require a vector
$v$ that corresponds either to large negative or to central rapidity,
where we define the rapidity of a spacelike vector as
\begin{align}
  \label{rap-def}
y_v = \frac{1}{2} \log \bigg| \frac{v^+}{v^-} \bigg| \,.
\end{align}
One cannot take the limit $y_v \to -\infty$, i.e.\ one cannot take $v$
lightlike in the parton density, since this would give divergences from
the region where gluons coupling to eikonal lines have small $\ell^+$ but
large $\ell^-$, i.e.\ from the region of large negative gluon rapidities
\cite{Collins:1981uk,Collins:2003fm,Collins:2011}.  The approximations for
soft gluons in section~\ref{sec:soft} require a vector $v$ with large
negative rapidity and a vector $w$ with large positive rapidity.  Again
one cannot take the limit where $y_v \to -\infty$ and $y_w \to +\infty$,
as we shall see explicitly in section~\ref{sec:soft-basic}.  However, this
limit can be taken for appropriate combinations of matrix elements, which
leads to important simplifications, see \cite{Collins:2007ph} and
\cite{Collins:2011}.

A further technical point is that the matrix elements discussed so far
include contributions from self energy graphs of Wilson lines and from
graphs where gluons are exchanged between different Wilson lines pointing
in the same direction (see e.g.\ figure~\ref{fig:wilson-self} below).  For
spacelike vectors $v$ and $w$ such graphs give infinite results already at
tree level, as shown in appendix~A of \cite{Bacchetta:2008xw}.  Such
graphs do not appear in the derivation of the factorization formula: as we
have seen in the two previous sections, Wilson lines appear when treating
gluon exchange between partons that have a large rapidity difference.  The
offending graphs cancel in the combination of matrix elements that appears
in the final factorization formula, but in the individual factors they
must be explicitly excluded.  (Only in the scheme of \cite{Collins:2011}
do these graphs already cancel in the parton distributions.)

Finally, the hard-scattering subgraphs have radiative corrections
themselves.  Since we require the produced bosons to have small transverse
momenta, there are only virtual corrections: radiation into the final
state can only be collinear or soft and is included in the collinear or
soft factors.  For Drell-Yan production at one-loop accuracy, one thus
only has the vertex correction to the quark-antiquark-photon three-point
function.  The regions of soft and collinear gluon momenta in the virtual
graphs have to be explicitly subtracted in the definition of the
hard-scattering cross section, in order to ensure that this factor is
dominated by large virtualities.  This removes in particular the
well-known soft divergences of the virtual graphs (which in the more
familiar case of inclusive observables cancel when real emission graphs
are added).

We expect that the above procedure can be generalized to the case of
double Drell-Yan production.  The division by vacuum expectation values of
the form \eqref{single-soft-fact} will be replaced by multiplication with
the inverse of the matrix \eqref{double-soft-fact} in color space.  We
leave it to future work to show that this can actually be done.  In the
remainder of this section, we will take a closer look at the elementary
building blocks of factorization, namely at the soft factor in
\eqref{double-soft-fact} and at the dependence of the proton matrix
elements of the operator \eqref{full-wilson-op} on the direction $v$ of
the Wilson lines.

To conclude this section we note that a factorization theorem for the
double Drell-Yan process also needs to provide a proper separation between
the production of the two gauge bosons by one or two hard-scattering
processes.  We will discuss this problem in
section~\ref{sec:splitting-X-sect}.


\subsection{The soft factor at small transverse distances}
\label{sec:soft-perturb}

If all three transverse distances $\tvec{y}, \tvec{z}_1, \tvec{z}_2$ are
small compared with a hadronic scale $\Lambda^{-1}$, the soft factor in
\eqref{double-soft-fact} is dominated by perturbative dynamics and can be
evaluated in perturbation theory.  In this section we compute the
short-distance form of this factor to leading order in the strong
coupling.

\begin{figure}
\begin{center}
\includegraphics[width=0.9\textwidth]{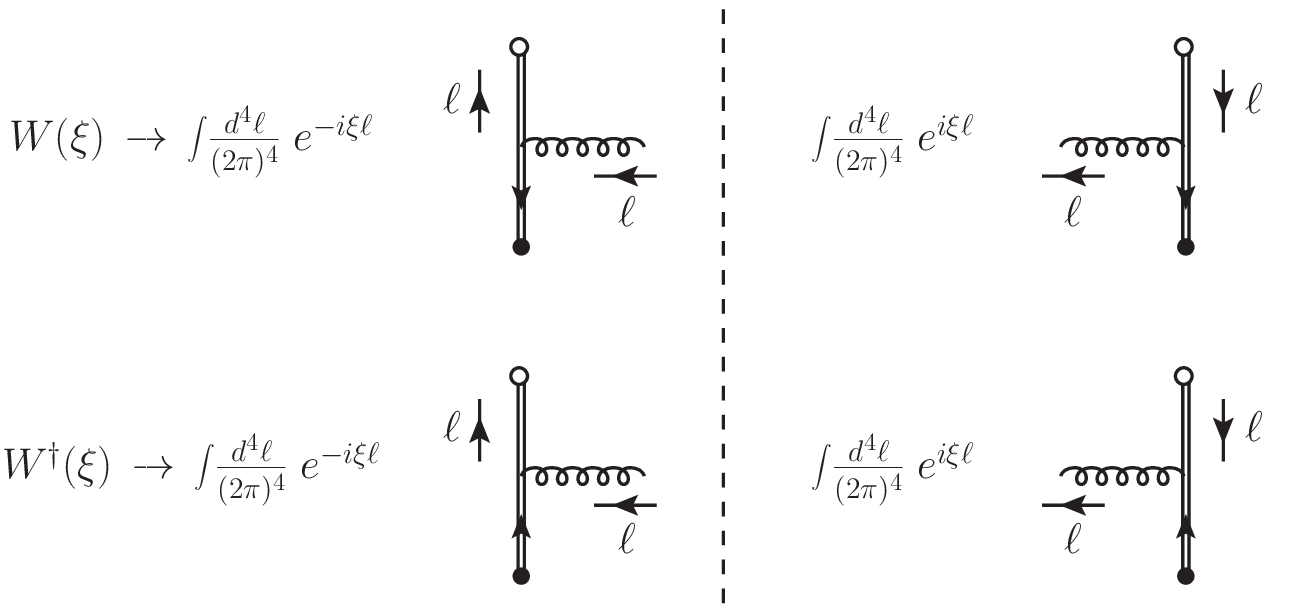}
\end{center}
\caption{\label{fig:wilson-rules} Feynman rules for the term of order $g$
  in the expansion of the Wilson lines defined in
  \protect\eqref{wilson-dy}.  The rules for lines to the left of the
  final-state cut (indicated by the dashed line) follow from
  \protect\eqref{wilson-rule-eq}, and those for lines to the right of the
  final-state cut are obtained by complex conjugation.}
\end{figure}

{}From \eqref{eikonal-fourier} and \eqref{dy-wilson-a} one obtains the
representation
\begin{align}
  \label{wilson-rule-eq}
ig \int_0^{\infty} d\lambda\, v A^a(\xi - \lambda v)\, t^a
 &= -ig t^a v^\alpha \int \frac{d^4 \ell}{(2\pi)^4}\;
    e^{-i \xi\ell}\, \frac{-i}{\ell v + i\epsilon}\, 
    \int d^4\zeta\; e^{i \zeta\ell} A^{a}_{\alpha}(\zeta)
\end{align}
for the exponent of the Wilson line $W(\xi; v)$.  Together with the
Feynman rules for eikonal lines and their coupling to gluons in
figure~\ref{fig:eikonal-rules} this gives a Feynman rule for the order $g$
term of the Wilson line to the left of the final-state cut, as shown in
figure~\ref{fig:wilson-rules}.

\subsubsection{The basic graphs}
\label{sec:soft-basic}

\begin{figure}
\begin{center}
\includegraphics[width=0.99\textwidth]{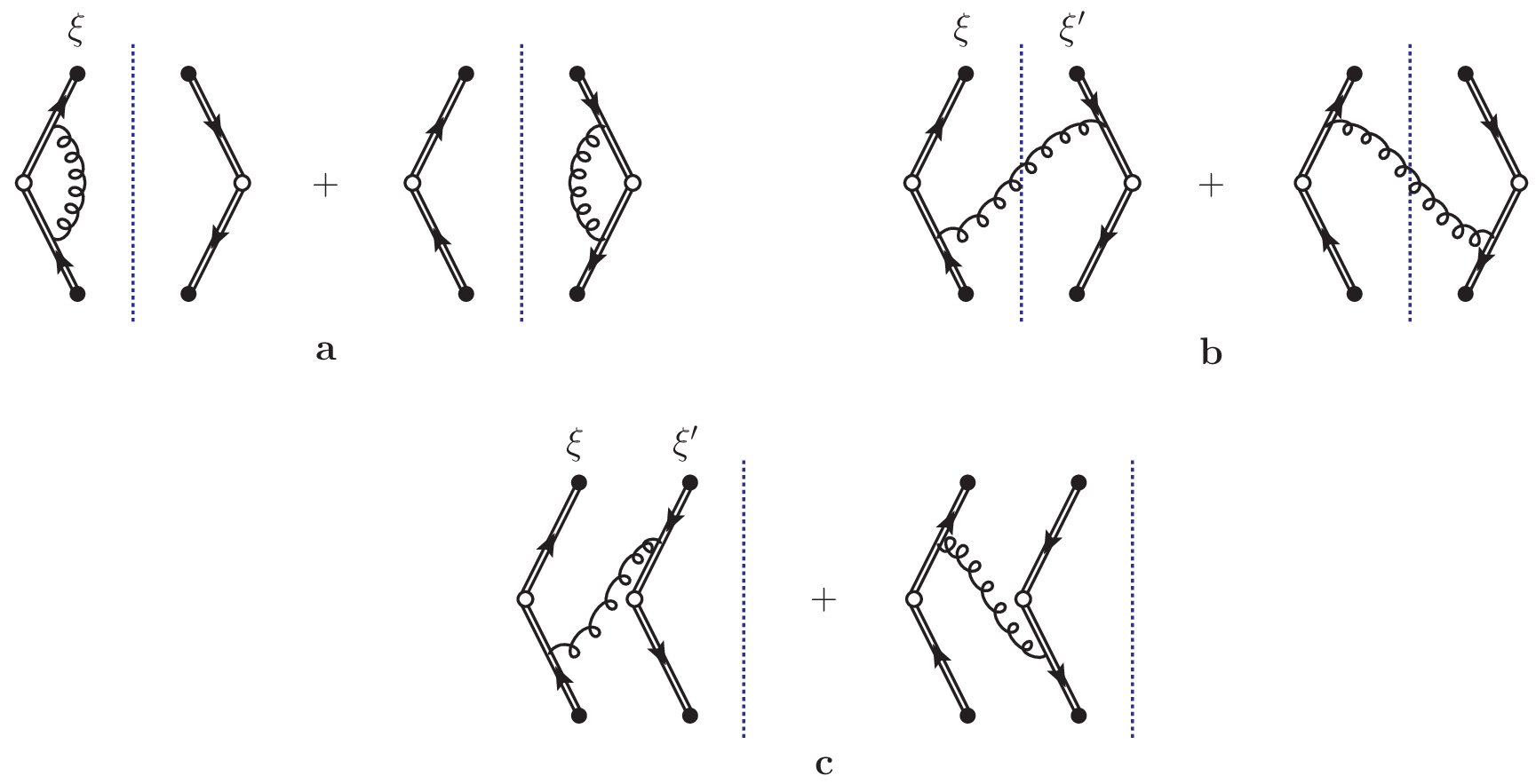}
\end{center}
\caption{\label{fig:soft-basic} Basic graphs contributing to the soft
  factor at order $\alpha_s$.}
\end{figure}

\begin{figure}
\begin{center}
\includegraphics[width=0.7\textwidth]{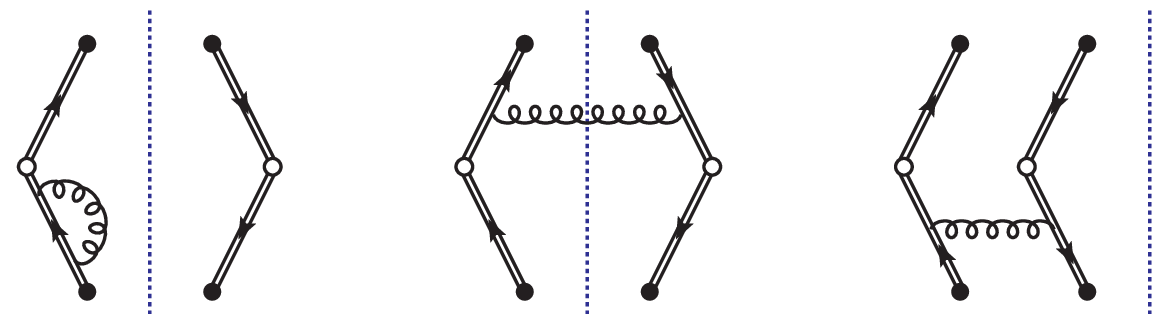}
\end{center}
\caption{\label{fig:wilson-self} Graphs with gluon exchange between
  eikonal lines having the same rapidity.  These graphs are excluded in
  the evaluation of soft factors, as discussed in the text.}
\end{figure}

The expansion at $\mathcal{O}(\alpha_s)$ of the soft factor
\eqref{double-soft-fact} involves the three types of graphs shown in
figure~\ref{fig:soft-basic}, which we now calculate.  Graphs a and b
already appear in the soft factor \eqref{single-soft-fact} for the single
Drell-Yan process, whereas graph c is specific for multiparton
interactions.  As discussed in the previous section, we discard graphs as
in figure~\ref{fig:wilson-self}, where gluons are exchanged between Wilson
lines pointing both along $v$ or both along $w$.

To regulate ultraviolet divergences, we work in $4-2\epsilon$ dimensions,
and to exhibit infrared divergences in individual graphs we use a small
gluon mass $\lambda$.  For the time being we omit the color matrices $t^a$
in the Feynman rules.  One finds that each pair of graphs in
figure~\ref{fig:soft-basic} gives the same result, so that in the
following we only list the expressions of the left-hand graphs and
multiply by two.

Before renormalization, the vertex correction graphs in
figure~\ref{fig:soft-basic}a give
\begin{align}
  \label{Ua-start}
U_{a} &= 2 \mu^{2\epsilon}\!
  \int \frac{d^{4-2\epsilon} \ell}{(2\pi)^{4-2\epsilon}}\;
  e^{i \xi_T \ell - i \xi_T \ell}\,
  \frac{i}{-\ell w + i\epsilon}\, (-ig w^\alpha)\,
  \frac{-i g_{\alpha\beta}}{\ell^2 - \lambda^2 + i\epsilon}\,
  (-ig v^\beta)\, \frac{-i}{\ell v + i\epsilon}
\nonumber \\
 &= -4 i \alpha_s\, \frac{v w}{v^+ w^+}\,
  \mu^{2\epsilon}\!
  \int \frac{d^{2-2\epsilon}\ms \tvec{\ell}}{(2\pi)^{2-2\epsilon}}\,
  \int_{-\infty}^{\infty} d\ell^+
  \int_{-\infty}^{\infty} \frac{d\ell^-}{2\pi}\,
  \frac{1}{2\ell^+ \ell^- - \tvec{\ell}^2 - \lambda^2 + i\epsilon}
\nonumber \\
 &\quad \times
  \frac{1}{\ell^- + \frac{w^-}{w^+}\ms \ell^+ - i\epsilon}\,
  \frac{1}{\ell^- + \frac{v^-}{v^+}\ms \ell^+ - i\epsilon} \,,
\end{align}
where we recall that the vectors $v$ and $w$ have zero transverse momenta
and satisfy $v^- >0$, $v^+ < 0$, $w^+ >0$, $w^- <0$.
For $\ell^+ < 0$ all poles in $\ell^-$ are on the same side of the real
axis, so that this region gives a zero contribution to the integral over
$\ell^-$.  For $\ell^+ > 0$ we close the integration contour in $\ell^-$
around the pole of the gluon propagator and obtain
\begin{align}
  \label{Ua-mid}
U_a &= - \alpha_s\, \frac{v w}{v^- w^-}\,
  \mu^{2\epsilon}\!
  \int \frac{d^{2-2\epsilon}\ms \tvec{\ell}}{(2\pi)^{2-2\epsilon}}\,
  \int_{0}^{\infty} d(\ell^+)^2
\nonumber \\
 &\quad \times
  \frac{1}{(\ell^+)^2 +
    \half \frac{w^+}{w^-}\, (\tvec{\ell}^2 + \lambda^2) + i\epsilon}
  \frac{1}{(\ell^+)^2 +
    \half \frac{v^+}{v^-}\, (\tvec{\ell}^2 + \lambda^2) + i\epsilon}
  \,.
\end{align}
As both poles in $(\ell^+)^2$ are on the same side of the real axis, one
can deform the integration contour to $-\infty < (\ell^+)^2 < 0$ and
obtains
\begin{align}
  \label{Ua-final}
U_a &= - 2\alpha_s\, \frac{v w}{v^- w^+ - v^+ w^-}\,
  \log \biggl( \frac{v^- w^+}{v^+ w^-} \biggr)\,
  \mu^{2\epsilon}\!
  \int \frac{d^{2-2\epsilon}\ms \tvec{\ell}}{(2\pi)^{2-2\epsilon}}\,
       \frac{1}{\tvec{\ell}^2 + \lambda^2}
\nonumber \\
 &= - \frac{\alpha_s}{2\pi}\;
  \frac{v^- w^+ + v^+ w^-}{v^- w^+ - v^+ w^-}\,
  \log \biggl( \frac{v^- w^+}{v^+ w^-} \biggr)\,
  \Gamma(\epsilon)\,
  \biggl( \frac{4\pi \mu^2}{\lambda^2} \biggr)^{\epsilon} \,.
\end{align}
We see that both vectors $v$ and $w$ must be chosen away from the light
cone, since taking even one of them lightlike gives an infinite result for
$U_a$.  With the definition \eqref{rap-def} for the rapidity of a
spacelike vector, we have
\begin{align}
  \label{vw-to-y}
\log \biggl( \frac{v^- w^+}{v^+ w^-} \biggr)
  &= 2 (y_w - y_v) \,,
&
\frac{v^- w^+ + v^+ w^-}{v^- w^+ - v^+ w^-}
  &= \tanh(y_w - y_v) \,.
\end{align}
We require a large rapidity difference, $|y_w - y_v| \gg 1$, i.e.\ $v^-
w^+ \gg v^+ w^-$.  This is satisfied by the choice $y_v \ll 1$ and $y_w
\gg 1$ made in section \ref{sec:soft}, but it also allows one of $v$ or
$w$ to have central rapidity, as long as the rapidity of the other one is
large (positive or negative).  We can then approximate $\tanh(y_w - y_v)
\approx 1$.
The expression \eqref{Ua-final} is ultraviolet divergent, and using
$\overline{\text{MS}}$ subtraction we obtain the renormalized result
\begin{align}
  \label{Sa-result}
S_a(\mu) &= \bigl[ U_a(\mu) \bigr]^{\text{ren}} 
  = - \frac{\alpha_s}{\pi}\,
    (y_w - y_v) \log\frac{\mu^2}{\lambda^2} \,.
\end{align}
For the graphs in figure \ref{fig:soft-basic}b, which describe gluon
emission into the final state, we have
\begin{align}
  \label{Ub}
U_{b} &= 2 \mu^{2\epsilon}\!
  \int \frac{d^{4-2\epsilon} \ell}{(2\pi)^{3-2\epsilon}}\;
  \theta(\ell^+)\, \delta(\ell^2 - \lambda^2)\,
  e^{i \xi_T^{} \ell - i \xi'_T \ell}\;
  \frac{i}{-\ell w + i\epsilon}\, (-ig w^\alpha)\,
   (-g_{\alpha\beta})\,
  (ig v^\beta)\, \frac{i}{- \ell v - i\epsilon}
\nonumber \\
 &= 4 \alpha_s\, \frac{v w}{v^+ w^+}\,
  \mu^{2\epsilon}\!
  \int \frac{d^{2-2\epsilon}\ms \tvec{\ell}}{(2\pi)^{2-2\epsilon}}\;
     e^{i (\tvec{\xi} - \tvec{\xi}')\ms \tvec{\ell}}
  \int_{0}^{\infty} d\ell^+\,
  \int_{-\infty}^{\infty} d\ell^-\,
  \delta(2\ell^+ \ell^- - \tvec{\ell}^2 - \lambda^2)
\nonumber \\
 &\quad \times
  \frac{1}{\ell^- + \frac{w^-}{w^+}\ms \ell^+ - i\epsilon}\,
  \frac{1}{\ell^- + \frac{v^-}{v^+}\ms \ell^+ - i\epsilon}
\nonumber \\
 &= \alpha_s\, \frac{v w}{v^- w^-}\,
  \mu^{2\epsilon}\!
  \int \frac{d^{2-2\epsilon}\ms \tvec{\ell}}{(2\pi)^{2-2\epsilon}}\;
     e^{i (\tvec{\xi} - \tvec{\xi}')\ms \tvec{\ell}}
  \int_{0}^{\infty} d(\ell^+)^2
  \nonumber \\
 &\quad \times
  \frac{1}{(\ell^+)^2 +
    \half \frac{w^+}{w^-}\, (\tvec{\ell}^2 + \lambda^2) + i\epsilon}\,
  \frac{1}{(\ell^+)^2 +
    \half \frac{v^+}{v^-}\, (\tvec{\ell}^2 + \lambda^2) + i\epsilon}
  \,.
\end{align}
Comparing with \eqref{Ua-mid} we observe that $U_a + U_b = 0$ at
$\tvec{\xi} = \tvec{\xi}'$.  For $\tvec{\xi} \neq \tvec{\xi}'$ the
integral over $\tvec{\ell}$ in \eqref{Ub} is ultraviolet finite and no
subtraction is needed before we set $\epsilon=0$.  We thus obtain
\begin{align}
  \label{Sb-result}
S_b(\tvec{z}) &\underset{\tvec{z} \neq \tvec{0}}{=}
  \bigl[ U_b(\tvec{z}) \bigr]^{\epsilon=0}
 = 4 \alpha_s\, (y_w - y_v)\,
  \int \frac{d^{2}\ms \tvec{\ell}}{(2\pi)^{2}}\;
     e^{i \tvec{z} \tvec{\ell}}
     \frac{1}{\tvec{\ell}^2 + \lambda^2}
 = \frac{\alpha_s}{\pi}\, (y_w - y_v)\,
    2 K_0\bigl( \lambda\ms |\tvec{z}| \bigr)
\nonumber \\
 & \underset{\lambda\to 0}{=}
  - \frac{\alpha_s}{\pi}\, (y_w - y_v)\,
    \log \frac{\lambda^2 \tvec{z}^2}{b_0^2} \,,
\end{align}
where $b_0 = 2 e^{-\gamma}$ and $\gamma$ is the Euler number.

The graphs in figure~\ref{fig:soft-basic}a and b give the full
$\mathcal{O}(\alpha_s)$ contribution for the vacuum matrix element $S_q$
defined in \eqref{single-soft-fact}, which appears in single Drell-Yan
production.  Let us evaluate this contribution as a side result.  The
color factor for all graphs is $N^{-1} \tr( t^a t^a ) = C_F$ in this case,
so that we have $S_{q}(\tvec{z}, \mu) = 1 + S(\tvec{z}, \mu) +
\mathcal{O}(\alpha_s^2)$ with
\begin{align}
  \label{single-soft-pert}
S(\tvec{z}, \mu) &= C_F \bigl[ S_a(\mu) + S_{b}(\tvec{z}) \bigr]
 = - \frac{\alpha_s}{\pi}\, C_F\, (y_w - y_v)\,
    \log \frac{\mu^2 \tvec{z}^2}{b_0^2}
\end{align}
for $\tvec{z} \neq \tvec{0}$.  Notice that the infrared divergences
regulated by a gluon mass $\lambda$ have cancelled in the sum over all
graphs, as is required for a perturbative evaluation of $S_{q}$.  For
$\tvec{z} = \tvec{0}$ the relation $U_a + U_b = 0$ ensures that the
condition $S_{q}(\tvec{0}) = 1$ does not receive radiative corrections, in
agreement with the general result discussed below
\eqref{single-soft-fact}.  This complete cancellation between real and
virtual corrections plays a crucial role for collinear factorization, see
section~\ref{sec:coll-fact}.
The fact that the limit $\tvec{z}\to 0$ of \eqref{single-soft-pert} is
infinite rather than zero is due to our use of modified minimal
subtraction.  For $\tvec{z} = \tvec{0}$ the ultraviolet divergences in
$U_a$ and $U_b$ cancel each other (as do the finite parts of the graphs),
so that there is no ultraviolet subtraction for their sum.  At $\tvec{z}
\neq \tvec{0}$, however, $U_a + U_b$ does require ultraviolet subtraction
since the first term is ultraviolet divergent whereas the second term is
not.

Finally, the graphs in figure~\ref{fig:soft-basic}c give
\begin{align}
  \label{Uc}
U_{c} &= 2 \mu^{2\epsilon}\!
  \int \frac{d^{4-2\epsilon} \ell}{(2\pi)^{4-2\epsilon}}\;
  e^{i \xi_T^{} \ell - i \xi'_T \ell}\,
  \frac{i}{-\ell w + i\epsilon}\, (-ig w^\alpha)\,
  \frac{-i g_{\alpha\beta}}{\ell^2 - \lambda^2 + i\epsilon}\,
  (-ig v^\beta)\, \frac{i}{\ell v + i\epsilon}
\nonumber \\
 &= 4 \alpha_s\, \frac{v w}{v^+ w^+}\,
  \mu^{2\epsilon}\!
  \int \frac{d^{2-2\epsilon}\ms \tvec{\ell}}{(2\pi)^{2-2\epsilon}}\;
     e^{i (\tvec{\xi} - \tvec{\xi}')\ms \tvec{\ell}}
  \int_{-\infty}^{\infty} d\ell^+\,
  \int_{-\infty}^{\infty} \frac{d\ell^-}{2\pi}\,
  \frac{i}{2\ell^+ \ell^- - \tvec{\ell}^2 - \lambda^2 + i\epsilon}
\nonumber \\
 &\quad \times
  \frac{1}{\ell^- + \frac{w^-}{w^+}\ms \ell^+ - i\epsilon}\,
  \frac{1}{\ell^- + \frac{v^-}{v^+}\ms \ell^+ - i\epsilon} \,.
\end{align}
For $\ell^+ < 0$ all poles in $\ell^-$ are on the same side of the real
axis and give a zero net result, whereas for $\ell^+ > 0$ the integral
over $\ell^-$ can be obtained from the residue of the gluon propagator.
This gives the same expression as in the second step of \eqref{Ub}.  The
graphs in figure~\ref{fig:soft-basic}c hence give the same result as those
in figure~\ref{fig:soft-basic}b:
\begin{align}
  \label{Sc-result}
S_{c}(\tvec{z}) &= S_{b}(\tvec{z}) \,.
\end{align}
Note that the graphs in figure~\ref{fig:soft-basic}b and c only differ by
the position of the final-state cut.  The result \eqref{Sc-result} shows
that this does not matter and provides a consistency check for the
omission of explicit time ordering in \eqref{soft-graph-mat}.  Namely,
time ordering of the two fields $A(\xi)$ and $A(\bar{\xi})$ associated
with the gluon propagator is required when the gluon does not cross the
final-state cut.  When it does cross the cut, no time ordering
prescription arises, however, since the two fields then have a complete
set of final states between them: $\sum_X A(\xi) |X \rangle\, \langle X|
A(\bar{\xi}) = A(\xi)\, A(\bar{\xi})$.

The identity \eqref{Sc-result} also holds before setting $\epsilon=0$ and
can then be used also at $\tvec{z} = \tvec{0}$.  This provides a simple
explanation of the relation $U_b(\tvec{0}) = - U_a$ discussed above.
Indeed, the equality $U_c(\tvec{0}) = - U_a$ is already evident from the
starting expressions \eqref{Ua-start} and \eqref{Uc}, which only differ by
the sign of the eikonal propagator $\pm i/(\ell v + i\epsilon)$.  This is
a direct consequence of the Feynman rules in
figures~\ref{fig:eikonal-rules}, since the upper eikonal lines to which
the gluon couples in figures~\ref{fig:soft-basic}a and c only differ by
the flow of color charge.

For later use we note that the graphs in figures~\ref{fig:soft-basic}b and
c change sign when the flow of color charge in one of their eikonal lines
is reversed, whereas those in figure~\ref{fig:soft-basic}a remain the
same.


\subsubsection{Soft factor associated with two-quark distributions}
\label{sec:soft-qq}

With these building blocks at hand we can construct the soft factor
associated with two-parton distributions in the cross section formula
\eqref{X-sect-soft-2}.  We first discuss the factor $S_{qq}$ in some
detail and give the results for $S_{q\bar{q}}$ and $S_I$ in the end.

\begin{figure}
\begin{center}
\includegraphics[width=0.87\textwidth]{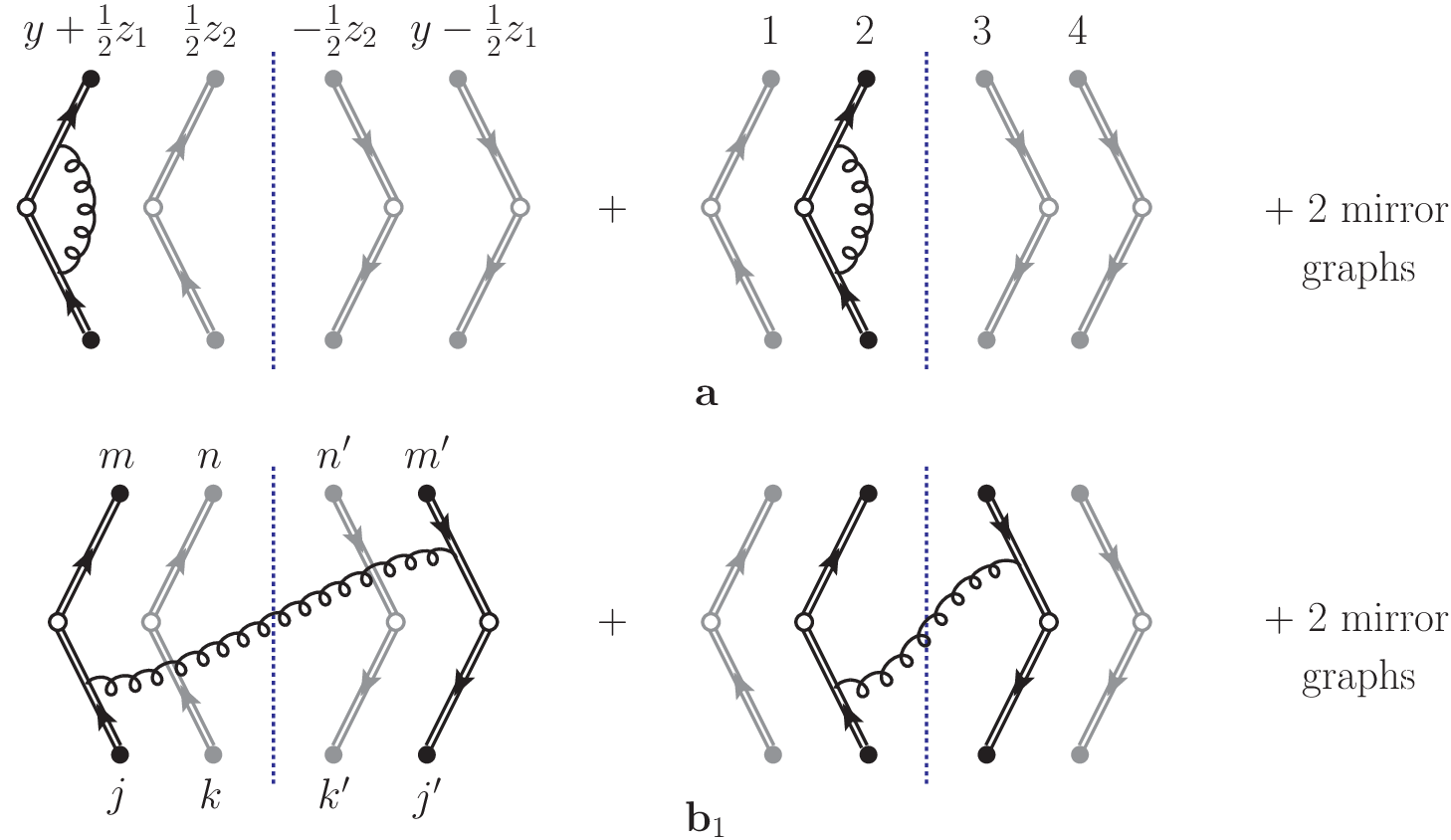} \\[0.8em]
\includegraphics[width=0.87\textwidth]{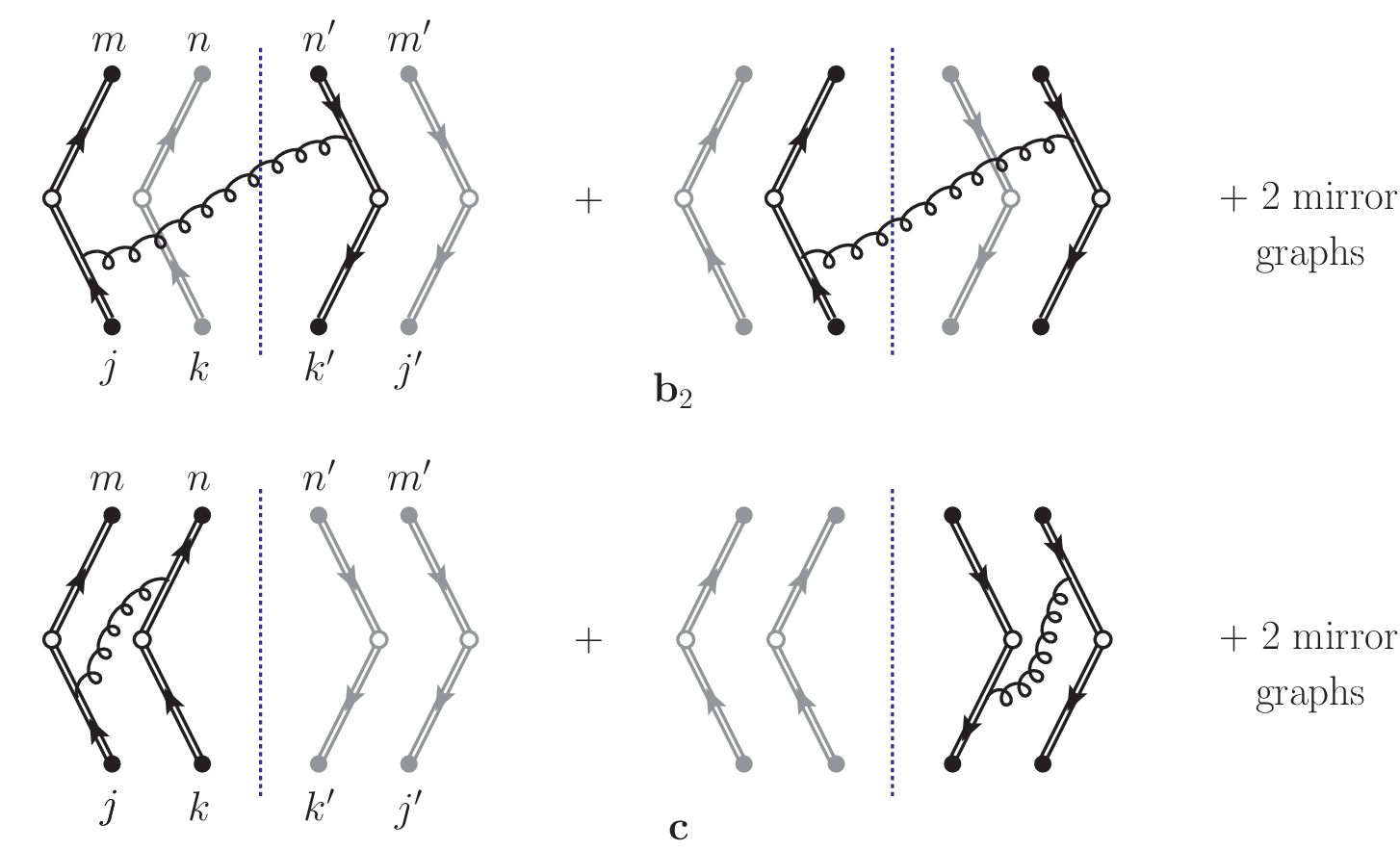}
\end{center}
\caption{\label{fig:soft-four-legs} Graphs for the soft factor associated
  with two-quark distributions.  The ``mirror graphs'' not shown are as in
  figure~\protect\ref{fig:soft-basic}.  The numbers, position arguments
  and color labels on the eikonal lines correspond to those of the quark
  lines in the distribution $F_{a_1,a_2}$, see
  figure~\protect\ref{fig:distrib}.}
\end{figure}

The graphs contributing to $S_{qq}$ are those in
figure~\ref{fig:soft-four-legs}.  In the graphs of
figure~\ref{fig:soft-four-legs}a and b$_{1}$ there is either no gluon
coupling to the pair $\{14 \}$ or no gluon coupling to the pair $\{23 \}$
of eikonal lines, so that the lines have the same color coupling at the
bottom and the top of the graph.  The color factor for the vertex
correction graphs~\ref{fig:soft-four-legs}a is $C_F$, independent of the
color flow along the eikonal lines and of the way in which their color
indices are coupled.  For figure~\ref{fig:soft-four-legs}b$_{1}$ we have
color factors
\begin{align}
\frac{1}{N^2}\, \tr( t^a\ms t^a )\, \tr \one
  &= C_F & & \text{for}~{}^{11\!}S_{qq} \,,
\nonumber \\
\frac{4}{N^2-1}\, \tr( t^a\ms t^c\ms t^b\ms t^c ) \, \tr(t^a\ms t^b)
  &= - \frac{1}{2N} & & \text{for}~{}^{88\!}S_{qq} \,,
\end{align}
where the prefactors $1/N^2$ and $4/(N^2-1)$ come from the analog of the
color decomposition \eqref{soft-color-decomp} for $S_{qq}$.  We note that
these color factors are the same as for the ladder graphs we shall discuss
in section~\ref{sec:ladders-color}.

By contrast, the graphs in figure~\ref{fig:soft-four-legs}b$_{2}$ and c
can change the color coupling of the eikonal lines.  As an illustration,
consider the first graph in figure~\ref{fig:soft-four-legs}b$_{2}$ with
the color indices at the bottom multiplied by octet matrices $t^b_{jj'}\,
t^b_{kk'}$.  The color structure is then
\begin{align}
  \label{octet-soft-1}
\bigl( t^a t^b \bigr)_{mm'}\, \bigl( t^b t^a \bigr)_{nn'} 
&= \frac{1}{2} \biggl[ \frac{1}{N}\, \delta^{ab} \delta_{mm'}
   + d^{abc}\ms t^{c}_{mm'} + i f^{abc}\ms t^{c}_{mm'} \biggr]
   \bigl( t^b t^a \bigr)_{nn'} 
\nonumber \\
&= \frac{1}{2} \biggl[ \frac{C_F}{N}\, \delta_{mm'}\ms \delta_{nn'}
      + \frac{N^2-4}{2N}\, t^a_{mm'}\ms t^a_{nn'}
      + \frac{N}{2}\, t^a_{mm'}\ms t^a_{nn'} \biggr]
\nonumber \\
&= \frac{N^2-1}{4 N^2}\, \delta_{mm'}\ms \delta_{nn'}
      + \biggl( C_F - \frac{1}{2N} \biggr)\, t^a_{mm'}\ms t^a_{nn'}
\end{align}
and we see that this graph contributes to both ${}^{81\!}S$ and
${}^{88\!}S$.  If the indices $jj'$ and $kk'$ are coupled to singlets,
then the graph is proportional to $t^a_{mm'}\, t^a_{nn'}$ and thus
contributes to ${}^{18\!}S$.  For the first graph in
figure~\ref{fig:soft-four-legs}c the color factors are analogous, with the
difference that one has
\begin{align}
  \label{octet-soft-2}
\bigl( t^a t^b \bigr)_{mm'}\, \bigl( t^a t^b \bigr)_{nn'} 
&= \frac{1}{2} \biggl[ \frac{C_F}{N}\, \delta_{mm'}\ms \delta_{nn'}
      + \frac{N^2-4}{2N}\, t^a_{mm'}\ms t^a_{nn'}
      - \frac{N}{2}\, t^a_{mm'}\ms t^a_{nn'} \biggr]
\nonumber \\
&= \frac{N^2-1}{4 N^2}\, \delta_{mm'}\ms \delta_{nn'}
      - \frac{1}{N}\, t^a_{mm'}\ms t^a_{nn'}
\end{align}
instead of \eqref{octet-soft-1}.  Putting everything together and using
the results of the previous section, we have
\begin{align}
  \label{S-matrix-qq1}
{}^{11\!}S_{qq} &= 1 + C_F
  \bigl[ 2 S_a(\mu) + S_b(\tvec{z}_1) + S_b(\tvec{z}_2) \bigr] \,,
\nonumber \\[0.3em]
{}^{88\!}S_{qq} &= 1 + 2 C_F S_a(\mu)
  - \frac{1}{2N}\, \bigl[ S_b(\tvec{z}_1) + S_b(\tvec{z}_2) \bigr]
\nonumber \\[0.3em]
&\quad 
 + \biggl( C_F - \frac{1}{2N} \biggr)\, \biggl[
    S_b\Bigl( \tvec{y} + \frac{\tvec{z}_1 + \tvec{z}_2}{2} \Bigr)
  + S_b\Bigl( \tvec{y} - \frac{\tvec{z}_1 + \tvec{z}_2}{2} \Bigr)
   \biggr]
\nonumber \\[0.2em]
&\quad 
 + \frac{1}{N}\, \biggl[
    S_c\Bigl( \tvec{y} + \frac{\tvec{z}_1 - \tvec{z}_2}{2} \Bigr)
  + S_c\Bigl( \tvec{y} - \frac{\tvec{z}_1 - \tvec{z}_2}{2} \Bigr)
   \biggr] \,,
\nonumber \\[0.3em]
{}^{18\!}S_{qq} = {}^{81\!}S_{qq}
&= \sqrt{\frac{C_F}{2N}}\; \biggl[
    S_b\Bigl( \tvec{y} + \frac{\tvec{z}_1 + \tvec{z}_2}{2} \Bigr)
  + S_b\Bigl( \tvec{y} - \frac{\tvec{z}_1 + \tvec{z}_2}{2} \Bigr)
\nonumber \\[0.2em]
&\qquad\qquad
  - S_c\Bigl( \tvec{y} + \frac{\tvec{z}_1 - \tvec{z}_2}{2} \Bigr)
  - S_c\Bigl( \tvec{y} - \frac{\tvec{z}_1 - \tvec{z}_2}{2} \Bigr)
  \biggr]  \,.
\end{align}
The terms $S_c$ come with an overall minus sign, because in
graph~\ref{fig:soft-four-legs}c one of the eikonal lines coupled to a
gluon has its color flow reversed compared with
graph~\ref{fig:soft-basic}c, which reverses the sign in the eikonal
propagator according to the Feynman rules in
figure~\ref{fig:eikonal-rules}.

The matrix elements in \eqref{S-matrix-qq1} are such that we can replace
$S_a$ and $S_b = S_c$ by their sum, which is infrared finite according to
\eqref{single-soft-pert}.  Using $S = C_F\ms (S_a + S_b)$ we have
\begin{align}
  \label{S-matrix-qq2}
\mathbf{S}_{qq}(\tvec{z}_i, \tvec{y}, \mu) &=
\bigl[ 1 + S(\tvec{z}_1, \mu) + S(\tvec{z}_2,\mu) \bigr]
\begin{pmatrix} 1 \ & 0 \\ 0 & \ 1 \end{pmatrix}
\nonumber \\
&\quad +
\begin{pmatrix} 0 & c\ms S_d \\[0.2em]
  c\ms S_d \quad & - (1+c^2) S_y - 2 c^2 S_d \end{pmatrix} \, ,
\end{align}
with a color factor
\begin{align}
  \label{c-color-factor}
c &= \frac{1}{\sqrt{2N C_F}} = \frac{1}{\sqrt{N^2 - 1}}
\end{align}
and linear combinations
\begin{align}
  \label{S-dy}
S_d(\tvec{z}_i, \tvec{y}) &= 
    S\Bigl( \tvec{y} + \frac{\tvec{z}_1 + \tvec{z}_2}{2}, \mu \Bigr)
  + S\Bigl( \tvec{y} - \frac{\tvec{z}_1 + \tvec{z}_2}{2}, \mu \Bigr)
\nonumber \\[0.1em]
&\quad
  - S\Bigl( \tvec{y} + \frac{\tvec{z}_1 - \tvec{z}_2}{2}, \mu \Bigr)
  - S\Bigl( \tvec{y} - \frac{\tvec{z}_1 - \tvec{z}_2}{2}, \mu \Bigr) \,,
\nonumber \\[0.2em]
S_y(\tvec{z}_i, \tvec{y}) &=
    S(\tvec{z}_1, \mu) + S(\tvec{z}_2, \mu)
  - S\Bigl( \tvec{y} + \frac{\tvec{z}_1 + \tvec{z}_2}{2}, \mu \Bigr)
  - S\Bigl( \tvec{y} - \frac{\tvec{z}_1 + \tvec{z}_2}{2}, \mu \Bigr) \,.
\end{align}
Note that the $\mu$ dependence of $S$ has canceled in $S_d$ and $S_y$.

\begin{figure}
\begin{center}
\includegraphics[width=0.32\textwidth]{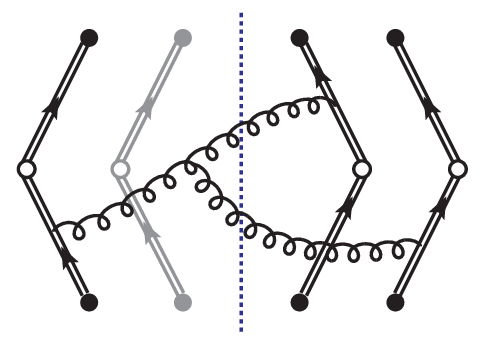}
\end{center}
\caption{\label{fig:soft-higher-order} A higher-order graph for the soft
  factor, which connects more than two eikonal lines.}
\end{figure}

At one-loop level the soft factor for two-quark distributions can thus be
expressed in terms of its analog $S$ for single-parton distributions.
This simplification will most likely no longer hold at higher orders in
$\alpha_s$, since one then has graphs like in
figure~\ref{fig:soft-higher-order}, where more than two eikonal lines are
connected.

It is easy to see that the soft factor $S_{q\bar{q}}(\tvec{z}_i,
\tvec{y})$ for quark-antiquark distributions is obtained from
$S_{qq}(\tvec{z}_i, \tvec{y})$ by replacing $\tvec{z}_2 \to -\tvec{z}_2$.
This corresponds to the interchange of the labels $2$ and $3$ in the
definitions of $F_{a_1,a_2}$ and $F_{a_1,\bar{a}_2}$ (see
\eqref{index-key} to \eqref{qqbar-mixed-F}) and uses the result that the
kernels $S_b$ and $S_c$ for gluons that cross or do not cross the
final-state cut are identical.

The soft factor multiplying the interference distributions is obtained
from the same graphs as those in figure~\ref{fig:soft-four-legs}, apart
from changes in the color labels and in the color flow of the eikonal
lines.  Making the appropriate replacements of position arguments in
$\mathbf{S}_{qq}$ we get
\begin{align}
  \label{S-matrix-int1}
{}^{11\!}S_I &= 1 + C_F
  \biggl[ 2 S_a(\mu)
   + S_b\Bigl( \tvec{y} + \frac{\tvec{z}_1 + \tvec{z}_2}{2} \Bigr)
   + S_b\Bigl( \tvec{y} - \frac{\tvec{z}_1 + \tvec{z}_2}{2} \Bigr)
   \biggr] \,,
\nonumber \\[0.3em]
{}^{88\!}S_I &= 1 + 2 C_F S_a(\mu)
  - \frac{1}{2N}\; \biggl[
     S_b\Bigl( \tvec{y} + \frac{\tvec{z}_1 + \tvec{z}_2}{2} \Bigr)
   + S_b\Bigl( \tvec{y} - \frac{\tvec{z}_1 + \tvec{z}_2}{2} \Bigr)
   \biggr]
\nonumber \\[0.3em]
&\quad 
 + \biggl( C_F - \frac{1}{2N} \biggr)\,  \biggl[
     S_c\Bigl( \tvec{y} + \frac{\tvec{z}_1 - \tvec{z}_2}{2} \Bigr)
   + S_c\Bigl( \tvec{y} - \frac{\tvec{z}_1 - \tvec{z}_2}{2} \Bigr)
  \biggr]
\nonumber \\[0.2em]
&\quad 
 + \frac{1}{N}\, \bigl[
     S_b(\tvec{z}_1) + S_b(\tvec{z}_2) \bigr] \,,
\nonumber \\[0.3em]
{}^{18\!}S_I = {}^{81\!}S_I
&= \sqrt{\frac{C_F}{2N}}\; \biggl[
     S_c\Bigl( \tvec{y} + \frac{\tvec{z}_1 - \tvec{z}_2}{2} \Bigr)
   + S_c\Bigl( \tvec{y} - \frac{\tvec{z}_1 - \tvec{z}_2}{2} \Bigr)
 - S_b(\tvec{z}_1) - S_b(\tvec{z}_2) \biggr] \,,
\end{align}
which can be rewritten as
\begin{align}
  \label{S-matrix-int2}
\mathbf{S}_I(\tvec{z}_1, \tvec{z}_2, \tvec{y}) &=
\bigl[ 1 + S(\tvec{z}_1, \mu) + S(\tvec{z}_2,\mu) \bigr]
\begin{pmatrix} 1 \ & 0 \\ 0 & \ 1 \end{pmatrix}
\nonumber \\
 &\quad -
\begin{pmatrix} S_y & c\ms (S_y + S_d) \\[0.3em]
   c\ms (S_y + S_d) \quad & (1-2c^2) (S_y + S_d) + c^2 S_d
\end{pmatrix} \,.
\end{align}


\subsection{Collins-Soper equation}
\label{sec:cs-general}

As is well-known, cross sections with measured transverse momenta
$|\tvec{q}_i| \sim q_T$ much smaller than the hardest scale $Q$ in the
process contain large logarithms in $q_T/Q$, which need to be summed to
all orders in $\alpha_s$ in order to have a perturbatively stable result.
A powerful method to resum these Sudakov logarithms is due to Collins,
Soper and Sterman (CSS) \cite{Collins:1984kg}.  This method uses
transverse-momentum-dependent factorization and is therefore, up to now,
limited to the production of color singlet particles, such as a Drell-Yan
lepton pair or a Higgs boson.  In this section we show how this formalism
extends to double Drell-Yan production, and we calculate the corresponding
Sudakov factor to next-to-leading logarithmic accuracy.

We begin by a brief account of the CSS method for processes with a single
hard scattering initiated by quark-antiquark annihilation.  As we
mentioned in section~\ref{sec:full-fact}, the Wilson lines in the
definition of transverse-momentum dependent parton distributions must be
taken along a direction $v$ with a finite rapidity.  Their dependence on
this rapidity is governed by the Collins-Soper (CS) equation
\cite{Collins:1981uk}, whose solution resums Sudakov logarithms to all
orders.  By Lorentz invariance, the distributions depend on $v$ via the
scalar parameter\footnote{To avoid the appearance of square roots, our
  definition~\protect\eqref{zeta-def} follows the convention of Ji et
  al.~\cite{Ji:2004wu} and differs from the one of Collins and Soper
  \cite{Collins:1981uk}, with $\zeta^2 |_{\text{here}} =
  \zeta_{\text{\protect\cite{Collins:1981uk}}}$.}
\begin{align}
  \label{zeta-def}
\zeta^2 &= \frac{(2 p\ms v)^2}{|v^2|} \,.
\end{align}
As discussed earlier, we require $v$ to be spacelike (although the
manipulations in the following would also work for timelike $v$, which is
the choice adopted in \cite{Ji:2004wu}).  In terms of the rapidities $y_v$
of $v$ and $y_p$ of the proton, we have
\begin{align}
\zeta &= 2 M \sinh(y_p - y_v)  \approx M\ms e^{y_p - y_v} \,,
&
\frac{\partial}{\partial\log\zeta}
 & = - \tanh(y_p - y_v)\, \frac{\partial}{\partial y_v}
 \approx - \frac{\partial}{\partial y_v}
\end{align}
for a right-moving proton, where $M$ is the proton mass.  The
approximations are valid for $y_v \ll y_p$, which we have seen to be
necessary in section \ref{sec:coll}.  This implies $\zeta\gg M$.

Following the original paper \cite{Collins:1981uk} we work here with
``unsubtracted parton distributions'' in the parlance of
\cite{Ji:2004wu,Idilbi:2004vb} and \cite{Collins:2011}, i.e.\ with matrix
elements of operators constructed from quark fields and Wilson lines along
$v$ as in \eqref{full-wilson-op}.  Subtractions for the soft momentum
region are not made in the parton distributions but in the soft factor
that appears in the cross section (see \cite{Collins:2007ph}).  The fields
in the operator are renormalized, so that the distributions depend on an
ultraviolet renormalization scale $\mu$.  Fourier transforming from
transverse momentum to position space, we then have single-quark
distributions $f(x, \tvec{z}; \zeta, \mu)$.  The $\mu$ dependence is given
by a homogeneous renormalization group equation~\cite{Collins:1981uk}
\begin{align}
  \label{tmd-rge}
\frac{d}{d\log\mu} f(x, \tvec{z}; \zeta, \mu)
 &= 2\gamma_q\bigl( \alpha_s(\mu) \bigr)\,
    f(x, \tvec{z}; \zeta, \mu) \,,
\end{align}
where $\gamma_q = 3\ms C_F \alpha_s /(4\pi) + \mathcal{O}(\alpha_s^2)$ can
be identified with the anomalous dimension of the quark field in the axial
gauge $v A = 0$ (where the Wilson lines reduce to unity).  In a covariant
gauge it corresponds to renormalization of the composite operator
$W(\xi)\ms q(\xi)$ or $\bar{q}(\xi)\ms W^\dagger(\xi)$, see
\eqref{full-wilson-op}.

\begin{figure}
\begin{center}
\includegraphics[width=0.99\textwidth]{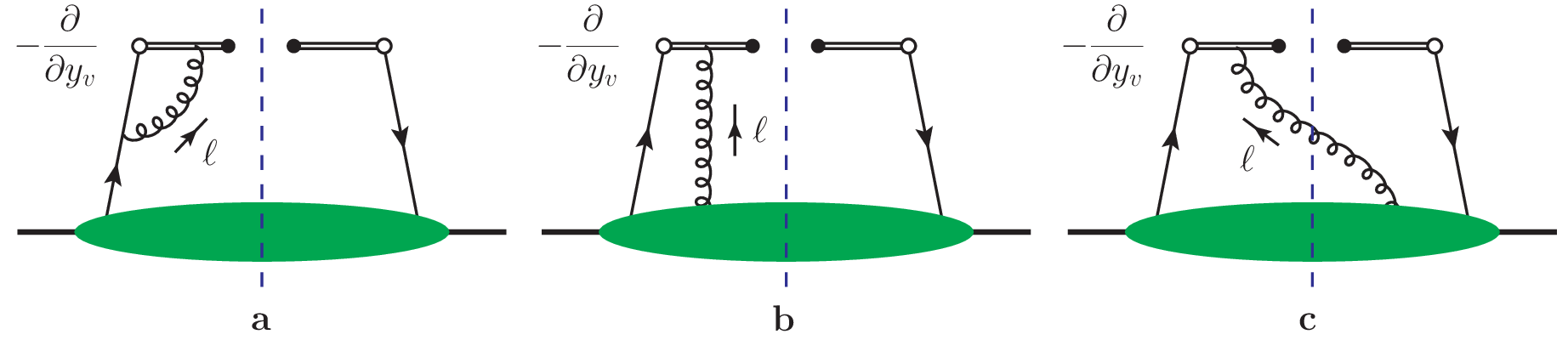}
\end{center}
\caption{\label{fig:cs-kernels} Graphs describing the dependence of a
  single parton distribution on the Wilson line rapidity $y_v$ at order
  $\alpha_s$.  The complex conjugate graphs are not shown.  Note that
  graph b includes the case where the gluon couples to the left quark
  line, i.e.\ graph a is included in b.}
\end{figure}

In terms of graphs, the dependence of $f$ on $v$ arises from the
propagators of eikonal lines and from their coupling to gluons, as is
obvious from the Feynman rules in figure~\ref{fig:eikonal-rules}.  A power
counting analysis shows that dominant contributions to $\partial
f/\partial\zeta$ come from regions where the momentum $\ell$ flowing
through eikonal lines is either hard or soft; contributions with $\ell$
collinear to the proton are power suppressed.  The only important graphs
where $\ell$ is a hard momentum have the form of a vertex correction shown
in figure~\ref{fig:cs-kernels}a (at higher orders this vertex graph
becomes dressed with further gluon and quark lines).  If the momentum
$\ell$ flows through spectator partons (see figure~\ref{fig:cs-kernels}b
and c), the hard region is power suppressed and only the region of soft
$\ell$ is important.

The contribution from the region of hard momenta $\ell$ reads
\begin{align}
\frac{\partial}{\partial\log\zeta}\, f(x, \tvec{z}; \zeta, \mu) 
\Big|_{\text{hard}}
 &= G(x \zeta/ \mu)\, f(x, \tvec{z}; \zeta, \mu) \,.
\end{align}
Since $\ell$ is hard, the kernel $G$ can be calculated perturbatively.
$G$ is extracted from subgraphs whose external lines are the quark on the
left of the final-state cut and the adjacent eikonal line, or the
corresponding two lines on the right of the cut.  As a result, $G$ is
independent of the transverse distance $\tvec{z}$ between the two quark
field operators.  Furthermore, it depends only on the longitudinal quark
momentum $xp$ rather than on the proton momentum $p$, so that the
dependence on $\zeta$ is via the combination $x\zeta = 2 (xp)\ms
v/\sqrt{|v^2|}$.  For dimensional reasons this parameter must be divided
by $\mu$, since $G$ is dominated by hard momenta and hence independent of
nonperturbative scales.
At leading order in $\alpha_s$, one obtains $G$ from the graph of
figure~\ref{fig:cs-kernels}a and the complex conjugate graph, with
subtractions made for the region of soft $\ell$.  Using
$\overline{\text{MS}}$ subtraction for the ultraviolet divergences, one
finds \cite{Collins:1981uk,Idilbi:2004vb}
\begin{align}
  \label{G-kernel}
G(x \zeta/ \mu) &= - \frac{\alpha_s}{\pi}\, C_F\, \biggl[
  \log\biggl( \frac{x^2\zeta^2}{\mu^2} \biggr) - 1 \biggr]
+ \mathcal{O}(\alpha_s^2) \,.
\end{align}

If the momentum carried by the eikonal lines and hence by the gluons they
couple to is soft, we have graphs with soft gluons coupling to parton
lines that move fast to the right.  We can then use the same procedure as
in section~\ref{sec:soft}, i.e.\ approximate the gluon coupling to the
right-moving particles and then use a Ward identity.  The gluons then
couple to eikonal lines with large positive rapidity, associated with a
vector $w$, with one eikonal line for each parton line attached to the
collinear subgraph.  This gives
\begin{align}
  \label{cs-q-soft}
\frac{\partial}{\partial\log\zeta}\, f(x, \tvec{z}; \zeta, \mu) 
\Big|_{\text{soft}}
 &= - \frac{\partial S_q^{v,w}(\tvec{z}, \mu)}{\partial y_v}\,
    \frac{1}{S_q^{v,w}(\tvec{z}, \mu)}\, f(x, \tvec{z}; \zeta, \mu) \,,
\end{align}
where we used $\partial/(\partial\log\zeta) = -\partial/(\partial y_v)$
and where we have explicitly displayed the dependence of $S_q$ defined in
\eqref{single-soft-fact} on the two vectors $v$ and $w$.  As shown in
\cite{Collins:2011}, one can take the limit of lightlike $w$ in
\eqref{cs-q-soft} and thus has
\begin{align}
  \label{cs-q}
\frac{\partial}{\partial\log\zeta}\, f(x, \tvec{z}; \zeta, \mu)
 &= \bigl[ G(x \zeta/ \mu) + K(\tvec{z}, \mu) \bigr]
    f(x, \tvec{z}; \zeta, \mu)
\end{align}
with
\begin{align}
K(\tvec{z}, \mu) = - \lim_{y_w\to \infty}
\frac{\partial S_q^{v,w}(\tvec{z}, \mu)}{\partial y_v}\,
    \frac{1}{S_q^{v,w}(\tvec{z}, \mu)} \,.
\end{align}
Having taken the limit of infinite $y_w$ the dependence of $K$ on $y_v$
has disappeared as well, since by Lorentz invariance $K$ could only depend
on $y_w - y_v$.
For large $\tvec{z}$ the kernel $K$ is dominated by nonperturbative
dynamics, just as $S_q$.  For sufficiently small $\tvec{z}$ we can however
use the perturbative expression of $S_q$ derived in the previous section.
From \eqref{single-soft-pert} we readily obtain
\begin{align}
  \label{K-pert}
K(\tvec{z}, \mu) &=
  - \frac{\alpha_s}{\pi}\, C_F\,\log \frac{\mu^2 \tvec{z}^2}{b_0^2} \,,
\end{align}
verifying that at lowest order $K$ is independent of $v$ and $w$.

An analysis of the ultraviolet divergences in $G$ and $K$ shows that they
satisfy renormalization group equations \cite{Collins:1981uk}
\begin{align}
  \label{kernels-rg}
\frac{d}{d\log\mu}\, G\bigl( x \zeta/ \mu, \alpha_s(\mu) \bigr)
 &= - \frac{d}{d\log\mu}\, K(\tvec{z}, \mu)
  = \gamma_K\bigl( \alpha_s(\mu) \bigr) \,,
\end{align}
so that the sum $G+K$ is independent of $\mu$.  At lowest order in
$\alpha_s$ this is readily verified from \eqref{G-kernel} and
\eqref{K-pert}, which also give the leading term
\begin{align}
  \label{gamma-K}
\gamma_K(\alpha_s)
&= \frac{2 \alpha_s}{\pi}\, C_F + \mathcal{O}(\alpha_s^2)
\end{align}
of the anomalous dimension.  The $\mathcal{O}(\alpha_s^2)$ term is also
known \cite{Collins:1984kg}.

We are now ready to generalize the CS equation to (unsubtracted) two-quark
distributions.  The contribution from hard eikonal momenta $\ell$ is again
given by vertex graphs as in figure~\ref{fig:cs-kernels}a, with one graph
for each of the four quark legs.  This gives just the sum of kernels
$G(x_1 \zeta/ \mu) + G(x_2 \zeta/ \mu)$.  As for the soft momentum region,
the argument leading to \eqref{cs-q-soft} can be repeated.  In
section~\ref{sec:soft} we have seen that the Ward identity for soft gluons
coupling to collinear lines in a two-parton distribution has a nontrivial
color structure.  We therefore now have a matrix equation in color space,
\begin{align}
  \label{cs-soft-mult}
\frac{\partial}{\partial\log\zeta}\,
 \begin{pmatrix}
  \sing{F}_{a_1,a_2}(x_i, \tvec{z}_i, \tvec{y}; \zeta, \mu) \\[0.2em]
  \oct{F}_{a_1,a_2}(x_i, \tvec{z}_i, \tvec{y}; \zeta, \mu)
 \end{pmatrix}
\Bigg|_{\text{soft}}
 &= \mathbf{K}_{qq}(\tvec{z}_i, \tvec{y}; \mu) \,
 \begin{pmatrix}
  \sing{F}_{a_1,a_2} (x_i, \tvec{z}_i, \tvec{y}; \zeta, \mu) \\[0.2em]
  \oct{F}_{a_1,a_2} (x_i, \tvec{z}_i, \tvec{y}; \zeta, \mu)
 \end{pmatrix}
\end{align}
with a kernel
\begin{align}
  \label{matrix-K-def}
\mathbf{K}_{qq}(\tvec{z}_i, \tvec{y}; \mu)
 &= - \lim_{y_w\to \infty}
   \frac{\partial \mathbf{S}_{qq}^{v,w}(\tvec{z}_i, \tvec{y}; \mu)}{%
      \partial y_v}\,
   \bigl[ \mathbf{S}_{qq}^{v,w}(\tvec{z}_i, \tvec{y}; \mu) \bigr]^{-1} \,.
\end{align}
Analogous equations hold for quark-antiquark and for interference
distributions, with kernels $\mathbf{K}$ constructed from the appropriate
soft factors $\mathbf{S}$.  Note that the kernels are sensitive to the
color charge of partons (i.e.\ to the difference between quarks and
antiquarks) but not to their polarization.  For $\mathbf{K}$ this follows
from the analogous property of $\mathbf{S}$, whereas for $G$ it follows
from parity invariance applied to the relevant subgraph with one external
quark and one eikonal line.
Putting everything together, we can write the CS equation for a two-quark
distribution into the form
\begin{align}
  \label{general-cs-eq}
\frac{d}{d\log\zeta}\, \begin{pmatrix}
  \sing{F} \\[0.2em]
  \oct{F} \end{pmatrix}
 &= \bigl[ G(x_1 \zeta, \mu) + G(x_2 \zeta, \mu)
         + K(\tvec{z}_1, \mu) + K(\tvec{z}_2, \mu) \bigr]
    \begin{pmatrix}
      \sing{F} \\[0.2em]
      \oct{F} \end{pmatrix}
\nonumber \\
 &\quad + \mathbf{M}(\tvec{z}_1, \tvec{z}_2, \tvec{y})
      \begin{pmatrix} \sing{F} \\[0.2em]
      \oct{F} \end{pmatrix}
\end{align}
with
\begin{align}
  \label{matrix-M-def}
\mathbf{M}(\tvec{z}_1, \tvec{z}_2, \tvec{y})
 &= \mathbf{K}(\tvec{z}_1, \tvec{z}_2, \tvec{y}; \mu)
  - \bigl[ K(\tvec{z}_1, \mu) + K(\tvec{z}_2,\mu) \bigr]
\begin{pmatrix} 1 \ & 0 \\ 0 & \ 1 \end{pmatrix} \,.
\end{align}
According to \eqref{kernels-rg} the $\mu$ dependence cancels between the
kernels $G$ and $K$ in \eqref{general-cs-eq}.  The matrix $\mathbf{M}$ is
independent of $\mu$ as well.  This can be traced back to the fact that
the only ultraviolet divergent graphs for $\mathbf{S}$ (and hence for
$\mathbf{K}$) have the form of a vertex correction as in
figure~\ref{fig:soft-four-legs}a.  As discussed in the previous section,
these graphs only contribute to the color diagonal elements ${}^{11\!}S$
and ${}^{88\!}S$ and are the same in both cases, since they are
insensitive to the color of the different eikonal lines.

Finally, the dependence of $F_{a_1, a_2}$, $F_{a_1, \bar{a}_2}$ etc.\ on
the renormalization scale is given by
\begin{align}
  \label{double-rge}
\frac{d}{d\log\mu} F(x_i, \tvec{z}_i, \tvec{y}; \zeta, \mu)
 &= 4\gamma_q\bigl( \alpha_s(\mu) \bigr)\,
    F(x_i, \tvec{z}_i, \tvec{y}; \zeta, \mu)
\end{align}
for both $\sing{F}$ and $\oct{F}$.  This is because ultraviolet
renormalization in $F$ is performed for individual operators $W(\xi)\ms
q(\xi)$ and $\bar{q}(\xi)\ms W^\dagger(\xi)$.


\subsubsection{General solution}
\label{sec:cs-solve}

Before solving \eqref{general-cs-eq} let us first consider the simpler
equation
\begin{align}
  \label{cs-simple}
\frac{d}{d\log\zeta}\, F(x_i, \tvec{z}_i, \tvec{y}; \zeta, \mu)
 &= \Bigl[ G\bigl( x_1\zeta/ \mu, \alpha_s(\mu) \bigr)
         + G\bigl( x_2\zeta/ \mu, \alpha_s(\mu) \bigr)
\nonumber \\
& \quad  + 2 K_{12}(\tvec{z}_1, \tvec{z}_2, \mu) \Bigr]
  F(x_i, \tvec{z}_i, \tvec{y}; \zeta, \mu) \,,
\end{align}
where we have abbreviated
\begin{align}
K_{12}(\tvec{z}_1, \tvec{z}_2, \mu)
 &= \frac{1}{2}\ms \Bigl[ K(\tvec{z}_1, \mu)
                        + K(\tvec{z}_2, \mu) \Bigr] \,.
\end{align}
We have included the argument of the running coupling in $G$ since this
will be needed shortly.  For the moment we do not assume that $K$ is given
by a perturbative expansion.  The solution of \eqref{cs-simple} can be
obtained by adapting the well-known solution of the CS equation for
single-parton distributions
\cite{Collins:1981uk,Collins:1984kg,DescotesGenon:2001hm}.  We have
\begin{align}
  \label{cs-simple-solution}
F(x_i, \tvec{z}_i, \tvec{y}; \zeta, \mu)
 &= \exp\bigl[- \mathcal{S}(x_1\zeta, \tvec{z}_1, \tvec{z}_2, \mu_0)
              - \mathcal{S}(x_2\ms\zeta, \tvec{z}_1, \tvec{z}_2, \mu_0)
    \bigr]\,
    F^{\mu_0}(x_i, \tvec{z}_i, \tvec{y}; \mu) \,,
\end{align}
where $F^{\mu_0}$ specifies the initial condition of evolution in $\zeta$.
The scale $\mu_0$ should be chosen such that $F^{\mu_0}$ does not depend
on widely disparate scales.  If this is not possible because $\tvec{z}_i$
and $\tvec{y}$ widely differ in size, further steps may be required in
order to resum all large logarithms.
We note that since $\mathcal{S}$ is $\mu$ independent (see below), the
$\mu$ dependence of $F^{\mu_0}$ is given by the same renormalization group
equation as in \eqref{double-rge}.
The Sudakov exponent in \eqref{cs-simple-solution} reads
\begin{align}
  \label{sudakov-1}
\mathcal{S}(x \zeta, \tvec{z}_1, \tvec{z}_2, \mu_0)
&= - \int_{\mu_0}^{x \zeta} \frac{d\zeta'}{\zeta'}
     \biggl[\ms G\bigl( x\zeta'/ \mu, \alpha_s(\mu) \bigr)
              + K_{12}(\tvec{z}_1, \tvec{z}_2, \mu) \biggr]
\end{align}
with $x$ equal to $x_1$ or $x_2$.  It is well-known from the solution
\begin{align}
  \label{cs-q-solution}
f(x, \tvec{z}; \zeta, \mu) &=
  \exp\bigl[- \mathcal{S}(x\zeta, \tvec{z}, \tvec{z}; \mu_0) \bigr]\,
  f^{\mu_0}(x, \tvec{z}; \mu)
\end{align}
of the CS equation \eqref{cs-q} for single-quark distributions.
We note that a more general set of solutions can be obtained by
multiplying $x\zeta$ with a constant $C_2$ in the upper integration limit
of \eqref{sudakov-1}; the initial condition $F^{\mu_0}$ in
\eqref{cs-simple-solution} then depends on that constant.  One can easily
restore the $C_2$ dependence of the expressions to follow, but for
simplicity we limit ourselves to the choice $C_2=1$ here.

The integrand of \eqref{sudakov-1} contains functions that depend on the
ratio of two large scales.  To make this dependence explicit, one uses the
renormalization group equation \eqref{kernels-rg} for $G$ and $K$.
Obviously, $K_{12}$ has the same $\mu$ dependence as $K$, so that the
$\mu$ dependence cancels between $G$ and $K_{12}$ in \eqref{cs-simple} and
\eqref{sudakov-1}.  Using \eqref{kernels-rg} one can rewrite
\begin{align}
  \label{kernel-sum-1}
& G\bigl( x \zeta'/ \mu, \alpha_s(\mu) \bigr)
       + K_{12}(\tvec{z}_1, \tvec{z}_2, \mu)
\nonumber \\
&\quad = G\bigl( 1, \alpha_s(x \zeta') \bigr)
       + K_{12}(\tvec{z}_1, \tvec{z}_2, \mu_0)
       - \int_{\mu_0}^{x \zeta'} \frac{d\mu'}{\mu'}\,
              \gamma_K\bigl( \alpha_s(\mu') \bigr) \,.
\end{align}
Inserting this into \eqref{sudakov-1} one can perform the integration over
$\zeta'$ for the terms containing $\gamma_K$ or $K_{12}$ and obtains
\begin{align}
  \label{sudakov-2}
\mathcal{S}(x \zeta, \tvec{z}_1, \tvec{z}_2, \mu_0)
&= \int_{\mu_0}^{x \zeta} \frac{d\mu}{\mu}
  \biggl[ \gamma_K\bigl( \alpha_s(\mu) \bigr) \log\frac{x \zeta}{\mu}
        - G\bigl( 1, \alpha_s(\mu) \bigr) \biggr]
\nonumber \\
&\quad
 - K_{12}(\tvec{z}_1, \tvec{z}_2, \mu_0)\, \log\frac{x \zeta}{\mu_0} \,.
\end{align}
The term with $\gamma_K$ in \eqref{sudakov-2} gives rise to the leading
double logarithm in $x\zeta/ \mu_0$, whereas the other terms give only
single logarithms.  Using \eqref{G-kernel} and \eqref{gamma-K} and
neglecting the running of $\alpha_s$ one has
\begin{align}
\mathcal{S}(x \zeta, \tvec{z}_1, \tvec{z}_2, \mu_0)
&= \frac{\alpha_s}{\pi}\, C_F \log^2\frac{x \zeta}{\mu_0}
   - \biggl[ \frac{\alpha_s}{\pi}\, C_F 
             + K_{12}(\tvec{z}_1, \tvec{z}_2, \mu_0) \biggr]
      \log\frac{x \zeta}{\mu_0}
 + \mathcal{O}(\alpha_s^2)
\end{align}
at leading order in $\alpha_s$.  A more precise expression is obtained by
rewriting $\int d\mu/\mu = \half \int d\alpha_s /\beta(\alpha_s)$, where
$\beta = d \alpha_s(\mu) /d\log \mu^2$.  After expanding
$1/\beta(\alpha_s)$ in $\alpha_s$, the integral in \eqref{sudakov-2} is
straightforward to evaluate for the one-loop expression \eqref{G-kernel}
of $G$ and the two-loop expression of $\gamma_K$ (given e.g.\ in
\cite{Collins:1984kg}).

The form \eqref{sudakov-2} is valid even if $K_{12}(\tvec{z}_1,
\tvec{z}_2, \mu)$ cannot be evaluated perturbatively because one or both
of $\tvec{z}_1$ and $\tvec{z}_2$ are large.  If both distances are small,
$K(\tvec{z}_i, \mu)$ and thus $K_{12}(\tvec{z}_1, \tvec{z}_2, \mu)$ is
given by a power series in $\alpha_s(\mu)$.  An alternative form of the
Sudakov exponent \cite{Collins:1984kg} is then obtained by rewriting
\eqref{kernel-sum-1} as
\begin{align}
  \label{kernel-sum-2}
& G\bigl( x \zeta'/ \mu, \alpha_s(\mu) \bigr)
  + K_{12}\bigl( \tvec{z}_1, \tvec{z}_2, \mu, \alpha(\mu) \bigr)
\nonumber \\
&\quad = G\bigl( 1, \alpha_s(x \zeta') \bigr)
       + K_{12}\bigl( \tvec{z}_1, \tvec{z}_2, \mu_0,
                      \alpha_s(x \zeta') \bigr)
       - \int_{\mu_0}^{x \zeta'} \frac{d\mu'}{\mu'}\,
         A\bigl( \tvec{z}_1, \tvec{z}_2, \mu_0, \alpha_s(\mu') \bigr)           
\end{align}
with
\begin{align}
A(\tvec{z}_1, \tvec{z}_2, \mu_0, \alpha_s)
 &= \gamma_K\bigl( \alpha_s \bigr)
           + 2\beta(\alpha_s)\, \frac{\partial}{\partial \alpha_s}
             K_{12}\bigl( \tvec{z}_1, \tvec{z}_2, \mu_0, \alpha_s \bigr) \,,
\end{align}
where we now distinguish between the explicit $\mu$ dependence of $K_{12}$
and the implicit dependence via the running coupling.  One then obtains
\begin{align}
  \label{sudakov-3}
\mathcal{S}(x \zeta, \tvec{z}_1, \tvec{z}_2, \mu_0)
&= \int_{\mu_0}^{x \zeta} \frac{d\mu}{\mu}
  \biggl[ A\bigl( \tvec{z}_1, \tvec{z}_2, \mu_0, \alpha_s(\mu) \bigr)
          \log\frac{x \zeta}{\mu}
        - G\bigl( 1, \alpha_s(\mu) \bigr)
\nonumber \\[0.3em]
&\qquad\qquad\quad
 - K_{12}\bigl( \tvec{z}_1, \tvec{z}_2, \mu_0, \alpha_s(\mu) \bigr)
 \biggr] \,,
\end{align}
where all perturbative functions are evaluated with $\alpha_s$ at the same
scale.  With the one-loop expression of $K$ in \eqref{K-pert} we have
\begin{align}
K_{12}(\tvec{z}_1, \tvec{z}_2, \mu, \alpha_s) =
- \frac{\alpha_s}{\pi}\, C_F
  \log\frac{\mu^2\, |\tvec{z}_1| \ms |\tvec{z}_2|}{b_0^2}
+ \mathcal{O}(\alpha_s^2) \,.
\end{align}
A natural choice for the starting scale of evolution in $\zeta$ is thus
\begin{align}
  \label{mu0-choice}
\mu_0^2 &= \frac{C_1}{|\tvec{z}_1| \ms |\tvec{z}_2|}
\end{align}
with a constant $C_1$ of order 1.  If one takes $C_1 = b_0^2$ then
$K_{12}$ vanishes.

It is now easy to write down the solution of the full CS equation
\eqref{general-cs-eq} for a two-parton distribution.  It is given by
\begin{align}
  \label{general-cs-solution}
\begin{pmatrix}
  \sing{F}(x_i, \tvec{z}_i, \tvec{y}; \zeta, \mu) \\[0.2em]
  \oct{F}(x_i, \tvec{z}_i, \tvec{y}; \zeta, \mu) \end{pmatrix}
&= \exp\bigl[- \mathcal{S}(x_1\zeta, \tvec{z}_1, \tvec{z}_2, \mu_0)
             - \mathcal{S}(x_2\ms\zeta, \tvec{z}_1, \tvec{z}_2, \mu_0)
       \bigr]
\nonumber \\
&\quad\times
   \exp\biggl[\ms \mathbf{M}(\tvec{z}_1, \tvec{z}_2, \tvec{y})\,
              \log\frac{\sqrt{x_1 x_2\rule{0pt}{1.5ex}}\, \zeta}{\mu_0}
       \ms\biggr]\,
\begin{pmatrix}
  \sing{F}^{\mu_0}(x_i, \tvec{z}_i, \tvec{y}; \mu) \\[0.2em]
  \oct{F}^{\mu_0}(x_i, \tvec{z}_i, \tvec{y}; \mu) \end{pmatrix}
\end{align}
with $\mathcal{S}$ given by \eqref{sudakov-2} for arbitrary values of
$\tvec{z}_i$ and by \eqref{sudakov-3} if both $\tvec{z}_1$ and
$\tvec{z}_2$ are small.  The logarithm in the second line has been chosen
such that it coincides with the one that multiplies $2 K_{12}$ when one
evaluates $- \mathcal{S}(x_1\zeta) - \mathcal{S}(x_2\zeta)$ from
\eqref{sudakov-2}.  Other choices are possible and lead to different
initial conditions $\sing{F}^{\mu_0}$ and $\oct{F}^{\mu_0}$.

Unless all distances $\tvec{z}_1$, $\tvec{z}_2$ and $\tvec{y}$ are small,
the matrix $\mathbf{M}$ cannot be calculated perturbatively and we cannot
further simplify the exponentiated matrix in \eqref{general-cs-solution}.
Nevertheless, \eqref{general-cs-solution} contains some important
information, since it gives the explicit form of the dependence on the
large scales $x_1 \zeta$ and $x_2 \zeta$.  In particular, we see that to
leading double logarithmic accuracy, where only squared logarithms of $x_1
\zeta /\mu_0$ and $x_2 \zeta /\mu_0$ are retained, the Sudakov factor for
two-quark distributions is the same for $\sing{F}$ and for $\oct{F}$ and
given by the product of the corresponding Sudakov factors for single-quark
densities with momentum fractions $x_1$ and $x_2$.  At next-to-leading
logarithmic accuracy, $\sing{F}$ and $\oct{F}$ mix under evolution in
$\zeta$, with the amount of mixing depending on the transverse distances
$\tvec{z}_1$, $\tvec{z}_2$ and $\tvec{y}$.

It should be possible to generalize the CS equation \eqref{general-cs-eq}
and its general solution \eqref{general-cs-solution} to the case of
multiparton distributions for more than two partons.  The same holds for
multi-gluon distributions, where the general structure will remain the
same but the kernels $G$ and $K$ will be different.


\subsubsection{Small transverse distances}
\label{sec:cs-perturb}

Let us now consider the situation when $\tvec{z}_1$, $\tvec{z}_2$ and
$\tvec{y}$ are all small enough to calculate $\mathbf{M}$ in perturbation
theory.  The kernels $K(\tvec{z}_i, \mu)$ in \eqref{general-cs-eq} are
then given by \eqref{K-pert} at leading order in $\alpha_s$, and the
Sudakov exponent $\mathcal{S}$ in the solution \eqref{general-cs-solution}
of the CS equation can be evaluated from \eqref{sudakov-3}.  It remains to
investigate the matrix $e^{L \mathbf{M}}$ in \eqref{general-cs-solution},
where we abbreviate
\begin{align}
 L &= \log\frac{\sqrt{x_1 x_2\rule{0pt}{1.5ex}}\, \zeta}{\mu_0} \,.
\end{align}
We treat the kernel $\mathbf{M}_{qq}$ for two-quark distributions $F_{a_1,
  a_2}$ in detail and discuss its analogs for quark-antiquark
distributions $F_{a_1, \bar{a}_2}$ and interference distributions $I_{a_1,
  a_2}$ later.

Using the definitions \eqref{matrix-K-def}, \eqref{matrix-M-def} and our
perturbative result \eqref{S-matrix-qq2} for $\mathbf{S}_{qq}$, we readily
find
\begin{align}
  \label{M-matrix-qq}
\mathbf{M}_{qq}(\tvec{z}_1, \tvec{z}_2, \tvec{y}) &=
\begin{pmatrix} 0 & c\ms K_d \\[0.2em]
  c\ms K_d \quad & - (1+c^2) K_y - 2 c^2 K_d \end{pmatrix}
\end{align}
with $c$ given in \eqref{c-color-factor} and
\begin{align}
  \label{K-dy-kernels}
K_d(\tvec{z}_i, \tvec{y}) &= 
    K\Bigl( \tvec{y} + \frac{\tvec{z}_1 + \tvec{z}_2}{2}, \mu \Bigr)
  + K\Bigl( \tvec{y} - \frac{\tvec{z}_1 + \tvec{z}_2}{2}, \mu \Bigr)
\nonumber \\[0.1em]
&\quad
  - K\Bigl( \tvec{y} + \frac{\tvec{z}_1 - \tvec{z}_2}{2}, \mu \Bigr)
  - K\Bigl( \tvec{y} - \frac{\tvec{z}_1 - \tvec{z}_2}{2}, \mu \Bigr)
\nonumber \\[0.2em]
K_y(\tvec{z}_i, \tvec{y}) &=
    K(\tvec{z}_1, \mu) + K(\tvec{z}_2, \mu)
  - K\Bigl( \tvec{y} + \frac{\tvec{z}_1 + \tvec{z}_2}{2}, \mu \Bigr)
  - K\Bigl( \tvec{y} - \frac{\tvec{z}_1 + \tvec{z}_2}{2}, \mu \Bigr)
\end{align}
in analogy to \eqref{S-dy}. Using the explicit form \eqref{K-pert} or the
renormalization group equation \eqref{kernels-rg} for $K$, we see that
$\mathbf{M}$ is $\mu$ independent, as we anticipated earlier.  We must of
course choose a scale in $\alpha_s$ when using the one-loop result
\eqref{K-pert} for evaluating $K_d$ and $K_y$, which gives rise to a
residual scale dependence of order $\alpha_s^2$.  The situation is the
same as for a physical (and hence formally $\mu$ independent) quantity
evaluated in fixed-order perturbation theory.  An appropriate scale of
$\alpha_s$ in $K_d$ and $K_y$ will be constructed from $\tvec{z}_i$ and
$\tvec{y}$.

Let $\mathbf{D}_{qq}$ be the diagonal matrix with the eigenvalues of
$\mathbf{M}_{qq}$ and let $\mathbf{E}_{qq}$ be the matrix whose columns
are the corresponding eigenvectors, i.e.
\begin{align}
\mathbf{M}_{qq} \ms \mathbf{E}_{qq}
  &= \mathbf{E}_{qq}\ms \mathbf{D}_{qq} &
\text{with}\
\mathbf{D}_{qq}
 &= \begin{pmatrix} d_+ & 0 \, \\ 0 \, & d_- \end{pmatrix} \,.
\end{align}
One then has
\begin{align}
e^{L\ms \mathbf{M}_{qq}}
 &= \mathbf{E}_{qq}^{}\;
    e^{L\ms \mathbf{D}_{qq}}\; \mathbf{E}_{qq}^{-1} \,.
\end{align}
The matrix \eqref{M-matrix-qq} has eigenvalues
\begin{align}
  \label{M-matrix-qq-evals}
d_\pm &= \frac{1}{2}\, \biggl[ - (1+c^2) K_y - 2 c^2 K_d \pm
    \sqrt{ \bigl( (1+c^2) K_y + 2 c^2 K_d \bigr)^2
         + \bigl( 2 c K_d \bigr)^2 } \;\biggr]
\end{align}
and a matrix of eigenvectors
\begin{align}
\mathbf{E}_{qq} &= \begin{pmatrix}
 - d_- \; & - \dfrac{d_+}{c K_d} \\[1em]
 c K_d \; & 1 \end{pmatrix} \,,
\end{align}
so that
\begin{align}
e^{L\ms \mathbf{M}_{qq}} &= \frac{1}{d_+ - d_-}
  \begin{pmatrix}
     d_+ e^{L d_-} - d_- e^{L d_+} &
     c K_d\ms \bigl( e^{L d_+} - e^{L d_-} \bigr) \\[0.5em]
     c K_d\ms \bigl( e^{L d_+} - e^{L d_-} \bigr) \quad &
     d_+ e^{L d_+} - d_- e^{L d_-}
  \end{pmatrix} \,.
\end{align}
Let us see how this matrix behaves for
\begin{align}
  \label{K-condition}
     |\ms c K_d \ms| & \ll |K_y| \,.
\end{align}
One can then Taylor expand the square root in \eqref{M-matrix-qq-evals}
and obtains
\begin{align}
  \label{qq-evectors-limit}
d_+ &= \frac{c^2}{1+c^2}\, \frac{K_d^2}{K_y}
     + \mathcal{O}\biggl( c^4\, \frac{K_d^3}{K_y^2} \biggr) \,,
&
d_- &= - (1+c^2) K_y + \mathcal{O}( c^2 K_d )
\end{align}
if $K_y>0$, whereas the role of $d_+$ and $d_-$ in
\eqref{qq-evectors-limit} is interchanged if $K_y<0$.  In both cases one
gets
\begin{align}
  \label{qq-matrix-approx}
e^{L\ms \mathbf{M}_{qq}} &\approx
\exp\biggl[\ms L\, \frac{1}{N^2}\, \frac{K_d^2}{K_y} \biggr]
  \begin{pmatrix}
    1 & \dfrac{1}{N b}\, \dfrac{K_d}{K_y}\,
        \bigl( 1 - e^{- L\ms b^2 K_y} \bigr) \\[1em]
        \dfrac{1}{N b}\, \dfrac{K_d}{K_y}\,
        \bigl( 1 - e^{- L\ms b^2 K_y} \bigr) \qquad &
    e^{- L\ms b^2 K_y} + \Bigl( \dfrac{1}{N b}\, \dfrac{K_d}{K_y} \Bigr)^2
  \end{pmatrix} \,,
\end{align}
where we have traded the color factor $c = 1/\sqrt{N^2-1}$ for
\begin{align}
b &= \frac{N}{\sqrt{N^2-1}} \,.
\end{align}

Since $c \sim 1/N$, the condition \eqref{K-condition} holds in the
large-$N$ limit.  Inserting \eqref{qq-matrix-approx} into
\eqref{general-cs-solution}, we see that $\sing{F}(\zeta)$ is then
controlled by the initial condition $\sing{F}^{\mu_0}$ because the
admixture from $\oct{F}^{\mu_0}$ is suppressed, although only by $1/N$.
Whether $\oct{F}(\zeta)$ is dominated by $\sing{F}^{\mu_0}$ or
$\oct{F}^{\mu_0}$ depends on whether the $1/N$ suppressed factor in the
lower row of \eqref{qq-matrix-approx} or the exponential $e^{- L\ms b^2
  K_y}$ is smaller.  In either case $\oct{F}(\zeta)$ is parametrically
smaller than $\sing{F}(\zeta)$.  To which extent the large-$N$ limit gives
a valid description of the physics for $N=3$ depends on the relative size
of $K_y$ and $K_d$, as well as the relative size of $\sing{F}^{\mu_0}$ and
$\oct{F}^{\mu_0}$.  This can only be decided by a more detailed analysis,
which we will not attempt here.

There is, however, a region of phase space where \eqref{K-condition} holds
beyond the large-$N$ limit.  From its definition \eqref{K-dy-kernels} we
see that $K_d$ vanishes if $\tvec{z}_1 = \tvec{z}_2 = \tvec{0}$.  In the
double-scattering process one has $|\tvec{z}_1| \sim |\tvec{z}_2| \sim
1/q_T$, so that the limit $|\tvec{z}_1|, |\tvec{z}_2| \ll |\tvec{y}|$ is
relevant in the region where $|\tvec{y}|$ is much larger than $1/q_T$.
Taylor expansion then gives
\begin{align}
K_d &= \frac{\alpha_s}{\pi}\, 2 C_F\, \biggl[\,
  \frac{2\ms (\tvec{y} \tvec{z}_1)\ms
             (\tvec{y} \tvec{z}_2)}{(\tvec{y}^2)^2}
- \frac{\tvec{z}_1 \tvec{z}_2}{\tvec{y}^2} \,\biggr]
+ \mathcal{O}\biggl( \frac{|\tvec{z}_i|^4}{|\tvec{y}|^4} \biggr) \,,
\nonumber \\[0.2em]
K_y &= \frac{\alpha_s}{\pi}\, 2 C_F\,
  \log\frac{\tvec{y}^2}{|\tvec{z}_1|\ms |\tvec{z}_2|}
+ \mathcal{O}\biggl( \frac{|\tvec{z}_i|^2}{\tvec{y}^2} \biggr) \,.
\end{align}
This implies
\begin{align}
\biggl| \frac{K_d}{K_y} \biggr| &\le
  \frac{|\tvec{z}_1|\ms |\tvec{z}_2|}{\tvec{y}^2}\,
  \biggl(\ms \log\frac{\tvec{y}^2}{|\tvec{z}_1|\ms |\tvec{z}_2|}
  \ms\biggr)^{-1}
+ \mathcal{O}\biggl( \frac{|\tvec{z}_i|^4}{|\tvec{y}|^4} \biggr)
\end{align}
for the factor in the off-diagonal elements of the matrix
\eqref{qq-matrix-approx}, whereas for the exponential factor we find
\begin{align}
  \label{exponential-p}
e^{-L\ms b^2 K_y} &\approx \biggl( 
  \frac{|\tvec{z}_1|\ms |\tvec{z}_2|}{\tvec{y}^2} \biggr)^{N p}
&
\text{with}\ \, p
 &= \frac{\alpha_s}{\pi}\, L = \frac{\alpha_s}{2\pi}\,
    \log\frac{|\tvec{z}_1|\ms |\tvec{z}_2|\, x_1 x_2\ms \zeta^2}{C_1} \,,
\end{align}
where we have chosen $\mu_0$ as in \eqref{mu0-choice}.  We thus find that
the octet admixture in the $\zeta$ evolution of $\sing{F}(\zeta)$ is power
suppressed by $|\tvec{z}_1|\ms |\tvec{z}_2| \big/ \tvec{y}^2$, and that
$\oct{F}(\zeta)$ is power suppressed compared with $\sing{F}(\zeta)$ by
\begin{align}
  \label{ratio-power-supp}
\frac{\oct{F}_{a_1,a_2}(x_i, \tvec{z}_i, \tvec{y}; \zeta, \mu)}{%
      \sing{F}_{a_1,a_2} (x_i, \tvec{z}_i, \tvec{y}; \zeta, \mu)}
\;\sim\;
\frac{\oct{F}_{a_1,a_2}^{\mu_0}(x_i, \tvec{z}_i, \tvec{y}; \mu)}{%
      \sing{F}_{a_1,a_2}^{\mu_0}(x_i, \tvec{z}_i, \tvec{y}; \mu)} \;\,
\biggl( \frac{|\tvec{z}_1|\ms |\tvec{z}_2|}{\tvec{y}^2} \biggr)^{%
  \min(1, N p)} \,.
\end{align}
In straightforward generalization of single Drell-Yan production
\cite{Collins:1981uk,Collins:2011}, an adequate choice of $\zeta$ in the
cross section for double hard scattering is $x_1 x_2\ms \zeta^2 \sim Q^2$.
Together with $|\tvec{z}_1|\ms |\tvec{z}_2| \sim 1/q_T^2$ this gives $p
\sim (\alpha_s/\pi) \log(Q/q_T)$.  Again, a more quantitative picture can
only be obtained by a detailed analysis.

The preceding results all rely on the validity of perturbation theory for
the soft CS kernel and thus require not only $\tvec{z}_1$ and $\tvec{z}_2$
but also $\tvec{y}$ to be perturbatively small.  We cannot draw any strict
conclusions about the case where $\tvec{z}_1$ and $\tvec{z}_2$ are small,
whereas $\tvec{y}$ is in the nonperturbative region.  However, we observe
that the power suppression parameter $|\tvec{z}_1|\ms |\tvec{z}_2| \big/
\tvec{y}^2$ becomes smaller rather than larger in this case.  One may thus
speculate that the general features of our analysis, namely the autonomous
$\zeta$ evolution of $\sing{F}$ and the suppression of $\oct{F}$ will
continue in the nonperturbative regime.

We conclude this section by noting that the CS equation and its solution
for quark-antiquark distributions $F_{a_1, \bar{a}_2}$ is readily obtained
from the previous results by replacing $\tvec{z}_2 \to -\tvec{z}_2$ and
that corresponding replacements are to be made for $F_{\bar{a}_1, a_2}$
and $F_{\bar{a}_1, \bar{a}_2}$.  This follows from the corresponding
property of the soft factor $\mathbf{S}_{q\bar{q}}$ discussed at the end
of section~\ref{sec:soft-qq}.


\paragraph{Interference distributions.}

The Collins-Soper equation for interference distributions
$\sing{I}_{a_1,\bar{a}_2}$ and $\oct{I}_{\bar{a}_1,\bar{a}_2}$ has the
same form as \eqref{general-cs-eq} with $F$ replaced by $I$.  The
appropriate kernel $\mathbf{M}_I$ in the perturbative regime follows from
$\mathbf{S}_I$ in \eqref{S-matrix-int2} and reads
\begin{align}
  \label{M-matrix-interf}
\mathbf{M}_I(\tvec{z}_1, \tvec{z}_2, \tvec{y}) &= {}-
\begin{pmatrix} K_y & c\ms (K_y + K_d) \\[0.3em]
   c\ms (K_y + K_d) \quad & (1-2c^2) (K_y + K_d) + c^2 K_d
\end{pmatrix}
\end{align}
with $K_d$ and $K_y$ given in \eqref{K-dy-kernels}.  This matrix has
eigenvalues
\begin{align}
d^{\ms\prime}_\pm &= \frac{1}{2}\, \biggl[ - (1-c^2)\ms (2 K_y + K_d)
  \pm \sqrt{1+c^2\rule{0pt}{1.8ex}}\,
      \sqrt{c^2\ms (2 K_y + K_d)^2 + K_d^2} \;\biggr]
\end{align}
and exponentiates to
\begin{align}
e^{L\ms \mathbf{M}_I} &= \frac{1}{d^{\ms\prime}_+ - d^{\ms\prime}_-}
  \begin{pmatrix}
     d^{\ms\prime}_+ e^{L d^{\ms\prime}_-}
   - d^{\ms\prime}_- e^{L d^{\ms\prime}_+}  &
     - c K_d\ms \bigl( e^{L d^{\ms\prime}_+}
     - e^{L d^{\ms\prime}_-} \bigr) \\[0.5em]
     - c K_d\ms \bigl( e^{L d^{\ms\prime}_+}
     - e^{L d^{\ms\prime}_-} \bigr) \quad &
     d^{\ms\prime}_+ e^{L d^{\ms\prime}_+}
   - d^{\ms\prime}_- e^{L d^{\ms\prime}_-}
  \end{pmatrix}
\nonumber \\[0.3em]
& \quad - \frac{K_y\ms (e^{L d^{\ms\prime}_+}
   - e^{L d^{\ms\prime}_-})}{d^{\ms\prime}_+ - d^{\ms\prime}_-}
  \begin{pmatrix} 1 & c \\[0.2em] c \; & -1 \end{pmatrix}     
\end{align}
For $|K_d| \ll K_y$, i.e.\ if $|\tvec{z}_1|, |\tvec{z}_2| \ll |\tvec{y}|$,
we can Taylor expand the eigenvalues as
\begin{align}
d^{\ms\prime}_+ &= 
 - \frac{N-2}{N-1}\, K_y + \mathcal{O}(K_d) \,,
&
d^{\ms\prime}_- &= 
 - \frac{N+2}{N+1}\, K_y + \mathcal{O}(K_d) \,.
\end{align}
The result simplifies if we use the orthogonal matrix
\begin{align}
\mathbf{U} &= \frac{1}{\sqrt{2N}}\,
\begin{pmatrix} \sqrt{N-1} \quad & - \sqrt{N+1} \\
                \sqrt{N+1} \quad & \quad \sqrt{N-1}
\end{pmatrix}
\end{align}
that implements the basis transformation from $\sing{I}, \oct{I}$ to the
combinations ${}^{\bar{3}\!}I, {}^{6\!}I$ introduced in
\eqref{sextet-def}.  We then have
\begin{align}
  \label{interf-matrix-approx}
\mathbf{U}\, e^{L\ms \mathbf{M}_I}\, \mathbf{U}^{\mathrm{\ms T}}
 &\approx \exp\biggl[ - L\, \frac{N-2}{N-1}\, K_y \biggr]\,
\begin{pmatrix}
  1 \quad &
  \dfrac{K_d}{4c K_y}\,
       \bigl(1 - e^{-L\ms K_y /C_F}\bigr) \\[0.5em]
  \dfrac{K_d}{4c K_y}\,
       \bigl(1 - e^{-L\ms K_y /C_F}\bigr) &
  e^{-L\ms K_y /C_F}
\end{pmatrix} \,.
\end{align}
Repeating the argument that led to \eqref{ratio-power-supp} we obtain
\begin{align}
  \frac{{}^{6\!}I(x_i, \tvec{z}_i, \tvec{y}; \zeta, \mu)}{%
  {}^{\bar{3}\!}I(x_i, \tvec{z}_i, \tvec{y}; \zeta, \mu)}
\;\sim\;
  \frac{{}^{6\!}I^{\mu_0}(x_i, \tvec{z}_i, \tvec{y}; \mu)}{%
  {}^{\bar{3}\!}I^{\mu_0}(x_i, \tvec{z}_i, \tvec{y}; \mu)} \;\,
\biggl( \frac{|\tvec{z}_1|\ms |\tvec{z}_2|}{\tvec{y}^2}
\biggr)^{\min(1, 2p)}
\end{align}
with $p$ given in \eqref{exponential-p}.  Likewise, comparing
\eqref{interf-matrix-approx} with \eqref{qq-matrix-approx} we find
\begin{align}
  \frac{{}^{\bar{3}\!}I(x_i, \tvec{z}_i, \tvec{y}; \zeta, \mu)}{%
  \sing{F}(x_i, \tvec{z}_i, \tvec{y}; \zeta, \mu)}
\;\sim\;
  \frac{{}^{\bar{3}\!}I^{\mu_0}(x_i, \tvec{z}_i, \tvec{y}; \mu)}{%
  \sing{F}^{\mu_0}(x_i, \tvec{z}_i, \tvec{y}; \mu)} \;\,
\biggl( \frac{|\tvec{z}_1|\ms |\tvec{z}_2|}{\tvec{y}^2}
\biggr)^{\frac{(N+1) (N-2)}{N}\ms p} \,.
\end{align}
For $|\tvec{z}_1|, |\tvec{z}_2| \ll |\tvec{y}|$ the sextet combination of
$I$ is hence suppressed compared with the antitriplet one, which in its
turn is small compared with the singlet combination $\sing{F}$.

We finally note that, in contrast to the case of $F$, the limit $|K_d|
\ll K_y$ just discussed does not give the same result as the large-$N$
limit.  In the latter one finds
\begin{align}
e^{L\ms \mathbf{M}_I} &\approx e^{-L K_y}\,
\begin{pmatrix} 1 &
  - \dfrac{1}{N}\, \Bigl( 1 + \dfrac{K_y}{K_d} \Bigr)\,
    \bigl( 1 - e^{-L\ms K_y /C_F} \bigr) \\[0.5em]
  - \dfrac{1}{N}\, \Bigl( 1 + \dfrac{K_y}{K_d} \Bigr)\,
    \bigl( 1 - e^{-L\ms K_y /C_F} \bigr) &
  e^{-L\ms K_y /C_F}
\end{pmatrix}
\end{align}
with relative corrections of order $K_y /(N K_d)$.  This expansion is
obviously not useful if one has $|K_d| \ll K_y$.  More generally, the
large-$N$ limit for the $\zeta$ evolution of $I$ will only be useful in
kinematical regions where $K_y /(N K_d)$ is small enough for $N=3$.


\subsection{Collinear factorization}
\label{sec:coll-fact}

The analysis in the previous sections was concerned with
transverse-momentum dependent (TMD) factorization, i.e.\ with cross
sections differential in transverse momenta that are small compared with
the large scale.  We now turn to collinear factorization, adequate for
cross sections with integrated transverse momenta.  We point out the main
changes compared with TMD factorization but do not work out the formalism
in detail.

As in previous sections, we first recapitulate the situation for single
Drell-Yan production.  The main changes compared with TMD factorization
are as follows.
\begin{itemize}
\item We recall from section~\ref{sec:soft} that, after a complex contour
  deformation that avoids the Glauber region, the effect of soft gluon
  exchange is described by a soft factor $S_q(\tvec{z})$.  Integration of
  the cross section \eqref{X-sect-soft} over $\tvec{q}$ sets $\tvec{z}$
  equal to zero in this factor.  Because the cancellation between real and
  virtual graphs gives $S_q(\tvec{0}) = 1$ as discussed in
  section~\ref{sec:soft-basic}, there is no net effect of soft gluons in
  the $q_T$ integrated cross section.

  With the elementary soft factor $S_q$ reduced to unity, subtractions of
  soft-gluon contributions as discussed in section~\ref{sec:full-fact} are
  not required, neither for the parton distributions nor for the
  hard-scattering subprocess.
\item Setting $\tvec{z}=\tvec{0}$ in the quark and antiquark distributions
  $f(x,\tvec{z})$, which is equivalent to integrating $f(x,\tvec{k})$ over
  $\tvec{k}$, gives rise to short distance singularities in addition to
  those that are removed by defining the distributions with renormalized
  quarks fields and Wilson lines.  The dependence on the ultraviolet
  subtraction scale $\mu$ is described by the well-known DGLAP evolution
  equations.
  
  The rapidity divergences (from gluons with small $\ell^+$ and large
  $\ell^-$) that prevent us from taking lightlike Wilson lines in the
  definition of $f(x,\tvec{z})$ cancel between real and virtual
  corrections when $\tvec{z}=\tvec{0}$ \cite{Collins:2003fm}.  Indeed, the
  relevant one-loop graphs are those in figure~\ref{fig:cs-kernels}b and c
  (without the derivative $-\partial/ \partial y_v$), and the
  approximation discussed in section~\ref{sec:cs-general} , which connects
  these graphs to the soft factor $S_q(\tvec{z})$, is valid for any
  $\ell^-$ as long as $\ell^+$ is small.
  
  Collinear parton distributions can hence be defined with lightlike
  Wilson lines and do not depend on a parameter $\zeta$.  Correspondingly,
  the $q_T$ integrated cross section is free of Sudakov logarithms.  In
  the operator definition of $f(x, \mu)$, the Wilson lines in
  $\bar{q}(-\half z)\ms W^\dagger(-\half z;v)$ and $W(\half z;v)\ms
  q(\half z)$ merge to a single Wilson line $W[-\half z, \half z]$, given
  by
\begin{align}
  \label{Wilson-finite}
W[\xi', \xi] &= \operatorname{P}
  \exp\biggl[ ig \int_0^{1} d\lambda\,
  A^{+ a}\bigl( \xi - \lambda\ms (\xi-\xi') \bigr)\, t^a \biggr] \,.
\end{align}
  The sections of the paths that go to infinity in $W^\dagger(-\half z;v)$
  and $W(\half z;v)$ have cancelled, and a path of finite length between
  $-\half z$ and $\half z$ remains.
\item The hard-scattering subprocess now receives radiative corrections
  not only from virtual graphs but also from real ones, since emission of
  partons with large transverse momenta in the final state is permitted
  once we do not fix the transverse momentum $q_T$ of the Drell-Yan
  photon.  As already mentioned, a subtraction for the soft-gluon region
  is not required in this case, in contrast to the situation for TMD
  factorization discussed in section~\ref{sec:full-fact}.  Subtractions
  are however needed for the regions where momenta are collinear to one of
  the partons entering the hard subprocess.  These subtractions must be
  performed in a way that matches the ultraviolet subtractions in the
  parton densities.  In particular, the $\mu$ dependence due to collinear
  subtractions in the hard subprocess has to cancel against the $\mu$
  dependence of the parton distributions in the cross section.
\end{itemize}

Let us now investigate the situation for double Drell-Yan production,
limiting ourselves to a one-loop analysis as we have done throughout the
preceding sections.  A key to understanding the role of soft gluons is to
set $\tvec{z}_i = \tvec{0}$ in $\mathbf{S}_{qq}(\tvec{z}_i, \tvec{y})$,
which results from integrating the cross section \eqref{X-sect-soft-2}
over $\tvec{q}_i$.  Our discussion in section~\ref{sec:soft-basic} implies
that $S_b(\tvec{0},\mu) = \bigl[ U_b(\tvec{0},\mu) \bigr]{}^{\text{ren}} =
- \bigl[ U_a(\mu) \bigr]{}^{\text{ren}} = - S_a(\mu)$ for the graphs in
figure~\ref{fig:soft-basic}.  Together with the relation $S_c(\tvec{z}) =
S_{\smash{b}}(\tvec{z})$, this turns our general one-loop result
\eqref{S-matrix-qq1} into
\begin{align}
  \label{soft-zero}
{}^{11\!}S_{qq} - 1
  &= {}^{18\!}S_{qq} = {}^{81\!}S_{qq} = 0 \,,
& 
{}^{88\!}S_{qq} -1
  &= 2 (1+c^2)\, S(\tvec{y}, \mu)
\end{align}
at $\tvec{z}_1 = \tvec{z}_2 = \tvec{0}$.  The one-loop contributions to
${}^{11\!}S_{qq}$ cancel between the vertex corrections
\ref{fig:soft-four-legs}a and the real graphs
\ref{fig:soft-four-legs}b$_{1}$, in full analogy with the case of $S_{q}$
discussed above.  In ${}^{18\!}S_{qq}$ and ${}^{81\!}S_{qq}$ we have a
cancellation between the real graphs \ref{fig:soft-four-legs}b$_{2}$ and
the virtual graphs \ref{fig:soft-four-legs}c.

We see from \eqref{soft-zero} that in the $q_T$ integrated cross section
for double Drell-Yan production the contributions from color singlet and
color octet distributions decouple from each other, and that they have a
different behavior concerning soft gluon exchange.  In the term with color
singlet distributions $\sing{F}$ we have a cancellation of soft gluon
effects, in full analogy to single Drell-Yan production.  Also, the graphs
for the hard-scattering subprocess are exactly as in the single Drell-Yan
process, with a cancellation of the soft-gluon region but with necessary
subtractions for the regions of collinear parton momenta.  From our above
discussion it follows that collinear two-parton distributions $\sing{F}$
can be defined with the same operators as their single-parton analogs,
with lightlike Wilson lines $W[y - \half z_1, y + \half z_1]$ and $W[-
\half z_2, \half z_2]$ between quark and antiquark fields, and that their
contribution to the $q_T$ integrated cross section is free of Sudakov
logarithms.  Like their single-parton counterparts, the distributions have
ultraviolet divergences; the scale dependence that follows from their
subtraction will be discussed in section~\ref{sec:evolution}.

The contribution of collinear color-octet distributions $\oct{F}$ is quite
different.  Because real and virtual graphs have different color factors,
soft gluon effects do not cancel between them, and their net effect is
described by ${}^{88\!}S_{qq}(\tvec{y})$.  As a consequence, the different
factors in the cross section formula require soft subtractions, as they do
in the case of measured transverse momenta.  Since the color indices of
$[\bar{q}(-\half z_2)\ms W^\dagger(-\half z_2;v) ]_{k'}$ and $[W(\half
z_2;v)\ms q(\half z_2)]_{k}$ are not contracted, the two Wilson lines do
not merge into a single one of finite length, and the same holds for their
analogs with arguments $y - \half z_1$ and $y + \half z_1$.  The vector
$v$ in the Wilson lines cannot be taken lightlike, so that collinear octet
distributions will depend on a parameter $\zeta$.  The resulting
Collins-Soper equation gives rise to Sudakov logarithms, which suppress
the color octet contribution to the $q_T$ integrated cross section.  This
important result was already obtained in \cite{Mekhfi:1988kj}, based on
the observation that in the hard-scattering subprocesses there is no
cancellation of the soft-gluon region.  An adequate scale $\mu_0$ for the
initial condition of the CS equation will in this case be a hadronic
scale, inverse to the typical distance $|\tvec{y}|$ between the two
scattering partons.

Let us finally take a look at the interference distributions
$I_{a_1,\bar{a}_2}$.  From \eqref{S-matrix-int1} we find a soft factor
\begin{align}
\mathbf{S}_I &=
\begin{pmatrix} 1 \ & 0 \\ 0 & \ 1 \end{pmatrix}
+ 2 S(\tvec{y}, \mu) 
\begin{pmatrix} 1 \ & c \\
                c \ \ & 1-2 c^2 \\
\end{pmatrix}
\end{align}
at $\tvec{z}_1 = \tvec{z}_2 = \tvec{0}$.  There is hence no cancellation
of soft-gluon effects, so that a formulation of collinear factorization
will in this case be similar to the one for the color octet distributions
$\oct{F}$ just discussed, with the additional complication of mixing
between the color singlet and octet channels.

\section{Some properties of two-quark and quark-antiquark distributions}
\label{sec:quark}

\subsection{Spin structure}
\label{sec:quark:spin}

Multiparton distributions have a nontrivial spin structure because the
polarizations of different partons can be correlated among themselves,
even in an unpolarized proton.  In the following two sections we first
investigate some general properties of spin correlations between two
quarks and then show that they have observable consequences in multiple
scattering cross sections.  We will encounter several examples for parton
spin correlations in section~\ref{sec:splitting-dist}.


\subsubsection{Spin decomposition}
\label{sec:spin-decomp}

Let us first take a closer look at the spin dependence of the two-quark
distributions $F_{a_1,a_2}$ introduced in \eqref{quark-mixed-F}, making
use of rotation and parity invariance.  We always assume that the hadron
is unpolarized, i.e.\ that the matrix element \eqref{quark-mixed-F} is
averaged over the hadron spin.
The simplest cases are the distributions
\begin{align}
  \label{VV}
F_{q, q} (x_i, \tvec{k}_i, \tvec{y}) &= f_{q, q} (x_1, x_2,
  \tvec{k}_1^2, \tvec{k}_2^2, \tvec{k}_1^{} \tvec{k}_2^{},
  \tvec{k}_1^{} \tvec{y}, \tvec{k}_2^{}\ms \tvec{y}, \tvec{y}^2) \,,
\nonumber \\
F_{\Delta q, \Delta q}(x_i, \tvec{k}_i, \tvec{y}) &=
f_{\Delta q, \Delta q}(x_1, x_2,
  \tvec{k}_1^2, \tvec{k}_2^2, \tvec{k}_1^{} \tvec{k}_2^{},
  \tvec{k}_1^{} \tvec{y}, \tvec{k}_2^{}\ms \tvec{y}, \tvec{y}^2) \,,
\end{align}
which are parity even, i.e.\ scalar functions.  By contrast, the
distributions $F_{q, \Delta q}$ and $F_{\Delta q, q}$ are parity odd,
i.e.\ pseudoscalar functions.  Their general form is
\begin{align}
  \label{VA}
F^{}_{q, \Delta q}(x_i, \tvec{k}_i, \tvec{y}) &=
  \epsilon^{jj'} \tvec{k}_1^j\ms \tvec{y}^{j'}\ms
    f_{q, \Delta q}^{\ms 1}
+ \epsilon^{jj'} \tvec{k}_2^j\ms \tvec{y}^{j'}\ms
    f_{q, \Delta q}^{\ms 2}
+ \epsilon^{jj'} \tvec{k}_1^j\ms \tvec{k}_2^{j'}\ms
    f_{q, \Delta q}^{\ms 3} \,,
\nonumber \\
F^{}_{\Delta q, q}(x_i, \tvec{k}_i, \tvec{y}) &=
  \epsilon^{jj'} \tvec{k}_1^j\ms \tvec{y}^{j'}\ms
    f_{\Delta q, q}^{\ms 1}
+ \epsilon^{jj'} \tvec{k}_2^j\ms \tvec{y}^{j'}\ms
    f_{\Delta q, q}^{\ms 2}
+ \epsilon^{jj'} \tvec{k}_1^j\ms \tvec{k}_2^{j'}\ms
    f_{\Delta q, q}^{\ms 3} \,,
\end{align}
where $f^{1}$, $f^{2}$ and $f^{3}$ are scalar functions with the same
arguments as in \eqref{VV}.  The scalar functions are in general neither
even nor odd in $\tvec{k}_i$ or $\tvec{y}$ since their dependence on
$\tvec{k}_1 \tvec{y}$, $\tvec{k}_2\ms \tvec{y}$ and $\tvec{k}_1^{}
\tvec{k}_2^{}$ is not constrained by symmetry.  Note that the three
two-dimensional vectors $\tvec{y}$, $\tvec{k}_1$ and $\tvec{k}_2$ are
linearly dependent, so that the three cross products in \eqref{VA} are
linearly dependent as well.  Expressing e.g.\ $\tvec{y}$ as a linear
combination of $\tvec{k}_1$ and $\tvec{k}_2$ one obtains
\begin{align}
  \label{lin-dependent}
\epsilon^{jj'} \tvec{k}_1^j\ms \tvec{y}^{j'} &=
  {}- \frac{\tvec{k}_1^2\, (\tvec{k}_2^{}\ms \tvec{y})
       - (\tvec{k}_1^{} \tvec{k}_2^{})\ms (\tvec{k}_1^{} \tvec{y})}{%
    \tvec{k}_1^2\ms \tvec{k}_2^2 - (\tvec{k}_1^{} \tvec{k}_2^{})^2}\;
  \epsilon^{jj'} \tvec{k}_1^j\ms \tvec{k}_2^{j'} \,,
\nonumber \\
\epsilon^{jj'} \tvec{k}_2^j\ms \tvec{y}^{j'} &=
  \frac{\tvec{k}_2^2\, (\tvec{k}_1^{} \tvec{y})
   - (\tvec{k}_1^{} \tvec{k}_2^{})\ms (\tvec{k}_2^{}\ms \tvec{y})}{%
    \tvec{k}_1^2\ms \tvec{k}_2^2 - (\tvec{k}_1^{} \tvec{k}_2^{})^2}\;
 \epsilon^{jj'} \tvec{k}_1^j\ms \tvec{k}_2^{j'}
\end{align}
and can thus write $F_{q, \Delta q}$ as $\epsilon^{jj'} \tvec{k}_1^j\ms
\tvec{k}_2^{j'}$ times a single scalar function.  However, that scalar
function is singular when $\tvec{k}_1$ and $\tvec{k}_2$ become collinear,
as is evident from the denominators in~\eqref{lin-dependent}.  To avoid
such artificial singularities, one can use \eqref{VA} if it is necessary
to make the appearance of $\epsilon^{jj'}$ explicit.

Using that $\half (1 \pm \gamma_5)$ projects on quarks with definite
helicity, one can readily identify the combinations of quark polarizations
that are described by the above functions.  In a schematic notation one
has
\begin{align}
F_{q, q} & \leftrightarrow
   q_1^+ q_2^+  +  q_1^- q_2^-  +  q_1^+ q_2^-  +  q_1^- q_2^+ \,,
\nonumber \\
F_{\Delta q, \Delta q} & \leftrightarrow
   q_1^+ q_2^+  +  q_1^- q_2^-  -  q_1^+ q_2^-  -  q_1^- q_2^+ \,,
\nonumber \\
F_{q, \Delta q} & \leftrightarrow
   q_1^+ q_2^+  -  q_1^- q_2^-  -  q_1^+ q_2^-  +  q_1^- q_2^+ \,,
\nonumber \\
F_{\Delta q, q} & \leftrightarrow
   q_1^+ q_2^+  -  q_1^- q_2^-  +  q_1^+ q_2^-  -  q_1^- q_2^+ \,,
\end{align}
where the superscript in $q^\pm$ denotes the quark helicity.  The
distribution $F_{\Delta q, \Delta q}$ thus describes the degree to which
the two quark helicities are aligned rather than antialigned, whereas
$F_{q, \Delta q}$ and $F_{\Delta q, q}$ describe the correlation between
the helicity of one of the quarks and one of the cross products in
\eqref{VA}.

To illustrate that spin correlations between two partons need not be
small, let us consider the simple case of a $SU(6)$ symmetric three-quark
wave function of the proton.  Its spin-flavor part reads
\begin{align}
\tfrac{1}{\sqrt{6}}\, \bigl( | u^+ u^- d^+ \rangle
  + | u^- u^+ d^+ \rangle - 2 | u^+ u^+ d^- \rangle \bigr) \,,
\end{align}
where $+$ and $-$ respectively indicate that the quark spin is aligned and
antialigned with the proton spin.  As is well known, this wave function
gives $\Delta u /u = 2/3$ and $\Delta d /d = -1/3$ for the longitudinal
polarization of $u$ and $d$ quarks, which reproduces at least the trend of
what is empirically found for the lowest $x$ moments of the polarized
quark densities.  For two-quark distributions one finds
\begin{align}
F_{\Delta u, \Delta u} / F_{u, u} &= 1/3 \,,
&
F_{\Delta u, \Delta d} / F_{u, d} &= -2/3
\end{align}
and thus an appreciable correlation between the longitudinal
polarizations of the quarks.

Of course, the study of a three-quark wave function tells us little about
partons with $x \sim 10^{-2}$ or smaller, which are of particular
relevance for LHC phenomenology.  To the extent that they are known,
polarized single-parton densities in this $x$ range are small compared
with their unpolarized counterparts, which means that there is only a weak
spin correlation between a small-$x$ quark and the proton as a whole.
This is not too surprising, given that a small-$x$ quark and the proton
are far apart in phase space.  It does however \emph{not} imply small spin
correlations between two quarks that have small but comparable momentum
fractions $x_1 \sim x_2$ and are thus closer in phase space.  How large
such correlations are is an important open question.

The distributions defined with one or two tensor operators
$\mathcal{O}_{\delta q}^j$ are associated with transverse quark
polarization, since $\half (\gamma^+ + s^j i\sigma^{j +} \gamma_5)$
projects on quarks with a transverse spin vector $s^j$.  We now discuss
their parametrization in terms of scalar or pseudoscalar functions.  Let
us begin with $F_{\Delta q, \delta q}^j$ and $F_{\delta q, \Delta q}^j$,
which transform like two-dimensional vectors.  They can hence be written
as a sum of three scalar functions that are respectively multiplied by
$\tvec{y}^j$, $\tvec{k}_1^j$ and $\tvec{k}_2^j$.  Only two of these
functions are independent because of the linear dependence of the three
vectors.  A minimal parametrization is obtained by taking $\tvec{y}^j$ and
\begin{align}
\tilde{\tvec{y}}^j = \epsilon^{jj'} \tvec{y}^{j'}
\end{align}
as basis vectors.  This gives
\begin{align}
  \label{TA}
F_{\Delta q, \delta q}^j(x_i, \tvec{k}_i, \tvec{y}) &=
  \tvec{y}^{j} M f_{\Delta q, \delta q} +
  \tilde{\tvec{y}}^j M g_{\ms\Delta q, \delta q} \,,
\nonumber \\
F_{\delta q, \Delta q}^j(x_i, \tvec{k}_i, \tvec{y}) &=
  \tvec{y}^{j} M f_{\delta q, \Delta q} +
  \tilde{\tvec{y}}^j M g_{\ms\delta q, \Delta q} \,,
\end{align}
where we have inserted the proton mass $M$ on the r.h.s.\ so that $f$ and
$g$ have the same mass dimension as $F$.  Here and in the following we
denote scalar functions by $f$ and pseudoscalar ones by $g$.  The latter
can be represented in the same way as $F_{q, \Delta q}$.  If one wants to
avoid pseudoscalar functions, one can replace the basis vector
$\tilde{\tvec{y}}^j$ by
\begin{align}
  \label{other-vector}
\tvec{k}{}_y^j &= (\tvec{k}_1 + \tvec{k}_2)^j
- \frac{(\tvec{k}_1 + \tvec{k}_2)\ms \tvec{y}}{\tvec{y}^2}\, \tvec{y}^j \,.
\end{align}
Since both $\tilde{\tvec{y}}$ and $\tvec{k}_y$ are orthogonal to
$\tvec{y}$, they must be proportional to each other, and explicitly one
finds
\begin{align}
\tilde{\tvec{y}}^j&= \frac{\tvec{y}^2\, 
  \epsilon^{ll'} (\tvec{k}_1 + \tvec{k}_2)^l \tvec{y}^{\,l'}}{%
  \tvec{y}^2\ms (\tvec{k}_1 + \tvec{k}_2)^2
  - [ (\tvec{k}_1 + \tvec{k}_2)\ms \tvec{y} \ms]^2\rule{0pt}{2.2ex}}\;
  \tvec{k}{}_y^j\,.
\end{align}
If one inserts this into \eqref{TA} then $\tvec{k}{}_y^j$ is multiplied by
scalar functions, which are however singular when $\tvec{k}_1 +
\tvec{k}_2$ and $\tvec{y}$ become collinear.  One could replace
$\tvec{k}_1 + \tvec{k}_2$ in \eqref{other-vector} by another linear
combination of $\tvec{k}_1$ and $\tvec{k}_2$, but this would only move the
singularities to a different part of phase space.

The distributions $F_{q, \delta q}^j$ and $F_{\delta q, q}^j$ transform
like axial vectors, so that one has
\begin{align}
  \label{TV}
F_{q, \delta q}^j(x_i, \tvec{k}_i, \tvec{y}) &=
  \tilde{\tvec{y}}^j M f_{q, \delta q} +
  \tvec{y}^{j} M g_{\ms q, \delta q} \,,
\nonumber \\
F_{\delta q, q}^j(x_i, \tvec{k}_i, \tvec{y}) &=
  \tilde{\tvec{y}}^j M f_{\delta q, q} +
  \tvec{y}^{j} M g_{\ms\delta q, q} \,.
\end{align}
A decomposition in terms of scalar functions can be obtained by replacing
$\tvec{y}^j$ with $\epsilon^{jj'}_{} \tvec{k}{}_y^j$.  The tensor
distribution $F_{\delta q, \delta q}^{jj'}$ can finally be written as
\begin{align}
  \label{TT}
F_{\delta q, \delta q}^{jj'}(x_i, \tvec{k}_i, \tvec{y}) &=
  \delta^{jj'}\! f^{}_{\delta q, \delta q}
+ \bigl( 2 \tvec{y}^{j} \tvec{y}^{j'} - \delta^{jj'} \tvec{y}^2 \bigr)\ms
  M^2 f^{\ms t}_{\delta q, \delta q}
\nonumber \\
& \quad
+ \bigl( \tvec{y}^{j} \tilde{\tvec{y}}^{j'}
       + \tilde{\tvec{y}}^j \tvec{y}^{j'} \bigr)\ms
  M^2 g^{\ms s}_{\ms\delta q, \delta q}
+ \bigl( \tvec{y}^{j} \tilde{\tvec{y}}^{j'}
       - \tilde{\tvec{y}}^j \tvec{y}^{j'} \bigr)\ms
  M^2 g^{\ms a}_{\ms\delta q, \delta q} \,.
\end{align}
Notice that the four basis tensors $t_{p}^{jj'}$ in this decomposition are
orthogonal to each other, i.e.\ $t_{p}^{jj'} t_{q}^{jj'} \propto
\delta_{pq}$ with $p,q= 1,2,3,4$.  Contracting \eqref{TT} with the
transverse polarization vectors $s_{1}^{j} s_{2}^{j'}$ of the quarks, we
see in particular that $f_{\delta q, \delta q}$ goes with $\tvec{s}_1
\tvec{s}_2$ and thus describes the correlation between the two transverse
quark spins.

In summary, we can represent the spin structure of $F_{a_1, a_2}(x_i,
\tvec{k}_i, \tvec{y})$ by eight scalar and eight pseudoscalar functions
for each combination of quark flavors.  The pseudoscalar functions can be
traded for scalar ones, which have however artificial singularities for
particular values of the vectors $\tvec{y}$, $\tvec{k}_1$ and
$\tvec{k}_2$.

If one integrates over transverse momenta to obtain collinear
distributions, one finds
\begin{align}
  \label{no-pseudoscalar}
F_{\Delta q, q}(x_i, \tvec{y}) &= F_{q, \Delta q}(x_i, \tvec{y}) = 0
\end{align}
because one cannot construct a pseudoscalar function with only one vector
$\tvec{y}$.  Likewise, the functions $g$ in \eqref{TA}, \eqref{TV} and
\eqref{TT} integrate to zero, so that we are left with
\begin{align}
  \label{collinear-decomp}
F_{q, q}(x_i, \tvec{y}) &= f_{q, q} \,,
&
F_{\Delta q, \Delta q}(x_i, \tvec{y}) &= f_{\Delta q, \Delta q} \,,
\nonumber \\
F_{q, \delta q}^j(x_i, \tvec{y}) &= \tilde{\tvec{y}}^j M
   f_{q, \delta q} \,,
&
F_{\Delta q, \delta q}^j(x_i, \tvec{y}) &= \tvec{y}^{j} M
   f_{\Delta q, \delta q} \,,
\nonumber \\
F_{\delta q, q}^j(x_i, \tvec{y}) &= \tilde{\tvec{y}}^j M
   f_{\delta q, q} \,,
&
F_{\delta q, \Delta q}^j(x_i, \tvec{y}) &= \tvec{y}^{j} M
   f_{\delta q, \Delta q} \,,
\nonumber \\
F_{\delta q, \delta q}^{jj'}(x_i, \tvec{y}) &=
  \delta^{jj'}\! f^{}_{\delta q, \delta q}
+ \bigl( 2 \tvec{y}^{j} \tvec{y}^{j'}
    - \delta^{jj'} \tvec{y}^2 \bigr)\ms
  M^2 f^{\ms t}_{\delta q, \delta q} \,. \hspace{-5em}
\end{align}
The eight functions $f$ on the right-hand side now depend on $x_1$, $x_2$
and $\tvec{y}^2$ and are obtained from their counterparts in \eqref{VV},
\eqref{TA}, \eqref{TV} and \eqref{TT} by integration over $\tvec{k}_1$
and $\tvec{k}_2$.

The above decompositions are given for distributions in a right-moving
proton.  For a left-moving proton one has to change the sign of
$\epsilon^{jj'}$ and hence of $\tilde{\tvec{y}}$ and of all pseudoscalar
functions, see our remark below \eqref{gluon-projectors}.  Analogous
decompositions can be written down for distributions $F_{a_1, \bar{a}_2}$,
$F_{\bar{a}_1, a_2}$, $F_{\bar{a}_1, \bar{a}_2}$ that involve antiquarks,
as well as for interference distributions $I_{a_1, \bar{a_2}}$ and
$I_{\bar{a}_1, a_2}$.


\paragraph{Symmetry properties}

The terms appearing in the decompositions \eqref{VV} to
\eqref{collinear-decomp} are consistent with rotation and parity
invariance.  Let us now discuss their symmetry properties.  Using that the
operators in \eqref{quark-bilinears} satisfy $\mathcal{O}_a^*(y_i, z_i) =
\mathcal{O}_a^{}(y_i, -z_i)$ one finds that the distributions $F_{a_1,
  a_2}(x_i, \tvec{k}_i, \tvec{y})$ are real valued,
\begin{align}
  \label{real-valued}
F^*_{a_1, a_2}(x_i, \tvec{k}_i, \tvec{y}) &=
   F^{}_{a_1, a_2}(x_i, \tvec{k}_i, \tvec{y}) \,.
\end{align}
For distributions that are purely defined in momentum or position space,
see \eqref{momentum-F} and \eqref{position-F}, this implies
\begin{align}
F^*_{a_1, a_2}(x_i, \tvec{k}_i, \tvec{r}_i) &=
   F^{}_{a_1, a_2}(x_i, \tvec{k}_i, -\tvec{r}_i) \,,
\nonumber \\
F^*_{a_1, a_2}(x_i, \tvec{z}_i, \tvec{y}) &=
   F^{}_{a_1, a_2}(x_i, -\tvec{z}_i, \tvec{y}) \,.
\end{align}
These functions are in general not real-valued, nor are the scalar or
pseudoscalar functions one can introduce to parameterize them in analogy
to \eqref{VV} to \eqref{TT}.

For the symmetry properties of parton distributions with respect to time
reversal, the Wilson lines appearing in their definition are essential.
As we argued in section~\ref{sec:coll}, multiparton distributions involve
the past pointing Wilson lines $W$ given in \eqref{wilson-dy}.  Upon time
reversal these turn into the future pointing Wilson lines $W'$ in
\eqref{wilson-sidis}.  A distribution is called $T$ even (odd) if it is
even (odd) under time reversal \emph{without} taking into account this
change of Wilson lines.  The time reversal invariance of strong
interactions thus implies that $T$ odd distributions are only nonzero
thanks to Wilson line effects; a prominent example from spin physics is
the Sivers distribution function \cite{Collins:2002kn}.  However, time
reversal does force distributions to vanish if they are $T$ odd and have
Wilson lines that are invariant under time reversal.  This is the case for
the Wilson lines along a finite lightlike path that appear in collinear
single-parton densities and in the collinear two-parton distributions
$\sing{F}$ in the color singlet sector, as we discussed in
section~\ref{sec:coll-fact}.  By contrast, collinear color octet
distributions $\oct{F}$, as well as interference distributions $\sing{I}$
and $\oct{I}$, have Wilson lines that do change under time reversal.

After these preliminaries we can now investigate the time reversal
properties of two-quark distributions.  We find
\begin{align}
  \label{time-reversal}
F_{a_1, a_2}^{W}(x_i, \tvec{k}_i, \tvec{y}) &=
  \eta_{a_1} \eta_{a_2} \,
  F_{a_1, a_2}^{W'}(x_i, \tvec{k}_i, -\tvec{y})
\end{align}
with sign factors $\eta_q = +1$ and $\eta_{\Delta q} = \eta_{\delta q} =
-1$, where the superscripts indicate the type of Wilson line in the matrix
element defining the distributions.  The relations \eqref{real-valued} to
\eqref{time-reversal} also hold for the distributions $F_{a_1,
  \bar{a}_2}$, $F_{\bar{a}_1, a_2}$, $F_{\bar{a}_1, \bar{a}_2}$ with
antiquarks and for the interference distributions $I_{a_1, \bar{a_2}}$,
$I_{\bar{a}_1, a_2}$.

Since the scalar functions parameterizing $F_{a_1, a_2}^{W}(x_i,
\tvec{k}_i, \tvec{y})$ are in general neither even nor odd in $\tvec{y}$,
they are not $T$ even or odd either.  The scalar functions that
parameterize the collinear distributions in \eqref{collinear-decomp} are
however even in $\tvec{y}$.  As a consequence, $F_{\Delta q,\delta
  q}^j(x_i, \tvec{y})$ and $\smash{F_{\delta q,\Delta q}^j(x_i,
  \tvec{y})}$ are $T$ odd and all other distributions in
\eqref{collinear-decomp} are $T$ even.  For the color singlet sector this
implies
\begin{align}
\sing{F}_{\Delta q,\delta q}^j(x_i, \tvec{y}) &=
  \sing{F}_{\delta q,\Delta q}^j(x_i, \tvec{y}) = 0 \,,
\end{align}
whereas the corresponding color-octet distributions $\oct{F}_{\Delta
  q,\delta q}^j(x_i, \tvec{y})$ and $\smash{\oct{F}_{\delta q,\Delta
    q}^j(x_i, \tvec{y})}$ can be nonzero due to the Wilson lines appearing
in their definitions.  Analogous statements hold for the corresponding
distributions with one or two antiquarks.
Collinear interference distributions $\sing{I}$ and $\oct{I}$ are not
restricted by time reversal invariance.


\subsubsection{Spin effects in gauge boson pair production}
\label{sec:boson-pairs}

In this section we show that the quark spin correlations discussed in the
previous section have observable consequences in multiparton interactions.
As we did earlier in this paper, we consider the production of a pair of
gauge bosons $\gamma$, $Z$ or $W$.  We include the decay of each boson
into a lepton pair, which carries information on the spin state of the
gauge boson.  While these processes have a rather small cross section,
they may be suited for experimental studies due to their clean final-state
signature.  We do not present a full analysis here, but highlight the
effects of selected parton spin correlations.

For simplicity we limit our attention to those distributions that do not
involve explicit vectors $\tvec{y}$ or $\tilde{\tvec{y}}$ on the r.h.s.\
of the decompositions in the previous section, i.e.\ to $F_{q,q}$,
$F_{\Delta q, \Delta q}$ and the term $\delta^{jj'}\! f_{\delta q, \delta
  q}^{}$ in $F_{\delta q, \delta q}^{jj'}$.  For definiteness we analyze
the graph of figure~\ref{fig:double-scatt}a, with two quarks emitted from
the right-moving proton and two antiquarks from the left-moving one.  We
approximate the transverse momenta $\tvec{q}_i$ of the bosons by zero when
calculating their production and decay, as deviations from this limit are
suppressed by powers of $q_T/Q$.
The partonic cross section for the production of a lepton pair can be
written as the product of a production tensor of the boson and a tensor
for its decay,
\begin{align}
  \label{prod-decay-tensor}
\hat{\sigma}_{a,\bar{a}} &=
  ( \hat{\sigma}_{a,\bar{a}} )^{\mu\mu'}\, D_{\mu\mu'} \,,
\end{align}
where $\mu$ is associated with the boson in the amplitude and $\mu'$ with
the one in the complex conjugate amplitude.  For unpolarized or
longitudinally polarized quarks one easily finds
\begin{alignat}{3}
  \label{bosons-long}
( \hat{\sigma}_{q, \bar{q}} )^{\mu\mu'}
 &= - ( \hat{\sigma}_{\Delta q, \Delta\bar{q}} )^{\mu\mu'}
&&= - A\ms g_\perp^{\mu\mu'} - B\ms i\epsilon_\perp^{\mu\mu'} \,,
\nonumber \\
( \hat{\sigma}_{\Delta q, \bar{q}} )^{\mu\mu'}
 &= - ( \hat{\sigma}_{q, \Delta\bar{q}} )^{\mu\mu'}
&&= - B\ms g_\perp^{\mu\mu'} - A\ms i\epsilon_\perp^{\mu\mu'} \,,
\end{alignat}
with coefficients $A$ and $B$ depending on kinematic variables and the
vector and axial-vector couplings of the gauge boson to the quark $q$.  In
the case of a photon one has $B=0$.  The transverse tensors
$g_\perp^{\mu\mu'}$ and $\epsilon_\perp^{\mu\mu'}$ have as nonzero
components $g_\perp^{11} = g_\perp^{22} = -1$ and $\epsilon_\perp^{12} =
-\epsilon_\perp^{21} = 1$.  From \eqref{bosons-long} it follows that the
overall cross section depends on the combinations
\begin{align}
& F_{q,q}\, F_{\bar{q},\bar{q}}
 + F_{\Delta q, \Delta q}\, F_{\Delta\bar{q}, \Delta\bar{q}}
& \text{and} &
& F_{q,q}\, F_{\Delta\bar{q}, \Delta\bar{q}}
 + F_{\Delta q, \Delta q}\, F_{\bar{q}, \bar{q}}\
\end{align}
of multiparton distributions.  The contraction of $g_\perp^{\mu\mu'}$ and
$\epsilon_\perp^{\mu\mu'}$ with the vector boson decay matrices results in
different angular distributions of the leptons.  If one integrates over
their angles then only the contribution from $g_\perp^{\mu\mu'}$ remains.
We thus find that nonzero values of $F_{\Delta q, \Delta q}$ and
$F_{\Delta\bar{q}, \Delta\bar{q}}$ modify both the total rate and the
lepton angular distribution compared with the contribution from the
unpolarized term $F_{q,q}\, F_{\bar{q},\bar{q}}$.

We now turn to transverse quark polarization.  In this case the production
tensor from each hard scattering depends also on the transverse indices
associated with the quark polarization.  We recall that the quark field
bilinears $\bar{q}\ms i\sigma^{+j} \gamma_5\ms q$ are chiral odd, so that
in the helicity basis they correspond to quarks or antiquarks with
opposite helicities in the scattering amplitude and its conjugate.  As a
result transverse quark polarization does not contribute to the production
of a $W$ boson.  Keeping only the term $\delta^{jj'}\! f_{\delta q,\delta
  q}$ in the decomposition \eqref{TT} of $F_{\delta q,\delta q}^{jj'}$ and
the corresponding term in the decomposition of
$F_{\delta\bar{q},\delta\bar{q}}^{kk'}$, one finds for the neutral bosons
$\gamma$ or $Z$
\begin{align}
  \label{bosons-trans}
\delta^{jj'} \delta^{kk'}\,
( \hat{\sigma}_{1,\ms \delta q, \delta\bar{q}}^{jk} \bigr)^{\mu\mu'}_{}\;
( \hat{\sigma}_{2,\ms \delta q, \delta\bar{q}}^{j'k'} \bigr)^{\nu\nu'}_{}
 &\;\propto\;
   g_\perp^{\mu\nu} g_\perp^{\mu'\nu'}
   + g_\perp^{\mu\nu'} g_\perp^{\mu'\nu}
   - g_\perp^{\mu\mu'} g_\perp^{\nu\nu'\phantom{\mu}} \,,
\end{align}
where the polarization indices $\mu,\mu'$ belong to the boson with
momentum $q_1$ and the indices $\nu,\nu'$ to the boson with momentum
$q_2$.  We observe that the polarization indices of the two bosons are
entangled in \eqref{bosons-trans}.  Contracting with the well-known boson
decay matrices, one obtains an azimuthal dependence like $\cos(2\varphi)$,
where $\varphi$ is the relative azimuthal angle between the two leptons
(as opposed to the antileptons).\footnote{%
  Since we are working in the approximation $\tvec{q}_1 = \tvec{q}_2 =
  \tvec{0}$, the azimuthal angles for the leptons of both boson decays are
  naturally defined w.r.t.\ the $z$ axis in the $pp$ center-of-mass.}
We thus obtain the important result that a correlation between transverse
quark and antiquark spins, as expressed by the distribution $f_{\delta
  q,\delta q}$ in \eqref{TT}, leads to a correlation between the decay
planes of the two produced bosons.

It is instructive to rewrite the production tensors in terms of boson
polarization vectors $\epsilon_{+} = - (0, 1, i, 0)\ms /\sqrt{2}$ and
$\epsilon_{-} = (0, 1, -i, 0)\ms /\sqrt{2}$, which respectively correspond
to angular momentum $+1$ or $-1$ along the $z$ axis.  One finds
\begin{align}
  \label{ge-pol}
- g_\perp^{\mu\mu'} &= \epsilon_{+}^\mu\, \epsilon_{+}^{*\mu'}
                     + \epsilon_{-}^\mu\, \epsilon_{-}^{*\mu'} \,,
&
- i\epsilon_\perp^{\mu\mu'} &= \epsilon_{+}^\mu\, \epsilon_{+}^{*\mu'}
                             - \epsilon_{-}^\mu\, \epsilon_{-}^{*\mu'}
\end{align}
and
\begin{align}
  \label{tau-pol}
\half \bigl(
   g_\perp^{\mu\nu} g_\perp^{\mu'\nu'}
   + g_\perp^{\mu\nu'} g_\perp^{\mu'\nu}
   - g_\perp^{\mu\mu'} g_\perp^{\nu\nu'} \bigr)
&= \epsilon_{+}^\mu\, \epsilon_{-}^{*\mu'}\,
   \epsilon_{-}^\nu\, \epsilon_{+}^{*\nu'}
 + \epsilon_{-}^\mu\, \epsilon_{+}^{*\mu'}\,
   \epsilon_{+}^\nu\, \epsilon_{-}^{*\nu'} \,.
\end{align}
We can easily understand why each boson is transversely polarized.  Recall
that a massless quark and antiquark can only annihilate into a vector
boson if their helicities are coupled to $\pm 1$.  Since we neglect the
transverse momentum of the bosons, their angular momentum along $z$ must
also be $\pm 1$.  The tensors $g_\perp^{\mu\mu'}$ and
$i\epsilon_\perp^{\mu\mu'}$ in \eqref{ge-pol} correspond to the same boson
polarization in amplitude and conjugate amplitude and thus do not give
rise to an azimuthal dependence in the leptonic decays, but they do give
different distributions in the polar angles of the leptons, or
equivalently in their rapidities.  By contrast, \eqref{tau-pol} involves
the interference between $J^z = 1$ and $-1$ for each of the bosons, which
readily translates into the $\cos(2\varphi)$ dependence already mentioned.

It is natural to expect that spin correlations between partons also lead
to angular correlations in the final state for other double-scattering
processes, such as the production of two dijets.  In this case two-parton
distributions involving linear gluon polarization can contribute as well.
We note that in the analysis of \cite{Berger:2009cm} uncorrelated dijet
planes were taken as a characteristic feature of the double-scattering
mechanism.  This is only adequate if parton spin correlations in the
proton are negligible.

\begin{figure}
\begin{center}
\includegraphics[width=0.65\textwidth]{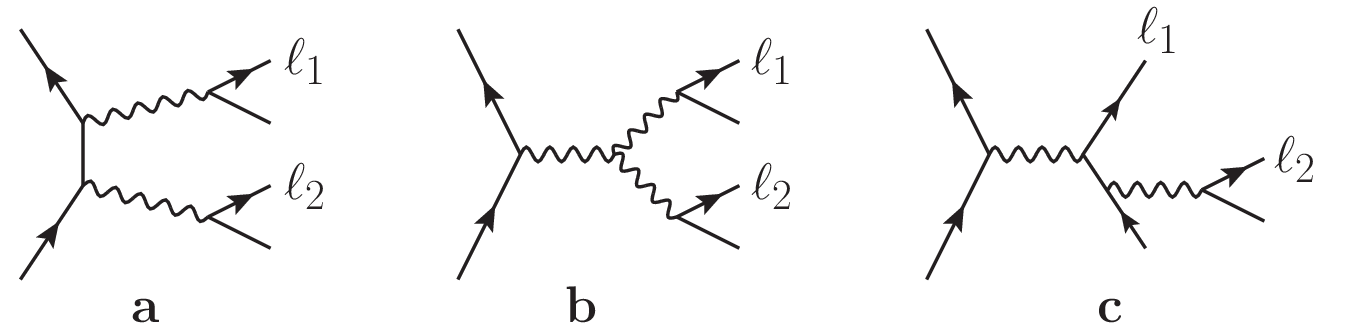}
\end{center}
\caption{\label{fig:four-leptons} Parton-level graphs for $q\bar{q}$
  annihilation into four leptons.  Further graphs are obtained by
  interchanging the leptons $\ell_1$ and $\ell_2$ together with the
  corresponding antileptons in a and c, and by reversing the charge flow
  of the central fermion line in c. Graph b requires at least one of the
  produced bosons to be a $W$.}
\end{figure}

Returning to four-lepton production, let us compare our results for double
hard scattering with the contribution from a single $q\bar{q}$
annihilation, remaining in kinematics where $\tvec{q}_1$ and $\tvec{q}_2$
can be neglected compared with $Q$.  The corresponding Feynman graphs
involve either quark exchange in the $t$ or $u$ channel, or an
intermediate boson in the $s$ channel in case one or both final-state
bosons are charged, see figure~\ref{fig:four-leptons}a and b.  In
addition, the four-lepton final state can be produced by graphs as in
figure~\ref{fig:four-leptons}c, where only one lepton pair comes from the
decay of a vector boson.  Such graphs were recently discussed and termed
``single-resonance graphs'' in \cite{Kom:2011nu}.  They should be taken
into account unless each lepton pair has an invariant mass inside the $Z$
or $W$ mass peaks.

The dependence of the single-scattering cross section on the azimuthal
angle $\varphi$ between the leptons can be deduced from symmetry arguments
since we set $\tvec{q}_1$ and $\tvec{q}_2$ to zero in the parton-level
process.  Let us assume that the initial $q\bar{q}$ pair has total angular
momentum $J^z = 1$ along the $z$ axis.  This must then hold for both the
scattering amplitude and its conjugate, since a single quark or antiquark
in an unpolarized proton is unpolarized.  One of the lepton pairs
originates from the decay of a gauge boson and is thus in a partial wave
with $J=1$.  The possible combinations of $(J_1^z, J_2^z)$ for the two
lepton pairs are thus $(2,-1)$, $(1,0)$, $(0,1)$ and $(-1,2)$, where the
first and last possibilities are only possible for single-resonance
graphs.  In the cross section this gives a $\varphi$ independent term and
a modulation with $\cos\varphi$ from all graphs, as well as modulations
with $\cos(2\varphi)$ and $\cos(3\varphi)$ from single-resonance graphs.
The same angular terms are obtained when the initial $q\bar{q}$ pair has
$J^z = -1$, whereas the configuration where the pair has $J^z = 0$
decouples in the hard-scattering graphs of figure~\ref{fig:four-leptons}.
In summary, the $\cos(2\varphi)$ modulation we found in the
double-scattering mechanism competes with single-scattering contributions
involving single-resonance graphs.\footnote{This was not realized in
  \protect\cite{Diehl:2011tt}, where single-resonance graphs were not
  taken into account.}
One may hope that the two sources of $\cos(2\varphi)$ dependence can be
separated by a more detailed analysis --- making for instance use of the
fact that the single-resonance graphs also give a $\cos(3\varphi)$ term
--- but this issue must be left to a dedicated study.


\subsection{Mellin moments and lattice calculations}
\label{sec:mellin}

If one takes Mellin moments in $x_1$ and $x_2$ of the color singlet
distributions $\sing{F}$, then the light-cone operators $\mathcal{O}_a$ in
their definition turn into local operators.  The corresponding moments of
single-parton densities can be calculated in lattice QCD, and we will now
investigate to which extent the same can be done for two-parton
distributions.  From our discussion in section~\ref{sec:coll-fact} it
follows that the Mellin moments of color octet distributions $\oct{F}$ do
not involve local operators because of their Wilson line structure.  We
therefore limit ourselves to the color singlet sector.

Using the relation \eqref{quark-antiquark} and its analogs for
$F_{a_1,\bar{a}_2}$ and $F_{\bar{a}_1, \bar{a}_2}$, one obtains double
Mellin moments
\begin{align}
  \label{mellin-moment}
M_{a_1,a_2}^{n_1, n_2}(\tvec{y}^2) &=
\int_0^1 dx_1^{}\, x_1^{n_1-1} \int_0^1 dx_2^{}\, x_2^{n_2-1}
\Bigl[ \sing{F}_{a_1,a_2}(x_1,x_2, \tvec{y})
 + (-1)^{n_1}\ms \sigma_{a_1} \sing{F}_{\bar{a}_1,a_2}(x_1,x_2, \tvec{y})
\nonumber \\
&\qquad\qquad
  + (-1)^{n_2}\ms \sigma_{a_2} \sing{F}_{a_1,\bar{a}_2}(x_1,x_2, \tvec{y})
  + (-1)^{n_1+n_2}\ms \sigma_{a_1} \sigma_{a_2} 
        \sing{F}_{\bar{a}_1,\bar{a}_2}(x_1,x_2, \tvec{y}) \Bigr]
\nonumber \\
&= \frac{1}{2}\ms (p^+)^{1-n_1-n_2} \int dy^- 
   \big\langle p \big|\, \mathcal{O}_{a_1}^{+\cdots+}(0)\,
        \mathcal{O}_{a_2}^{+\cdots+}(y) \big| p \big\rangle_{y^+ = 0}
\end{align}
with $\sigma_{q} = \sigma_{\delta q} = +1$ and $\sigma_{\Delta q} = -1$.
For each label $\delta q$ the moments $M_{a_1, a_2}^{n_1, n_2}$ and the
corresponding operator on the r.h.s.\ carry an additional transverse
Lorentz index not displayed in \eqref{mellin-moment}.  On the r.h.s.\ we
have the twist-two quark operators familiar from the operator product
expansion,
\begin{align}
\mathcal{O}^{\mu_1 \cdots \mu_n}_{q}(y) &=
  \operatorname*{T}_{(\mu_1 \cdots \mu_n)}
  \operatorname*{S}_{(\mu_1 \cdots \mu_n)}
  \bar{q}(y)\ms \gamma^{\mu_1}\ms
                i\lrD^{\mu_2}(y) \cdots i\lrD^{\mu_n}(y)\, q(y) \,,
\nonumber \\
\mathcal{O}^{\mu_1 \cdots \mu_n}_{\Delta q}(y) &=
  \operatorname*{T}_{(\mu_1 \cdots \mu_n)}
  \operatorname*{S}_{(\mu_1 \cdots \mu_n)}
  \bar{q}(y)\ms \gamma^{\mu_1} \gamma_5\,
                i\lrD^{\mu_2}(y) \cdots i\lrD^{\mu_n}(y)\, q(y) \,,
\nonumber \\
\mathcal{O}^{\lambda \mu_1 \cdots \mu_n}_{\delta q}(y) &=
  \operatorname*{T}_{(\lambda \mu_1 \cdots \mu_n)}
  \operatorname*{A}_{(\lambda \mu_1)}
  \operatorname*{S}_{(\mu_1 \cdots \mu_n)}
  \bar{q}(y)\ms i \sigma^{\lambda \mu_1} \gamma_5\,
                i\lrD^{\mu_2}(y) \cdots i\lrD^{\mu_n}(y)\, q(y) \,,
\end{align}
where $\lrD^\mu(y) = \half (\rpartial^\mu - \lpartial^\mu) + i g A^\mu(y)$
is the antisymmetrized covariant derivative.  $\operatorname{S}$,
$\operatorname{A}$ and $\operatorname{T}$ respectively denotes
symmetrization, antisymmetrization and subtraction of traces in the
indicated indices.
It is understood that the composite operators $\mathcal{O}_{a_1}(y)$ and
$\mathcal{O}_{a_2}(0)$ in \eqref{mellin-moment} are each renormalized in a
standard manner, e.g.\ in the $\overline{\text{MS}}$ scheme.  As long as
they are taken at a finite spacelike distance $y$, no further
ultraviolet divergences appear.  The operators $\mathcal{O}_{q}$ and
$\mathcal{O}_{\Delta q}$ mix of course with their gluon counterparts
$\mathcal{O}_{g}$ and $\mathcal{O}_{\Delta g}$ under renormalization.  We
shall return to the renormalization of collinear multiparton distributions
in section~\ref{sec:evolution}.

Let us rewrite the r.h.s.\ of \eqref{mellin-moment} in a manifestly
covariant form.  We first introduce the covariant decomposition
\begin{align}
  \label{covar-decomp}
 \big\langle p \big|\,
   \mathcal{O}_{q}^{\nu_1 \cdots \nu_{n_2}}(0)\,
   \mathcal{O}_{q}^{\mu_1 \cdots \mu_{n_1}}(y) \big| p \big\rangle
&= 2\ms p^{\nu_1} \cdots p^{\nu_{n_2}}\,
        p^{\mu_1} \cdots p^{\mu_{n_1}}\,
   \langle \mathcal{O}_{q}^{n_2}\,
           \mathcal{O}_{q}^{n_1} \rangle(py, y^2)
 + \cdots \,,
\end{align}
where the ellipsis represents terms with uncontracted vectors $y^{\mu}$
and terms involving the metric tensor $g^{\mu\nu}$.  The reduced matrix
element $\langle \mathcal{O}_{q}^{n_2}\, \mathcal{O}_{q}^{n_1} \rangle$
can only depend on the invariants $py$ and $y^2$.  We then choose a frame
where $\tvec{p} = \tvec{0}$ and $y^+ = 0$, so that $py = p^+ y^-$ and $y^2
= -\tvec{y}^2$.  This allows us to write \eqref{mellin-moment} in the
desired form
\begin{align}
  \label{mellin-covar}
M_{q,q}^{n_1, n_2}(- {y}^2) &= \int d(py)\;
  \langle \mathcal{O}_{q}^{n_2}\,
          \mathcal{O}_{q}^{n_1} \rangle(py, y^2)  \,.
\end{align}
A corresponding representation is readily obtained for $M_{\Delta q,\Delta
  q}^{n_1, n_2}$.  For one or two polarization labels $\delta q$ the
analogs of \eqref{covar-decomp} involve the tensor structures in the
decomposition \eqref{collinear-decomp} of the two-quark distributions.

The matrix element in \eqref{covar-decomp} and its counterparts with
polarized quarks can be evaluated on a lattice in Euclidean spacetime if
one chooses $y^0=0$.  This is rather similar to lattice studies of
transverse-momentum dependent single-quark distributions
\cite{Hagler:2009mb,Musch:2010ka}, with the main difference that the
operators taken at different spacetime points are single quark fields in
that case, whereas they are gauge invariant bilinear operators here.  The
restriction $y^0=0$ entails
\begin{align}
\frac{(py)^2}{-y^2} &=
   \frac{(\mvec{p} \mvec{y})^2}{\mvec{y}^2}  \;\le\; \mvec{p}^2 \,,
\end{align}
where $\mvec{p}$ and $\mvec{y}$ denote the spacelike three-vectors.  Thus,
the integral over all $py$ at fixed $y^2$ on the r.h.s.\ of
\eqref{mellin-covar} can unfortunately not be evaluated from results on a
discrete Euclidean lattice, where the maximal momentum is fixed by the
lattice spacing.  This is completely analogous to the single-quark case,
as discussed in \cite{Musch:2010ka}.  Despite this limitation of
principle, we hope that lattice data in a certain range of $py$ and $y^2$
will in the future provide genuinely nonperturbative information about the
behavior of multi-parton distributions.

We note that a lattice calculation has been reported in
\cite{Alexandrou:2008ru} for the correlation function of two vector
currents at equal time in a proton at rest.  This corresponds to setting
$n_1=n_2=1$ and $py=0$ in \eqref{covar-decomp}.  The reduced matrix
element $\langle \mathcal{O}_{a_2}^{n_2}\, \mathcal{O}_{a_1}^{n_1}
\rangle$ at $py = 0$ is directly related to an integral of the two-quark
correlation function $\Phi(k_1, k_2, r)$ defined in \eqref{Phi-quarks}.
The relative plus-momentum $r^+$ is integrated over in that case, rather
than being set to zero as required for the distributions $F(x_i,
\tvec{k}_i, \tvec{r})$ that appear in double hard-scattering cross
sections.


\subsection{Relation with generalized parton distributions}
\label{sec:gpd-connection}

In section~\ref{sec:reduct} we derived an approximate relation between
multi- and single-parton distributions in a model theory with scalar
partons.  We now extend this relation to distributions for two quarks or
antiquarks, taking into account the different combinations of fermion
number and color.  For the time being we neglect aspects related to the
proton spin, which will be discussed in section~\ref{sec:gpd-spin}.  The
distributions we will deal with are
\begin{align}
  \label{six-distributions}
\sing{F}_{a_1,a_2} &=
  \mat{(\bar{q}_3\ms \Gamma_{a_2}\ms q_2)\,
       (\bar{q}_4\ms \Gamma_{a_1}\ms q_1)} \,,
&
\sing{\tilde{F}}_{a_1,a_2} &=
  \mat{(\bar{q}_{4}\ms \Gamma_{a_2}\ms q_{2})\,
       (\bar{q}_{3}\ms \Gamma_{a_1}\ms q_{1})} \,,
\nonumber \\[0.2em]
\sing{F}_{a_1,\bar{a}_2} &=
  \mat{(\bar{q}_2\ms \Gamma_{\bar{a}_2}\ms q_3)\,
       (\bar{q}_4\ms \Gamma_{a_1}\ms q_1)} \,,
&
\sing{\tilde{F}}_{a_1,\bar{a}_2} &=
  \mat{(\bar{q}_{4}\ms \Gamma_{\bar{a}_2}\ms q_{3})\,
       (\bar{q}_{2}\ms \Gamma_{a_1}\ms q_{1})} \,,
\nonumber \\[0.2em]
\sing{I}_{a_1,\bar{a}_2} &=
  \mat{(\bar{q}_2\ms \Gamma_{\bar{a}_2}\ms q_4)\,
       (\bar{q}_3\ms \Gamma_{a_1}\ms q_1)} \,,
&
\sing{\tilde{I}}_{a_1,\bar{a}_2} &=
  \mat{(\bar{q}_3\ms \Gamma_{\bar{a}_2}\ms q_4)\,
       (\bar{q}_2\ms \Gamma_{a_1}\ms q_1)} \,.
\end{align}

The general result \eqref{impact-product-2} for $n$ scalar partons readily
carries over to the two-quark distributions $\sing{F}$:
\begin{align}
  \label{F-result}
\sing{F}_{a_1,a_2}(x_i, \tvec{z}_i, \tvec{y}) &\approx
\int d^2\tvec{b}\;
  f_{a_2}\bigl( x_2, \tvec{z}_2; \tvec{b} + \half x_1 \tvec{z}_1 \bigl)\,
  f_{a_1}\bigl( x_1, \tvec{z}_1;
                \tvec{b} + \tvec{y} - \half x_2 \tvec{z}_2 \bigr) \,,
\end{align}
where the impact parameter dependent single-quark distributions $f_a(x,
\tvec{z}; \tvec{b})$ are defined in analogy to the scalar case in
\eqref{impact-dist-1} and \eqref{impact-dist-2}.  Setting $\tvec{z}_1 =
\tvec{z}_2 = \tvec{0}$ in \eqref{F-result}, one obtains collinear
distributions on both sides and has the probability interpretation
represented in figure~\ref{fig:impact-reduct}.
As a counterpart to \eqref{mom-product-1} one can transform the relation
\eqref{F-result} into transverse-momentum space, where it reads
\begin{align}
  \label{F-result-mom}
\sing{F}_{a_1,a_2}(x_i, \tvec{k}_i, \tvec{r}) &\approx
  f_{a_2}\bigl( x_2, \tvec{k}_2 - \half x_2\ms \tvec{r}; - \tvec{r} \bigl)\,
  f_{a_1}\bigl( x_1, \tvec{k}_1 - \half x_1\ms \tvec{r}; \tvec{r} \bigr)
\end{align}
with distributions $f_a(x, \tvec{k}; \tvec{\Delta})$ defined in analogy to
\eqref{mom-dist}.  Integrating over $\tvec{k}_i$ we obtain the relation
recently given in \cite{Blok:2010ge}.
Relations analogous to \eqref{F-result} and \eqref{F-result-mom} are
obtained for $\sing{F}_{a_1,\bar{a}_2}$ by replacing the label $a_2$ with
$\bar{a}_2$ on both sides.

To reduce $\sing{\tilde{F}}$ to single-particle distributions we could
repeat our earlier derivation that started with \eqref{state-insertion}.
We find it more convenient to work in the transverse-momentum rather than
impact parameter representation.  We insert a complete set $|X \rangle$ of
intermediate states between the color singlet operators $(\bar{q}_{4}\ms
\Gamma_{a_2}\ms q_{2})$ and $(\bar{q}_{3}\ms \Gamma_{a_1}\ms q_{1})$ and
assume that single-proton intermediate states dominate.  We then have
\begin{align}
  \label{mom-skewed-0}
\sing{\tilde{F}}_{a_1,a_2}(x_i, \tvec{k}_i, \tvec{r})
&= \sum_X \,
2 p^+\!\! \int dy^-\ms d^2\tvec{y}\; e^{i \tvec{y} \tvec{r}}\;\,
\biggl[\, \prod_{i=1}^2
       \int \frac{dz_i^-\, d^2\tvec{z}_i^{}}{(2\pi)^3}\;
       e^{i x_i^{} z_i^- p^+ -i \tvec{z}_i^{} \tvec{k}_i^{}}
\biggr]\,
\nonumber \\
&\quad \times
  \big\langle p^+, \tvec{p} \,\big|\,
    \mathcal{O}_{a_2}(y, z_2) \,\big|\, X \big\rangle \,
  \big\langle X \,\big|\,
    \mathcal{O}_{a_1}(y, z_1) \,\big|\, p^+, \tvec{p}
  \big\rangle \,\Big|_{\tvec{p} = \tvec{0}}
 \phantom{\int}
\nonumber \\
& \approx
\int \frac{dp'^+\, d^2\tvec{p}'}{2 p'^+\ms (2\pi)^3}\;
2 p^+\!\! \int dy^-\ms d^2\tvec{y}\; e^{i \tvec{y} \tvec{r}}\;\,
\biggl[\, \prod_{i=1}^2
       \int \frac{dz_i^-\, d^2\tvec{z}_i^{}}{(2\pi)^3}\;
       e^{i x_i^{} z_i^- p^+ -i \tvec{z}_i^{} \tvec{k}_i^{}}
\biggr]\,
\nonumber \\
&\quad \times
  \big\langle p^+, \tvec{p} \,\big|\,
    \mathcal{O}_{a_2}(y, z_2) \,\big|\, p'^+, \tvec{p}' \big\rangle \,
  \big\langle p'^+, \tvec{p}' \,\big|\,
    \mathcal{O}_{a_1}(y, z_1) \,\big|\, p^+, \tvec{p}
  \big\rangle \,\Big|_{\tvec{p} = \tvec{0}} \,.
 \phantom{\int}
\end{align}
After an appropriate shift of the position arguments in the bilinear field
operators and a change of integration variables from $y, z_1, z_2$ to $u_0
= \half (z_1 - z_2)$, $u_1 = \half (z_1 + z_2) + y$ and $u_2 = \half (z_1
+ z_2) - y$ this gives
\begin{align}
  \label{mom-skewed-1}
\sing{\tilde{F}}_{a_1,a_2}(x_i, \tvec{k}_i, \tvec{r})
& \approx \frac{p^+}{p'^+}
\int dp'^+\, d^2\tvec{p}'\,
\biggl[\, \prod_{i=0}^2
       \int \frac{du_i^-\, d^2\tvec{u}_i^{}}{(2\pi)^3}
\biggr]\;
e^{iu_0^{\smash{-}} (\ms p' - [1 - x_1^{} + x_2^{}]\, p\ms )^+
 - i\tvec{u}_0^{} \ms
      (\tvec{p}' - \tvec{k}_1^{} + \tvec{k}_2^{})}
\nonumber \\
&\quad \times e^{i (u_1^{\smash{-}} + u_2^{\smash{-}})
                 (x_1^{} + x_2^{})\ms p^+ /2
   - i \tvec{u}_1^{} (\tvec{k}_1^{} + \tvec{k}_2^{} - \tvec{r}) /2
   - i \tvec{u}_2^{} (\tvec{k}_1^{} + \tvec{k}_2^{} + \tvec{r}) /2}
 \phantom{\int}
\nonumber \\
&\quad \times
  \big\langle p^+, \tvec{p} \,\big|\,
    \mathcal{O}_{a_2}(0, u_2) \,\big|\, p'^+, \tvec{p}' \big\rangle \,
  \big\langle p'^+, \tvec{p}' \,\big|\,
    \mathcal{O}_{a_1}(0, u_1) \,\big|\, p^+, \tvec{p}
  \big\rangle \,\Big|_{\tvec{p} = \tvec{0}} \,.
 \phantom{\int} \hspace{-2em}
\end{align}
The integrations over $u_0^{\smash{-}}$ and $\tvec{u}_0^{}$ fix the
momentum $p'$ of the intermediate state.  In particular, its plus-momentum
is $p'^+ = (1 - x_1 + x_2)\ms p^+$, which reflects that the operator
$\bar{q}_{3}\ms \Gamma_{a_1}\ms q_{1}$ describes the emission of a quark
with plus-momentum $x_1\ms p^+$ and the reabsorption of a quark with
plus-momentum $x_2\ms p^+$.  The matrix elements in the approximation
\eqref{mom-skewed-1} are thus given by \emph{generalized} parton
distributions (GPDs), which play a prominent role in the description of
hard exclusive processes, see
\cite{Mueller:1998fv,Ji:1996ek,Radyushkin:1997ki} and the reviews
\cite{Goeke:2001tz,Diehl:2003ny,Belitsky:2005qn}.  To evaluate the
unapproximated form in \eqref{mom-skewed-0} one would need the
corresponding matrix elements for all transitions $p\to X$.  This is
obviously impractical, although for selected transitions to single
baryons, e.g.\ for $p\to \Delta(1232)$, some information can be obtained
\cite{Goeke:2001tz,Belitsky:2005qn}.  GPDs are defined by
\begin{align}
  \label{gpd-mom}
f_a(x,\xi, \tvec{k}; \tvec{p}, \tvec{p'}) &=
  \int \frac{dz^-}{2\pi}\, e^{i x z^- P^+}
  \int \frac{d^2\tvec{z}}{(2\pi)^2}\, e^{-i \tvec{k} \tvec{P}}\;
  \big\langle p'^+, \tvec{p}' \,\big|\, \mathcal{O}_a(0, z)
  \,\big|\, p^+, \tvec{p} \big\rangle \,,
\end{align}
where $P = \half (p + p')$ and $\Gamma_a$ is one of the matrices in
\eqref{Gamma-twist-2}.  The parameter
\begin{align}
\xi &= \frac{p^+ - p'^+}{p^+ + p'^+}
\end{align}
is often called skewness.  One finds that $\tvec{k}$ is the average
transverse momentum of the two quark legs and $x$ their average
plus-momentum divided by the average plus-momentum $P^+$ of the proton
states.  For ease of notation we do not indicate the polarization states
of the protons, which are in general different.  A parameterization of the
matrix elements \eqref{gpd-mom} for spin $1/2$ hadrons in terms of scalar
functions can be found in \cite{Meissner:2009ww}.  Invariance under the
transverse boost specified by $\tvec{v}\to \tvec{v} - (v^+ /P^+)\,
\tvec{P}$ gives the relation $f_a(x,\xi, \tvec{k}; \tvec{p}, \tvec{p'}) =
f_a(x,\xi, \tvec{k} - x \tvec{P}; -\half \tvec{\Delta}, \half
\tvec{\Delta})$ with $\tvec{\Delta} = (1+\xi)\ms \tvec{p}' - (1-\xi)\ms
\tvec{p}$.  Abbreviating
\begin{align}
f_a(x,\xi, \tvec{k}; \tvec{\Delta})
 &= f_a(x,\xi, \tvec{k}; -\half \tvec{\Delta}, \half \tvec{\Delta})
\end{align}
we can thus rewrite the relation \eqref{mom-skewed-1} as
\begin{align}
  \label{tilde-F-result}
\sing{\tilde{F}}_{a_1,a_2}(x_i, \tvec{k}_i, \tvec{y})
& \,\approx\, \frac{1+\xi}{1-\xi}\;
  f_{a_2}\bigl( x,-\xi, \half (\tvec{k}_+ + \tvec{r} + x \tvec{k}_-);
               (1+\xi)\ms \tvec{k}_- \bigr)\,
\nonumber \\[0.2em]
&\hspace{3.1em} \times
  f_{a_1}\bigl( x, \xi, \half (\tvec{k}_+ - \tvec{r} + x \tvec{k}_-);
               - (1+\xi)\ms \tvec{k}_- \bigr) \,,
\end{align}
where $\tvec{k}_{\pm} = \tvec{k}_1 \pm \tvec{k}_2$ and
\begin{align}
  \label{x-xi-1}
x   &= \frac{x_1+x_2}{2-x_1+x_2} \,, &
\xi &= \frac{x_1-x_2}{2-x_1+x_2} \,.
\end{align}
In complete analogy one derives
\begin{align}
  \label{I-result}
\sing{I}_{a_1,\bar{a}_2}(x_i, \tvec{k}_i, \tvec{r})
& \,\approx\, \frac{1+\xi}{1-\xi}\;
  f_{\bar{a}_2}\bigl( x,-\xi, \half (\tvec{k}_+ + \tvec{r} + x \tvec{k}_-);
               (1+\xi)\ms \tvec{k}_- \bigr)\,
\nonumber \\[0.2em]
&\hspace{3.1em} \times
  f_{a_1}\bigl( x, \xi, \half (\tvec{k}_+ - \tvec{r} + x \tvec{k}_-);
               - (1+\xi)\ms \tvec{k}_- \bigr) \,,
\end{align}
where the generalized parton distributions $f_{\bar{a}}$ for antiquarks
are defined by replacing $\mathcal{O}_a(0, z)$ with
$\mathcal{O}_{\bar{a}}(0, z)$ in \eqref{gpd-mom}.  One finds
\begin{align}
f_{\bar{a}}(x, \xi, \tvec{k}; \tvec{\Delta}) &=
  \sigma_{a}\, f_{a}(-x, \xi, -\tvec{k}; \tvec{\Delta})
\end{align}
with the same sign factors $\sigma_{q} = \sigma_{\delta q} = +1$ and
$\sigma_{\Delta q} = -1$ that appeared in \eqref{quark-antiquark}.  We
also note that generalized parton distributions with positive and negative
skewness parameter are easily related to each other by taking the complex
conjugate of \eqref{gpd-mom}.

For the distributions $\sing{\tilde{F}}_{a_1,\bar{a}_2}$ and
$\sing{\tilde{I}}_{a_1,\bar{a}_2}$ we obtain
\begin{align}
  \label{tilde-I-result}
\sing{\tilde{F}}_{a_1,\bar{a}_2}(x_i, \tvec{k}_i, \tvec{r})
& \,\approx\, \frac{1+\xi}{1-\xi}\;
  f_{\bar{a}_2}\bigl( -x,-\xi,
               - \half (\tvec{k}_- + \tvec{r} + x \tvec{k}_+);
               (1+\xi) \tvec{k}_+ \bigr)\,
\nonumber \\[0.2em]
&\hspace{3.1em} \times
  f_{a_1}\bigl( x, \xi, \half (\tvec{k}_- - \tvec{r} + x \tvec{k}_+);
               - (1+\xi)\ms \tvec{k}_+ \bigr) \,
\nonumber \\[1em]
\sing{\tilde{I}}_{a_1,\bar{a}_2}(x_i, \tvec{k}_i, \tvec{r})
& \,\approx\, \frac{1+\xi}{1-\xi}\;
  f_{\bar{a}_2}\bigl( x,-\xi, \half (\tvec{k}_- + \tvec{r} + x \tvec{k}_+);
               (1+\xi) \tvec{k}_+ \bigr)\,
\nonumber \\[0.2em]
&\hspace{3.1em} \times
  f_{a_1}\bigl( x, \xi, \half (\tvec{k}_- - \tvec{r} + x \tvec{k}_+);
               - (1+\xi)\ms \tvec{k}_+ \bigr) \,,
\end{align}
where we have again $\tvec{k}_{\pm} = \tvec{k}_1 \pm \tvec{k}_2$ but now
\begin{align}
  \label{x-xi-2}
x   &= \frac{x_1-x_2}{2-x_1-x_2} \,, &
\xi &= \frac{x_1+x_2}{2-x_1-x_2} \,.
\end{align}
In this case we have $|x|\le \xi$, which describes the emission of a
quark-antiquark pair.  Again, this could be anticipated from
figure~\ref{fig:distrib} since now the parton lines combined to color
singlets are $\{ 12 \}$ and $\{ 34 \}$, with each pair being on the same
side of the final-state cut in the double parton distribution.

An important difference between the approximations for $\sing{\tilde{F}}$,
$\sing{I}$, $\sing{\tilde{I}}$ and the one for $\sing{F}$ given in
\eqref{F-result} is that the generalized distributions on the r.h.s.\ of
\eqref{tilde-F-result}, \eqref{I-result} and \eqref{tilde-I-result} do not
reduce to collinear functions if we integrate over $\tvec{k}_1$ and
$\tvec{k}_2$, because these momenta appear in their fourth arguments.

The preceding derivations can easily be extended to distributions
describing interference between different quark flavors.  The
distributions corresponding to figure~\ref{fig:interference} are defined
with bilinear operators $\bar{d} \ms\Gamma\ms u$ or $\bar{u} \ms\Gamma\ms
d$.  For a proton target, the ground state in the sum over intermediate
states inserted between the two operators is then a neutron.  Isospin
symmetry relates the resulting matrix elements to matrix elements in the
proton: $\langle n | \bar{d}\, \Gamma\ms u | p \rangle = \langle p |
\bar{u}\ms \Gamma\ms d | n \rangle = \langle p | \bar{u}\ms \Gamma\ms u |
p \rangle - \langle p | \bar{d}\, \Gamma\ms d | p \rangle$.  Under the
assumption of $SU(3)$ flavor symmetry one can derive similar relations for
distributions involving strange quarks
\cite{Goeke:2001tz,Belitsky:2005qn}.

Although the relation between multiparton distributions and GPDs is an
approximation whose accuracy is not easy to estimate (and although our
current knowledge of GPDs is far less advanced than that of ordinary
parton densities) this relation provides opportunities to obtain
information about multiple interactions that is hard to get by other
means.  One example are the different interference distributions discussed
above, which are so far entirely unknown.  Perhaps even more important is
that GPDs give rather direct information about the distribution of single
partons in the impact parameter $\tvec{b}$, which is Fourier conjugate to
a transverse momentum transfer $\tvec{\Delta}$ that can be measured in
physical processes.  This is in stark contrast to the interparton distance
$\tvec{y}$ in two-parton distributions, which appears as an integration
variable in cross section formulae like \eqref{X-sect-qq} and is not
directly related to observable kinematic quantities.  We already mentioned
in section~\ref{sec:sigma-eff} that studies of GPDs give evidence for a
correlation between the longitudinal momentum and the impact parameter of
partons in the proton.  For values of the momentum fraction where such a
correlation is strong, it is hardly plausible that there should be no
correlation between $x_1, x_2$ and $\tvec{y}$ in two-parton distributions,
even if there were important corrections to approximations like
\eqref{F-result}.


\subsubsection{Spin correlations}
\label{sec:gpd-spin}

We now take a closer look at the role of the proton spin in the
approximate relation between double and single parton distributions, which
we have glossed over up to now.  For our purpose, a suitable choice to
describe the spin state of a proton is the light-cone helicity $\lambda =
\pm \half$, which is equal to the usual helicity in a frame where the
proton plus-momentum $p^+$ tends to infinity (see e.g.\
\cite{Soper:1972xc} or \cite[section~3.5.1]{Diehl:2003ny}).  We denote the
corresponding momentum eigenstates by $|p^+, \tvec{p}, \lambda \rangle$.

When inserting intermediate proton states between the two color singlet
operators in a two-parton distribution for an unpolarized proton, we
schematically have
\begin{align}
  \label{proton-helicity-sum}
& \frac{1}{2}\ms \sum_{\lambda}
  \big\langle p^+, \tvec{p}, \lambda \,\big|\,
  \mathcal{O}_{a_2}\, \mathcal{O}_{a_1} \,\big|\,
  p^+, \tvec{p}, \lambda \big\rangle
\nonumber \\
&\quad \approx \frac{1}{2}\ms \sum_{\lambda,\lambda'}
\int \frac{dp'^+\, d^2\tvec{p}'}{2 p'^+\ms (2\pi)^3}\;
  \big\langle p^+, \tvec{p}, \lambda \,\big|\,
    \mathcal{O}_{a_2} \,\big|\,
    p'^+, \tvec{p}', \lambda' \big\rangle\,
  \big\langle p'^+, \tvec{p}', \lambda' \,\big|\,
    \, \mathcal{O}_{a_1} \,\big|\,
    p^+, \tvec{p}, \lambda \big\rangle
\end{align}
or an analogous relation with states $|p^+, \tvec{b}, \lambda \rangle$ of
definite transverse position.  The sum over states on the r.h.s.\ thus
includes single-parton matrix elements where the proton helicity differs
in the bra and the ket state.  Corresponding sums over polarization states
should hence be inserted in the relations \eqref{F-result},
\eqref{tilde-F-result}, \eqref{I-result} and \eqref{tilde-I-result}.

To discuss the implications of this observation, let us focus on the
collinear distribution $\sing{F}(x_i, \tvec{r})$.  At the level of matrix
elements we have a relation
\begin{align}
  \label{F-helicity}
\sing{F}_{a_1,a_2}(x_i, \tvec{r}) &\approx
\frac{1}{2}\ms \sum_{\lambda,\lambda'}
  f_{a_2}^{\lambda,\lambda'}(x_2; - \tvec{r})\,
  f_{a_1}^{\lambda',\lambda}(x_1; \tvec{r}) \,,
\end{align}
where the superscripts $\lambda$ and $\lambda'$ of $f_a$ denote the proton
helicities as in \eqref{proton-helicity-sum} and an average over proton
helicities is understood in $\sing{F}$.  A standard decomposition of the
spin dependence of generalized parton distributions involves two
distributions $H$ and $E$ for unpolarized quarks and two distributions
$\tilde{H}$ and $\tilde{E}$ for longitudinally polarized quarks.  The
distribution $\tilde{E}$ does not contribute in the case of zero skewness
$\xi=0$ we are dealing with here.  Using the conventions and the matrix
elements for definite proton light-cone helicity in eq.~(54) of
\cite{Diehl:2003ny}, we have
\begin{align}
  \label{unpol-gpds}
f_{q}^{++}(x,\tvec{r}) &= H^q(x,0, -\tvec{r}^2) \,,
&
f_{q}^{--}(x,\tvec{r}) &= H^q(x,0, -\tvec{r}^2) \,,
\nonumber \\[0.2em]
f_{q}^{-+}(x,\tvec{r}) 
  &= \frac{\tvec{r}^1 + i \tvec{r}^2}{2M}\, E^q(x,0, -\tvec{r}^2) \,,
&
f_{q}^{+-}(x,\tvec{r})
  &= - \frac{\tvec{r}^1 - i \tvec{r}^2}{2M}\, E^q(x,0, -\tvec{r}^2)
\intertext{and}
\label{pol-gpds}
f_{\Delta q}^{++}(x,\tvec{r}) &= \tilde{H}^q(x,0, -\tvec{r}^2) \,,
&
f_{\Delta q}^{--}(x,\tvec{r}) &= - \tilde{H}^q(x,0, -\tvec{r}^2) \,,
\nonumber \\[0.2em]
f_{\Delta q}^{-+}(x,\tvec{r}) &= 0 \,,
&
f_{\Delta q}^{+-}(x,\tvec{r}) &= 0 \,,
\end{align}
where $M$ is the proton mass and $H^q(x,\xi,t)$, $E^q(x,\xi,t)$ and
$\tilde{H}^q(x,\xi,t)$ are the usual GPDs defined in \cite{Diehl:2003ny}.
$H^q$ and $\tilde{H}^q$ are the respective generalizations of the
unpolarized and longitudinally polarized quark densities $q$ and $\Delta
q$.  Changing the basis of the proton spin states, one can see that $E^q$
is related to unpolarized quarks in a transversely polarized proton
\cite{Burkardt:2002hr}.  Inserting \eqref{unpol-gpds} into
\eqref{F-helicity}, we get
\begin{align}
  \label{H-E-result}
\sing{F}_{q,q}(x_i, \tvec{r}) &\approx
 H^q(x_2, 0, -\tvec{r}^2)\, H^q(x_1,0, -\tvec{r}^2)
 + \frac{\tvec{r}^2}{4 M^2}\,
   E^q(x_2, 0, -\tvec{r}^2)\, E^q(x_1,0, -\tvec{r}^2) \,.
\end{align}
The term with $H$ corresponds to the simplest approximation of the
two-parton distribution as a product of single-parton distributions,
whereas the one with $E$ appears in addition.  $E$ describes a correlation
between the position of a single quark and the proton spin, and
\eqref{H-E-result} shows how such a correlation may lead to a correlation
between two quarks in an unpolarized proton.  It is difficult to say
whether this correction term alone already provides an improved
approximation of $\sing{F}_{q,q}(x_i, \tvec{r})$, but one may take it as
an indicator for the possible departure from a simple factorized ansatz
that neglects all correlations.  In cases where the correction from $E$ is
large, it is plausible to expect that the factorized ansatz involving only
$H$ will fail.

Applying the same method to the distribution $\sing{F}_{\Delta q, \Delta
  q}$, we obtain from \eqref{pol-gpds}
\begin{align}
  \label{H-tilde-result}
\sing{F}_{\Delta q, \Delta q}(x_i, \tvec{r}) &\approx
   \tilde{H}(x_2,0, -\tvec{r}^2)\, \tilde{H}(x_1,0, -\tvec{r}^2) \,.
\end{align}
Again one should be cautious regarding the validity of this approximation.
For the region of small but similar $x_1$ and $x_2$ we already argued in
section~\ref{sec:spin-decomp} that one may well have sizeable correlations
between the longitudinal polarization of two quarks, even if there is
little correlation between the longitudinal polarizations of one quark and
the proton as a whole.  Conversely, in kinematics where the product of
quark-proton spin correlations in \eqref{H-tilde-result} is sizeable it
seems natural to assume that quark-quark spin correlations are sizeable as
well.

\section{Perturbatively large transverse momentum}
\label{sec:high-qt}

So far we have treated multiple interactions as a two-scale problem, in
which 
the virtualities $q_1^2, q_2^2 \sim Q^2$ define a large scale whereas the
transverse momenta $|\tvec{q}_1|$, $|\tvec{q}_2|$ and the scale $\Lambda$
of nonperturbative interactions are treated as small.  We now make a
distinction between the different scales previously treated as small,
requiring $|\tvec{q}_1| \sim |\tvec{q}_2| \sim q_T$ to be large compared
with the hadronic scale $\Lambda$.  We thus have a three-scale problem
characterized by the hierarchy
\begin{align}
  \Lambda \ll q_T \ll Q \,.
\end{align}

Large $\tvec{q}{}_i$ implies that at least some of the transverse parton
momenta $\tvec{k}_i$ and $\bar{\tvec{k}}_i$ must be large.  The occurrence
of partons with large transverse momentum $k_T$ can be thought of as
resulting from the perturbative splitting of partons with low $k_T$, which
leads to a factorization formula for transverse-momentum dependent parton
distributions in terms of a hard-scattering kernel and collinear
distributions.  This significantly adds predictive power since collinear
distributions depend on fewer variables than $k_T$ dependent ones.

\begin{figure}
\begin{center}
\includegraphics[width=0.62\textwidth]{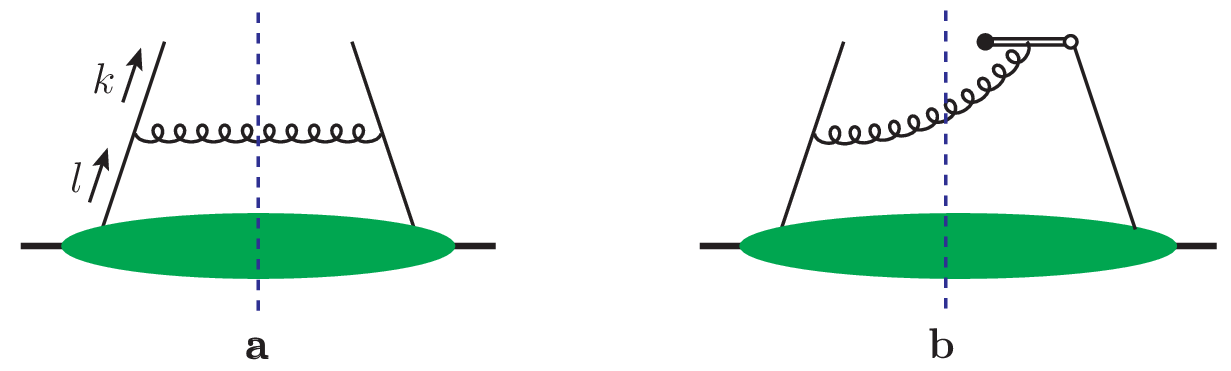}
\end{center}
\caption{\label{fig:simple-ladder} Graphs for a single-parton distribution
  $f(x,\tvec{k}$) at perturbatively large $\tvec{k}$.  Here and in the
  following it is understood that lines emerging from the lower blob have
  virtualities of order $\Lambda \ll |\tvec{k}|$.  The eikonal line in
  graph b results from the Wilson lines in the definition of $f(x,
  \tvec{k})$, see section~\protect\ref{sec:coll}.}
\end{figure}

Example graphs for the case of a single-quark distribution are shown in
figure~\ref{fig:simple-ladder}.  The description based on such graphs was
extensively used for spin effects and azimuthal correlations in Drell-Yan
production \cite{Ji:2006ub,Ji:2006vf,Ji:2006br} and semi-inclusive DIS in
\cite{Ji:2006br,Koike:2007dg,Bacchetta:2008xw}, building on the seminal
work in \cite{Collins:1981uk,Collins:1984kg}.

This description carries over to the case of two-parton distributions and
is discussed in section~\ref{sec:ladders}.  In the subsequent sections we
investigate a competing mechanism for the generation of high transverse
momentum, in which the two partons with momentum fractions $x_1$ and $x_2$
originate from the perturbative splitting of a single parton.  We will see
that this mechanism has profound consequences for the theoretical
description of multiple interactions.


\subsection{Ladder graphs at large \texorpdfstring{$\tvec{y}$}{y}}
\label{sec:ladders}

\begin{figure}[b]
\begin{center}
\includegraphics[width=0.99\textwidth]{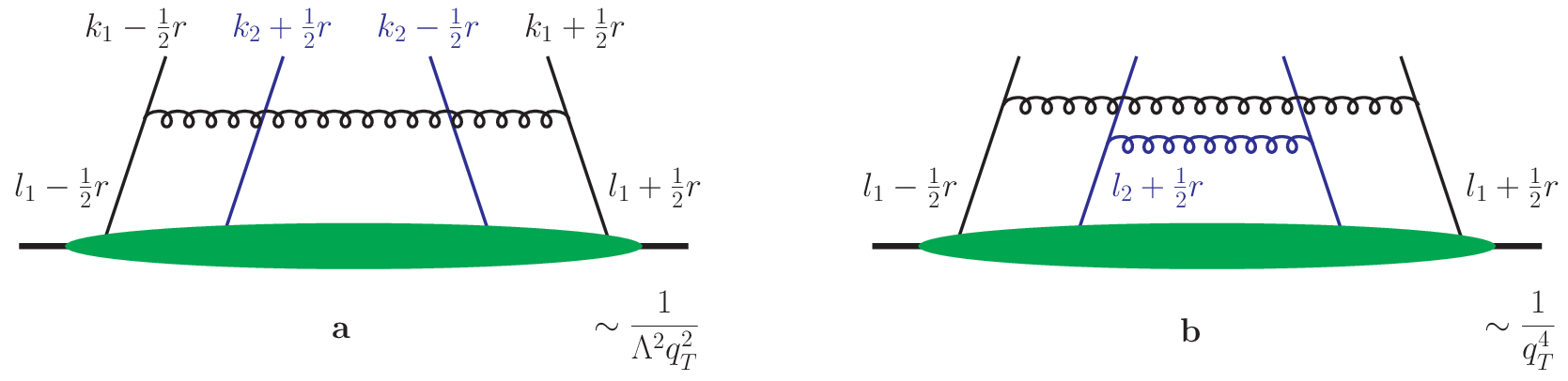}
\end{center}
\caption{\label{fig:ladders} Ladder graphs for the region of small
  $\tvec{r}$ with large $\tvec{k}_1$ (a) or large $\tvec{k}_1$ and
  $\tvec{k}_2$ (b).  The power behavior refers to
  $F(x_i,\tvec{k}_i,\tvec{r})$ and is discussed in the text.  Here and in
  the following we omit the dashed line that indicates the final-state cut
  as in figure~\protect\ref{fig:simple-ladder}.}
\end{figure}

The single and double ladder graphs in figure~\ref{fig:ladders} are
natural generalizations of the ladder graph for a single parton density in
figure~\ref{fig:simple-ladder}a.  In the following we concentrate on these
ladder graphs, bearing in mind that in a covariant gauge there are further
graphs with eikonal lines as in figure~\ref{fig:simple-ladder}b.  Those do
not change the conclusions we obtain for the ladder graphs.  Indeed, they
are absent in the axial gauge $v A = 0$, where the Wilson lines in the
definition of parton distributions reduce to unity (apart from pieces at
infinity, as discussed at the end of section~\ref{sec:coll}).

When interpreting the graphs in figures~\ref{fig:simple-ladder} and
\ref{fig:ladders} it is important to bear in mind that they represent a
separation of dynamics at different scales, with lines attached to the
lower blob having virtualities of order $\Lambda$, whereas propagators in
the upper part of the graphs are for virtualities of the order of the
large transverse momentum $q_T$.  An important feature of
figure~\ref{fig:ladders} is that no hard gluons are exchanged between the
parton lines that have different momentum fractions $x_1$ and $x_2$ at the
top of the graphs.  The requirement that both $l - \half r$ and $l + \half
r$ have small virtualities forces $|\tvec{r}|$ to be of order $\Lambda$,
which translates into interparton distances $|\tvec{y}|$ of hadronic size
in the Fourier transformed distributions $F(x_i, \tvec{k}_i, \tvec{y})$.


\subsubsection{Power behavior}
\label{sec:ladders-power}

Let us first investigate the general power behavior associated with ladder
graphs.  In the following we refer to $q_T$ as the hard scale (compared
with $\Lambda$), keeping in mind that $q_T$ is still much smaller than
$Q$.

We proceed in a similar way as in section~\ref{sec:power-counting}.  In
particular, we use the modified parton correlation functions $\Phi'_n$,
which contain a factor $1 /\sqrt{l^+}$ for each quark or antiquark of
momentum $l$ and in which pairs of quark fields are contracted with a
Dirac matrix $\Gamma_{a}$ from \eqref{Gamma-twist-2}.  For the transition
from $k$ to $m$ partons in the $t$ channel, we correspondingly use
hard-scattering coefficients $V'_{k\to m}$ that include a factor
$\sqrt{l^+}$ for each incoming quark or antiquark and a factor $1
/\sqrt{l^+}$ for each outgoing one.  Spinor indices in $V'$ are contracted
with an appropriate matrix $\half \gamma^+$, $\half \gamma^+ \gamma_5$,
$\half i \sigma^{+ j} \gamma_5$ for outgoing lines and with $\half
\gamma^-$, $\half \gamma_5 \gamma^-$, $\half i \sigma^{- j} \gamma_5$ for
incoming ones.  $V'$ is invariant under a boost along $z$ and thus can
only depend on the scale $q_T$ but not on $Q$ (cf.\ the corresponding
argument for $\Phi'$ in section~\ref{sec:power-counting}).  One thus
obtains
\begin{align}
  \label{V-scaling}
V'_{k\to m} &\sim q_T^{4 - k - 3m} \,,
\end{align}
as one can easily check for the example graphs below.  Note that compared
with the hard-scattering amplitudes in \eqref{T-scaling} we now have $3m$
instead of $m$ because $V'$ includes the propagators of the outgoing
partons, as is appropriate for the calculation of parton distributions.

The power behavior for the single-ladder graph in figure~\ref{fig:ladders}a
can be obtained from
\begin{align}
  \label{single-ladder}
F(x_i, \tvec{k}_i, \tvec{r}) 
\big|_{\text{fig.~\protect\ref{fig:ladders}a}}
 &= p^+ k_1^+ k_2^+ \int dr^-\, dk_1^-\, dk_2^-\, d^4 l_1^{}\;
  V'_{2\to 2}\, \Phi'_{4}
\nonumber \\
 &\approx  p^+ k_1^+ k_2^+ \int dk_1^-\, dl_1^+\,
  V'_{2\to 2} \int dr^-\, dk_2^-\, dl_1^-\, d^2\tvec{l}_1^{}\, \Phi'_4 \,.
\end{align}
The factor $p^+$ and the integrations over minus-momenta come from the
definition of $F$, whereas $k_1^+ k_2^+$ compensates the corresponding
factors in $V'$ and $\Phi'$.  It is understood that $V'$ includes a
$\delta$ function for each parton line going across the final-state cut.
This does not affect the power counting, since one could first consider
the hard-scattering amplitude without cut and then take the appropriate
discontinuity in the $s$ channel.
The momenta $k_2 \pm \half r$ and $l_1 \pm \half r$ attach to the parton
distribution at the bottom of the graph and hence have virtualities of
order $\Lambda$, whereas $k_1 \pm \half r$ emerges from the hard
scattering and hence has virtuality of order $q_T$.  As a result, the
momentum components $k_2^-, l_1^-, r^- \sim \Lambda^2 /p^+$ and
$|\tvec{l}_1| \sim \Lambda$ are small and can be neglected in the
hard-scattering kernel $V'$.  We used this when rearranging the order of
integrations in the second step.  By contrast, the large components $k_1^-
\sim q_T^2 /p^+$ and $k_1^+, k_2^+, l_1^+ \sim p^+$ are to be kept in
$V'$.  For the power behavior we obtain
\begin{align}
  \label{single-ladder-scaling}
F(x_i, \tvec{k}_i, \tvec{r}) 
\big|_{\text{fig.~\protect\ref{fig:ladders}a}}
 & \sim \alpha_s \times
     q_T^2 \times q_T^{-4} \times (\Lambda^2)^4 \times \Lambda^{-10}
   = \alpha_s \times \frac{1}{\Lambda^2\ms q_T^2}
\end{align}
with \eqref{Phi-scaling} and \eqref{V-scaling}.  We recognize the
$1/q_T^2$ behavior that is characteristic of the splitting of one parton
(the incoming quark) into two partons (the outgoing quark and the gluon).
The power behavior for the double-ladder graph in
figure~\ref{fig:ladders}b is obtained by the same type of analysis:
\begin{align}
  \label{double-ladder-scaling}
F(x_i, \tvec{k}_i, \tvec{r})
  \big|_{\text{fig.~\protect\ref{fig:ladders}b}}
 &\approx p^+ k_1^+ k_2^+ 
  \int dk_1^-\, dl_1^+\, V'_{2\to 2}\, 
  \int dk_2^-\, dl_2^+\, V'_{2\to 2}\, 
  \int dr^-\, dl_1^-\, dl_2^-\, d^2\tvec{l}_1\, d^2\tvec{l}_2\,
       \Phi'_{4}
\nonumber \\
 &\sim \alpha_s^2 \times
    (q_T^2 \times q_T^{-4})^2 \times (\Lambda^2)^5 \times \Lambda^{-10} 
  = \alpha_s^2 \times \frac{1}{q_T^4} \,.
\end{align}
If all transverse momenta are small, the distribution $F(x_i, \tvec{k}_i,
\tvec{r})$ scales of course like $\Lambda^{-4}$.

\begin{figure}
\begin{center}
\includegraphics[width=0.88\textwidth]{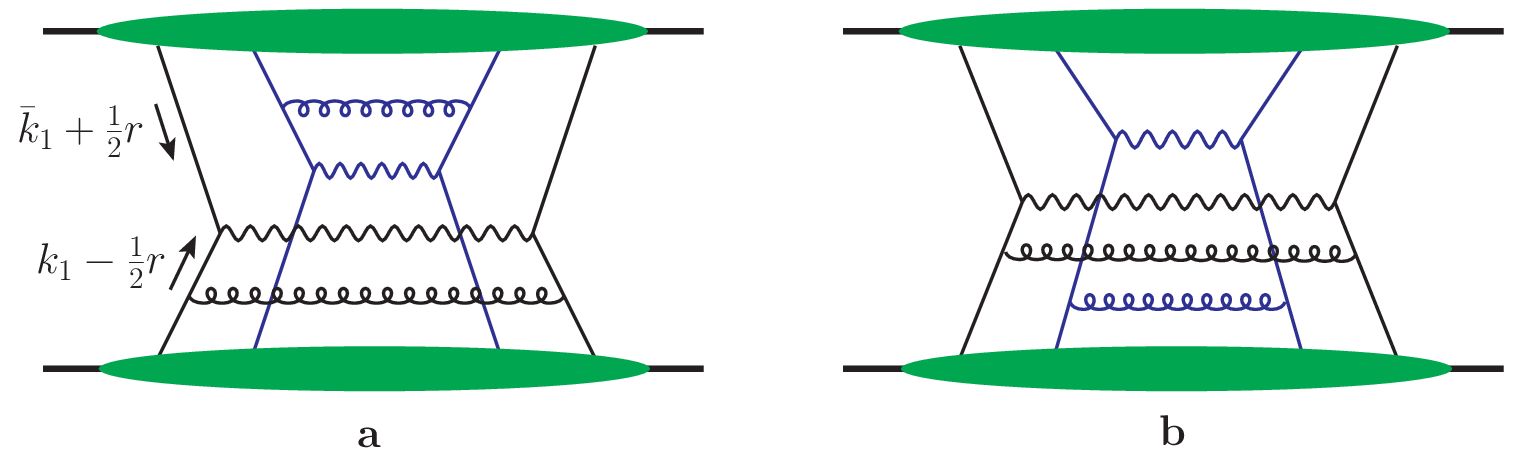}
\end{center}
\caption{\label{fig:ladder-Xsect} Ladder graphs contributing to the cross
  section for boson pair production at large $q_T$.}
\end{figure}

Let us now see how the power behavior of the two-parton distributions
translates into the power behavior of the cross section
\begin{align}
  \label{X-sect-again}
\frac{s^2\ms d\sigma}{\prod_{i=1}^2 dx_i\, d\bar{x}_i\, d^2\tvec{q}{}_i}
  \bigg|_{\text{figs.~\protect\ref{fig:ladder-Xsect}a,b}}
&\propto
\biggl[\, \prod_{i=1}^{2} \, s\ms \hat{\sigma}_i(x_i \bar{x}_i s) \biggr]\,
\biggl[\, \prod_{i=1}^{2} 
     \int d^2\tvec{k}_i\, d^2\bar{\tvec{k}}_i\;
     \delta^{(2)}(\tvec{q}{}_i - \tvec{k}_i - \bar{\tvec{k}}_i) \biggr]
\nonumber \\
& \quad\times
\int d^2\tvec{r}\,
  F(x_i, \tvec{k}_i, \tvec{r})\,
  F(\bar{x}_i, \bar{\tvec{k}}_i, - \tvec{r}) \,,
\end{align}
where we have omitted numerical factors as well as labels for parton
species, spin and color.  We have multiplied the cross section with $s^2$
for convenience, since this gives factors $s\ms \hat{\sigma}(x_i \bar{x}_i
s)$ of order $1$ on the r.h.s.

To have both large $\tvec{q}_1$ and $\tvec{q}_2$ requires at lowest order
in $\alpha_s$ either a single-ladder graph in the distribution for each
colliding proton, or a double-ladder graph in one of the distributions
with no hard gluons in the other, as shown in
figure~\ref{fig:ladder-Xsect}.  In both cases one has $F(x_i, \tvec{k}_i,
\tvec{r})\, F(\bar{x}_i, \bar{\tvec{k}}_i, - \tvec{r}) \sim \alpha_s^2
/(\Lambda^{4}\ms q_T^{4})$ for the product of distributions, and the
integration volume $d^2\tvec{k}_1\, d^2\bar{\tvec{k}}_1\;
\delta^{(2)}(\tvec{q}{}_1 - \tvec{k}_1 - \bar{\tvec{k}}_1)$ is of order
$\Lambda^2$ since $\bar{\tvec{k}}_1^{}$ and thus $\tvec{k}_1^{} -
\tvec{q}_1^{}$ are restricted to be of size $\Lambda$.  Similarly, one
finds $d^2\tvec{k}_2\, d^2\bar{\tvec{k}}_2\; \delta^{(2)}(\tvec{q}{}_2 -
\tvec{k}_2 - \bar{\tvec{k}}_2) \sim \Lambda^2$ in both cases, so that the
overall power behavior is
\begin{align}
  \label{ladders-X-sect}
\frac{s^2\ms d\sigma}{\prod_{i=1}^2 dx_i\, d\bar{x}_i\, d^2\tvec{q}{}_i}
  \,\bigg|_{\text{figs.~\protect\ref{fig:ladder-Xsect}a,b}}  
 &\sim \alpha_s^2 \times
    (\Lambda^2)^3 \times \Bigl( \frac{1}{\Lambda^2\ms q_T^2} \Bigr)^2
  = \alpha_s^2\times \frac{\Lambda^2}{q_T^4} \,.
\end{align}
By similar arguments one finds that the power behavior remains the same at
higher order in $\alpha_s$, when one can have more than two ladders in the
graphs for the cross section.  Any decrease by a factor $\Lambda^2/q_T^2$
in the product $F(x_i, \tvec{k}_i, \tvec{r})\, F(\bar{x}_i,
\bar{\tvec{k}}_i, - \tvec{r})$ is compensated by an increase from
$\Lambda^2$ to $q_T^2$ in the integration volume over the transverse
parton momenta $\tvec{k}_i$ or $\bar{\tvec{k}}_i$.

\subsubsection{Factorization formulae}
\label{sec:ladders-fact}

Let us now investigate the structure of the factorization formulae for
ladder graphs.  Still omitting spin and color indices for the moment, we
write our result \eqref{single-ladder} as
\begin{align}
  \label{single-ladder-again}
F(x_i, \tvec{k}_i, \tvec{r}) 
  \big|_{\text{fig.~\protect\ref{fig:ladders}a}}
 = \int_{k_1^+}^{(p-k_2)^+}\,
 & \frac{dl_1^+}{l_1^+}\,
   \biggl[ k_1^+ \int dk_1^-\, V'_{2\to 2}  \biggr]
\nonumber \\
 & \times
  \biggl[ p^+ l_1^+ k_2^+ \int dr^-\, dl_1^-\, dk_2^-\, d^2\tvec{l}_1^{}\,
          \Phi'_4(l_1^{}, k_2^{}, r) \biggr]_{r^+ = 0} \,.
\end{align}
The lower limit on the $l_1^+$ integration reflects that the gluon with
momentum $l_1-k_1$ crosses the final-state cut and hence cannot have
negative plus-momentum.  Up to a numerical factor, the second expression
in square brackets is just the transverse-momentum dependent double-parton
distribution $F(u_1, x_2, \tvec{l}_1,\tvec{k}_2, \tvec{r})$ with $u_1^{} =
l_1^+ /p^+$.  The first factor in square brackets is invariant under
boosts along $z$ and can thus only depend on $\tvec{k}_1$ and $l_1^+
/k_1^+ = u_1^{} /x_1^{}$.  We can write it as a numerical factor times
$\tvec{k}_1^{-2}\ms P\bigl( u_1^{}/x_1^{} , \tvec{k}_1^{} \bigr)$, where
$P$ is dimensionless.  If $P$ is a scalar it depends on $\tvec{k}_1^{}$
only via $\tvec{k}_1^2$ and only because there are other dimensionful
variables $\zeta$ and $\mu$, which we have not displayed for ease of
writing.  For certain parton polarizations, $P$ is a tensor with
transverse indices and can hence depend on the components of
$\tvec{k}_1^{}$, as discussed below.

We can finally Fourier transform \eqref{single-ladder-again} from
$\tvec{r}$ to $\tvec{y}$ and then have
\begin{align}
  \label{single-ladder-y-diff}
F(x_1,x_2, \tvec{k}_1, \tvec{k}_2, \tvec{y})
  \big|_{\text{fig.~\protect\ref{fig:ladders}a}}
 &= \frac{1}{\pi \tvec{k}_1^2}\,
    \int_{x_1}^{1-x_2}\, \frac{du_1}{u_1}\,
    P\Bigl( \frac{x_1}{u_1} , \tvec{k}_1 \Bigr)
    \int d^2\tvec{l}_1\, F(u,x_2, \tvec{l}_1,\tvec{k}_2, \tvec{y}) \,,
\end{align}
where the factor $1/\pi$ has been chosen for convenience.  Inserting this
and its analog for $F(\bar{x}_1,\bar{x}_2, \bar{\tvec{k}}_1,
\bar{\tvec{k}}_2, \tvec{y})$ in the cross section formula
\eqref{X-sect-mixed}, we have
\begin{align}
  \label{single-ladders-intermed}
& \frac{d\sigma}{\prod_{i=1}^2 dx_i\, d\bar{x}_i\, d^2\tvec{q}{}_i}
  \bigg|_{\text{fig.~\protect\ref{fig:ladder-Xsect}a}}
 = \frac{1}{C}\, 
\hat{\sigma}_1(x_1 \bar{x}_1 s)\, \hat{\sigma}_2(x_2 \bar{x}_2 s)\, 
\int d^2\bar{\tvec{k}}_1\, d^2\tvec{k}_2\;
\frac{1}{\pi\ms (\tvec{q}_1^{} - \bar{\tvec{k}}_1^{})^2 \rule{0pt}{2.3ex}}\,
\frac{1}{\pi\ms (\tvec{q}_2^{} - \tvec{k}_2^{})^2 \rule{0pt}{2.3ex}}
\nonumber \\[0.2em]
& \qquad\times
  \int_{x_1}^{1-x_2}\, \frac{du_1}{u_1}\, P\Bigl( \frac{x_1}{u_1},
                        \tvec{q}_1^{} - \bar{\tvec{k}}_1^{} \Bigr)\,
  \int_{\bar{x}_2}^{1-\bar{x}_1}\, \frac{d\bar{u}_2}{\bar{u}_2}\,
         P\Bigl( \frac{\bar{x}_2}{\bar{u}_2},
         \tvec{q}_2^{} - \tvec{k}_2^{} \Bigr)
\nonumber \\[0.2em]
& \qquad \times \int d^2\tvec{y}
    \int d^2\tvec{l}_1\, F(u_1,x_2, \tvec{l}_1,\tvec{k}_2, \tvec{y})
    \int d^2\bar{\tvec{l}}_2\,
    F(\bar{x}_1,\bar{u}_2, \bar{\tvec{k}}_1,\bar{\tvec{l}}_2, \tvec{y}) \,,
\end{align}
where we have used the $\delta$ function constraints in
\eqref{X-sect-mixed} to eliminate $\tvec{k}_1$ and $\bar{\tvec{k}}_2$.  We
can now approximate $\tvec{q}_1^{} - \bar{\tvec{k}}_1^{} \approx
\tvec{q}_1^{}$ and $\tvec{q}_2^{} - \tvec{k}_2^{} \approx \tvec{q}_2^{}$,
after which the integrations over $\bar{\tvec{k}}_1$ and $\tvec{k}_2$ only
concern the double-parton distributions, which are then integrated over
both transverse-momentum arguments.  The result is
\begin{align}
  \label{single-ladders-X}
& \frac{d\sigma}{\prod_{i=1}^2 dx_i\, d\bar{x}_i\, d^2\tvec{q}{}_i}
  \bigg|_{\text{fig.~\protect\ref{fig:ladder-Xsect}a}}
 = \frac{1}{C}\, 
\hat{\sigma}_1(x_1 \bar{x}_1 s)\, \hat{\sigma}_2(x_2 \bar{x}_2 s)\, 
\frac{1}{\pi \tvec{q}_1^2}\, \frac{1}{\pi \tvec{q}_2^2}\,
  \int_{x_1}^{1-x_2}\, \frac{du_1}{u_1}\,
       P\Bigl( \frac{x_1}{u_1}, \tvec{q}_1 \Bigr)\,
\nonumber \\[0.1em]
& \qquad \times
  \int_{\bar{x}_2}^{1-\bar{x}_1}\, \frac{d\bar{u}_2}{\bar{u}_2}\,
         P\Bigl( \frac{\bar{x}_2}{\bar{u}_2}, \tvec{q}_2 \Bigr)
  \int d^2\tvec{y}\,
    F(u_1,x_2, \tvec{y})
    F(\bar{x}_1,\bar{u}_2, \tvec{y}) \,.
\end{align}
Only collinear two-parton distributions appear on the r.h.s., so that the
relation \eqref{single-ladder-y-diff} is only needed in the form
\begin{align}
  \label{single-ladder-y}
\int d^2\tvec{k}_2\, F(x_1,x_2, \tvec{k}_1, \tvec{k}_2, \tvec{y})
  \big|_{\text{fig.~\protect\ref{fig:ladders}a}}
 &= \frac{1}{\pi \tvec{k}_1^2}\,
    \int_{x_1}^{1-x_2}\, \frac{du_1}{u_1}\,
    P\Bigl( \frac{x_1}{u_1} , \tvec{k}_1 \Bigr)\,
    F(u, x_2, \tvec{y}) \,.
\end{align}
Repeating the preceding arguments for the double-ladder graph, one obtains
\begin{align}
  \label{double-ladder-y}
F(x_1,x_2, \tvec{k}_1, \tvec{k}_2, \tvec{y}) 
  \big|_{\text{fig.~\protect\ref{fig:ladders}b}}
 = \frac{1}{\pi \tvec{k}_1^2}\, &
   \frac{1}{\pi \tvec{k}_2^2}\, 
   \int_{x_1}^{1-x_2}\, \frac{du_1}{u_1}\,
         P\Bigl( \frac{x_1}{u_1} , \tvec{k}_1 \Bigr)
\nonumber \\
& \quad \times
    \int_{x_2}^{1-u_1}\, \frac{du_2}{u_2}\,
         P\Bigl( \frac{x_1}{u_2} , \tvec{k}_2 \Bigr)\,
     F(u_1,u_2, \tvec{y})
\end{align}
from the result \eqref{double-ladder-scaling} and
\begin{align}
  \label{double-ladder-X}
& \frac{d\sigma}{\prod_{i=1}^2 dx_i\, d\bar{x}_i\, d^2\tvec{q}{}_i}
  \bigg|_{\text{fig.~\protect\ref{fig:ladder-Xsect}b}}
 = \frac{1}{C}\, 
\hat{\sigma}_1(x_1 \bar{x}_1 s)\, \hat{\sigma}_2(x_2 \bar{x}_2 s)\, 
\frac{1}{\pi \tvec{q}_1^2}\, \frac{1}{\pi \tvec{q}_2^2}\,
  \int_{x_1}^{1-x_2}\, \frac{du_1}{u_1}\,
       P\Bigl( \frac{x_1}{u_1}, \tvec{q}_1 \Bigr)\,
\nonumber \\[0.1em]
& \qquad \times
  \int_{x_2}^{1-u_1}\, \frac{du_2}{u_2}\,
       P\Bigl( \frac{x_2}{u_2}, \tvec{q}_2 \Bigr)
  \int d^2\tvec{y}\,
    F(u_1,u_2, \tvec{y}) F(\bar{x}_1,\bar{x}_2, \tvec{y})
\end{align}
for the contribution of figure~\ref{fig:ladder-Xsect}b to the cross
section, where again only collinear two-parton distributions appear.

The analog of \eqref{single-ladder-y} for single-parton distribution reads
\begin{align}
  \label{simple-ladder-high-kt}
f(x, \tvec{k}) 
  \big|_{\text{fig.~\protect\ref{fig:simple-ladder}}}
 &= \frac{1}{\pi \tvec{k}^2}\,
    \int_{x}^1\, \frac{du}{u}\,
    P\Bigl( \frac{x}{u} , \tvec{k} \Bigr)\, f(u) \,.
\end{align}
At order $\alpha_s$ the kernel $P(x/u)$ is just the usual DGLAP splitting
function, up to terms proportional to $\delta(1-x/u)$ which will be
discussed shortly.  To see this, let us consider the $k_T$ integrated
parton density defined with a naive cutoff,
\begin{align}
f(x; \mu) & \overset{\text{naive}}{=} \int d^2\tvec{k}\;
   \theta(\mu^2 - \tvec{k}^2)\, f(x,\tvec{k}^2) \,.
\end{align}
Using that the transverse-momentum dependent density depends only on the
square of $\tvec{k}$ we then have
\begin{align}
\mu^2\ms \frac{d}{d\mu^2}\, f(x; \mu)
  \overset{\text{naive}}{=} \pi \mu^2 \, f(x, \tvec{k}^2=\mu^2) \,,
\end{align}
and comparing with the DGLAP equation for the l.h.s.\ we can identify the
kernel in \eqref{simple-ladder-high-kt} as the familiar splitting
function.

The preceding argument is oversimplified in two respects.  Firstly, the
calculation of $f(x, \tvec{k})$ for large $\tvec{k}$ only involves real
graphs like those in figure~\ref{fig:simple-ladder} at leading order in
$\alpha_s$, because to obtain a parton with large $\tvec{k}$ one needs a
recoiling parton in the final state.  (Higher-order graphs can include
virtual loops, so that our argument cannot be applied any more.)  By
contrast, the evolution equation for the collinear parton density $f(x)$,
which is integrated over all $\tvec{k}$, involves both real and virtual
graphs at order $\alpha_s$.  The latter give contributions proportional to
$\delta(1-x/u)$ to the DGLAP splitting kernels, which are absent from
$P(x/u)$ in \eqref{simple-ladder-high-kt}.  Secondly, the distribution
$f(x, \tvec{k})$ must be defined with non-lightlike Wilson lines as
discussed in section~\ref{sec:factorization}, which leads to a dependence
on the parameter $\zeta$ defined in \eqref{zeta-def}.  Since the collinear
distribution $f(x)$ has no such dependence, it is the kernel in
\eqref{simple-ladder-high-kt} that must depend on $\zeta$.  The explicit
calculation in \cite{Bacchetta:2008xw} shows that the $\zeta$ dependence
of $P(x/u)$ comes with a factor $\delta(1-x/u)$, which in the light of our
discussion in section~\ref{sec:full-fact} is plausible if one observes
that the point $x=u$ in \eqref{simple-ladder-high-kt} corresponds to
infinite negative gluon rapidity in figure~\ref{fig:simple-ladder}.

Let us mention that there is an analog of \eqref{simple-ladder-high-kt}
for the distribution $f(x,\tvec{z})$ at small transverse distance
$\tvec{z}$.  The corresponding hard-scattering kernel differs again from
$P(x/u)$ by terms proportional to $\delta(1-x/u)$.  This is because
$f(x,\tvec{z})$ is given by an integral $\int d^2\tvec{k}\, e^{i \tvec{z}
  \tvec{k}}\, f(x,\tvec{k})$ over all $\tvec{k}$, so that already at order
$\alpha_s$ virtual graphs appear in addition to real ones.

We now turn our attention to the role of color in two-parton distributions
at high transverse momentum, which we have glossed over so far.  Since the
graphs we are discussing do not connect parton lines with different
momentum fractions $x_1$ and $x_2$, the color coupling of the
distributions on the left and on the right of the factorization formulae
\eqref{single-ladder-y} and \eqref{double-ladder-y} are the same.
For distributions $\sing{F}(x_i, \tvec{k}_i, \tvec{y})$ in the color
singlet channel, the kernels $P$ in \eqref{single-ladder-y} and
\eqref{double-ladder-y} coincide with the one in
\eqref{simple-ladder-high-kt} at least at leading order in $\alpha_s$,
since the relevant hard-scattering graphs to be calculated are identical.
The leading-order kernels for the color octet distributions $\oct{F}(x_i,
\tvec{k}_i, \tvec{y})$ differ by an overall color factor from those for
$\sing{F}(x_i, \tvec{k}_i, \tvec{y})$, which is the subject of the next
section.  Note that, unlike their color singlet analogs, the collinear
color octet distributions appearing on the r.h.s.\ of
\eqref{single-ladder-y} and \eqref{double-ladder-y} depend on $\zeta$ as
discussed in section~\ref{sec:coll-fact}.  Given our discussion of the
cancellation or non-cancellation of soft contributions between real and
virtual graphs in section~\ref{sec:coll-fact}, we expect that the kernels
for the position space distributions $\sing{F}(x_i, \tvec{z}_i, \tvec{y})$
and $\oct{F}(x_i, \tvec{z}_i, \tvec{y})$ at small $\tvec{z}_i$ will differ
by more than an overall color factor already at leading order.  A
systematic investigation of this is left to future work.

The power counting in section~\ref{sec:ladders-power} and the discussion
in the present section do not depend on whether the parton lines in the
ladder graphs are quarks or gluons.  As is well-known for single parton
distributions, a quark with high transverse momentum can originate from a
gluon with low transverse momentum and vice versa.  The corresponding
elementary ladder graphs are shown in figure~\ref{fig:ladder-color} below.

We now discuss the spin structure of the ladder graphs.  As we have seen
in sections~\ref{sec:spin:quarks} and \ref{sec:spin:gluons}, there are
three polarization combinations for each quark or gluon in a two-parton
distribution, which we can choose as unpolarized ($q, g$), longitudinally
polarized $(\Delta q, \Delta g)$ and transversely $(\delta q)$ or linearly
$(\delta g)$ polarized.  The possible transitions between these
combinations in the factorization formulae \eqref{single-ladder-y} and
\eqref{double-ladder-y} are restricted by symmetries.  Transverse quark
polarization $\delta q$ is described by a chiral-odd operator $\bar{q}\ms
i\sigma^{+j} \gamma_5\, q$ and the chirality of light quarks is conserved
in hard-scattering subprocesses, so that the only transitions for
transversely polarized quarks are of the form $\delta q \to \delta q$.  In
the longitudinally polarized sector one has all possible transitions
between $\Delta q$ and $\Delta g$ on the left and on the right of
\eqref{single-ladder-y} and \eqref{double-ladder-y}, with transitions to
other polarizations being forbidden by parity invariance.  Likewise, one
has all possible transitions between unpolarized quarks and gluons.  In
addition, ladder graphs allow the transitions $g\to \delta g$ and $q\to
\delta g$ from unpolarized collinear distributions to linearly polarized
gluons at high $\tvec{k}$, as has been observed in the study of single
Higgs production in \cite{Nadolsky:2007ba,Catani:2010pd}.  Since $\delta
g$ corresponds to a helicity difference of two units between the gluon on
the left and the gluon the right of the final-state cut, there is no
collinear distribution for a single linearly polarized gluon in a proton,
so that transitions $\delta g \to g$ and $\delta g \to q$ played no role
in \cite{Nadolsky:2007ba,Catani:2010pd}.  However, one finds that the
corresponding hard-scattering kernels are nonzero and hence allow these
transitions for two-parton distributions.  One thus has all possible
transitions between $q, g$ and $\delta g$.

For distributions that involve polarizations $\delta q$ or $\delta g$ the
kernels $P$ in \eqref{single-ladder-y} and \eqref{double-ladder-y} are
tensors with transverse Lorentz indices, constructed from $\delta^{jl}$
and from the large transverse momentum $\tvec{k}_1$ or $\tvec{k}_2$ in the
ladder.  Explicit calculation at order $\alpha_s$ shows that the kernel
$P_{\delta q\ms \delta q}^{jl}$ for the transition from $F_{\delta q,
  a}^l$ to $F_{\delta q, a}^j$ (with arbitrary $a$) is proportional
to~$\delta^{jl}$.  As a result, ladder graphs do not generate the
distributions $g^{\ms s}_{\ms\delta q, \delta q}$ and $g^{\ms
  a}_{\ms\delta q, \delta q}$ in the decomposition \eqref{TT} of
$\smash{F_{\delta q, \delta q}^{jj'}(x_i, \tvec{k}_i, \tvec{y})}$, given
that they come with tensors that are absent in the collinear distributions
$\smash{F_{\delta q, \delta q}^{l\ms l'}(x_i, \tvec{y})}$ according to
\eqref{collinear-decomp}.  This adds to the predictive power of the
perturbative mechanism at large transverse momenta.  For transitions
involving linear gluon polarization we find kernels
\begin{align}
  \label{tensor-kernels}
P_{\delta g\ms \delta g}^{jj',l\ms l'}
 &\propto \tau^{jj',l\ms l'}_{} \,,
&
P_{\delta g\ms g}^{ll'} &\propto P_{\delta g\ms q}^{ll'}
  \propto P_{g\ms \delta g}^{l\ms l'} \propto P_{q\ms \delta g}^{l\ms l'}
  \propto 2 \tvec{k}_i^l \tvec{k}_i^{l'} - \delta^{l\ms l'} \tvec{k}_i^2 \,,
\end{align}
where the transition from $\smash{F_{\delta g, a}^{l\ms l'}}$ to
$\smash{F_{\delta g, a}^{jj'}}$ is described by $\smash{P_{\delta g\ms
    \delta g}^{jj',l\ms l'}}$, the transition from $\smash{F_{g, a}^{}}$
to $\smash{F_{\delta g, a}^{l\ms l'}}$ by $\smash{P_{\delta g\ms g}^{l\ms
    l'}}$ etc., and where $\tvec{k}_i$ is $\tvec{k}_1$ or $\tvec{k}_2$.
The second tensor in \eqref{tensor-kernels} is symmetric and traceless and
describes two units of orbital angular momentum, which compensates the
mismatch of helicities in the transitions $g\to \delta g$, $q\to \delta
g$, $\delta g\to g$ and $\delta g\to q$.  For later use we note that terms
involving this tensor vanish by rotation invariance when
\eqref{single-ladder-y-diff} or~\eqref{double-ladder-y} is integrated over
$\tvec{k}_1$ and $\tvec{k}_2$.

The representation of the cross section derived in this section is based
on a two-step procedure.  In the first step we have used factorization to
separate the annihilation processes $q\bar{q} \to V$ into vector bosons
$V$ with mass or virtuality of order $Q$ from transverse-momentum
dependent two-parton distributions, in which the largest scale is $q_T$.
In a second step, we have used factorization to compute these
distributions in terms of hard-scattering processes at scale $q_T$ and
distributions that reflect physics at a hadronic scale $\Lambda$, where it
turned out that in the cross section we only need the latter distributions
integrated over $\tvec{k}_1$ and $\tvec{k}_2$.

An alternative procedure is to first use factorization to represent the
graphs in figure~\ref{fig:ladder-Xsect} as the product of collinear
two-parton distributions and inclusive hard-scattering processes $q\bar{q}
\to V + X$, where at lowest order in $\alpha_s$ the unobserved system $X$
consists of just one gluon.  In a second step one can then simplify the
corresponding hard-scattering kernels by taking the limit $q_T \ll Q$ we
are interested in.  The relation between these two procedures has been
studied in detail for single Drell-Yan production or for semi-inclusive
deep inelastic scattering in \cite{Bacchetta:2008xw} and
\cite{Ji:2006ub,Ji:2006vf,Ji:2006br,Koike:2007dg}.  An important property
of the procedure using transverse-momentum dependent distributions in a
first step is that it permits the resummation of Sudakov logarithms with
the method of Collins, Soper and Sterman~\cite{Collins:1984kg}.


\subsubsection{Color factors and quark-gluon transitions}
\label{sec:ladders-color}

\begin{figure}
\begin{center}
\includegraphics[width=0.65\textwidth]{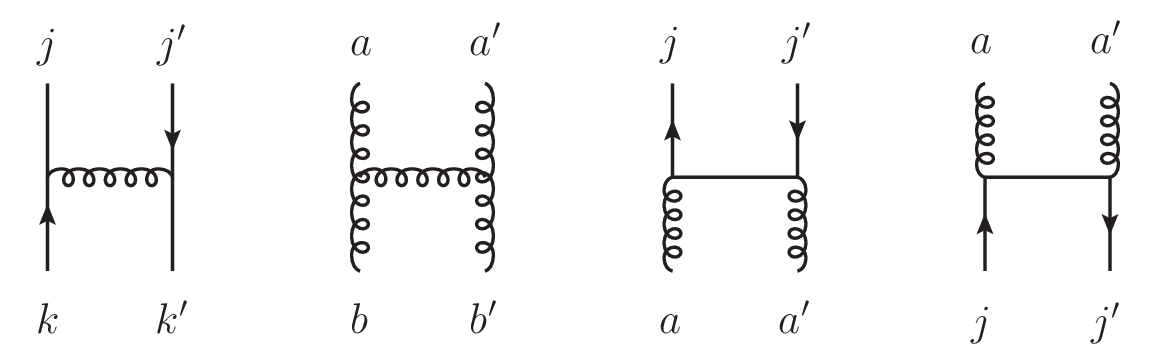}
\end{center}
\caption{\label{fig:ladder-color} Elementary ladder graphs for transitions
  between quarks and gluons.}
\end{figure}

Let us now compute the color factors for the transitions from collinear to
transverse-momentum dependent two-parton distributions, restricting
ourselves to the leading order in $\alpha_s$.  For this it is sufficient
to consider ladder graphs for the hard scattering, because graphs with
eikonal lines as in figure~\ref{fig:simple-ladder}b can be eliminated by
choosing the gauge $v A = 0$.
Multiplying the color structure of the ladder graphs in
figure~\ref{fig:ladder-color} with the appropriate color matrix for the
incoming partons, one obtains the following color transitions:
\begin{align}
  \label{color-transitions}
\sing{q} &\to \sing{q}: &
 t^a_{jk}\, t^a_{k'j'}\, \delta_{kk'}^{} &= C_F\, \delta_{jj'} \,,
   \phantom{\frac{1}{1}}
\nonumber \\
\oct{q} &\to \oct{q}: &
 t^a_{jk}\, t^a_{k'j'}\, t^c_{kk'} &= - \frac{1}{2N}\, t^c_{jj'} \,,
\nonumber \\
\sing{g} &\to \sing{g}: &
 f^{abd}\, f^{a'b'd}\, \delta^{bb'} &= N\, \delta^{aa'} \,,
  \phantom{\frac{1}{1}}
\nonumber \\
\octA{g} &\to \octA{g}: &
 f^{abd}\, f^{a'b'd}\, f^{cbb'} &= \frac{N}{2}\, f^{caa'} \,,
\nonumber \\
\octS{g} &\to \octS{g}: &
 f^{abd}\, f^{a'b'd}\, d^{cbb'} &= \frac{N}{2}\, d^{caa'} \,,
\nonumber \\
\sing{g} &\to \sing{q}: &
 t^a_{jk}\, t^{a'}_{kj'}\, \delta^{aa'} &= \frac{N^2-1}{2N}\, \delta_{jj'} \,,
\nonumber \\
\octA{g} &\to \oct{q}: &
 t^a_{jk}\, t^{a'}_{kj'}\, (-i f^{caa'}) &= \frac{N}{2}\, t_{jj'}^c \,,
\nonumber \\
\octS{g} &\to \oct{q}: &
 t^a_{jk}\, t^{a'}_{kj'}\, d^{caa'} &= \frac{N^2-4}{2N}\, t_{jj'}^c \,,
\nonumber \\
\sing{q} &\to \sing{g}: &
 t^{a'}_{j'k}\, t^{a}_{kj}\, \delta_{jj'}^{} &= \frac{1}{2}\, \delta^{aa'} \,,
\nonumber \\
\oct{q} &\to \octA{g},\, \octS{g}: &
 t^{a'}_{j'k}\, t^{a}_{kj}\, t^c_{jj'} &=
   - \frac{1}{4}\, if^{caa'} + \frac{1}{4}\, d^{caa'} \,.
\end{align}
Reversing the fermion lines in the ladder graphs changes the order of
multiplication for the $t$ matrices on the l.h.s.\ of the above relations.
This leads to a change of sign on the r.h.s.\ for the transitions
$\octA{g}\to \oct{q}$ and $\oct{q}\to \octA{g}$, whereas transitions
between $\oct{q}$ and $\octS{g}$ or between singlets are unchanged.  This
implies that the difference of distributions for $q$ and $\bar{q}$ does
not mix with gluons in the singlet or the symmetric octet channel but
does mix with gluons coupled to an antisymmetric octet.

In the cases where there is mixing, the relation \eqref{single-ladder-y}
for a double-parton distribution at large $\tvec{k}_1$ becomes a matrix
equation, which can be written as
\begin{align}
  \label{ladder-matrix}
\int d^2\tvec{k}_2\; F^J(x_1,x_2, \tvec{k}_1, \tvec{k}_2, \tvec{y})
 &= \frac{1}{\pi \tvec{k}_1^2}\, \sum_{J'}
    \int_{x_1}^{1-x_2} \frac{du_1}{u_1}\,
    P^{JJ'}\Bigl( \frac{x_1}{u_1} \Bigr)\,
    F^{J'}(u_1,x_2, \tvec{y}) \,,
\end{align}
where the arguments of $F$ and $P$ are as in \eqref{single-ladder-y}.  For
polarizations $\delta q$ and $\delta g$ the distributions and kernels
carry tensor indices, which were discussed in the previous section and
will be omitted here.  The matrix structure in \eqref{ladder-matrix} can
be generalized to the relation \eqref{double-ladder-y} for double-ladder
graphs, but the resulting expressions are rather cluttered with indices
and will not be given here.

In the color singlet sector one has
\begin{align}
\sing{F}^{J} &= 
\begin{pmatrix}
  \label{singlet-matrix}
  \sum_{q} \, [ \sing{F}_{q,a} + \sing{F}_{\bar{q},a} ] \\[0.3em]
  \sing{F}_{g,a} \\[0.3em]
  \sing{F}_{\delta g,a}
\end{pmatrix} \,,
&
\sing{P}^{JJ'} &=
\begin{pmatrix}
C_F P_{qq\phantom{\ms\delta}} & n_F P_{qg}
                              & n_F P_{q\ms \delta g} \\[0.3em]
C_F P_{gq\phantom{\ms\delta}} & N P_{gg\phantom{\ms\delta}}
                      & N P_{g\ms \delta g\phantom{\delta}} \\[0.3em]
C_F P_{\delta g\ms q} & N P_{\delta g\ms g}
                      & N P_{\delta g\ms \delta g}
\end{pmatrix} \,,
\end{align}
where $n_F$ is the number of quark flavors and where the second parton $a$
may be an unpolarized or polarized quark, antiquark or gluon.\footnote{%
  We note that in \protect\cite{Diehl:2011tt} the possibility of
  transitions between $q, g$ and $\delta g$ was overlooked, and only the
  mixing between $q$ and $g$ was considered.}
To obtain the color factors for the off-diagonal elements one must take
into account the prefactors in the definitions \eqref{quark-color-decomp},
\eqref{gluon-color-decomp} and \eqref{qg-color-decomp}.  In the upper left
$2\times 2$ submatrix of $\sing{P}$ we recognize the structure of the
mixing matrix in the usual DGLAP equations.  Mixing in the symmetric octet
sector involves the vectors
\begin{align}
\octS{F}^{J} &= 
\begin{pmatrix}
  \sum_{q} \, [ \oct{F}_{q,a} + \oct{F}_{\bar{q},a} ] \\[0.3em]
  \octS{F}_{g,a} \\[0.3em]
  \octS{F}_{\delta g, a}
\end{pmatrix} \,,
&
\octA{F}^{J} &= 
\begin{pmatrix}
  \sum_{q} \, [ \oct{F}_{q,a} - \oct{F}_{\bar{q},a} ] \\[0.3em]
  \octA{F}_{g,a} \\[0.3em]
  \octA{F}_{\delta g, a}
\end{pmatrix} \,.
\end{align}
If $a$ indicates a gluon, one should replace $\oct{F}_{q,a} +
\oct{F}_{\bar{q},a}$ by $\octS{F}_{q,a} + \octS{F}_{\bar{q},a}$ and
$\oct{F}_{q,a} - \oct{F}_{\bar{q},a}$ by $\octA{F}_{q,a} -
\octA{F}_{\bar{q},a}$.  The splitting matrices now read
\begin{align}
  \label{S-octet-matrix}
\octS{P}^{JJ'} &=
\begin{pmatrix}
-\frac{1}{2N}\ms P_{qq} &
 \sqrt{\frac{N^2-4}{2 (N^2-1)}}\; n_F P_{qg} \quad &
 \sqrt{\frac{N^2-4}{2 (N^2-1)}}\; n_F P_{q\ms \delta g} \\[0.5em]
 \sqrt{\frac{N^2-1}{8}}\ms \sqrt{\frac{N^2-4}{N^2}}\,
    P_{gq\phantom{\ms\delta}} &
 \frac{N}{2} P_{gg\phantom{\ms\delta}} &
 \frac{N}{2} P_{g\ms \delta g\phantom{\delta}} \\[0.5em]
 \sqrt{\frac{N^2-1}{8}}\ms \sqrt{\frac{N^2-4}{N^2}}\, P_{\delta g\ms q} &
 \frac{N}{2} P_{\delta g g} &
 \frac{N}{2} P_{\delta g\ms \delta g}
\end{pmatrix}
\intertext{and}
  \label{A-octet-matrix}
\octA{P}^{JJ'} &=
\begin{pmatrix}
-\frac{1}{2N}\ms P_{qq} &
 \sqrt{\frac{N^2}{2 (N^2-1)}}\; n_F P_{qg} \quad &
 \sqrt{\frac{N^2}{2 (N^2-1)}}\; n_F P_{q\ms \delta g} \\[0.5em]
 \sqrt{\frac{N^2-1}{8}}\,
    P_{gq\phantom{\ms\delta}} &
 \frac{N}{2} P_{gg\phantom{\ms\delta}} &
 \frac{N}{2} P_{g\ms \delta g\phantom{\delta}} \\[0.5em]
 \sqrt{\frac{N^2-1}{8}}\, P_{\delta g\ms q} &
 \frac{N}{2} P_{\delta g g} &
 \frac{N}{2} P_{\delta g\ms \delta g}
\end{pmatrix} \,.
\end{align}
There is no mixing for the combinations $\sum_q \,[ \sing{F}_{q,a} -
  \sing{F}_{\bar{q},q} ]$, nor for the difference of distributions for two
quark flavors, nor for matrix elements where the quark flavors differ on
the two sides of the final-state cut as in figure~\ref{fig:interference}.
In these cases the behavior at large $\tvec{k}_1$ is described by
\eqref{single-ladder-y} with $P$ replaced by $C_F P_{qq}$ for color
singlet and by $- \frac{1}{2N}\ms P_{qq}$ for color octet combinations.

Experience with the usual parton densities tells us that gluons quickly
dominate over sea quarks as one goes to momentum fractions below $0.1$
(except possibly if one considers very low factorization scales).  One
therefore expects that for typical values of $x_1$ and $x_2$ at LHC,
two-parton distributions at high transverse momentum are dominated by
those combinations that can originate from gluons in
\eqref{ladder-matrix}.
 
Comparing \eqref{singlet-matrix} with \eqref{S-octet-matrix} and
\eqref{A-octet-matrix} we find that the color factors are always smaller
in the octet channels than in the singlet channel, with the biggest
suppression occurring for $P_{qq}$.  In the large-$N$ limit the singlet
matrix $\sing{P}$ has one eigenvalue $\frac{N}{2} P_{qq}$ and two
eigenvalues with color factors $N$ for the submatrix in the $g, \delta g$
sector.  Both $\octS{P}$ and $\octA{P}$ have the same two eigenvalues, but
with color factors $\frac{N}{2}$ instead of $N$, and another eigenvalue of
order $1$.  One can hence expect a dominance of color singlet
distributions for sufficiently large transverse momentum, which would
significantly simplify the theoretical description and the phenomenology
of multiple interactions.  How strong the suppression of color octet
channels is in given kinematics should, however, be studied quantitatively
before drawing strong conclusions.

We have also calculated the color factors for higher color representations
of gluon distributions, restricting ourselves to $N=3$ as we did in
\eqref{gluon-decomp-8}.  Mixing with quark distributions is of course not
possible in this case.  We find that the color factors for decuplet and
antidecuplet distributions are zero, so that ladder graphs do not admit
these color combinations, at least not at leading order in $\alpha_s$.
For the $27$ representation we obtain
\begin{align}
{}^{27\!}{F}^{J} &= 
\begin{pmatrix}
  \label{27-matrix}
  {}^{27\!}{F}_{g,a} \\[0.3em]
  {}^{27\!}{F}_{\delta g,a}
\end{pmatrix} \,,
&
{}^{27\!}{P}^{JJ'} &\underset{N=3}{=}\; -
\begin{pmatrix}
P_{gg\phantom{\ms\delta}} & P_{g\ms \delta g\phantom{\delta}} \\[0.3em]
P_{\delta g\ms g}         & P_{\delta g\ms \delta g}
\end{pmatrix} \,,
\end{align}
where $a=g$ or $\delta g$.  The color factor is equal to $-1$ and thus
smaller in magnitude than the factors $N$ or $\half N$ we have in the
singlet and octet sectors, respectively.

The color factors we have obtained agree with those given in
\cite{Bukhvostov:1985rn}, provided that one restores a missing factor
$\sqrt{2/N}$ in the expression of $P_{8f}$ in eq.~(54b) of that
paper.\footnote{The projector $P_0$ in \protect\cite{Bukhvostov:1985rn}
  appears only for $SU(N)$ with $N>3$ \protect\cite{Lipatov:2012pc}.}
Our color factors for transitions in the gluon sector are also in
agreement with eq.~(A.6) in \cite{Bartels:1993ih}.

The splitting matrices for longitudinally polarized quarks and gluons in
\eqref{ladder-matrix} are obtained from the upper left submatrices for
unpolarized quarks and gluons in \eqref{singlet-matrix} to
\eqref{A-octet-matrix} by changing the kernels but keeping the color
factors.  Likewise, the splitting kernel $P_{\delta q\ms \delta q}$ for
transverse quark polarization comes with the same color factors as
$P_{qq}$.  In both cases, it is again the color singlet sector that has
the largest color factors and will therefore be enhanced at high
transverse parton momentum.

Let us finally consider ladder graphs for the quark-antiquark interference
distributions $I$ represented in figure~\ref{fig:distrib}c.  The color
independent part of the splitting kernel is different for distributions
$I$ and $F$ because of their different spin structure (in one case a gluon
is exchanged between a quark and an antiquark line and in the other case
between two quark lines), but we shall not pursue this issue further here.
We can, however, easily determine the color structure of the graphs.  For
definiteness, consider the exchange of a gluon between the two lines with
momentum fraction $x_1$ and color indices $j$, $j'$.  The color
decomposition that remains invariant under this exchange is the one in
\eqref{interf-color-decomp-alt}, since
\begin{align}
t^{a}_{jl}\, t^{a}_{j'l'}\, \bigl( \delta_{lk'}\ms \delta_{l'k}
         + \delta_{lk}\ms \delta_{l'k'} \bigr)
&= t^{a}_{jk'}\, t^{a}_{j'k} + t^{a}_{jk}\, t^{a}_{j'k'}
= \frac{N-1}{2N}\,
    \bigl( \delta_{jk'}\ms \delta_{j'k}
         + \delta_{jk}\ms \delta_{j'k'} \bigr)
\nonumber \\
t^{a}_{jl}\, t^{a}_{j'l'}\, \bigl( \delta_{lk'}\ms \delta_{l'k}
         - \delta_{lk}\ms \delta_{l'k'} \bigr)
&= t^{a}_{jk'}\, t^{a}_{j'k} - t^{a}_{jk}\, t^{a}_{j'k'}
= - \frac{N+1}{2N}\,
    \bigl( \delta_{jk'}\ms \delta_{j'k}
         - \delta_{jk}\ms \delta_{j'k'} \bigr)
\end{align}
For $N=3$ we thus find a color factor $\frac{1}{3}$ if the quarks are
coupled to a sextet and $-\frac{2}{3}$ if they are coupled to an
antitriplet.  Both factors are smaller than $C_F = \frac{4}{3}$ in the
color singlet channel.


\subsection{Parton splitting at high transverse parton momenta}
\label{sec:parton-splitting}

We now turn to another mechanism that generates large transverse momenta
in multiparton distributions: the perturbative splitting of one parton
into two partons, both of which subsequently take part in hard-scattering
processes.  This mechanism turns out to be enhanced by powers of $q_T
/\Lambda$ in the cross section.  In addition, it leads to conceptual
issues concerning the very notion of multiparton interactions.  In the
following sections we derive several results about the splitting
mechanism, but we will be left with a number of open questions for future
research.


\subsubsection{Power behavior}
\label{sec:splitting-power}

\begin{figure}
\begin{center}
\includegraphics[width=0.95\textwidth]{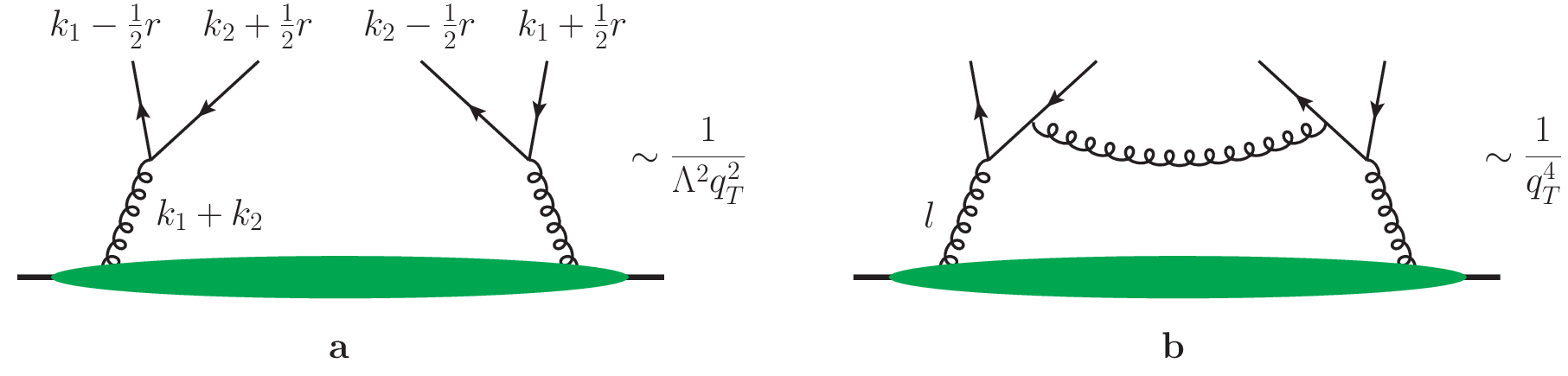}
\end{center}
\caption{\label{fig:2-4} Graphs for a quark-antiquark distribution that
  involve the perturbative splitting of one into two partons on both sides
  of the final-state cut.  The power behavior refers to
  $F(x_i,\tvec{k}_i,\tvec{r})$ and is discussed in the text.}
\end{figure}

There are a multitude of graphs involving the splitting of one parton into
two partons, and in order to assess their importance we use power counting
as our first guiding principle.
Simple examples for parton splitting are shown in figure~\ref{fig:2-4}.
They allow all transverse momenta $\tvec{k}_1$, $\tvec{k}_2$ and
$\tvec{r}$ to be large.  This leads to large virtualities for $k_1$, $k_2$
and $r$, so that their minus components are large as well,
\begin{align}
  \label{2-4-kin}
\tvec{k}_1 \sim \tvec{k}_2 \sim \tvec{r} &\sim q_T \,, &
k_1^- \sim k_2^- \sim r^-                &\sim q_T^2 /p^+ \,. &
\end{align}
At lowest order in $\alpha_s$ one has disconnected graphs as in figure
\ref{fig:2-4}a, which leads to the kinematic constraint
\begin{align}
  \label{2-4a-constraint}
\tvec{k}_1 + \tvec{k}_2 &\sim \Lambda \,, &
k_1^- + k_2^-           &\sim \Lambda^2 /p^+ \,. &
\end{align}
We will see shortly that this constraint plays a special role when
two-parton distributions are combined to calculate a cross section.
The power behavior of graph \ref{fig:2-4}a can be determined using the
same method as in section~\ref{sec:ladders-power}, and we have
\begin{align}
  \label{2-4a}
F(x_i, \tvec{k}_i, \tvec{r})
  \big|_{\text{fig.~\protect\ref{fig:2-4}a}}
 &= p^+ k_1^+ k_2^+ \int dr^-\ms dk_1^-\;
    \big| V'_{1\to 2} \big|^2_{k_2^- = -k_1^-,\,
                               \tvec{k}_2^{} = -\tvec{k}_1^{}}
    \int dk_2^-\, \Phi'_2(k_1^{} + k_2^{})
\nonumber \\
 &\sim \alpha_s \times
    q_T^{4} \times (q_T^{-3})^2 \times \Lambda^2 \times \Lambda^{-4}
  = \alpha_s \times \frac{1}{\Lambda^2\ms q_T^2} \,,
\end{align}
where the power counting of integration volumes is $dr^- \sim dk_1^- \sim
q_T^2 /p^+$ and $dk_2^- \sim \Lambda^2 /p^+$ in order to fulfill the
constraint \eqref{2-4a-constraint}.  Notice that in $V'$ one should set
$k_2^- = -k_1^-$ because the difference of $k_2^-$ and $-k_1^-$ is
negligible compared with $k_1^-$.  Likewise, one should set
$\tvec{k}_2^{}$ equal to $-\tvec{k}_1^{}$ in $V'$.  The factor $\int
dk_2^-\, \Phi'_2(k_1^{} + k_2^{})$ in \eqref{2-4a} is proportional to the
transverse-momentum dependent distribution of a gluon in the proton.

Starting at order $\alpha_s^2$ one has connected graphs as in
figure~\ref{fig:2-4}b.  The restrictions \eqref{2-4a-constraint} are then
lifted, and we obtain a power behavior
\begin{align}
  \label{2-4b}
F(x_i, \tvec{k}_i, \tvec{r})
  \big|_{\text{fig.~\protect\ref{fig:2-4}b}}
 &= p^+ k_1^+ k_2^+ \int dl^+\, dr^-\, dk_1^-\, dk_2^-\;
    V'_{2\to 4}\, \int dl^-\, d^2\tvec{l}\, \Phi'_2
\nonumber \\
 &\sim \alpha_s^2 \times
    q_T^{6} \times q_T^{-10} \times \Lambda^4 \times \Lambda^{-4}
  = \alpha_s^2 \times \frac{1}{q_T^4} \,.
\end{align}

\begin{figure}
\begin{center}
\includegraphics[width=0.67\textwidth]{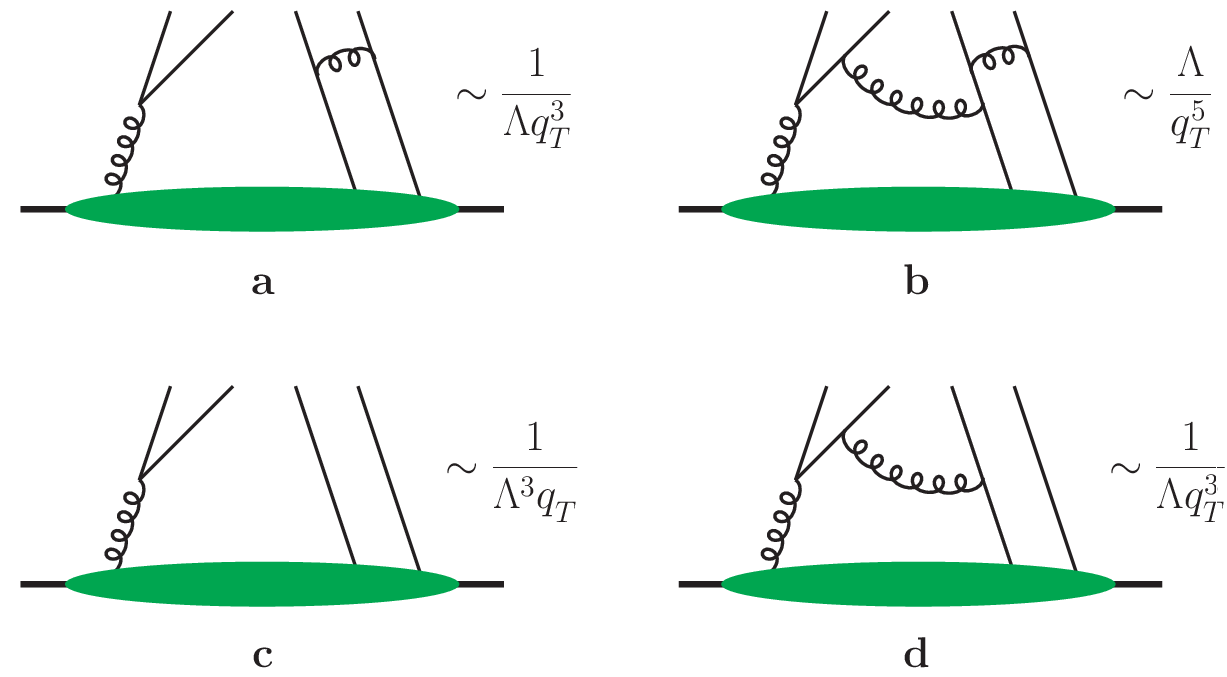}
\end{center}
\caption{\label{fig:3-4} Example graphs for the transition of three to
  four partons in the $t$ channel and the corresponding power behavior of
  $F(x_i, \tvec{k}_i, \tvec{r})$.  They involve the splitting of one into
  two partons only on one side of the final-state cut.}
\vspace{1.5em}
\begin{center}
\includegraphics[width=0.67\textwidth]{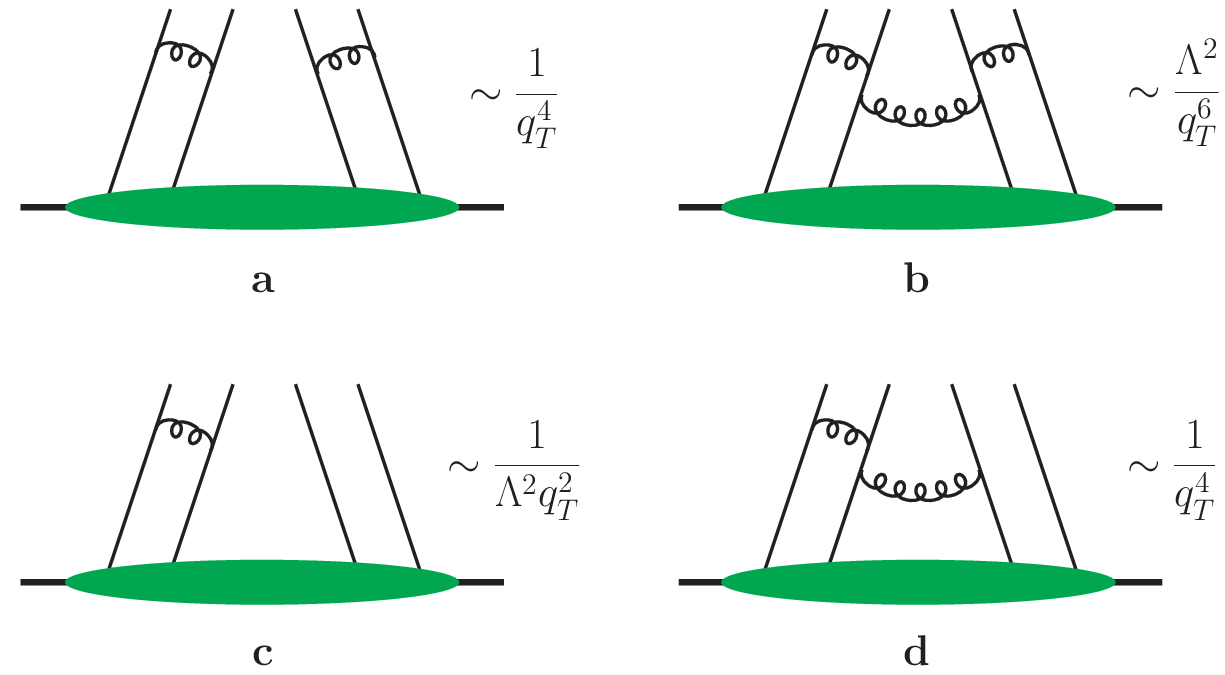}
\end{center}
\caption{\label{fig:4-4} As figure~\protect\ref{fig:3-4}, but for the
  transition of four to four partons in the $t$ channel.}
\end{figure}

\begin{figure}
\begin{center}
\includegraphics[width=0.67\textwidth]{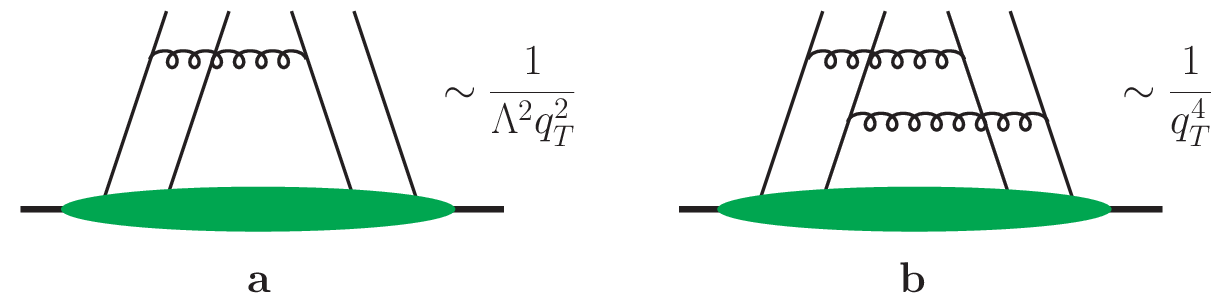}
\end{center}
\caption{\label{fig:4-4-alt} Ladder graphs with the kinematic constraint
  $|\tvec{k}_1 - \tvec{k}_2| \sim \Lambda$ and the resulting power
  behavior of $F(x_i, \tvec{k}_i, \tvec{r})$.}
\end{figure}

The graphs just discussed describe the transition from two to four partons
in the $t$ channel.  Let us compare them with transitions starting from
three or four partons.  The corresponding graphs admit a variety of
topologies, and we shall not give a comprehensive treatment here.
Important examples are shown in figures~\ref{fig:3-4} and \ref{fig:4-4}.
They are subject to different kinematic restrictions:
\begin{itemize}
\item graphs \ref{fig:3-4}a and \ref{fig:4-4}a require $|\tvec{k}_1 +
  \tvec{k}_2| \sim \Lambda$ and are thus analogous to \ref{fig:2-4}a,
\item in analogy to \ref{fig:2-4}b, graphs \ref{fig:3-4}b and
  \ref{fig:4-4}b produce partons with unconstrained transverse momenta,
\item in graphs \ref{fig:3-4}d and \ref{fig:4-4}d one must have
  $|\tvec{k}_1 + \half \tvec{r}| \sim \Lambda$ since the rightmost quark
  line is disconnected,
\item graphs \ref{fig:3-4}c and \ref{fig:4-4}c are subject to both
  constraints $|\tvec{k}_1 + \tvec{k}_2| \sim \Lambda$ and $|\tvec{k}_1 +
  \half \tvec{r}| \sim \Lambda$.
\end{itemize}
The power behavior of the resulting two-parton distributions can be
obtained by the same methods as previously and is given in the figures.
We see that within a given kinematic group, the graph with the smallest
number of partons initiating the hard scattering is dominant by power
counting, i.e.\ two partons in cases a and b, and three partons in cases c
and d.  The graph with the leading power behavior also has the lowest
power of $\alpha_s$.
We note that for graphs starting with four partons there are topologies
leading to different kinematic constraints, such as those in
figure~\ref{fig:4-4-alt}.  We shall not discuss these in the following.

So far we have assumed that the transverse momenta $\tvec{k}_1$,
$\tvec{k}_2$, $\tvec{r}$ are all large.  However, the graphs we have
discussed remain under perturbative control in more restricted kinematics
as well.  In graph \ref{fig:2-4}a for instance, we need large transverse
momenta for the four upper parton lines, i.e.\ large $\tvec{k}_1 \pm \half
\tvec{r}$ (recall that $\tvec{k}_2 \approx -\tvec{k}_1$ for this graph).
This allows either $\tvec{r}$ or $\tvec{k}_1$ to be small, as long as the
other is large.  Both configurations will be important in our further
discussion.  The power behavior we have derived above remains unchanged in
those kinematic regions, as we shall see explicitly in
section~\ref{sec:splitting-dist}.

\paragraph{Cross section.}

\begin{figure}
\begin{center}
\includegraphics[width=0.78\textwidth]{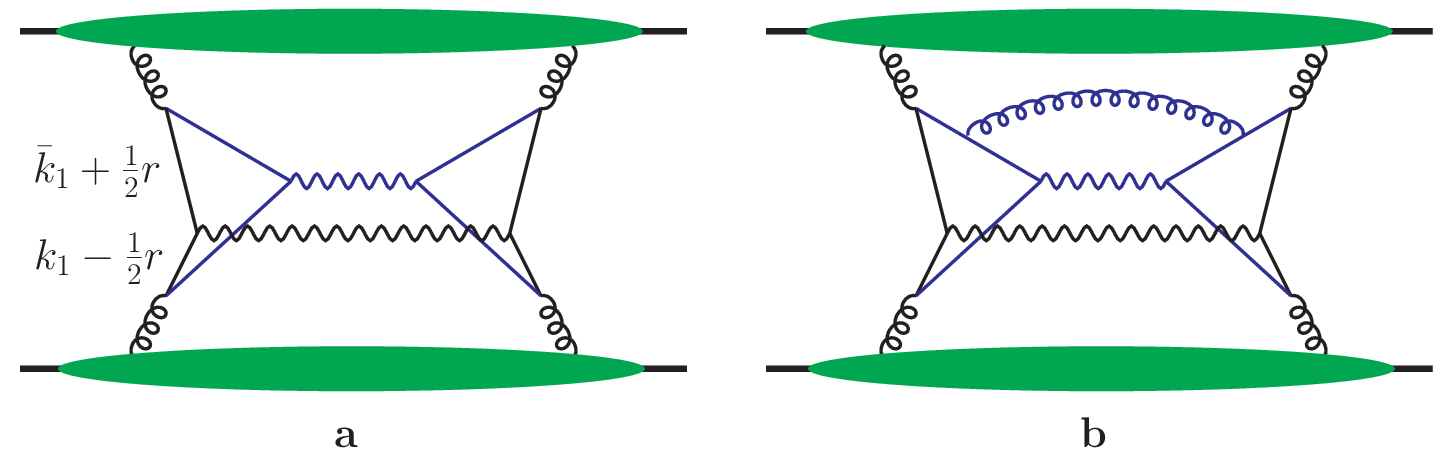}
\end{center}
\caption{\label{fig:2-4-X} Graphs for the cross section with two-to-four
  parton transitions in the $t$ channel for both colliding protons.  Graph
  a only contributes when $|\tvec{q}_1 + \tvec{q}_2| \sim \Lambda$.}
\vspace{1.5em}
\begin{center}
\includegraphics[width=0.73\textwidth]{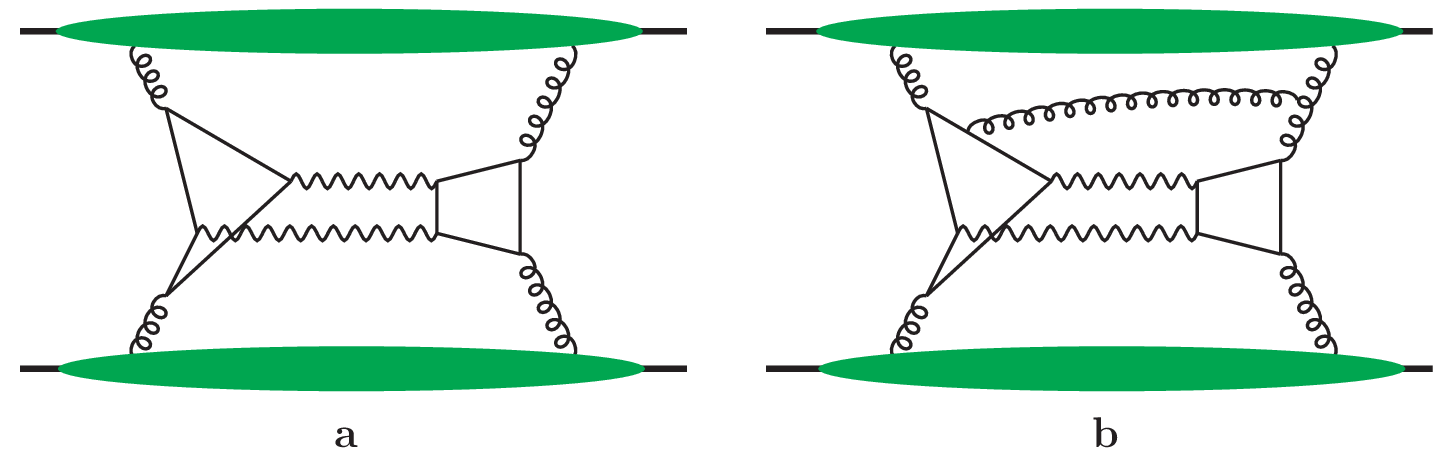}
\end{center}
\caption{\label{fig:2-3-X} Graphs for the cross section with two-to-three
  parton transitions in the $t$ channel for both protons.}
\vspace{1.5em}
\begin{center}
\includegraphics[width=0.54\textwidth]{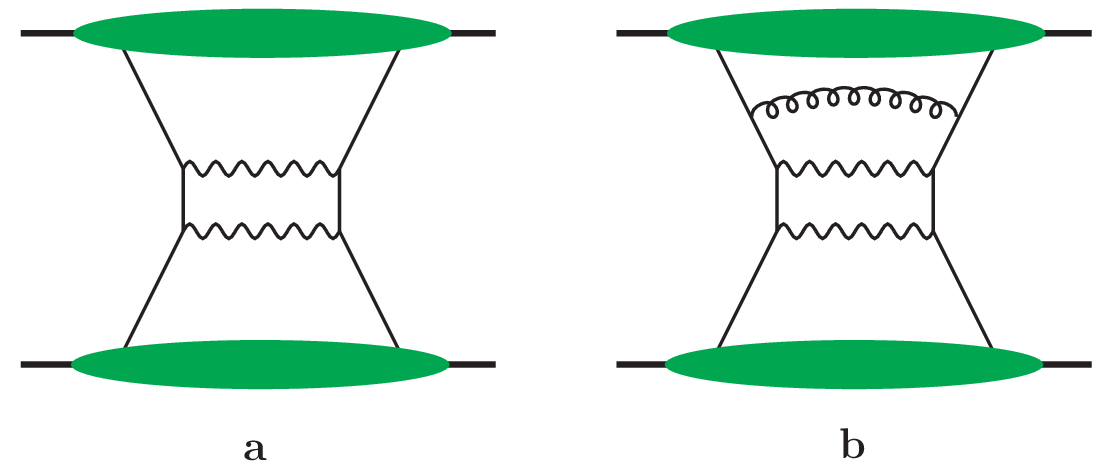}
\end{center}
\caption{\label{fig:simple-ladder-X} Graphs for the cross section with a
  single hard scattering.}
\end{figure}

Let us now see how the different contributions to multiparton
distributions enter the cross section for large $\tvec{q}_1$ and
$\tvec{q}_2$.
Taking the lowest-order parton splitting contribution of
figure~\ref{fig:2-4}a for both protons, we arrive at the graph in
figure~\ref{fig:2-4-X}a.  Both $\tvec{k}_1 + \tvec{k}_2$ and
$\bar{\tvec{k}}_1 + \bar{\tvec{k}}_2$ are restricted to be of order
$\Lambda$ in this case, which implies
\begin{align}
  \label{qt-constraint}
\tvec{q}_1^{} + \tvec{q}_2^{} = 
  \tvec{k}_1^{} + \bar{\tvec{k}}_1^{}
+ \tvec{k}_2^{} + \bar{\tvec{k}}_2^{}
\sim \Lambda \,.
\end{align}
In other words, the produced bosons must be almost back to back in
transverse momentum.  To determine the power behavior of the cross
section, we note that the integration element $d^2\tvec{k}_1\,
d^2\bar{\tvec{k}}_1\; \delta^{(2)}(\tvec{q}{}_1 - \tvec{k}_1 -
\bar{\tvec{k}}_1)$ scales like $q_T^2$ since $\tvec{k}_1$ can be freely
chosen of size $q_T$.  Once this choice is made, $\tvec{k}_2$ can only
differ from $-\tvec{k}_1$ by an amount of order $\Lambda$, so that
$d^2\tvec{k}_2\, d^2\bar{\tvec{k}}_2\; \delta^{(2)}(\tvec{q}{}_2 -
\tvec{k}_2 - \bar{\tvec{k}}_2)$ scales like $\Lambda^2$.  The
corresponding constraint on $\bar{\tvec{k}}_1 + \bar{\tvec{k}}_2$ is then
automatically fulfilled by virtue of \eqref{qt-constraint}.  With
$d^2\tvec{r} \sim q_T^2$ and the scaling behavior \eqref{2-4a} the cross
section formula \eqref{X-sect-again} then gives
\begin{align}
  \label{2-4a-X}
\frac{s^2\ms d\sigma}{\prod_{i=1}^2 dx_i\, d\bar{x}_i\, d^2\tvec{q}{}_i}
  \,\bigg|_{\text{fig.~\protect\ref{fig:2-4-X}a}}^{%
            |\tvec{q}_1^{} + \tvec{q}_2^{}| \,\sim\, \Lambda}
 &\sim \alpha_s^2 \times
       (q_T^2)^2 \times \Lambda^2 \times
       \Bigl( \frac{1}{\Lambda^2\ms q_T^2} \Bigr)^2
  = \alpha_s^2\times \frac{1}{\Lambda^2} \,.
\end{align}
Going one order higher in $\alpha_s$, one has graphs as in
figure~\ref{fig:2-4-X}b with a connected two-to-four parton transition on
one side.  The constraint \eqref{qt-constraint} is then lifted, and
$\tvec{q}_1$ and $\tvec{q}_2$ can be chosen independently.  The
integration elements in the cross section formula scale as in the previous
case, but due to the stronger falloff in $q_T$ in \eqref{2-4b}, one now
has
\begin{align}
  \label{2-4b-X}
\frac{s^2\ms d\sigma}{\prod_{i=1}^2 dx_i\, d\bar{x}_i\, d^2\tvec{q}{}_i}
  \,\bigg|_{\text{fig.~\protect\ref{fig:2-4-X}b}}  
 &\sim \alpha_s^3 \times
       (q_T^2)^2 \times \Lambda^2 \times \frac{1}{q_T^2}
       \times \frac{1}{\Lambda^2\ms q_T^2}
  = \alpha_s^3\times \frac{1}{q_T^2} \,.
\end{align}
At yet higher order in $\alpha_s$ one obtains the same power behavior if
both two-parton distributions contain a connected two-to-four parton
transition: the extra factor $\Lambda^2 /q_T^2$ from the two-parton
distribution is compensated by an increase from $\Lambda^2$ to $q_T^2$ in
the loop phase space, since both $\tvec{k}_1$ and $\tvec{k}_2$ can then be
chosen independently of order $q_T$.

We note that both \eqref{2-4a-X} and \eqref{2-4b-X} contribute at the same
power of $\Lambda^2 /q_T^2$ if one integrates the cross section over
$\tvec{q}_1$ and $\tvec{q}_2$ in a region of size $q_T$.  This is because
the contribution \eqref{2-4a-X} has a restricted phase space of order
$d^2\tvec{q}_1\, d^2\tvec{q}_2 \sim \Lambda^2\ms q_T^2$.  In the
differential cross section, however, the contribution \eqref{2-4a-X} gives
a peak in the distribution of $\tvec{q}_1 + \tvec{q}_2$, which is enhanced
not only by a power of $\alpha_s$ but also by $q_T^2 /\Lambda^2$.

There are more contributions to the cross section with the same power
behavior as the one we have just encountered.  We recall from our
discussion in section~\ref{sec:power-counting} that in the differential
cross section, double parton scattering has the same power behavior as the
interference of two hard scatters in the amplitude with a single hard
scatter in the conjugate amplitude (see graph \ref{fig:power-beh-2}a).  If
the two partons initiating the two hard scatters in the amplitude come
from the splitting of a single parton, we have graphs like in
figure~\ref{fig:2-3-X}.\footnote{The single hard scattering to the right
  of the final-state cut proceeds through a loop in our example, because
  gluons have no direct coupling with electroweak gauge bosons.  Other
  processes, like the production of two dijets, can proceed already at
  tree level.  The powers of $\alpha_s$ in our example are thus not
  representative of the generic case, whereas powers of $\Lambda /q_T$
  are.}
For graph \ref{fig:2-3-X}a one finds the same scaling behavior
\eqref{2-4a-X} as for graph \ref{fig:2-4-X}a, and for graph
\ref{fig:2-3-X}b one finds the same scaling behavior \eqref{2-4b-X} as for
graph \ref{fig:2-4-X}b.

The same power behavior is again found for the case where the two gauge
bosons are produced in a single hard scatter, both in the amplitude and
its conjugate.  The corresponding cross section formula can be found in
\eqref{single-hard-diff}.  For incoming gluons the hard scattering
proceeds through a loop as on the r.h.s.\ of graphs~\ref{fig:2-3-X} a and
b, whereas for incoming quarks or antiquarks one has graphs as those in
figure~\ref{fig:simple-ladder-X}.  If $|\tvec{q}_1 + \tvec{q}_2| \sim
\Lambda$ then one needs no parton exchange of virtuality $q_T$ in any of
the parton densities, and one immediately finds the same power behavior as
in \eqref{2-4a-X}.  If $|\tvec{q}_1 + \tvec{q}_2| \sim q_T$ then at lowest
order in $\alpha_s$ one parton distribution has large transverse momentum
generated by a ladder graph as shown in fig.~\ref{fig:simple-ladder-X}b.
According to \eqref{simple-ladder-high-kt} one has $f(x,\tvec{k}) \sim
\alpha_s /q_T^2$ for $|\tvec{k}| \sim q_T$, and for the cross section one
obtains the same power behavior as in \eqref{2-4b-X}.

Adapting our discussion at the end of section~\ref{sec:ladders-fact} we
see that the graph in figure~\ref{fig:simple-ladder-X}b can be calculated
either in terms of transverse-momentum dependent parton densities, one of
which involves a ladder graph, or in terms of collinear parton
distributions and the parton-level process $q\bar{q}\to V_1\ms V_2 + g$,
where $V_1$ and $V_2$ denote the produced vector bosons.  The result is
the same in both cases.

In a similar way, figures~\ref{fig:2-4-X} and \ref{fig:2-3-X} can be
interpreted as graphs for two-boson production by a single hard scattering
process at one-loop level, namely by $gg\to V_1\ms V_2$ for graphs~a and
$gg\to V_1\ms V_2 + g$ for graphs~b.  The quark lines in each loop are
then typically off-shell by order~$Q$, which is the hard scale set by the
final state.  Note that this differs from the case when one interprets the
same graphs as representing double hard scattering
(figure~\ref{fig:2-4-X}) or the interference of double and single hard
scattering (figure~\ref{fig:2-3-X}).  In that case the quark lines in the
loops (except for those on the r.h.s.\ of graphs~\ref{fig:2-3-X}a and b)
are understood to have typical virtualities of order $q_T$, which allows
one to treat them as incoming on-shell partons in the tree-level
subprocesses $q\bar{q}\to V$, whose large scale is $Q$.  Detailed
inspection of the quark loops in figure~\ref{fig:2-4-X} shows that they
receive contributions with the same scaling behavior in $q_T/Q$ from the
two regions where all quark virtualities are either of order~$q_T$ or of
order $Q$.\footnote{By contrast, there is no kinematic region where all
  quark lines in the loops on the r.h.s.\ of graphs~\ref{fig:2-3-X}a or b
  are off shell by order $q_T$.  This is easily seen by analyzing the flow
  of large plus and minus momenta.}
One thus obtains the same power behavior for the graphs in each of the two
interpretations just discussed.  The interpretation in terms of a single
hard-scattering process producing two gauge bosons for
graphs~\ref{fig:2-4-X}a, \ref{fig:2-3-X}a and \ref{fig:simple-ladder-X}a
and two gauge bosons plus a gluon for graphs~\ref{fig:2-4-X}b,
\ref{fig:2-3-X}b and \ref{fig:simple-ladder-X}b makes it clear that each
group of graphs has the same scaling behavior, respectively given by
\eqref{2-4a-X} and \eqref{2-4b-X}.

In section~\ref{sec:ladders-power} we found that ladder graphs as in
figure~\ref{fig:ladder-Xsect} contribute to the scaled cross section with
a power $\Lambda^2 /q_T^4$, with no distinction between the cases where
$\tvec{q}_1 + \tvec{q}_2$ is of order $\Lambda$ or $q_T$.  This means that
the contributions of figures~\ref{fig:2-4-X}, \ref{fig:2-3-X} and
\ref{fig:simple-ladder-X} are enhanced over the ladder graphs by $q_T^2
/\Lambda^2$ for $|\tvec{q}_1 + \tvec{q}_2| \sim q_T$ and by $q_T^4
/\Lambda^4$ for $|\tvec{q}_1 + \tvec{q}_2| \sim \Lambda$.  As we already
discussed in section~\ref{sec:small-x}, one can however expect that at
small $x_1$ and $x_2$ this enhancement is counteracted by the stronger
rise of the ladder graphs with decreasing parton momentum fractions, since
the ladder contributions involve two-parton distributions, whereas the
graphs in figures~\ref{fig:2-4-X}, \ref{fig:2-3-X} and
\ref{fig:simple-ladder-X} depend on single-parton distributions.  Whether
this small-$x$ enhancement is more important than powers of $q_T^2
/\Lambda^2$ cannot be determined on generic grounds, so that one will want
to keep both types of contribution in a flexible theoretical description.

\begin{figure}
\begin{center}
\includegraphics[width=0.99\textwidth]{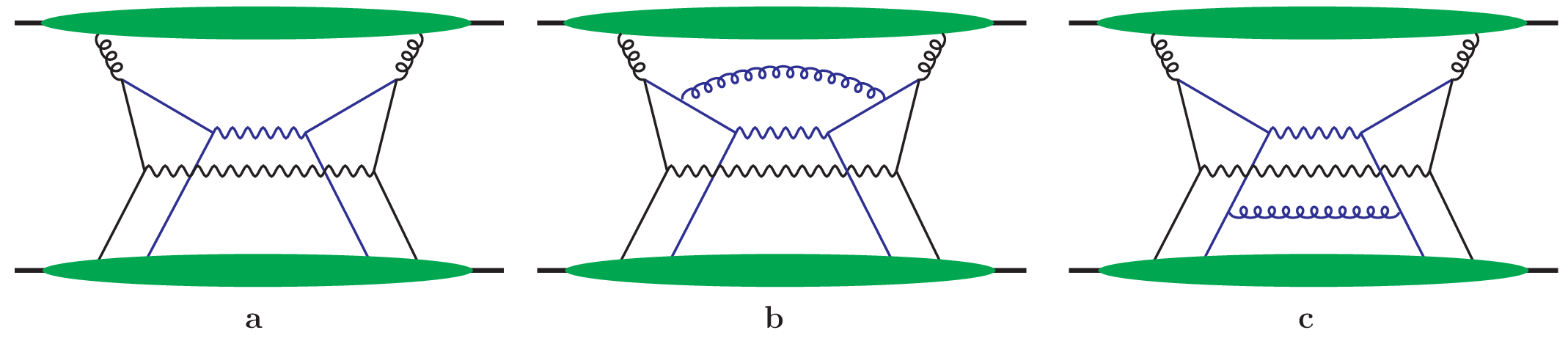}
\end{center}
\caption{\label{fig:high-mixed} Contribution of the perturbative
  transition from two to four quarks in one proton, for the region of
  small $\tvec{r}$.}
\vspace{1.5em}
\begin{center}
\includegraphics[width=0.99\textwidth]{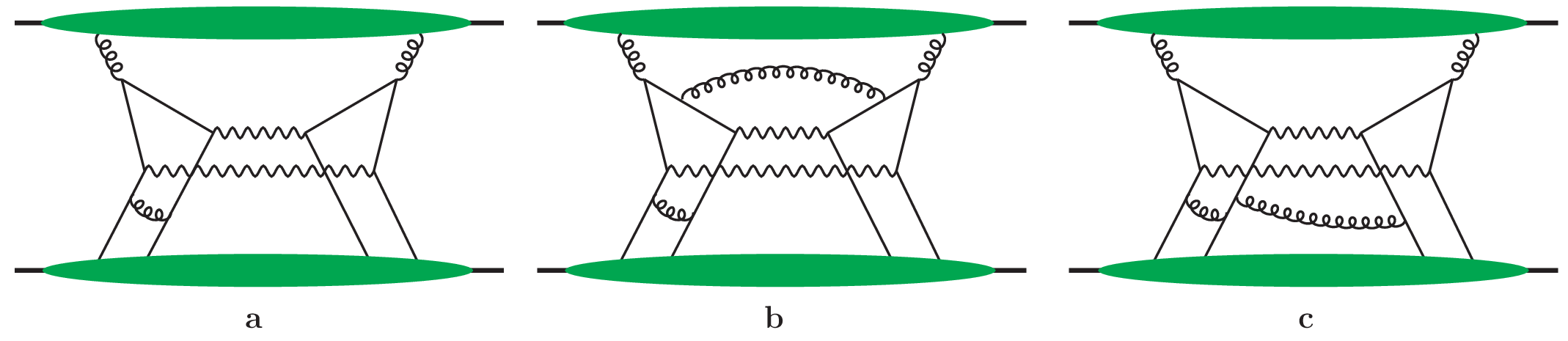}
\end{center}
\caption{\label{fig:high-mix-2} As figure~\protect\ref{fig:high-mixed} but
  for the region of perturbatively large $\tvec{r}$.}
\end{figure}

With this in mind, we now turn our attention to the graphs in
figures~\ref{fig:high-mixed} and \ref{fig:high-mix-2}, which involve
parton splitting and thus single-parton distributions for one proton but a
two-parton distribution for the other.  In the graphs of
figure~\ref{fig:high-mixed} the two-parton distributions in one proton
force $\tvec{r}$ to be of order $\Lambda$, whereas in
figure~\ref{fig:high-mix-2} an additional gluon exchanged between partons
with momentum fraction $x_1$ and $x_2$ allows $\tvec{r}$ to be of order
$q_T$.

The corresponding power behavior is readily obtained from our results for
the relevant parton distributions (given in figures~\ref{fig:ladders},
\ref{fig:2-4} and \ref{fig:4-4}) and the available loop phase space in
each graph.  We find
\begin{align}
  \label{mixed-X}
\frac{s^2\ms d\sigma}{\prod_{i=1}^2 dx_i\, d\bar{x}_i\, d^2\tvec{q}{}_i}
  \,\bigg|_{\text{fig.~\protect\ref{fig:high-mixed}a}}^{%
            |\tvec{q}_1^{} + \tvec{q}_2^{}| \,\sim\, \Lambda}
 &\sim \alpha_s\times \frac{1}{q_T^2} \,,
\nonumber \\[0.2em]
\frac{s^2\ms d\sigma}{\prod_{i=1}^2 dx_i\, d\bar{x}_i\, d^2\tvec{q}{}_i}
  \,\bigg|_{\text{figs.~\protect\ref{fig:high-mixed}b,c}} 
 &\sim \alpha_s^2\times \frac{\Lambda^2}{q_T^4}
\end{align}
and the analogous scaling behavior with an extra power of $\alpha_s$ for
the graphs in figure~\ref{fig:high-mix-2}.  For $|\tvec{q}_1 + \tvec{q}_2|
\sim q_T$, we thus find the same behavior as for the ladder graphs in
figure~\ref{fig:ladder-Xsect}, which involve however two two-parton
distribution in the cross section and therefore have a stronger small-$x$
enhancement.  In the region $|\tvec{q}_1 + \tvec{q}_2| \sim \Lambda$ we
have an extra power of $q_T^2 /\Lambda^2$, as in the other parton
splitting contributions discussed so far.

\begin{figure}
\begin{center}
\includegraphics[width=0.73\textwidth]{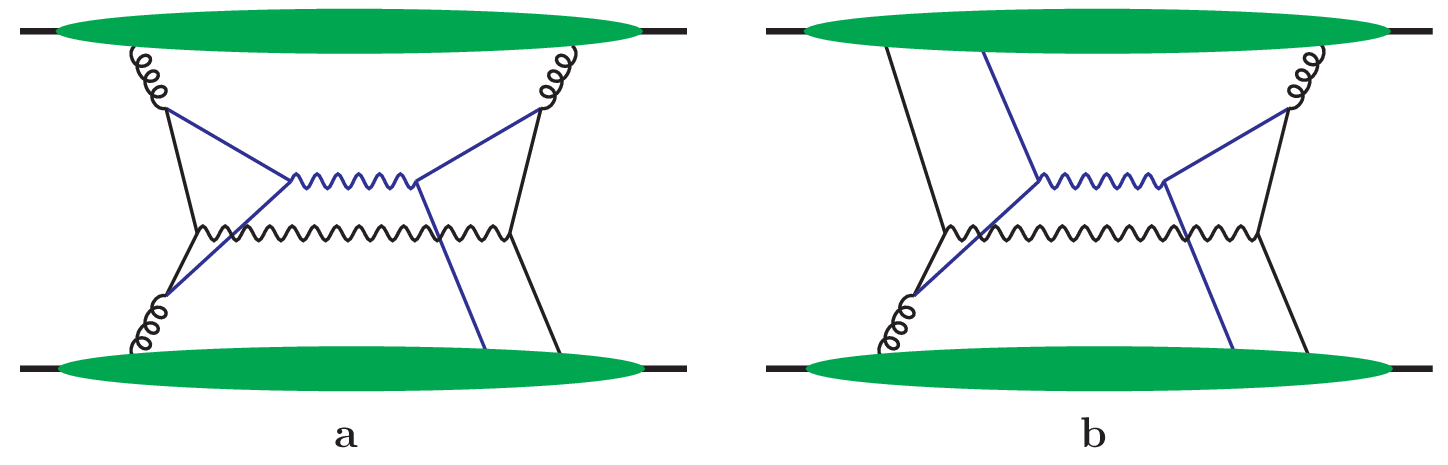}
\end{center}
\caption{\label{fig:3-4-X} Graphs with three partons in the $t$ channel
  for one proton (a) or for both (b).}
\end{figure}

In figure~\ref{fig:3-4} we have graphs initiated by proton matrix elements
with three partons in the $t$ channel.  Examples for their contribution to
the cross sections are given in figure~\ref{fig:3-4-X}.  They behave as
\begin{align}
  \label{3-4-Xsect}
\frac{s^2\ms d\sigma}{\prod_{i=1}^2 dx_i\, d\bar{x}_i\, d^2\tvec{q}{}_i}
  \,\bigg|_{\text{fig.~\protect\ref{fig:3-4-X}a}}^{%
            |\tvec{q}_1^{} + \tvec{q}_2^{}| \,\sim\, \Lambda}
 &\sim \alpha_s^{3/2}\times \frac{1}{\Lambda\ms q_T} \,,
\nonumber \\[0.2em]
\frac{s^2\ms d\sigma}{\prod_{i=1}^2 dx_i\, d\bar{x}_i\, d^2\tvec{q}{}_i}
  \,\bigg|_{\text{fig.~\protect\ref{fig:3-4-X}b}}^{%
            |\tvec{q}_1^{} + \tvec{q}_2^{}| \,\sim\, \Lambda}
 &\sim \alpha_s\times \frac{1}{q_T^2} \,.
\end{align}
The contribution from graph~\ref{fig:3-4-X}a is thus suppressed compared
with the one from graph~\ref{fig:2-4-X}a, although only by a power of
$\Lambda /q_T$, which corresponds to the loss of one power $\Lambda /q_T$
between the splitting graphs initiated by two or three partons in the $t$
channel (cf.\ figures~\ref{fig:2-4}a and \ref{fig:3-4}c).  Likewise,
graph~\ref{fig:3-4-X}b is suppressed by $\Lambda^2 /q_T^2$ compared with
graph~\ref{fig:2-4-X}a, thus having the same power behavior as graph
\ref{fig:high-mixed}a, but lacking the small-$x$ enhancement of the
latter.  An analogous situation is found for graphs that are like in
figure~\ref{fig:3-4-X} but have an extra gluon across the final-state cut
(constructed e.g.\ from graphs~\ref{fig:2-4}b or \ref{fig:3-4}d) and thus
contribute to the region $|\tvec{q}_1^{} + \tvec{q}_2^{}| \sim q_T$.  To
the extent that one power of $\Lambda /q_T$ is a small enough suppression
parameter and that the small-$x$ enhancement of four-parton matrix
elements is important, one can hence neglect contributions involving
three-parton matrix elements.  Our results for the power behavior of the
different contributions are collected in table~\ref{tab:splitting-power}.

\begin{table}
  \renewcommand{\arraystretch}{1.3}
\begin{center}
\begin{tabular}{cccc} \hline
partons & example graphs & \multicolumn{2}{c}{power behavior} \\
($t$ channel) & & $|\tvec{q}_1+\tvec{q}_2| \sim \Lambda$
 & $|\tvec{q}_1+\tvec{q}_2| \sim q_T$ \\
\hline
 $4 \times 4$ & \protect\ref{fig:ladder-Xsect}
  & $\Lambda^2 /q_T^4$ & $\Lambda^2 /q_T^4$ \\
 $4 \times 2$ & \protect\ref{fig:high-mixed}, \protect\ref{fig:high-mix-2}
  & $1/q_T^2$ & $\Lambda^2 /q_T^4$ \\
 $2 \times 2$ & \protect\ref{fig:2-4-X}, \protect\ref{fig:2-3-X},
 \protect\ref{fig:simple-ladder-X} & $1 /\Lambda^2$ & $1/q_T^2$ \\
 $2 \times 3$ & \protect\ref{fig:3-4-X}a
  & $1/(\Lambda\ms q_T)$ & $\Lambda /q_T^3$ \\
 $3 \times 3$ & \protect\ref{fig:3-4-X}b
  & $1/q_T^2$ & $\Lambda^2 /q_T^4$ \\
\hline
\end{tabular}
\end{center}
\caption{\label{tab:splitting-power} Power behavior to the scaled cross
  section $s^2 \ms d\sigma / (dx_1\ms dx_2\ms d\bar{x}_1\ms d\bar{x}_2\ms
  d^2\tvec{q}{}_1\ms d^2\tvec{q}{}_2)$ from various contributions.  An
  entry $m\times n$ in the first column means that the cross section
  involves matrix elements with $m$ and $n$ partons in the  $t$ channel
  for the first and the second proton, respectively.}
\end{table}

The contributions in the first three rows of the table were recently
investigated in \cite{Blok:2011bu}.  It
was pointed out in that work that the $2\times 4$ contribution in our
table has a further enhancement compared with the $4\times 4$ term, which
is due to the fact that the latter involves the product $F(x_i,
\tvec{r})\ms F(\bar{x}_i, -\tvec{r})$ of two distributions that decrease
with $\tvec{r}$, whereas the former involves only one factor $F(x_i,
\tvec{r})$ multiplied by the perturbative splitting contribution that is
approximately $\tvec{r}$ independent for $|\tvec{r}| \sim \Lambda$.

To compare our results with those in \cite{Blok:2011bu} we note that
$|\tvec{q}_1 + \tvec{q}_2| \sim \delta'$ was required to be in the
perturbative domain in that work, whereas we treat it as comparable to a
soft scale.  Our power counting results apply to this case as well as far
as the $q_T$ behavior is concerned, if one understands $\Lambda$ as either
$|\tvec{q}_1 + \tvec{q}_2|$ or a generically soft scale, without the
ability to distinguish between them.  What is important for our results is
the hierarchy $\Lambda \ll q_T \ll Q$, which in the notation of
\cite{Blok:2011bu} reads $\delta' \ll \delta \ll Q$.  The fact that in
\cite{Blok:2011bu} four-jet production rather than the double Drell-Yan
process was studied does not prevent us from a comparison since, as we
pointed out earlier, our power counting results hold independently of the
particular hard-scattering processes.  We agree with \cite{Blok:2011bu}
that the $2\times 4$ and the $4\times 4$ contributions to the cross
section respectively behave like $1/q_T^2$ and $1/q_T^4$, and that the
$2\times 2$ contribution does not have a $1/q_T^2$ behavior but depends
logarithmically on $q_T$.  However, the authors of \cite{Blok:2011bu}
write that the $2\times 2$ term is \emph{comparable} to the $2\times 4$
and $4\times 4$ terms.  We emphasize that the $2\times 2$ contribution
goes like $1/\Lambda^2$ in the scaled cross section and is therefore
\emph{power enhanced} compared with the other two contributions for $q_T
\gg \Lambda$.  This comes out of our power counting analysis and is
confirmed by explicit calculation, see \eqref{splitting-X-scale} below.
What can potentially make the $2\times 4$ and $4\times 4$ terms more
important is their small-$x$ behavior, as we noted above.


\subsubsection{Splitting in two-parton distributions}
\label{sec:splitting-dist}

After the general analysis in section~\ref{sec:splitting-power} we now
investigate splitting contributions to two-parton distributions in more
detail.  We begin with the graph in figure~\ref{fig:2-4}a, which describes
the splitting process $g\to q\bar{q}$.

From the color structure of the graph we readily find that it gives rise
to color octet distributions that are suppressed compared with the color
singlet ones by a factor
\begin{align}
  \label{splitting-color}
\frac{\oct{F}_{a_1,\bar{a}_2}}{\sing{F}_{a_1,\bar{a}_2}}
\,\bigg|_{g\to q\bar{q}}
&= {}- \frac{1}{\sqrt{N^2-1}} \,.
\end{align}
The color singlet distributions are given by
\begin{align}
  \label{splitting-1}
& \sing{F}_{a_1,\bar{a}_2}(x_i, \tvec{k}_i, \tvec{r})
  \,\Big|_{g\to q\bar{q}}
  = \frac{4\pi\alpha_s}{(2\pi)^5}\, \frac{1}{2}\;
    2p^+\! \int dr^-\ms dk_1^-\, dk_2^-\; \Phi^g_{\alpha\beta}(k_1+k_2)
\nonumber \\
 &\qquad \times
\tr \biggl[\ms \Gamma_{a_1}\,
 \frac{({k}_1 - \half {r}) \gamma}{(k_1 - \half r)^2 + i\epsilon}
 \,\gamma^\alpha\,
 \frac{({k}_2 + \half {r}) \gamma}{(k_2 + \half r)^2 + i\epsilon}
\nonumber \\
 &\qquad\quad \; \times
 \,\Gamma_{\bar{a}_2}\,
 \frac{({k}_2 - \half {r}) \gamma}{(k_2 - \half r)^2 - i\epsilon}
 \,\gamma^\beta\,
 \frac{({k}_1 + \half {r}) \gamma}{(k_1 + \half r)^2 - i\epsilon}
\,\biggr]_{k_2^- = - k_1^-,\,
          \tvec{k}_2^{\phantom{-}} = - \tvec{k}_1^{\phantom{-}}} \,,
\end{align}
where $\alpha$ and $\beta$ are polarization indices of the gluon
potentials in the correlation function~$\Phi^g$, whose definition follows
from \eqref{Phi-gluons}.  $\Phi^g$ is already summed over the gluon color
indices, and the corresponding trace over color matrices has given a
factor $\half$.  As discussed in section~\ref{sec:spin:gluons}, $\alpha$
and $\beta$ are restricted to be transverse at leading-power accuracy.
The second and third line of \eqref{splitting-1} represent the hard part
of the process, where we can neglect the difference between the transverse
and minus-components of $k_1$ and $k_2$, see \eqref{2-4a}.  We introduce
\begin{align}
k &= \half (k_1 - k_2) \,,
&
\kappa &= k_1 + k_2
\end{align}
and change integration variables from $k_1^-$ and $k_2^-$ to $k^-$ and
$\kappa^-$.  The integration over $\kappa^-$ only concerns the gluon
correlation function $\Phi^g$, which can be decomposed as
\cite{Mulders:2000sh}
\begin{align}
x p^+ \!\! \int d\kappa^- 
   \Phi^{g, jj'}(\kappa) \bigg|_{\kappa^+ = x p^+}
&= \frac{1}{2}\ms \delta^{jj'} f_1^g(x, \tvec{\kappa})
 + \frac{2 \tvec{\kappa}^j \tvec{\kappa}^{j'}
         - \delta^{jj'}\! \tvec{\kappa}^2}{4 M^2}\,
   h_1^{\perp g}(x, \tvec{\kappa}) \,,
\end{align}
where $M$ denotes the proton mass.  In terms of the operators introduced
in \eqref{gluon-bilinears} we have
\begin{align}
f_1^g(x, \tvec{\kappa}) &= \frac{1}{x p^+}
   \int \frac{dz^- d^2\tvec{z}}{(2\pi)^3}\,
     e^{i x z^- p^+ -i \tvec{z} \tvec{\kappa}}\,
   \big\langle p \big|\, \mathcal{O}_{g}(0, z)  \big| p \big\rangle \,,
\nonumber \\
\frac{2 \tvec{\kappa}^j \tvec{\kappa}^{j'}
      - \delta^{jj'}\! \tvec{\kappa}^2}{4 M^2}\,
h_1^{\perp g}(x, \tvec{\kappa})
 &= \frac{1}{x p^+} \int \frac{dz^- d^2\tvec{z}}{(2\pi)^3}\,
   e^{i x z^- p^+ -i \tvec{z} \tvec{\kappa}}\,
   \big\langle p \big|\, \mathcal{O}_{\delta g}^{jj'}(0, z)
   \big| p \big\rangle \,.
\end{align}
$f_1^g$ is the usual transverse-momentum dependent density of gluons,
whereas the gluon Boer-Mulders function $h_1^{\perp g}$ describes linearly
polarized gluons and is essentially unknown at present (see
\cite{Boer:2010zf,Qiu:2011ai,Boer:2011kf} for processes where this
distribution could be studied).
Writing the product of propagator denominators in \eqref{splitting-1} as
\begin{align}
& \frac{1}{2x_1\ms p^+ \bigl(k - \half r\bigr)^-
            - \bigl(\tvec{k} - \half \tvec{r}\bigr)^2 + i\epsilon}\,
  \frac{1}{2x_2\ms p^+ \bigl(k - \half r\bigr)^-
            + \bigl(\tvec{k} - \half \tvec{r}\bigr)^2 - i\epsilon}
\nonumber \\
\times \; &
  \frac{1}{2x_2\ms p^+ \bigl(k + \half r\bigr)^-
            + \bigl(\tvec{k} + \half \tvec{r}\bigr)^2 + i\epsilon}\,
  \frac{1}{2x_1\ms p^+ \bigl(k + \half r\bigr)^-
            - \bigl(\tvec{k} + \half \tvec{r}\bigr)^2 - i\epsilon}
\end{align}
we see that the integrations over $r^-$ and $k^-$ can conveniently be
performed using the theorem of residues, after a change of variables to
$(k - \half r)^-$ and $(k + \half r)^-$.  Performing the fermion trace, we
finally obtain
\begin{align}
  \label{splitting-result}
& \sing{F}_{a_1,\bar{a}_2}(x_i, \tvec{k}_i, \tvec{r})
   \,\Big|_{g\to q\bar{q}}
= \frac{\alpha_s}{4\pi^2}\, \Biggl[\ms
  \frac{f_1^g(x_1+x_2, \tvec{\kappa})}{x_1+x_2}\;
    T_{a_1,\bar{a}_2}^{l\ms l'}\biggl( \frac{x_1}{x_1+x_2} \biggr)
  + \frac{2 \tvec{\kappa}^m \tvec{\kappa}^{m'}
        - \delta^{mm'} \tvec{\kappa}^2}{2 M^2}\,
\nonumber \\[0.2em]
&\quad \times
   \frac{h_1^{\perp g}(x_1+x_2, \tvec{\kappa})}{x_1+x_2}\;
    U_{a_1,\bar{a}_2}^{l\ms l'mm'}\biggl( \frac{x_1}{x_1+x_2} \biggr)
\ms\Biggr]\,
  \frac{\bigl( \tvec{k} + \half \tvec{r} \bigr)^{l}
        \bigl( \tvec{k} - \half \tvec{r} \bigr)^{l'}}{%
           \bigl( \tvec{k} + \half \tvec{r} \bigr)^2
           \bigl( \tvec{k} - \half \tvec{r} \bigr)^2} \,.
\end{align}
With the abbreviation $\bar{u} = 1 - u$ the kernels read
\begin{align}
  \label{Tkernels}
T_{q,\bar{q}}^{l\ms l'}(u)
 &= - T_{\Delta q, \Delta\bar{q}}^{l\ms l'}(u)
  = \delta^{l\ms l'}\ms (u^2 + \bar{u}^2) \,,
\nonumber \\[0.2em]
T_{\Delta q, \bar{q}}^{l\ms l'}(u)
 &= - T_{q, \Delta\bar{q}}^{l\ms l'}(u) \hspace{1.2ex}
  = i\epsilon^{l\ms l'}\ms (u - \bar{u}) \,,
\nonumber \\[0.2em]
\bigl[\ms T_{\delta q, \delta\bar{q}}^{l\ms l'}(u) \ms\bigr]{}^{jj'}
 &= - 2\ms \delta^{l\ms l'}\ms \delta^{jj'}_{} u \bar{u}
\end{align}
and
\begin{align}
  \label{Ukernels}
U_{q,\bar{q}}^{l\ms l'mm'}(u)
 &= - U_{\Delta q, \Delta\bar{q}}^{l\ms l'mm'}(u)
  = - 2 \tau^{l\ms l',mm'}\ms u \bar{u} \,,
\nonumber \\[0.2em]
\bigl[\ms U_{\delta q,\delta \bar{q}}^{l\ms l'mm'}(u) \ms\bigr]{}^{jj'}
 &= 2 \tau^{l\ms l',j'm'}\ms \delta^{jm}\ms u
    + 2 \tau^{l\ms l',jm'}\ms \delta^{j'm}\ms \bar{u}
    - 2 \tau^{l\ms l',mm'}\ms \delta^{jj'} u \bar{u} \,,
\end{align}
where $j$ and $j'$ are the indices of the Dirac matrices
$i\sigma^{+j}\gamma_5$ in the definition of the distributions.  The
kernels $T$ and $U$ not listed in \eqref{Tkernels} or \eqref{Ukernels} are
zero.  Note that $\tvec{k} - \half\tvec{r}$ is half the difference between
the transverse parton momenta $\tvec{k}_1 - \half\tvec{r}$ and $\tvec{k}_2
+ \half\tvec{r}$ on the left of figure~\ref{fig:2-4}a, whereas $\tvec{k} +
\half\tvec{r}$ is half the corresponding momentum difference on the right.
To ensure that the quark lines with momenta $k_1 \pm \half r$ and $k_2 \pm
\half r$ in figure~\ref{fig:2-4}a are far off-shell it is sufficient that
\emph{one} of the transverse momenta $\tvec{r}$ and $\tvec{k}$ is large,
as already mentioned earlier.  This implies that \eqref{splitting-result}
describes the large $\tvec{r}$ behavior of $F_{a_1,\bar{a}_2}(x_i,
\tvec{k}_i, \tvec{r})$ for small $\tvec{k}_1$ and $\tvec{k}_2$, as well as
its behavior for large $\tvec{k}$ at small $\tvec{r}$.

We see that the short-distance splitting process gives rise to a rich spin
structure, with all chiral even two-parton distributions being nonzero.
The relations $F_{q,\bar{q}} = - F_{\Delta q, \Delta\bar{q}}$ and
$F_{\Delta q, \bar{q}} = - F_{q, \Delta\bar{q}}$ reflect that the
perturbative gluon splitting leads to a 100\% correlation between the
helicities of the quark and antiquark: if the quark has positive helicity
the antiquark has negative one, and vice versa.  For values of $u$ around
$\half$, the transverse spin correlation encoded in $F_{\delta q,
  \delta\bar{q}}$ is as large as the unpolarized distribution $F_{q,q}$.

The splitting contributions to other two-parton distributions are obtained
in close analogy to the case we have just discussed, and in the following
we only give the relevant starting expressions and results.  A reader not
interested in the details may skip forward to the paragraph after
equation~\eqref{splitting-g-gg-higher}.

The graph in figure~\ref{fig:2-4}a also contributes to the interference
distributions $I_{a_1,\bar{a}_2}$, with the same ratio
$\oct{I}_{a_1,\bar{a}_2} \big/\, \sing{I}_{a_1,\bar{a}_2} = -1 \big/
\sqrt{N^2 -1}$ of octet and singlet distributions as in
\eqref{splitting-color}.  The expression for $\sing{I}_{a_1,\bar{a}_2}$
can be obtained from the one in \eqref{splitting-1} by interchanging
$({k}_2 - \half {r}) \gamma$ and $({k}_1 + \half {r}) \gamma$ in the
fermion trace.  The result has the same structure as in
\eqref{splitting-result}, with the kernels $T_{a_1, \bar{a}_2}$ replaced
by
\begin{align}
  \label{Vkernels}
V_{q,\bar{q}}^{l\ms l'}(u)
 &= - V_{\Delta q, \Delta\bar{q}}^{l\ms l'}(u)
  = - 2\ms \delta^{l\ms l'}\ms u \bar{u} \,,
\nonumber \\[0.2em]
\bigl[\ms V_{\delta q, \delta\bar{q}}^{l\ms l'}(u) \ms\bigr]{}^{jj'}
 &= \delta^{l\ms l'} \delta^{jj'}_{}\ms (u^2 + \bar{u}^2)
   + (\delta^{jl} \delta^{j'l'} - \delta^{jl'} \delta^{j'l})\ms
      (u - \bar{u})
\end{align}
and the kernels $U_{a_1, \bar{a}_2}$ replaced by
\begin{align}
  \label{Wkernels}
W_{q,\bar{q}}^{l\ms l'mm'}(u)
 &= - W_{\Delta q, \Delta\bar{q}}^{l\ms l'mm'}(u)
  = \tau^{l\ms l',mm'}\ms (u^2 + \bar{u}^2) \,,
\nonumber \\[0.2em]
W_{\Delta q,\bar{q}}^{l\ms l'mm'}(u)
 &= - W_{q, \Delta\bar{q}}^{l\ms l'mm'}(u)
  = \tau^{l\ms l',mn}\, i\epsilon^{m'n}\ms (u - \bar{u}) \,,
\nonumber \\[0.2em]
\bigl[\ms W_{\delta q,\delta \bar{q}}^{l\ms l'mm'}(u) \ms\bigr]{}^{jj'}
 &= - \bigl( \delta^{jl'} \delta^{j'm} \delta^{lm'} 
           + \delta^{j'l} \delta^{jm} \delta^{l'm'} \bigr)\ms u
    - \bigl( \delta^{jl} \delta^{j'm} \delta^{l'm'}
           + \delta^{j'l'} \delta^{jm} \delta^{lm'} \bigr)\ms \bar{u}
\nonumber \\
 &\quad
     + \tau^{jj',mm'} \delta^{l\ms l'}
     + \tau^{ll',mm'} \delta^{jj'}\ms (u^2 + \bar{u}^2) \,.
\end{align}
All other kernels are zero.  We see that the splitting contribution to the
interference distributions $I_{a_1, \bar{a}_2}$ is generically of the same
size as for the distributions $F_{a_1, \bar{a}_2}$.

We now turn to the analog of figure~\ref{fig:2-4}a for the splitting
process $q\to g q$.  This graph (not shown here for brevity) involves
propagators for the outgoing gluons and requires a choice of gauge.  If we
work in the light-cone gauge $A n = A^+ = 0$ with $n = (1,0,0,-1)
/\sqrt{2}$, then the gluon propagator has a numerator
\begin{align}
D^{\alpha\beta}(\ell) &= - g^{\alpha\beta}
  + \frac{n^\alpha \ell^\beta + \ell^\alpha n^\beta}{\ell^+}
\end{align}
and the $q\to g q$ splitting contribution to quark-gluon distributions
reads
\begin{align}
  \label{splitting-q-gq-1}
& \sing{F}_{a_1,a_2}(x_i, \tvec{k}_i, \tvec{r}) \,\Big|_{q\to g q}
 = \frac{4\pi\alpha_s}{(2\pi)^5}\; C_F\,
    (x_1\ms p^+)\, 2p^+\! \int dr^-\ms dk_1^-\, dk_2^-
\nonumber \\
 &\quad \times
\frac{D_{\alpha j}(k_1 - \half r)}{(k_1 - \half r)^2 + i\epsilon}\
\Pi_{a_1}^{jj'}\;
\frac{D_{j' \beta}(k_1 + \half r)}{(k_1 + \half r)^2 - i\epsilon}\,
\nonumber \\
&\quad \times
\tr \biggl[\ms \gamma^\beta\,
  \frac{({k}_2 - \half {r}) \gamma}{(k_2 - \half r)^2 - i\epsilon}
\,\Gamma_{\bar{a}_2}\,
  \frac{({k}_2 + \half {r}) \gamma}{(k_2 + \half r)^2 + i\epsilon}
\,\gamma^\alpha\,
\Phi^q(k_1 + k_2)
\,\biggr]_{k_2^- = - k_1^-,\,
          \tvec{k}_2^{\phantom{-}} = - \tvec{k}_1^{\phantom{-}}} \,.
\end{align}
Since $j$ is a transverse index, the numerator factor of the first gluon
propagator simplifies to $- g_{\alpha j} + n_{\alpha} (k_1 - \half r)_{j}
/(k_1 - \half r)^+$.  If we work in covariant gauge instead, these two
terms correspond to the first two terms of the gluon field strength
$G^{+j} = \partial^+ A^j - \partial^j A^+ + \mathcal{O}(g)$ in the
operator definition of the quark-gluon distribution.  An analogous
statement holds for the second gluon propagator.
The expression \eqref{splitting-q-gq-1} involves the quark correlation
function $\Phi^q$ for an unpolarized proton, for which one has
\begin{align}
\int d\kappa^-\, \Phi^q(\kappa) \bigg|_{\kappa^+ = x p^+}
 &= \frac{1}{2}\ms \gamma^- f_1^q(x,\tvec{\kappa})
    + \frac{1}{2}\ms i\sigma^{j -}\gamma_5\,
      \frac{\epsilon^{jj'} \tvec{\kappa}^{j'}}{M}\,
      h_1^{\perp q}(x,\tvec{\kappa})
\end{align}
to leading-twist accuracy, or equivalently
\begin{align}
f_1^q(x, \tvec{\kappa}) &=
   \int \frac{dz^- d^2\tvec{z}}{(2\pi)^3}\,
     e^{i x z^- p^+ -i \tvec{z} \tvec{\kappa}}\,
   \big\langle p \big|\, \mathcal{O}_{q}(0, z)  \big| p \big\rangle \,,
\nonumber \\
\frac{\epsilon^{jj'} \tvec{\kappa}^{j'}}{M}\,
h_1^{\perp q}(x, \tvec{\kappa})
 &= \int \frac{dz^- d^2\tvec{z}}{(2\pi)^3}\,
   e^{i x z^- p^+ -i \tvec{z} \tvec{\kappa}}\,
   \big\langle p \big|\, \mathcal{O}_{\delta q}^{j}(0, z)
   \big| p \big\rangle \,.
\end{align}
The $q\to g q$ splitting process gives rise to all possible color
couplings in the quark-gluon distribution in \eqref{qg-color-decomp}, with
color factors
\begin{align}
  \label{splitting-q-gq-color}
\frac{\octS{F}_{a_1,a_2}}{\sing{F}_{a_1,a_2}} \,\bigg|_{q\to g q}
&= \sqrt{\frac{N^2-4}{2}} \,,
&
\frac{\octA{F}_{a_1,a_2}}{\sing{F}_{a_1,a_2}} \,\bigg|_{q\to g q}
&= {}- \frac{N}{\sqrt{2}} \,.
\end{align}
Contrary to the case of $g\to q\bar{q}$ analyzed above, the splitting
mechanism now favors color octet distributions over color singlet ones.
Evaluating \eqref{splitting-q-gq-1} we obtain
\begin{align}
  \label{splitting-q-gq-2}
& \sing{F}_{a_1,a_2}(x_i, \tvec{k}_i, \tvec{r}) \,\Big|_{q\to g q}
= \frac{\alpha_s}{2\pi^2}\, C_F\, \Biggl[\ms
  \frac{f_1^q(x_1+x_2, \tvec{\kappa})}{x_1+x_2}\;
    T_{a_1,a_2}^{l\ms l'}\biggl( \frac{x_1}{x_1+x_2} \biggr)
\nonumber \\[0.2em]
&\quad
 + \frac{\epsilon^{mm'} \tvec{\kappa}^{m'}}{M}\,
   \frac{h_1^{\perp q}(x_1+x_2, \tvec{\kappa})}{x_1+x_2}\;
     U_{a_1,a_2}^{l\ms l'm}\biggl( \frac{x_1}{x_1+x_2} \biggr)
\ms\Biggr]\,
 \frac{\bigl( \tvec{k} + \half \tvec{r} \bigr)^{l}
        \bigl( \tvec{k} - \half \tvec{r} \bigr)^{l'}}{%
           \bigl( \tvec{k} + \half \tvec{r} \bigr)^2
           \bigl( \tvec{k} - \half \tvec{r} \bigr)^2}
\end{align}
with
\begin{align}
  \label{Tkernels-q-gq}
T_{g,q}^{l\ms l'}(u) &= \delta^{l\ms l'}\ms (1+\bar{u}^2) /u \,,
&
T^{l\ms l'}_{\Delta g, \Delta q}(u) &= \delta^{l\ms l'}\, (1+\bar{u}) \,,
\nonumber \\ 
T^{l\ms l'}_{\Delta g, q}(u) &= -i \epsilon^{l\ms l'}\, (1+\bar{u}^2) /u \,,
&
T^{l\ms l'}_{g, \Delta q}(u) &= -i \epsilon^{l\ms l'}\, (1+\bar{u}) \,,
\nonumber \\
\bigl[\ms T^{l\ms l'}_{\delta g, q}(u) \ms\bigr]{}^{jj'}
 &= 2 \tau^{l\ms l',\ms jj'} \bar{u}/u
\end{align}
and
\begin{align}
\bigl[\ms U_{g, \delta q}^{l\ms l'm}(u) \ms\bigr]{}^{k}
  &= 2\ms \delta^{l\ms l'} \delta^{km}\, \bar{u} /u \,,
\qquad\qquad\qquad
\bigl[\ms U_{\Delta g, \delta q}^{l\ms l'm}(u) \ms\bigr]{}^{k}
   = - 2\ms i \epsilon^{l\ms l'} \delta^{km}\, \bar{u} /u \,,
\nonumber \\
\bigl[\ms U_{\delta g, \delta q}^{l\ms l'm}(u) \ms\bigr]{}^{jj',k}
  &= \tau^{jj',ml} \delta^{kl'} + \tau^{jj',ml'} \delta^{kl}
   - (\tau^{jj',kl} \delta^{ml'} + \tau^{jj',kl'} \delta^{ml}) \ms \bar{u}
   - \tau^{jj',km} \delta^{l\ms l'}\ms u
\nonumber \\
  &\quad
   + 2 \tau^{jj',l\ms l'} \delta^{km}\ms \bar{u} /u \,.
\end{align}
All other kernels are zero, in particular $F_{\delta g, \Delta q}$ is not
generated by the splitting mechanism at leading order in $\alpha_s$.
Analogous results can be derived for the splitting $\bar{q} \to g
\bar{q}$.

The splitting of one gluon into two gives a contribution to two-gluon
distributions, which reads
\begin{align}
  \label{splitting-g-gg-1}
& \sing{F}_{a_1,a_2}(x_i, \tvec{k}_i, \tvec{r}) \,\Bigl|_{g\to gg}
  = \frac{4\pi\alpha_s}{(2\pi)^5}\, N\,
    (x_1\ms p^+)\, (x_2\ms p^+)\, 2p^+\! \int dr^-\ms dk_1^-\, dk_2^-\;
    \Phi^g_{\alpha\beta}(k_1 + k_2)
  \,\Big|_{\text{fig.~\protect\ref{fig:2-4}a}}
\nonumber \\
 &\quad \times \biggl[ \ms
\frac{D_{\mu j}(k_1 - \half r)}{(k_1 - \half r)^2 + i\epsilon}\;
\Pi_{a_1}^{jj'}\;
\frac{D_{j' \mu'}(k_1 + \half r)}{(k_1 + \half r)^2 - i\epsilon}\;
\frac{D_{\nu k}(k_2 + \half r)}{(k_2 + \half r)^2 + i\epsilon}\;
\Pi_{a_2}^{kk'}\;
\frac{D_{k' \nu'}(k_2 - \half r)}{(k_2 - \half r)^2 - i\epsilon}\,
\nonumber \\
&\quad\; \times
  \Bigl( g^{\mu'\nu'} (k_1 - k_2 + r)^\beta
       - g^{\beta\mu'} (2 k_1 + k_2 + \half r)^{\nu'}
       + g^{\beta\nu'} (k_1 + 2 k_2 - \half r)^{\mu'} \Bigr)
\phantom{\biggl[ \biggr]}
\nonumber \\
&\quad\; \times
  \Bigl( g^{\mu\nu} (k_1 - k_2 - r)^\alpha 
       - g^{\alpha\mu} (2 k_1 + k_2 - \half r)^\nu
       + g^{\alpha\nu} (k_1 + 2 k_2 + \half r)^\mu \Bigr)
\ms\biggr]_{\substack{k_2^- = - k_1^- \\[0.1ex]
            \tvec{k}_2 = - \tvec{k}_1}}
\end{align}
in the gauge $A^+ = 0$.  Evaluating this expression, we obtain a result
with the same structure as for $g\to q\bar{q}$ in
\eqref{splitting-result},
\begin{align}
  \label{splitting-g-gg-2}
& \sing{F}_{a_1,a_2}(x_i, \tvec{k}_i, \tvec{r}) \,\Big|_{g\to gg}
= \frac{\alpha_s}{2\pi^2}\, N\, \Biggl[\ms
  \frac{f_1^g(x_1+x_2, \tvec{\kappa})}{x_1+x_2}\;
    T_{a_1,a_2}^{l\ms l'}\biggl( \frac{x_1}{x_1+x_2} \biggr)
 + \frac{2 \tvec{\kappa}^m \tvec{\kappa}^{m'}
      - \delta^{mm'} \tvec{\kappa}^2}{2 M^2}\,
\nonumber \\[0.2em]
&\quad \times
  \frac{h_1^{\perp g}(x_1+x_2, \tvec{\kappa}) }{x_1+x_2}\;
    U_{a_1,a_2}^{l\ms l'mm'}\biggl( \frac{x_1}{x_1+x_2} \biggr)
\ms\Biggr]\,
  \frac{\bigl( \tvec{k} + \half \tvec{r} \bigr)^{l}
        \bigl( \tvec{k} - \half \tvec{r} \bigr)^{l'}}{%
           \bigl( \tvec{k} + \half \tvec{r} \bigr)^2
           \bigl( \tvec{k} - \half \tvec{r} \bigr)^2}
\end{align}
with
\begin{align}
  \label{Tkernels-g-gg}
T_{g,g}^{l\ms l'}(u) &= 2 \delta^{l\ms l'}\ms
     (u/\bar{u} + \bar{u}/u + u \bar{u}) \,,
&
T_{\Delta g, \Delta g}^{l\ms l'}(u)
   &= 2 \delta^{l\ms l'}\ms (2 - u \bar{u}) \,,
\nonumber \\
T_{g, \Delta g}^{l\ms l'}(u)
   &= -2 i\epsilon^{l\ms l'}\ms (2\bar{u} + u /\bar{u}) \,,
&
\bigl[\ms T_{g, \delta g}^{l\ms l'}(u) \ms\bigr]{}^{kk'}
   &= 2 \tau^{l\ms l',kk'}\, u /\bar{u} \,,
\nonumber \\
\bigl[\ms T_{\delta g, \delta g}^{l\ms l'}(u) \ms\bigr]{}^{jj',kk'}
   &= \delta^{l\ms l'}\ms \tau^{jj',kk'} u \bar{u}
\end{align}
and
\begin{align}
U_{g,g}^{l\ms l'mm'}(u) &= - U_{\Delta g,\Delta g}^{l\ms l'mm'}(u)
    = 2 \tau^{ll',mm'} u \bar{u} \,,
\nonumber \\
\bigl[\ms U_{g, \delta g}^{l\ms l'mm'}(u) \ms\bigr]{}^{kk'}
   &= \delta^{l\ms l'} \tau^{kk',mm'}\, \bar{u}/u \,,
\qquad\qquad\qquad
\bigl[\ms U_{\Delta g, \delta g}^{l\ms l'mm'}(u) \ms\bigr]{}^{kk'}
    = - i\epsilon^{ll'} \tau^{kk',mm'}\, \bar{u} /u \,,
\nonumber \\[0.1em]
\bigl[\ms U_{\delta g, \delta g}^{l\ms l'mm'}(u) \ms\bigr]{}^{jj',kk'}
   &= \tau^{l\ms l',kk'}\, \tau^{mm',jj'}\, u /\bar{u}
    + \tau^{l\ms l',jj'}\, \tau^{mm',kk'}\, \bar{u} /u
    + \tau^{l\ms l',mm'}\, \tau^{jj',kk'}\, u \bar{u}
\nonumber \\
 &\quad + \tau^{lm,jj'}\, \tau^{l'm',kk'}
        + \tau^{l'm,jj'}\, \tau^{lm',kk'}
\nonumber \\
 &\quad
  - \bigl( \tau^{jj',mn}\, \tau^{nl',kk'}\, \delta^{lm'}
         + \tau^{jj',mn}\, \tau^{nl,kk'}\, \delta^{l'm'}
    \bigr)\ms u
\nonumber \\
 &\quad
  - \bigl( \tau^{kk',mn}\, \tau^{nl',jj'}\, \delta^{lm'}
         + \tau^{kk',mn}\, \tau^{nl,jj'}\, \delta^{l'm'}
    \bigr)\ms \bar{u} \,.
\end{align}
The kernels $T_{\Delta g,g}$, $T_{\delta g,g}$, $U_{\delta g,g}$ and
$U_{\delta g,\Delta g}$ are respectively obtained from $T_{g,\Delta g}$,
$T_{g,\delta g}$, $U_{g,\delta g}$ and $U_{\Delta g,\delta g}$ by
interchanging $u \leftrightarrow \bar{u}$ and the appropriate indices.
The remaining kernels are zero.
For the different color combinations we find
\begin{align}
  \label{splitting-g-gg-color}
\frac{\octS{F}_{a_1,a_2}}{\sing{F}_{a_1,a_2}} \,\bigg|_{g\to gg}
&= {}- \frac{\octA{F}_{a_1,a_2}}{\sing{F}_{a_1,a_2}} \,\bigg|_{g\to gg}
 = \frac{\sqrt{N^2-1}}{2} \underset{N=3}{=} \sqrt{2} \,,
\end{align}
where as in the case $q\to g q$ color octet distributions are enhanced
over color singlet ones.  The factors for the higher color representations
in the case $N=3$ are
\begin{align}
 \label{splitting-g-gg-higher}
{}^{10\!}F_{a_1,a_2} \,\big|_{g\to gg}
&= {}^{\overline{10}\!}{F}_{a_1,a_2} \,\big|_{g\to gg} = 0 \,,
&
\frac{{}^{27\!}{F}_{a_1,a_2}}{\sing{F}_{a_1,a_2}} \,\bigg|_{g\to gg}
 = - \sqrt{3} \,.
\end{align}
The $27$ representation is hence even more strongly enhanced than the two
color octet combinations.  Decuplet and antidecuplet distributions are not
generated by perturbative splitting at lowest order.  We recall that this
was also the case for the ladder graphs discussed in
section~\ref{sec:ladders-color}.

We see that the perturbative splitting mechanism gives rise to a multitude
of two-parton distributions at high transverse momentum, which we have
collected in table~\ref{tab:splitting}.  As the comparison of
\eqref{splitting-result}, \eqref{splitting-q-gq-2} and
\eqref{splitting-g-gg-2} shows, a common feature of all channels is the
dependence on the transverse momenta $\tvec{k}$ and $\tvec{r}$.

\begin{table}
\centering
\begin{tabular}{cccccccccc} \hline
 & $F_{q,\bar{q}}$ & $F_{\Delta q,\Delta\bar{q}}$
 & $F_{\Delta q,\bar{q}}$ & $F_{q,\Delta\bar{q}}$
 & $F_{\delta q, \delta\bar{q}}$
 & $F_{\delta q,\bar{q}}$ & $F_{\delta q,\Delta\bar{q}}$
 & $F_{q,\delta\bar{q}}$  & $F_{\Delta q,\delta\bar{q}}$ \\
\hline
$f^g_1$         & $\times$ & $\times$ & $\times$ & $\times$
                & $\times$ & & & & \\
$h^{\perp g}_1$ & $\times$ & $\times$ & & & $\times$ & & & & \\
\hline
 & \\
\hline
 & $I_{q,\bar{q}}$ & $I_{\Delta q,\Delta\bar{q}}$
 & $I_{\Delta q,\bar{q}}$ & $I_{q,\Delta\bar{q}}$
 & $I_{\delta q,\delta\bar{q}}$
 & $I_{\delta q,\bar{q}}$ & $I_{\delta q,\Delta\bar{q}}$
 & $I_{q,\delta\bar{q}}$  & $I_{\Delta q,\delta\bar{q}}$ \\
\hline
$f^g_1$         & $\times$ & $\times$ & & & $\times$ & & & & \\
$h^{\perp g}_1$ & $\times$ & $\times$ & $\times$ & $\times$
                & $\times$ & & & & \\
\hline
 & \\
\hline
 & $F_{g,q}$ & $F_{\Delta g,\Delta q}$
 & $F_{\Delta g, q}$ & $F_{g,\Delta q}$
 & $F_{\delta g,\delta q}$
 & $F_{\delta g,q}$ & $F_{\delta g,\Delta q}$
 & $F_{g,\delta q}$ & $F_{\Delta g,\delta q}$ \\
\hline
$f^q_1$         & $\times$ & $\times$ & $\times$ & $\times$ & & $\times$ & \\
$h^{\perp q}_1$ & & & & & $\times$ & & & $\times$ & $\times$ \\
\hline
 & \\
\hline
 & $F_{g,g}$ & $F_{\Delta g,\Delta g}$
 & $F_{\Delta g, g}$ & $F_{g,\Delta g}$
 & $F_{\delta g,\delta g}$
 & $F_{\delta g,g}$ & $F_{\delta g,\Delta g}$
 & $F_{g,\delta g}$ & $F_{\Delta g,\delta g}$ \\
\hline
$f^g_1$         & $\times$ & $\times$ & $\times$ & $\times$
                & $\times$ & $\times$ & & $\times$ & \\
$h^{\perp g}_1$ & $\times$ & $\times$ & & & $\times$ & $\times$
                & $\times$ & $\times$ & $\times$ \\
\hline
\end{tabular}
\caption{\label{tab:splitting} Overview of the two-parton distributions
  that receive nonzero contributions from perturbative splitting of a
  single quark or gluon.  A cross  indicates a nonzero contribution at
  order $\alpha_s$.  Not shown are entries for $F_{q,g}$ and its
  analogs with polarization, which are like those for $F_{g,q}$ and its
  polarized counterparts.}
\end{table}


\paragraph{Position space.}

The Feynman graphs for the splitting contributions are naturally evaluated
in momentum representation.  We now transform our results to position
space.  We restrict our attention to the splitting $g\to q\bar{q}$ since
the other distributions can be treated in close analogy.
Using the relation
\begin{align}
\frac{1}{2\pi} \int d^2\tvec{k}\; e^{i \tvec{k} \tvec{z}}\,
 \frac{\tvec{k}^l}{\tvec{k}^2}
  &= i\ms \frac{\tvec{z}^l}{\tvec{z}{\rule{0pt}{1.6ex}}^2}
\end{align}
one can easily transform \eqref{splitting-result} to impact parameter
space,
\begin{align}
  \label{splitting-impact}
& \sing{F}_{a_1,\bar{a}_2}(x_i, \tvec{z}_i, \tvec{y})
   \,\Big|_{g\to q\bar{q}}
= \frac{\alpha_s}{4\pi^2}\, \Biggl[\ms
  \frac{f_1^g(x_1+x_2, \tvec{\zeta})}{x_1+x_2}\;
    T_{a_1,\bar{a}_2}^{l\ms l'}\biggl( \frac{x_1}{x_1+x_2} \biggr)\,
  - \frac{4 \tvec{\zeta}^m \tvec{\zeta}^{m'}
          - 2 \delta^{mm'} \tvec{\zeta}^2}{M^2}
\nonumber \\[0.2em]
&\quad \times
 \biggl( \frac{\partial}{\partial \tvec{\zeta}^2} \biggr)^2\,
  \frac{h_1^{\perp g}(x_1+x_2, \tvec{\zeta})}{x_1+x_2}\;
    U_{a_1,\bar{a}_2}^{l\ms l'mm'}\biggl( \frac{x_1}{x_1+x_2} \biggr)
\ms\Biggr]\,
  \frac{\bigl( \tvec{y} + \half \tvec{z} \bigr)^{l}
        \bigl( \tvec{y} - \half \tvec{z} \bigr)^{l'}}{%
           \bigl( \tvec{y} + \half \tvec{z} \bigr)^2
           \bigl( \tvec{y} - \half \tvec{z} \bigr)^2} \,,
\end{align}
where
\begin{align}
\tvec{z} &= \tvec{z}_1 - \tvec{z}_2 \,,
&
\tvec{\zeta} &= \half (\tvec{z_1} + \tvec{z}_2) \,.
\end{align}
$f_1^g(x,\tvec{\zeta})$ and $h_1^{\perp g}(x,\tvec{\zeta})$ are the
Fourier transforms of the transverse-momentum dependent parton densities
in \eqref{splitting-result}.  The distribution in \eqref{splitting-impact}
is singular at $\tvec{y} - \half\tvec{z}_1 = - \half\tvec{z}_2$ and at
$\tvec{y} + \half\tvec{z}_1 = \half\tvec{z}_2$, i.e.\ at the points where
in the operator definition of $F(x_i, \tvec{z}_i, \tvec{y})$ either the
two fields on the left or the two fields on the right of the final-state
cut are taken at equal transverse positions (see \eqref{index-key}).

The form of the distributions $F_{a_1,\bar{a}_2}(x_i, \tvec{k}_i,
\tvec{y})$ in the mixed representation is slightly more involved.  For
convenience we introduce the function
\begin{align}
  \label{D-function}
D(\tvec{k}, \tvec{r}) &=
\frac{\bigl( \tvec{k} + \half \tvec{r} \bigr)
        \bigl( \tvec{k} - \half \tvec{r} \bigr)}{%
           \bigl( \tvec{k} + \half \tvec{r} \bigr)^2
           \bigl( \tvec{k} - \half \tvec{r} \bigr)^2}
 = \frac{1}{2} \int_{-1}^1 dt\,
   \frac{\tvec{k}^2 - \frac{1}{4} \tvec{r}^2}{%
     \bigl( \tvec{k}^2 + \frac{1}{4} \tvec{r}^2 + t\ms \tvec{k}\tvec{r}
     \bigr)^2} \,,
\end{align}
which appears in $F_{q,\bar{q}}$, $F_{\Delta q, \Delta\bar{q}}$ and
$F_{\delta q, \delta\bar{q}}$.  Its Fourier transform can be written as
\begin{align}
D(\tvec{k}, \tvec{y})
&=\int \frac{d^2\tvec{r}}{(2\pi)^2}\; e^{i \tvec{r}\tvec{y}}\,
  D(\tvec{k}, \tvec{r})
\nonumber \\
&= \frac{1}{2} \int \frac{d^2\tvec{r}}{(2\pi)^2}
    \int_{-1}^{1} dt \int_{0}^\infty d\alpha\, \alpha\,
    \Bigl( \tvec{k}^2 - \frac{1}{4} \tvec{r}^2 \Bigr)\;
    e^{\ms i \tvec{r}\tvec{y} - \alpha \left(
       \tvec{k}^2 + \frac{1}{4} \tvec{r}^2 + t\ms \tvec{k}\tvec{r}
       \right)} \,,
\end{align}
where the Gaussian integral over $\tvec{r}$ is straightforward to perform.
Using the representation
\begin{align}
  \label{MacDonald}
\frac{1}{2} \int_0^\infty d\alpha\, \alpha^{p-1}\,
\exp\biggl[ -\alpha a^2 - \frac{z^2}{4 \alpha} \biggr]
 &= \frac{1}{a^{2p}}\, \biggl( \frac{za}{2} \biggr)^p\, K_p(az)
\end{align}
of the MacDonald functions we obtain
\begin{align}
D(\tvec{k}, \tvec{y})
= \frac{1}{\pi} \int_{-1}^{1} dt\, e^{-2i t\ms \tvec{k}\tvec{y}}\,
 & \Bigl[\, \bigl( 2it\ms \tvec{k}\tvec{y} - 1 \bigl)\,
      K_0\Bigl( 2 \sqrt{1-t^2}\, |\tvec{k}| |\tvec{y}| \Bigr)
\nonumber \\
 & + 2\sqrt{1-t^2}\, |\tvec{k}| |\tvec{y}| \,
      K_1\Bigl( 2\sqrt{1-t^2}\, |\tvec{k}| |\tvec{y}| \Bigr) \,\Bigr] \,.
\end{align}
For the factor appearing in $F_{q,\Delta\bar{q}}$ and $F_{\Delta q,
  \bar{q}}$ one finds in a similar fashion
\begin{align}
\tilde{D}(\tvec{k}, \tvec{y})
&= \int \frac{d^2\tvec{r}}{(2\pi)^2}\; e^{i \tvec{r}\tvec{y}}\;
   \frac{i\epsilon^{l\ms l'}
        \bigl( \tvec{k} + \half \tvec{r} \bigr)^{l}
        \bigl( \tvec{k} - \half \tvec{r} \bigr)^{l'}}{%
           \bigl( \tvec{k} + \half \tvec{r} \bigr)^2
           \bigl( \tvec{k} - \half \tvec{r} \bigr)^2}
\nonumber \\
&= \epsilon^{l\ms l'} \tvec{k}^l \tvec{y}^{l'}\,
   \frac{2}{\pi} \int_{-1}^{1} dt\,  e^{-2i t\ms \tvec{k}\tvec{y}}\,
     K_0\Bigl( 2 \sqrt{1-t^2}\, |\tvec{k}| |\tvec{y}| \Bigr) \,.
\end{align}
For small $\tvec{k}^2 \tvec{y}^2$ we can approximate the MacDonald
functions and perform the integral over $t$, which gives
\begin{align}
D(\tvec{k}, \tvec{y}) &=
  \frac{1}{\pi} \Bigl[\ms \log( 4\ms\tvec{k}^2 \tvec{y}^2 ) + 2\gamma
      + \mathcal{O}( \tvec{k}^2 \tvec{y}^2 ) \ms\Bigr] \,,
\nonumber \\
\tilde{D}(\tvec{k}, \tvec{y}) &=
  \epsilon^{l\ms l'} \tvec{k}^l \tvec{y}^{l'}
  \frac{2}{\pi} \Bigl[\ms \log( 4\ms\tvec{k}^2 \tvec{y}^2 ) + 2\gamma - 2
      + \mathcal{O}( \tvec{k}^2 \tvec{y}^2 ) \ms\Bigr] \,,
\end{align}
where $\gamma$ is the Euler number.  In the short-distance limit
$\tvec{y}^2 \to 0$ we thus have a logarithmic divergence in $D(\tvec{k},
\tvec{y})$ and hence in the distributions $F_{q,\bar{q}}(x_i, \tvec{k}_i,
\tvec{y})$, $F_{\Delta q, \Delta\bar{q}}(x_i, \tvec{k}_i, \tvec{y})$ and
$F_{\delta q, \delta\bar{q}}(x_i, \tvec{k}_i, \tvec{y})$.


\subsubsection{Contribution to the cross section}
\label{sec:splitting-X-sect}

We now investigate how the splitting contribution to quark-antiquark
distributions in figure~\ref{fig:2-4}a enters in the cross section for
double hard scattering, as shown in figure~\ref{fig:2-4-X}a.
Concentrating on the factors that depend on transverse momenta, we have
with the cross section formula \eqref{X-sect-momentum} and the
distributions from \eqref{splitting-result}
\begin{align}
 \label{splitting-X-mom}
& \biggl[\, \prod_{i=1}^{2} 
     \int d^2\tvec{k}_i\, d^2\bar{\tvec{k}}_i\;
     \delta^{(2)}(\tvec{q}{}_i - \tvec{k}_i - \bar{\tvec{k}}_i) \biggr]
\int \frac{d^2\tvec{r}}{(2\pi)^2}\,
  F_{a_1,\bar{a}_2}(x_i, \tvec{k}_i, \tvec{r})\, 
  F_{\bar{a}_1,a_2}(\bar{x}_i, \bar{\tvec{k}}_i, - \tvec{r})
  \,\bigg|_{g\to q\bar{q}}
\nonumber \\[0.1em]
&\quad \propto
\int d^2\tvec{\kappa}\, d^2\bar{\tvec{\kappa}}\,
     \delta^{(2)}(\tvec{\kappa} + \bar{\tvec{\kappa}}
                  - \tvec{q}_1 - \tvec{q}_2)\,
  \bigl[ f_1^g(x_1+x_2, \tvec{\kappa})\, 
         f_1^g(\bar{x}_1+\bar{x}_2, \bar{\tvec{\kappa}}) + \ldots \,\bigr]
\nonumber \\
&\qquad \times
\int d^2\tvec{r}\, d^2\tvec{k}\, d^2\bar{\tvec{k}}\;
  \delta^{(2)}(\tvec{k} + \bar{\tvec{k}} - \tvec{q}) \;
  \frac{\bigl( \tvec{k} - \half\tvec{r} \bigr)^{l}
        \bigl( \tvec{k} + \half\tvec{r} \bigr)^{l'}}{%
        \bigl( \tvec{k} - \half\tvec{r} \bigr)^2
        \bigl( \tvec{k} + \half\tvec{r} \bigr)^2} \,
  \frac{\bigl( \bar{\tvec{k}} + \half\tvec{r} \bigr)^{m}
        \bigl( \bar{\tvec{k}} - \half\tvec{r} \bigr)^{m'}}{%
        \bigl( \bar{\tvec{k}} + \half\tvec{r} \bigr)^2
        \bigl( \bar{\tvec{k}} - \half\tvec{r} \bigr)^2}
\end{align}
with $\tvec{q} = \half ( \tvec{q}_1 - \tvec{q}_2 )$, where the ellipsis
stands for terms involving the gluon Boer-Mulders functions $h_1^{\perp
  g}$.  Changing integration variables to $\tvec{k}_{+} = \tvec{k} +
\half\tvec{r}$ and $\tvec{k}_{-} = \tvec{k} - \half\tvec{r}$, we can
rewrite the last line as
\begin{align}
  \label{splitting-X-int}
\int d^2\tvec{k}_{+}\,
\frac{\tvec{k}_{+}^{l} (\tvec{k}_{+}^{} - \tvec{q})^{m}}{%
      \tvec{k}_{+}^2 (\tvec{k}_{+}^{} - \tvec{q}){\rule{0pt}{1.6ex}}^2} \;
\int d^2\tvec{k}_{-}\, 
\frac{\tvec{k}_{-}^{l'} (\tvec{k}_{-}^{} - \tvec{q})^{m'}}{%
      \tvec{k}_{-}^2 (\tvec{k}_{-}^{} - \tvec{q}){\rule{0pt}{1.6ex}}^2} \,.
\end{align}
Each integral is infrared finite but has a logarithmic divergence at large
$\tvec{k}_\pm$.
This logarithmic divergence also appears if we use the impact parameter
space representation \eqref{splitting-impact}.  According to
\eqref{X-sect-position} the cross section is then proportional to
\begin{align}
  \label{splitting-X-impact}
& \int d^2\tvec{z}_1\, d^2\tvec{z}_2\, d^2\tvec{y}\;
  e^{- i \tvec{q}_1^{} \tvec{z}_1^{} - i \tvec{q}_2^{} \tvec{z}_2^{}}\,
  F_{a_1,\bar{a}_2}(x_i, \tvec{z}_i, \tvec{y})\, 
  F_{\bar{a}_1,a_2}(\bar{x}_i, \tvec{z}_i, \tvec{y})
  \,\bigg|_{g\to q\bar{q}}
\nonumber \\
&\quad \propto
\int d^2\tvec{\zeta}\; e^{- i\ms (\tvec{q}_1 + \tvec{q}_2)\ms \tvec{\zeta}}\,
  \bigl[ f_1^g(x_1+x_2, \tvec{\zeta})\,
         f_1^g(\bar{x}_1+\bar{x}_2, \tvec{\zeta}) + \ldots \,\bigr]
\nonumber \\
&\qquad \times
\int d^2\tvec{z}\, d^2\tvec{y}\; e^{- i \tvec{q} \tvec{z}}\;
  \frac{\bigl( \tvec{y} - \half\tvec{z} \bigr)^{l}
        \bigl( \tvec{y} - \half\tvec{z} \bigr)^{m}}{%
        \bigl( \tvec{y} - \half\tvec{z} \bigr)^4} \,
  \frac{\bigl( \tvec{y} + \half\tvec{z} \bigr)^{l'}
        \bigl( \tvec{y} + \half\tvec{z} \bigr)^{m'}}{%
        \bigl( \tvec{y} + \half\tvec{z} \bigr)^4} \,.
\end{align}
The last line diverges logarithmically for $\tvec{y} = \half\tvec{z}$ and
$\tvec{y} = -\half\tvec{z}$.  At these points one respectively has
$\tvec{y} - \half\tvec{z}_1 = - \half\tvec{z}_2$ and $\tvec{y} +
\half\tvec{z}_1 = \half\tvec{z}_2$, so that the singularities correspond
to configurations where partons are at the same transverse position,
either to the right or to the left of the final-state cut.

To understand the origin of this ultraviolet divergence, we go back to the
graph in figure~\ref{fig:2-4-X}a.  As mentioned in
section~\ref{sec:splitting-power} this graph receives leading
contributions from two kinematic regions.  In the first region, the
virtualities and transverse momenta of the quarks are of order $q_T$ and
thus much smaller than $Q$, whereas in the second region they are of order
$Q$.  The approximations that are necessary to derive factorization for
double hard scattering are only valid in the first region.  However, the
integrand in \eqref{splitting-X-int} does not decrease fast enough with
$\tvec{k}_\pm = \half (\tvec{k}_1 - \tvec{k}_2 \pm \tvec{r})$ to suppress
the second region, so that the factorization formula
\eqref{X-sect-momentum} requires a suitable regularization in order to
remove contributions from that region.  A corresponding statement holds in
the position space formulation.

A simple way to regularize the cross section formula in impact parameter
space is to impose a lower cutoff $1/ \mu^2$ on $(\tvec{y} +
\half\tvec{z})^2$ and $(\tvec{y} - \half\tvec{z})^2$.  The integral in the
last line of \eqref{splitting-X-impact} then becomes
\begin{align}
  \label{fix-by-cutoff}
\pi \biggl[\, \delta^{lm} \int_{|\tvec{q}| /\mu}^\infty
              \frac{dw}{w}\, J_0(w)
 + \biggl( \delta^{lm} - \frac{2 \tvec{q}^{l} \tvec{q}^m}{\tvec{q}^2}
   \biggr)\, \frac{\mu}{|\tvec{q}|}\,
             J_1\biggl( \frac{|\tvec{q}|}{\mu} \biggr)
\,\biggr]
\end{align}
times the same expression with $l\to l'$ and $m\to m'$.  The integral in
\eqref{fix-by-cutoff} behaves like $\log\bigl( \mu /|\tvec{q}| \bigr)$ for
$\mu\gg |\tvec{q}|$.  The $\mu$ dependence of the cross section obtained
in this way must cancel when one adds the contribution from
figure~\ref{fig:2-4-X}a in the region of transverse loop momenta of order
$Q$.  That region is naturally associated with single hard scattering as
discussed in section~\ref{sec:splitting-power}.  At this point, one must
obviously be careful to avoid double counting between the parts of the
graph that one associates with single or with double hard scattering.  The
analogous double counting problem in multijet production has been pointed
out in \cite{Cacciari:2009dp}.

To use a cutoff in \eqref{splitting-X-impact} is of course rather ad hoc,
and there should be better ways to construct a consistent factorization
scheme in which the formula for double-parton scattering has a controlled
ultraviolet behavior and in which the double counting problem is properly
taken care of.  One may for instance think of subtracting the perturbative
splitting contribution of figure~\ref{fig:2-4} at large momenta or small
transverse distances in the definition of the two-parton distributions, so
that graphs like in figure~\ref{fig:2-4-X} are not included in double hard
scattering at all.  To solve this issue is a nontrivial task and must be
left to future work.

We already remarked that the integrals in \eqref{splitting-X-int} are
finite in the infrared.  This is due to the numerator factors and can be
understood in simple physical terms, as noted in the detailed analysis
given in \cite{Gaunt:2011xd}.  The points where one of the four momenta
$\tvec{k}_+$, $\tvec{k}_+ - \tvec{q}$, $\tvec{k}_-$ or $\tvec{k}_- -
\tvec{q}$ vanishes correspond to configurations where one of the four
$g\to q\bar{q}$ splitting processes in figure~\ref{fig:2-4-X}a proceeds in
strictly collinear kinematics.  The amplitude for the collinear splitting
$g\to q\bar{q}$ is zero because an on-shell gluon has helicity $\pm 1$,
whereas the helicities of $q$ and $\bar{q}$ add up to zero due to
chirality conservation for massless quarks.

Referring to the end of section~\ref{sec:splitting-power} we finally
determine the dependence of \eqref{splitting-X-mom} on $q_T$ and on
$\Lambda$.  With $|\tvec{q}_1 + \tvec{q}_2| \sim \Lambda$ the second line
scales like $1 /\Lambda^2$, and with the behavior of the third line just
discussed we find
\begin{align}
 \label{splitting-X-scale}
\frac{s^2\ms d\sigma}{\prod_{i=1}^2 dx_i\, d\bar{x}_i\, d^2\tvec{q}{}_i}
  \bigg|_{\text{fig.~\protect\ref{fig:2-4-X}a}}
& \sim \frac{1}{\Lambda^2} \, \log^2 \frac{\mu^2}{q_T^2} \,,
\end{align}
where $\mu$ is an ultraviolet cutoff much larger than $q_T$.


\subsection{Parton splitting in collinear distributions}

The results of the previous section are relevant not only for
transverse-momentum dependent two-parton distributions but also for
collinear ones.  As we have seen, collinear two-parton distributions
appear in transverse-momentum integrated cross sections and in cross
sections at perturbatively large $q_T$ via the ladder graphs discussed in
section~\ref{sec:ladders}.  Since $\tvec{k}_1$ and $\tvec{k}_2$ are not
fixed in collinear distributions, the splitting contributions we computed
in section~\ref{sec:splitting-dist} are relevant for $F(x_i, \tvec{r})$ at
large $\tvec{r}$ and, after Fourier transform, for $F(x_i, \tvec{y})$ at
small $\tvec{y}$.

\subsubsection{Ultraviolet behavior}
\label{sec:splitting-collinear}

Integrating \eqref{splitting-result} over $\tvec{k}_1$ and $\tvec{k}_2$,
i.e.\ over $\tvec{k}$ and $\tvec{\kappa}$, one formally obtains
\begin{align}
  \label{splitting-mom-int}
\biggl[\, \prod_{i=1}^2 \int {d^2\tvec{k}_i} \,\biggr]
\, \sing{F}_{a_1,\bar{a}_2}(x_i, \tvec{k}_i, \tvec{r})
   \,\Big|_{g\to q\bar{q}}
&= \frac{\alpha_s}{4\pi^2}\,
  \frac{1}{x_1+x_2}\, f_1^g(x_1+x_2)\;
  T_{a_1,\bar{a}_2}^{l\ms l'}\biggl( \frac{x_1}{x_1+x_2} \biggr)\,
\nonumber \\
&\quad \times
\int {d^2\tvec{k}}\;
  \frac{\bigl( \tvec{k} + \half \tvec{r} \bigr)^{l}
        \bigl( \tvec{k} - \half \tvec{r} \bigr)^{l'}}{%
           \bigl( \tvec{k} + \half \tvec{r} \bigr)^2
           \bigl( \tvec{k} - \half \tvec{r} \bigr)^2} \,,
\end{align}
where the integration over $\tvec{\kappa}$ gives the collinear gluon
distribution $f_1^g(x_1+x_2)$, whereas the term with $h_1^{\perp g}$
disappears due to rotation invariance.  In the case where
$T_{a_1,\bar{a}_2}^{l\ms l'} \propto \delta^{l\ms l'}$, i.e.\ for
$F_{q,\bar{q}}$, $F_{\Delta q, \Delta\bar{q}}$ and $F_{\delta q,
  \delta\bar{q}}$, the integral over $\tvec{k}$ is ultraviolet divergent.
The corresponding integrals of $F_{q, \Delta\bar{q}}$ and $F_{\Delta
  q,\bar{q}}$ are proportional to $\epsilon^{l\ms l'}\ms \tvec{r}^{l}
\tvec{r}^{l'}$ and hence vanish, as they must according to the constraint
\eqref{no-pseudoscalar} from parity invariance.  An analogous discussion
can be given for the interference distributions $I_{a_1,\bar{a}_2}$ and
for the distributions resulting from the splitting processes $q\to g q$ or
$g\to g g$.  In all cases, contributions going with the Boer-Mulders
functions $h_1^{\perp q}$ or $h_1^{\perp g}$ vanish after integration over
$\tvec{\kappa}$ and one is left with contributions from the unpolarized
distributions $f_1^q$ or $f_1^g$.

With the kernels $T$ or $V$ given in \eqref{Tkernels}, \eqref{Vkernels},
\eqref{Tkernels-q-gq} and \eqref{Tkernels-g-gg} we find that the splitting
mechanism generates nonzero collinear two-parton distributions
\begin{align}
  \label{coll-dist-split}
& F_{q,\bar{q}} \,, F_{\Delta q, \Delta\bar{q}} \,,
  F_{\delta q, \delta\bar{q}} \,,
&&
I_{q,\bar{q}} \,, I_{\Delta q, \Delta\bar{q}} \,,
  I_{\delta q, \delta\bar{q}} \,,
&&
F_{g,q} \,, F_{\Delta g, \Delta q} \,, F_{\delta g, q} \,,
\nonumber \\
& F_{g,g} \,, F_{\Delta g, \Delta g} \,, F_{g, \delta g} \,,
  F_{\delta g, \delta g} \,,
\end{align}
as well as the distributions obtained by interchanging the first and
second subscripts in \eqref{coll-dist-split} or by replacing quarks with
antiquarks in $F_{g,q}$ and its polarized counterparts.  With the
exception of $F_{\delta g, \Delta q}$ (and $F_{\Delta q, \delta g}$,
$F_{\delta g, \Delta\bar{q}}$, $F_{\Delta\bar{q}, \delta g}$) these are
indeed all collinear distributions that are allowed by parity invariance
and that are chiral even.
For the distributions depending on polarization indices we have
\begin{align}
F_{\delta q, \delta\bar{q}}^{jj'}
  \propto I_{\delta q, \delta\bar{q}}^{jj'}
& \propto \delta^{jj'} \,,
&
F_{\delta g, \delta g}^{jj',kk'} &\propto \tau^{jj',kk'} \,,
\nonumber \\
F_{\delta g, q}^{jj'} \propto F_{g, \delta g}^{jj'}
& \propto 2 \tau^{jj', l\ms l'} \tvec{r}^l \tvec{r}^{l'}
  =\, 2 \tvec{r}^j\ms \tvec{r}^{j'} - \delta^{jj'} \tvec{r}^2 \,.
\end{align}

To further investigate the ultraviolet divergence mentioned below
\eqref{splitting-mom-int} we focus on $F_{q,\bar{q}}$ for definiteness.
Since $D(\tvec{k}, \tvec{r})$ in \eqref{D-function} falls off as
$1/\tvec{k}^2$ for fixed $\tvec{r}$ and as $1/\tvec{r}^2$ for fixed
$\tvec{k}$, one obtains logarithmic divergences if one integrates over one
or both of these variables.  To regulate these divergences one may work in
$4 - 2\epsilon$ dimensions.  The result for $F_{q,\bar{q}}(x_i,
\tvec{k}_i, \tvec{r})$ is then the same as in \eqref{splitting-result}
with a modified kernel
\begin{align}
  \label{epsilon-kernel}
T_{q,\bar{q}}^{l\ms l'}(u; \epsilon)
 &= \delta^{l\ms l'}\, \bigl[\ms u^2 + (1-u)^2 - \epsilon \ms\bigr] 
     \big/ (1-\epsilon)
\end{align}
times a power of $(2\pi)^{\epsilon}$ we need not specify here.
Integrating over both transverse momenta and changing integration
variables to $\tvec{k}_{+} = \tvec{k} + \half \tvec{r}$ and $\tvec{k}_{-}
= \tvec{k} - \half \tvec{r}$, one obtains
\begin{align}
  \label{twist-four-integral}
\int d^{2-2\epsilon}\tvec{r}\, d^{2-2\epsilon}\tvec{k}\,
D(\tvec{k}, \tvec{r})
  &= \int d^{2-2\epsilon}\tvec{k}_{+}\,
          \frac{\tvec{k}_{+}^{l}}{\tvec{k}_{+}^2}
     \int d^{2-2\epsilon}\tvec{k}_{-}\,
          \frac{\tvec{k}_{-}^{l}}{\tvec{k}_{-}^2} \,,
\end{align}
which is zero due to rotation invariance.  We note that integrating over
$\tvec{k}_1$, $\tvec{k}_2$ and $\tvec{r}$ puts all four fields in the
matrix element defining $F_{q,\bar{q}}$ at the same transverse position,
so that one obtains a twist-four operator.  If \eqref{twist-four-integral}
were not zero but finite after subtraction of the logarithmically
divergent pieces, the graph in figure~\ref{fig:2-4}a would contribute to
the scale evolution of a twist-four distribution.  The vanishing of
\eqref{twist-four-integral} thus reflects the fact that distributions of
twist four and of twist two (the collinear gluon distribution in
\eqref{splitting-mom-int}) do not mix under evolution.  The same zero
result is obtained in any regularization scheme that respects rotational
invariance.

Integrating over $\tvec{k}$ at fixed nonzero $\tvec{r}$ and using the
integral representation in \eqref{D-function}, one obtains
\begin{align}
  \label{split-r}
\mu^{2\epsilon} \int d^{2-2\epsilon}\tvec{k}\; D(\tvec{k}, \tvec{r})
&= \pi^{1-\epsilon}\,
   \frac{\Gamma^2(1-\epsilon)}{\Gamma(1-2\epsilon)}\,
   \Gamma(\epsilon)\,
   \biggl( \frac{\tvec{r}^2}{\mu^2} \biggr)^{-\epsilon}
\nonumber \\
&= \pi\ms \biggl[ \frac{1}{\epsilon} + \log\frac{\mu^2}{\tvec{r}^2}
                + \text{const.} + \mathcal{O}(\epsilon) \biggr] \,.
\end{align}
This contains an ultraviolet pole and an associated logarithm of the
renormalization scale~$\mu^2$.  The value of the constant is not of
relevance for our discussion.  If $\tvec{r} = 0$ then the integral on the
l.h.s.\ is scaleless and therefore vanishes.  To isolate the ultraviolet
singularity in that case, one can for instance give a small mass to the
quarks.  Up to corrections of order $m^2$ this leads to
\begin{align}
\mu^{2\epsilon} \int d^{2-2\epsilon}\tvec{k}\,
   \frac{1}{\tvec{k}^2 + m^2}
 &= \pi^{1-\epsilon}\, \Gamma(\epsilon)\,
   \biggl( \frac{m^2}{\mu^2} \biggr)^{-\epsilon} \,.
\end{align}
The ultraviolet pole and the associated logarithm are hence the same as
for nonzero $\tvec{r}$.

If one defines the collinear distribution $F_{q,\bar{q}}(x_i, \tvec{r}) =
\int d^2\tvec{k}_1\, d^2\tvec{k}_2\, F_{q,\bar{q}}(x_i, \tvec{k}_i,
\tvec{r})$ in the $\overline{\text{MS}}$ scheme, the above $1/\epsilon$
pole is subtracted, together with a constant.  To leading order in
$\alpha_s$ one finds for the scale dependence\footnote{
  We note that both $\alpha_s$ and the gluon distribution $f^g$ in
  \protect\eqref{splitting-mom-int} also have a scale dependence, which
  becomes relevant at order $\alpha_s^2$ in the evolution equation.}
\begin{align}
  \label{evol-splitting}
\frac{d}{d \log\mu^2}\, 
  F_{q,\bar{q}}(x_i, \tvec{r})
  \,\Big|_{g\to q\bar{q}}
&= \frac{1}{x_1+x_2}\, f_1^g(x_1+x_2)\;
   P_{q,g}\biggl( \frac{x_1}{x_1+x_2} \biggr) \,,
\end{align}
where
\begin{align}
  \label{full-Pqg}
P_{q,g}(u) = \frac{\alpha_s}{2\pi}\, \frac{u^2 + (1-u)^2}{2}
\end{align}
is the familiar DGLAP splitting function (now including a color factor
$T_R = 1/2$, unlike the function $P_{qg}$ we used in
section~\ref{sec:ladders-color}).  We come back to this in the next
section.  Let us note that with the results in \eqref{splitting-q-gq-2},
\eqref{Tkernels-q-gq} and \eqref{splitting-g-gg-2}, \eqref{Tkernels-g-gg}
we obtain relations analogous to \eqref{evol-splitting} for $F_{g, q}$ and
$F_{g, g}$.  On the r.h.s.\ of these relations we respectively find the
DGLAP splitting functions $P_{g,q}(u)$ and $P_{g,g}(u)$, except for terms
proportional to $\delta(1-u)$ in $P_{g,g}(u)$.

Quite interestingly, the situation changes if we consider
$F_{q,\bar{q}}(x_i, \tvec{y})$ instead of $F_{q,\bar{q}}(x_i, \tvec{r})$.
The Fourier transform of $(\tvec{r}^2)^{-\epsilon}$ in $2-2\epsilon$
transverse dimensions is
\begin{align}
\int d^{2-2\epsilon}\tvec{r}\; e^{-i \tvec{r} \tvec{y}}\,
    (\tvec{r}^2)^{-\epsilon}
 &= 4^{1-2\epsilon}\ms \pi^{1-\epsilon}\,
    \frac{\Gamma(1-2\epsilon)}{\Gamma(\epsilon)}\,
    (\tvec{y}^2)^{-1+2\epsilon} \,,
\end{align}
which can be seen by writing $(\tvec{r}^2)^{-\epsilon} =
\Gamma^{-1}(\epsilon) \int_{0}^\infty d\alpha\, \alpha^{\epsilon-1}\,
e^{-\alpha \tvec{r}^2}$, performing the integral over $\tvec{r}$ and then
the one over $\alpha$.  The factor $\Gamma(\epsilon)$ responsible for the
ultraviolet divergence in \eqref{split-r} is thus canceled if one Fourier
transforms from $\tvec{r}$ to $\tvec{y}$, and the result is finite for
$\epsilon=0$,
\begin{align}
  \label{unsubtracted-Fy}
\int \frac{d^2\tvec{k}\, d^2\tvec{r}}{(2\pi)^2}\;
 e^{-i \tvec{r} \tvec{y}}\, D(\tvec{k}, \tvec{r})
 &=  \frac{1}{\tvec{y}^2} \,.
\end{align}
The $1/\tvec{y}^2$ behavior can be obtained directly in $4$ dimensions by
setting $\tvec{z} = \tvec{0}$ in \eqref{splitting-impact}.

We thus find that $F_{q,\bar{q}}(x_i, \tvec{r})$ requires an ultraviolet
subtraction for the graph in figure~\ref{fig:2-4}a, whereas
$F_{q,\bar{q}}(x_i, \tvec{y})$ does not.  Let us see what we obtain if we
define a modified $\tvec{y}$ dependent distribution as the Fourier
transform of the ultraviolet subtracted distribution $F_{q,\bar{q}}(x_i,
\tvec{r})$,
\begin{align}
  \label{modif-F-y}
F^{\text{mod}}_{q,\bar{q}}(x_i, \tvec{y}) &=
  \int \frac{d^2\tvec{r}}{(2\pi)^2}\; e^{-i \tvec{r}\tvec{y}}\;
            F_{q,\bar{q}}(x_i, \tvec{r}) \,.
\end{align}
We have
\begin{align}
  \label{subtracted-Fr}
F_{q,\bar{q}}(x_i, \tvec{r})
  \,\Big|_{g\to q\bar{q}}
 &= f(x_1,x_2)\, \pi \log\frac{\mu^2}{\tvec{r}^2} + g(x_1,x_2) \,,
\\
  \label{unsubtracted-Fy-2} 
F_{q,\bar{q}}(x_i, \tvec{y})
  \,\Big|_{g\to q\bar{q}}
 &= f(x_1,x_2)\, \frac{1}{\tvec{y}^2} \,,
\end{align}
where the explicit expressions of $f$ and $g$ are easily obtained but not
relevant for our discussion.  As shown in appendix~\ref{app:fourier} the
Fourier transform of \eqref{subtracted-Fr} gives
\begin{align}
  \label{Fourier-log}
F^{\text{mod}}_{q,\bar{q}}(x_i, \tvec{y})
  \,\Big|_{g\to q\bar{q}}
&=  f(x_1,x_2)\, \biggl[ \frac{1}{\tvec{y}^2} \biggr]_{+\ms (\mu)}
  + g(x_1,x_2)\, \delta^{(2)}(\tvec{y}) \,,
\end{align}
where
\begin{align}
\biggl[ \frac{1}{\tvec{y}^2} \biggr]_{+\ms (\mu)}
 &= \underset{\varepsilon\to 0}{\lim} \,\biggl[
    \frac{1}{\tvec{y}^2}\, \theta(\tvec{y}^2 - \varepsilon)
  - \delta^{(2)}(\tvec{y}) \int d^2\tvec{y}'\, \frac{1}{\tvec{y}'^2}\,
      \theta(\tvec{y}'^2 - \varepsilon)\,
      \theta(b_0^2 - \mu^2 \tvec{y}'^2)
    \biggr]
\end{align}
with $b_0 = 2 e^{-\gamma}$.  We thus find the same $\tvec{y}$ dependence
in \eqref{unsubtracted-Fy-2} and \eqref{Fourier-log}, up to terms
concentrated at the singular point $\tvec{y} = \tvec{0}$.  This is not
surprising since the ultraviolet divergent term in \eqref{split-r} is
independent of $\tvec{r}$.

In section~\ref{sec:ladders-power} we have shown that the contribution of
ladder graphs at large $\tvec{y}$ to the cross section involves an
integral
\begin{align}
  \label{y-coll-integral}
\int d^2\tvec{y}\,
    F(u_i, \tvec{y}) F(\bar{u}_i, \tvec{y})
\end{align}
according to \eqref{single-ladders-X} and \eqref{double-ladder-X}.  Since
the collinear two-parton distributions behave like $1/\tvec{y}^2$ at small
$\tvec{y}$, the above integral has a linear divergence for small
$\tvec{y}^2$ and is hence not defined as it stands.  A corresponding
linear divergence is found if one Fourier transforms from $\tvec{y}$ to
$\tvec{r}$,
\begin{align}
  \label{r-coll-integral}
\int d^2\tvec{r}\,
    F(u_i, \tvec{r}) F(\bar{u}_i, -\tvec{r}) \,,
\end{align}
where according to \eqref{subtracted-Fr} the distributions behave like
$\log(\tvec{r}^2 /\mu^2)$ for large $\tvec{r}$.  The ultraviolet
subtraction already included in the definition of $F(u_i, \tvec{r})$ is
hence \emph{not} sufficient to render the integral in
\eqref{r-coll-integral} finite.

The reason for the unphysical divergences in \eqref{y-coll-integral} and
\eqref{r-coll-integral} is that the cross section formulae containing
these integrals have been derived for the region where $1/\tvec{y}^2$ or
$\tvec{r}^2$ is much smaller than $q_T^2$.  We thus encounter a similar
problem as in section~\ref{sec:splitting-X-sect}, with the difference that
the divergence to be regulated is now linear instead of logarithmic.  To
make the cross section formulae \eqref{single-ladders-X} and
\eqref{double-ladder-X} well-defined, one must either remove or suppress
the $\tvec{y}$ integral in the region where $|\tvec{y}|$ is not large
compared with $1/q_T$, or one must define $F(x_i,\tvec{y})$ such that in
this region the contribution from perturbative splitting as in
figure~\ref{fig:2-4} is subtracted.  Along with such a procedure, one must
provide a prescription for evaluating the splitting contribution at small
$|\tvec{y}|$ in such a way that there is no double counting, as discussed
in section~\ref{sec:splitting-X-sect}.

Integrating the cross section \eqref{X-sect-qq} over $\tvec{q}_1$ and
$\tvec{q}_2$, we readily obtain the integral in \eqref{y-coll-integral}
with $u_i = x_i$ and $\bar{u}_i = \bar{x}_i$.  The discussion of the
previous paragraph carries over to that case, with the difference that the
requirement for the validity of the cross section formula is then
$|\tvec{y}| \gg 1/Q$ instead of $|\tvec{y}| \gg 1/q_T$.  For the
corresponding momentum integral \eqref{r-coll-integral} with $u_i = x_i$
and $\bar{u}_i = \bar{x}_i$ one must require $|\tvec{r}| \ll Q$ instead of
$|\tvec{r}| \ll q_T$.  We note that in \cite{Ryskin:2011kk} it was
proposed to regulate this integral by imposing an upper cutoff $\tvec{r}^2
< \min(q_1^2, q_2^2)$.  By itself this is clearly insufficient to obtain a
reliable result, since the contribution from $\tvec{r}^2$ outside that
region is large and needs to be evaluated as well.


\subsubsection{Scale evolution}
\label{sec:evolution}

Let us now investigate the scale evolution of collinear two-parton
distributions.  We focus on the color-singlet combinations $\sing{F}$,
which are most closely related with single-parton densities as we already
saw in section~\ref{sec:coll-fact}.  For definiteness we consider the
quark-antiquark distribution $\sing{F}_{q,\bar{q}}$, which we studied
extensively in the previous section.  The generalization to other parton
and polarization combinations is straightforward.

The dependence on the scale $\mu$ of collinear parton distributions arises
from the regularization and subtraction of ultraviolet divergences in
their definition.  This involves divergences from self-energy graphs
(which also occur in transverse-momentum dependent distributions and can
be expressed in terms of suitable $Z$ factors) and divergences from
regions of large transverse parton momenta.  For a single-parton
distribution the contribution from the high-transverse-momentum tail was
already discussed in section~\ref{sec:ladders-fact}.

The scale dependence in the collinear distributions $\sing{F}(x_i,
\tvec{y})$ arises from self-energy graphs and in addition from the ladder
graphs in figure~\ref{fig:ladders}, which according to
\eqref{single-ladder-y} and \eqref{double-ladder-y} give rise to
ultraviolet divergent integrals $\int d^2\tvec{k}_1\, d^2\tvec{k}_2\;
\sing{F}(x_i, \tvec{k}_i, \tvec{y})$ unless one performs suitable
subtractions.  Since the ladder and self-energy graphs have exactly the
same structure as for single-parton distributions, the corresponding
evolution equation reads
\begin{align}
  \label{hom-dglap}
\frac{d}{d\log\mu^2}\, \sing{F}_{q, \bar{q}}(x_1,x_2, \tvec{y})
 &= \sum_{b_1 = q,g}\,
    \int_{x_1}^{1-x_2} \frac{du_1}{u_1}\,
       P_{q,b_1}\biggl(\frac{x_1}{u_1}\biggr)\,
         \sing{F}_{b_1, \bar{q}}(u_1,x_2, \tvec{y})
\nonumber \\
 &+ \sum_{b_2 = \bar{q},g}\,
    \int_{x_2}^{1-x_1} \frac{du_2}{u_2}\,
       P_{\bar{q},b_2}\biggl(\frac{x_2}{u_2}\biggr)\,
         \sing{F}_{q, b_2}(x_1,u_2, \tvec{y})
\end{align}
for a quark-antiquark distribution.  The splitting functions $P_{a,b}$ now
include the contributions from virtual corrections, unlike the
corresponding kernels in section~\ref{sec:ladders-color}.  Note that the
labels $b_1$ and $b_2$ do not take the value $\delta g$ here, because the
corresponding kernels $P_{q, \delta g}$ and $P_{\bar{q}, \delta g}$ vanish
due to rotation invariance, see our remark below \eqref{tensor-kernels}.

The evolution equation \eqref{hom-dglap} has the structure of a usual
DGLAP equation for each parton.  The corresponding operator appearing in
the distribution $F_{q,\bar{q}}(x_i, \tvec{y})$ is
\begin{align}
  \label{twice-twist-2}
& \Bigl\{\, \bar{q}\bigl( -\half z_2 \bigr)\, W[-\half z_2, \half z_2]\,
       \Gamma_{\bar{q}}\, q\bigl( \half z_2 \bigr)
\ms\Bigr\}^{\text{ren}, \mu}_{z_2^+ = 0,\ms \tvec{z}_2^{} = \tvec{0}}
\nonumber \\
\times\; & \Bigl\{\, \bar{q}\bigl( y-\half z_1 \bigr)\,
       W[y-\half z_1, y+\half z_1]\,
       \Gamma_{q}\ms q\bigl( y+\half z_1 \bigr)
    \ms\Bigr\}^{\text{ren}, \mu}_{z_1^+ = y^+ = 0,\,
       \tvec{z}_1^{} = \tvec{0}} \,,
\end{align}
where the Wilson line $W[\xi',\xi]$ is defined in \eqref{Wilson-finite}
and where $\{\, \ldots \,\}^{\text{ren}, \mu}$ indicates that each
bilinear operator $\bar{q}\ms W q$ is renormalized at scale $\mu$ in the
same way as for single-parton distributions.  As long as the transverse
distance $\tvec{y}$ between the two bilinear operators remains finite, no
further ultraviolet divergences appear, and one has the product of two
renormalized twist-two operators.  As remarked earlier in the literature,
one may choose different renormalization scales $\mu_1$ and $\mu_2$ for
the two operators, which appears useful when one has two hard-scattering
processes with rather different hard scales.  The separate evolution
equations in $\mu_1$ and $\mu_2$ are then simply the usual ones with a
single DGLAP kernel.

For the collinear distributions $\sing{F}_{a_1,\tvec{a}_2}(x_i, \tvec{r})$
that depend on the relative momentum $\tvec{r}$ the situation is
different, as we have seen in the previous section.  The splitting graph
in figure~\ref{fig:2-4}a and higher-order corrections as in
figure~\ref{fig:2-4}b give rise to additional ultraviolet divergences.
Their subtraction leads to an inhomogeneous term in the evolution
equation.  At leading order in $\alpha_s$ one has
\begin{align}
  \label{inhom-dglap}
\frac{d}{d\log\mu^2}\, \sing{F}_{q, \bar{q}}(x_1,x_2, \tvec{r})
 &= \sum_{b_1 = q,g}\,
    \int_{x_1}^{1-x_2} \frac{du_1}{u_1}\,
       P_{q,b_1}\biggl(\frac{x_1}{u_1}\biggr)\,
         \sing{F}_{b_1, \bar{q}}(u_1,x_2, \tvec{r})
\nonumber \\
 &+ \sum_{b_2 = \bar{q},g}\,
    \int_{x_2}^{1-x_1} \frac{du_2}{u_2}\,
       P_{\bar{q},b_2}\biggl(\frac{x_2}{u_2}\biggr)\,
         \sing{F}_{q, b_2}(x_1,u_2, \tvec{r})
\nonumber \\
 &+ \frac{1}{x_1+x_2}\,
       P_{q,g}\biggl( \frac{x_1}{x_1+x_2} \biggr)\,
       f_1^g(x_1+x_2) \,,
\end{align}
where the extra term follows from \eqref{evol-splitting}.  At higher
orders in $\alpha_s$, the inhomogeneous term will involve a convolution
integral, as can be anticipated from the graph in figure~\ref{fig:2-4}b.
The appearance of the extra term in the evolution equation can also be
understood in the impact parameter representation by writing
$F_{q,\bar{q}}(x_i, \tvec{r})$ as a Fourier transform
\begin{align}
F_{q,\bar{q}}(x_i, \tvec{r}; \mu) &=
  \biggl[\, \int d^2\tvec{y}\, e^{-i \tvec{r}\tvec{y}}\,
    F_{q,\bar{q}}(x_i, \tvec{y}; \mu) \,\biggr]^{\text{ren}, \mu} \,.
\end{align}
Since $F_{q,\bar{q}}(x_i, \tvec{y})$ has a $1/\tvec{y}^2$ singularity at
small $\tvec{y}$, the integral over this variable is logarithmically
divergent and requires a subtraction in addition to those already made in
$F_{q,\bar{q}}(x_i, \tvec{y})$.  We have indicated this extra subtraction
by $[\, \ldots \,]^{\text{ren}, \mu}$.

For the distribution
\begin{align}
F_{q,\bar{q}}(x_i; \mu) & \underset{\text{def}}{=} F_{q,\bar{q}}(x_i,
\tvec{r}=\tvec{0}; \mu) = 
  \biggl[\, \int d^2\tvec{y}\, F_{q,\bar{q}}(x_i, \tvec{y}; \mu)
  \,\biggr]^{\text{ren}, \mu}
\end{align}
the evolution equation \eqref{inhom-dglap} has long been known in the
literature, see \cite{Kirschner:1979im,Shelest:1982dg,Snigirev:2003cq} and
the recent detailed study \cite{Gaunt:2009re}.  We wish to comment in this
context on an ansatz that is often made in phenomenological studies, in
which the $\tvec{y}$ dependent two-parton distributions are written as
\begin{align}
  \label{factor-ansatz}
F(x_i, \tvec{y}; \mu) = f(\tvec{y})\, F(x_i; \mu) \,,
\end{align}
where $f(\tvec{y})$ is a smooth function normalized as $\int d^2\tvec{y}\,
f(\tvec{y}) = 1$.  A typical choice for $f(\tvec{y})$ is e.g.\ a Gaussian
or a sum of Gaussians.  This type of ansatz is obviously inconsistent if
$F(x_i, \tvec{y}; \mu)$ is defined from the product \eqref{twice-twist-2}
of twist-two operators, since the $\mu$ dependence of the l.h.s.\ is then
given by the homogeneous evolution equation whereas the $\mu$ dependence
on the r.h.s.\ is governed by the inhomogeneous evolution equation
\eqref{inhom-dglap}.  If one instead defines the $\tvec{y}$ dependent
distribution as the Fourier transform of $F(x_i, \tvec{r})$ as in
\eqref{modif-F-y} then the ansatz \eqref{factor-ansatz} is consistent
regarding evolution since by construction $F^{\text{mod}}(x_i, \tvec{y};
\mu)$ evolves as in \eqref{inhom-dglap}.  We do however not think that
this procedure is satisfactory.  As we have seen in \eqref{Fourier-log},
$F(x_i, \tvec{y})^{\text{mod}}$ differs from $F(x_i, \tvec{y})$ only by
terms proportional to $\delta^{(2)}(\tvec{y})$, and such terms do not
appear in the ansatz \eqref{factor-ansatz}, which is smooth and finite at
$\tvec{y} = \tvec{0}$.  In more physical terms, we recall that the
inhomogeneous term in the evolution equation \eqref{inhom-dglap} has its
origin in the $1/\tvec{y}^2$ behavior of $F(x_i, \tvec{y})$ at short
distances, which is not described by \eqref{factor-ansatz}.

We have seen in section~\ref{sec:splitting-X-sect} that this
short-distance behavior prevents us from using either $F(x_i, \tvec{y})$
or $F^{\text{mod}}(x_i, \tvec{y})$ in the double-scattering factorization
formula as it stands.  An ansatz like \eqref{factor-ansatz} with a smooth
function $f(\tvec{y})$ does not have this problem and may be regarded as
modeling a $\tvec{y}$ distribution where the perturbative splitting
contribution that gives rise to the $1/\tvec{y}^2$ singularity has been
removed.  Since the ansatz is ad hoc, one cannot say which evolution
equation should then be used on both sides of \eqref{factor-ansatz}.  Our
discussion suggests that the homogeneous form \eqref{hom-dglap} may be
more appropriate, at least for values $\tvec{y}$ of typical hadronic size,
which are of course most important when the ansatz is used in the
factorization formula.  With this choice, one also retains consistency
with respect to evolution if one makes the additional ansatz $F(x_i,\mu) =
f(x_1,\mu)\ms f(x_2,\mu)$, as is often done.  To find a systematic
solution that treats both splitting and non-splitting contributions in a
consistent manner remains a task for future work.

For the reasons discussed in section~\ref{sec:coll-fact}, the evolution of
color octet distributions $\oct{F}$ differs from the one of $\sing{F}$,
and these differences have not yet been worked out in detail.  However,
the issues discussed in the present section affect $\oct{F}$ in the same
way as $\sing{F}$, given that both the ladder graphs in
figure~\ref{fig:ladders} and the splitting graph in figure~\ref{fig:2-4}a
differ only by overall factors between the singlet and octet channels.
They hence give rise to the same logarithmic divergences when the relevant
transverse momenta are integrated over.  One may therefore expect that,
once a solution of the above problems for singlet distributions is found,
it will be possible to adapt it to the octet sector.

\section{Conclusions}
\label{sec:conclude}

We have investigated several aspects of multiparton interactions in QCD.
Such interactions can contribute to hadron-hadron collisions whenever one
has a final state with several groups of particles for which the vector
sum of transverse momenta is small compared with the large scale $Q$ that
characterizes the process.  As we have shown in
sections~\ref{sec:sing-vs-mult} and \ref{sec:power-counting}, multiple
interactions are then \emph{not} power suppressed in $1/Q$ compared with
the mechanism where these groups of particles are produced in a single
hard scattering.  Examples are the production of two lepton pairs
originating from the decay of two vector bosons with low transverse
momenta, or the production of two dijet pairs that are approximately
back-to-back.  For small parton momentum fractions $x$, which are typical
of collisions at the LHC, multiple hard scattering can even be enhanced
because one expects multiparton distributions to rise faster with
decreasing $x$ than single-parton densities, as we argued in
section~\ref{sec:small-x}.

Given the importance of transverse momenta in the final state, we have
given a factorization formula for multiple hard scattering in terms of
multiparton distributions that depend on the transverse momenta of the
partons.  Such a formula can be fully derived for lowest-order Feynman
graphs and generalizes the more familiar description in terms of collinear
(i.e.\ transverse-momentum integrated) multiparton distributions given in
the literature \cite{Paver:1982yp,Mekhfi:1983az,Calucci:2009ea}.  A
physically intuitive interpretation is obtained if one expresses the cross
section in a mixed representation, in which the multiparton distributions
depend on the average transverse momentum of the partons and on their
average transverse distance from each other, where the ``average'' refers
to the scattering amplitude of the process and its complex conjugate.
These distributions have the structure of Wigner functions.

The simple picture just sketched is however complicated by the presence of
correlation and interference effects, some of which have been pointed out
earlier in the literature \cite{Mekhfi:1985dv}.  The spin and color of the
partons described by a multiparton distribution can be correlated, and
such correlations change the overall rate of multiple interactions.
Two-quark distributions allow two color couplings, which we classified as
color singlet and color octet, whereas for gluons a number of color
couplings appear in addition to the color singlet one, see
section~\ref{sec:color}.  In section~\ref{sec:boson-pairs} we have shown
that spin correlations can also affect the distribution of particles in
the final state, using four-lepton production as an example.  Further
contributions to the cross section can come from interference effects in
fermion number or in quark flavor (figures~\ref{fig:double-scatt}c and
\ref{fig:interference}) and from the interference between single and
multiple hard scattering (figure~\ref{fig:power-beh-2}a).  One can however
expect that these interference effects will not benefit from the small-$x$
enhancement of multiple interactions mentioned above (although in the case
of interference between single and multiple scattering the situation is
not entirely settled as explained in section~\ref{sec:small-x}).
Regarding ``rescattering contributions'' of the type shown in
figure~\ref{fig:rescatter}a, we have shown that their evaluation in terms
of two sequential scattering processes with on-shell external partons is
inappropriate and that, when calculated properly, such contributions are
suppressed by powers of $1/Q$.

How large the above correlations and interference effects are remains an
important open question, both for the phenomenology of multiple
interactions and from the point of view of hadron structure.  The
possibility to study moments of multiparton distributions on the lattice
as explained in section~\ref{sec:mellin}, as well as the approximate
relations with generalized parton distributions we derived in
sections~\ref{sec:reduct} and \ref{sec:gpd-connection} provide two
possible avenues to investigate these issues further.

A proper factorization formula in QCD requires much more than an analysis
of the lowest-order Feynman graphs contributing to the process in
question.  In section~\ref{sec:factorization} we have taken first steps
towards a factorization proof for double hard scattering in terms of
transverse-momentum dependent distributions.  Our investigation only
applies to processes where each hard scatter produces color-singlet
particles, given the limitation of our current understanding for single
hard-scattering processes \cite{Rogers:2010dm}.  For definiteness we have
restricted our analysis to the double Drell-Yan process.  We have shown
how collinear and soft gluon exchange at order $\alpha_s$ can be arranged
into Wilson lines, which are basic building blocks in the construction of
an all-order factorization formula.  We have also seen that at this order
soft-gluon effects cancel in factorization formulae that involve collinear
two-parton distributions in the color singlet sector, whereas they do not
cancel in the color octet sector.  In section~\ref{sec:coll-soft} we have
listed the many issues that remain to be clarified and worked out for a
full factorization proof.  The most critical questions are probably
whether one can show that the effect of soft gluons in the Glauber region
cancels in the cross section and whether the double counting problem
mentioned below can be solved in a satisfactory way.

Our calculation of soft-gluon effects at leading order in $\alpha_s$ also
allows us to investigate the structure of Sudakov logarithms in the double
Drell-Yan process, extending the method of Collins, Soper and Sterman
\cite{Collins:1984kg}.  We find that the leading double logarithms are
given by the product of the corresponding Sudakov factors for each single
scattering process, whereas beyond this approximation soft gluon effects
connect the two hard scatters in a nontrivial way.  In the region where
all transverse parton momenta are large and the transverse distance
$\tvec{y}$ between the two partons is small compared to a hadronic scale,
we find that Sudakov effects favor the color singlet coupling in two-quark
distributions.  If this result could be generalized to large $\tvec{y}$,
it would provide a valuable simplification.

In generic kinematics, the description of multiple interactions involves a
multitude of terms, with many unknown distributions that describe
correlation effects already in the case of double hard scattering (not to
speak of the case with three or more scatters).  The predictive power of
the theory is increased in the region where the net transverse momentum
$q_T$ for each final state produced by a hard scattering is large compared
with the scale $\Lambda$ of nonperturbative interactions (while still
being small compared with the scale $Q$ characterizing the hard-scattering
processes).  Apart from the possible simplification due to Sudakov effects
just mentioned, the transverse-momentum dependent multiparton
distributions can then be computed in terms of collinear distributions and
a hard scattering at scale $q_T$.  The generation of high transverse
momenta can proceed by ladder graphs as in figure~\ref{fig:ladders}, and
we find that the color factors of these graphs favor the color singlet
coupling in two-parton distributions.

A different mechanism is shown in figure~\ref{fig:2-4}, where one parton
splits into two partons that subsequently take part in a hard scatter.  By
explicit calculation at order $\alpha_s$ we find that this splitting
mechanism generates a multitude of spin correlations between the two
emerging partons.  For $q\bar{q}$ distributions the color singlet coupling
is preferred, whereas for $qg$ and $gg$ distributions the opposite is the
case.  Contributions from ladder graphs and from parton splitting graphs
compete with each other in the double scattering cross section.  An
overview is given in table~\ref{tab:splitting-power}, where we see that
compared with splitting graphs the contribution of ladder graphs is
suppressed by powers of $\Lambda /q_T$.  On the other hand, the splitting
graphs lack the small-$x$ enhancement discussed earlier, so that one
cannot decide on generic grounds which mechanism is more important in
given kinematics.  Finally, we find that splitting contributions require a
modification of the formalism outlined so far, because they increase so
strongly for decreasing interparton distance $\tvec{y}$ that one obtains
divergent integrals when inserting them into the factorization formulae.
This is closely related with the problem that graphs like in
figure~\ref{fig:2-4-X}a can either be interpreted as representing double
hard scattering with parton splitting in each two-parton distribution, or
as representing a single hard-scattering process at two-loop level.  A
consistent factorization scheme must ensure that there is no double
counting of this graph in different kinematic regions.  A satisfactory
solution of these problems remains to be found, and as we argued in
section~\ref{sec:evolution} such a solution will also have consequences on
the evolution equation for collinear multiparton distributions.

In summary, we find that a systematic description of multiparton
interactions in QCD involves a considerable degree of complexity, but that
there are several elements that hint at possible simplifications.  More
work is required to work out these simplifications and to put the theory
on firmer ground.

\appendix

\section{Two-dimensional Fourier transform of the logarithm}
\label{app:fourier}

In this appendix we prove the relation
\begin{align}
  \label{Fourier-log-again}
\int \frac{d^2\tvec{r}}{4\pi}\; e^{i \tvec{r} \tvec{y}}\,
  \log\frac{\mu^2}{\tvec{r}^2}
&=  \underset{\varepsilon\to 0}{\lim} \,\biggl[
    \frac{1}{\tvec{y}^2}\, \theta(\tvec{y}^2 - \varepsilon)
  - \delta^{(2)}(\tvec{y}) \int d^2\tvec{y}'\, \frac{1}{\tvec{y}'^2}\,
      \theta(\tvec{y}'^2 - \varepsilon)\,
      \theta(b_0^2 - \mu^2 \tvec{y}'^2)
    \biggr]
\end{align}
with $b_0 = 2 e^{-\gamma}$, which we used in \eqref{Fourier-log}.  To this
end we integrate the relation over a test function, which must be
differentiable and decrease sufficiently fast for $\tvec{y}^2 \to \infty$.
We have
\begin{align}
\mathcal{I} &= \int d^2\tvec{y}\; f(\tvec{y})
               \int \frac{d^2\tvec{r}}{4\pi}\;
     e^{i \tvec{r} \tvec{y}}\, \log\frac{\mu^2}{\tvec{r}^2}
 = \int d^2\tvec{y}\; f(\tvec{y}) \int \frac{d^2\tvec{r}}{4\pi}\,
   \biggl[ \frac{1}{i}\ms \frac{\partial}{\partial\tvec{y}^j}\,
     \frac{\tvec{r}^j}{\tvec{r}^2}\;
     e^{i \tvec{r} \tvec{y}} \biggr]\, \log\frac{\mu^2}{\tvec{r}^2}
\nonumber \\
&= \int d^2\tvec{y}\;
   \biggl[ i\ms \frac{\partial}{\partial\tvec{y}^j}\, f(\tvec{y}) \biggr]
   \int \frac{d^2\tvec{r}}{4\pi}\, e^{i \tvec{r} \tvec{y}}\,
   \frac{\tvec{r}^j}{\tvec{r}^2}\, \log\frac{\mu^2}{\tvec{r}^2} \,.
\end{align}
The integral over $\tvec{r}$ is convergent and gives
\begin{align}
\int \frac{d^2\tvec{r}}{4\pi}\, e^{i \tvec{r} \tvec{y}}\,
   \frac{\tvec{r}^j}{\tvec{r}^2}\, \log\frac{\mu^2}{\tvec{r}^2}
&= \frac{i}{2}\; \frac{\tvec{y}^j}{|\tvec{y}|}
   \int_{0}^{\infty} dr\,
    J_1\bigl( r |\tvec{y}| \bigr) \log\frac{\mu^2}{\tvec{r}^2}
 = \frac{i}{2}\; \frac{\tvec{y}^j}{\tvec{y}^2}\,
    \log\frac{\mu^2 \tvec{y}^2}{b_0^2} \,,
\end{align}
which leads to
\begin{align}
\mathcal{I} &= - \frac{1}{2} \int d^2\tvec{y}\;
     \frac{\tvec{y}^j}{\tvec{y}^2}\,
     \log\frac{\mu^2 \tvec{y}^2}{b_0^2}\;
     \frac{\partial}{\partial\tvec{y}^j}\, f(\tvec{y})
\nonumber \\
 &= - \frac{1}{2} \hspace{-1ex}
    \int\limits_{\mu^2 \tvec{y}^2 < b_0^2} \hspace{-2.4ex} d^2\tvec{y}\;
     \frac{\tvec{y}^j}{\tvec{y}^2}\,
     \log\frac{\mu^2 \tvec{y}^2}{b_0^2}\;
     \frac{\partial}{\partial\tvec{y}^j}\,
        \bigl[ f(\tvec{y}) - f(\tvec{0}) \bigr]
    - \frac{1}{2} \hspace{-1ex}
     \int\limits_{\mu^2 \tvec{y}^2 > b_0^2} \hspace{-2.4ex} d^2\tvec{y}\;
     \frac{\tvec{y}^j}{\tvec{y}^2}\,
     \log\frac{\mu^2 \tvec{y}^2}{b_0^2}\;
     \frac{\partial}{\partial\tvec{y}^j}\, f(\tvec{y}) \,.
\end{align}
The integration region has been split in such a way that integration by
parts does not give any boundary term, so that one has
\begin{align}
\mathcal{I} &= \int\limits_{\mu^2 \tvec{y}^2 < b_0^2}
               \hspace{-2ex} d^2\tvec{y}\;
     \frac{f(\tvec{y}) - f(\tvec{0})}{\tvec{y}^2}
   + \hspace{-1.5ex}
     \int\limits_{\mu^2 \tvec{y}^2 > b_0^2} \hspace{-2ex} d^2\tvec{y}\;
     \frac{f(\tvec{y})}{\tvec{y}^2}
 = \int d^2\tvec{y}\; \frac{f(\tvec{y})
      - \theta(b_0^2 - \mu^2 \tvec{y}^2)\, f(\tvec{0})}{\tvec{y}^2} \,.
\end{align}
For an alternative derivation of the result in this form (with test
functions depending on~$\tvec{y}^2$) we refer to eqs.~(129), (133) and
(141) in \cite{Bozzi:2005wk}.  Further rewriting
\begin{align}
\mathcal{I} &= \underset{\varepsilon\to 0}{\lim}
     \int\limits_{\tvec{y}^2 > \varepsilon} \hspace{-1ex} d^2\tvec{y}\;
     \frac{f(\tvec{y})
       - \theta(b_0^2 - \mu^2 \tvec{y}^2)\, f(\tvec{0})}{\tvec{y}^2}
\end{align}
we can separate the terms with $f(\tvec{y})$ and with $f(\tvec{0})$ and
thus obtain \eqref{Fourier-log-again}.


\acknowledgments

We gratefully acknowledge discussions with J.~Bartels, J.~Bl\"umlein,
D.~Boer, V.~Braun, S.~Brodsky, F.~Ceccopieri, S.~Dawson, Yu.~Dokshitzer,
J.~Gaunt, Ph.~H\"agler, T.~Kasemets, F.~Krauss, P.~Kroll, L.~Lipatov,
Z.~Nagy, S.~Pl\"atzer, T.~Rogers, D.~Soper, R.~Venugopalan and
W.~Vogelsang.
A part of the calculations for this work was done using FORM
\cite{Vermaseren:2000nd}, and the figures were produced with JaxoDraw
\cite{Binosi:2003yf}.


\phantomsection

\newpage

\section*{Changes in arXiv version 3 compared with version 2}

\begin{description}
\item[equations \eqref{Ua-final} and \eqref{vw-to-y}:] Terms $v^+ v^-$
  have been changed into $v^+ w^-$.
\item[equation \eqref{mellin-moment}:] The factor $2$ on the r.h.s.\ has
  been changed into $1/2$.
\item[equation \eqref{splitting-mom-int}:] The factors $(2\pi)^2$ have
  been omitted on both sides.
\item[equation \eqref{epsilon-kernel}:] The expression has been divided by
  $(1-\epsilon)$, and an additional factor is mentioned after the
  equation.
\end{description}


\addcontentsline{toc}{section}{References}

\begin{thebibliography}{999}

\bibitem{Landshoff:1975eb}
  P.~V.~Landshoff, J.~C.~Polkinghorne, D.~M.~Scott,
  Phys.\ Rev.\  {\bf D12} (1975)  3738.

\bibitem{Landshoff:1978fq}
  P.~V.~Landshoff, J.~C.~Polkinghorne,
  Phys.\ Rev.\  {\bf D18} (1978)  3344.


\bibitem{Humpert:1983pw}
  B.~Humpert,
  Phys.\ Lett.\  {\bf B131} (1983)  461.

\bibitem{Humpert:1984ay}
  B.~Humpert, R.~Odorico,
  Phys.\ Lett.\  {\bf B154} (1985)  211.

\bibitem{Ametller:1985tp}
  L.~Ametller, N.~Paver, D.~Treleani,
  Phys.\ Lett.\  {\bf B169} (1986)  289.

\bibitem{Mangano:1988sq}
  M.~L.~Mangano,
  Z.\ Phys.\  {\bf C42} (1989)  331.

\bibitem{DelFabbro:2002pw}
  A.~Del Fabbro, D.~Treleani,
  Phys.\ Rev.\  {\bf D66} (2002)  074012
  [hep-ph/0207311].

\bibitem{Domdey:2009bg}
  S.~Domdey, H.~J.~Pirner, U.~A.~Wiedemann,
  Eur.\ Phys.\ J.\ {\bf C65} (2010) 153
  [arXiv:0906.4335].

\bibitem{Berger:2009cm}
  E.~L.~Berger, C.~B.~Jackson, G.~Shaughnessy,
  Phys.\ Rev.\  D {\bf 81} (2010) 014014
  [arXiv:0911.5348].

\bibitem{Drees:1996rw}
  M.~Drees, T.~Han,
  Phys.\ Rev.\ Lett.\  {\bf 77} (1996)  4142
  [hep-ph/9605430].


\bibitem{Goebel:1979mi}
  C.~Goebel, F.~Halzen, D.~M.~Scott,
  Phys.\ Rev.\  {\bf D22 } (1980)  2789.

\bibitem{Halzen:1986ue}
  F.~Halzen, P.~Hoyer, W.~J.~Stirling,
  Phys.\ Lett.\  {\bf B188} (1987)  375.

\bibitem{Kom:2011nu}
  C.~H.~Kom, A.~Kulesza, W.~J.~Stirling,
  Eur.\ Phys.\ J.\ C {\bf 71} (2011) 1802
  [arXiv:1109.0309].


\bibitem{Kom:2011bd}
  C.~H.~Kom, A.~Kulesza, W.~J.~Stirling,
  Phys.\ Rev.\ Lett.\  {\bf 107} (2011)  082002
  [arXiv:1105.4186].

\bibitem{Baranov:2011ch}
  S.~P.~Baranov, A.~M.~Snigirev, N.~P.~Zotov,
  Phys.\ Lett.\  {\bf B705} (2011)  116
  [arXiv:1105.6276].

\bibitem{Novoselov:2011ff}
  A.~Novoselov,
  [arXiv:1106.2184].


\bibitem{Godbole:1989ti}
  R.~M.~Godbole, S.~Gupta, J.~Lindfors,
  Z.\ Phys.\  {\bf C47 } (1990)  69.

\bibitem{Eboli:1997sv}
  O.~J.~P.~Eboli, F.~Halzen, J.~K.~Mizukoshi,
  Phys.\ Rev.\  {\bf D57} (1998)  1730
  [hep-ph/9710443].

\bibitem{DelFabbro:1999tf}
  A.~Del Fabbro, D.~Treleani,
  Phys.\ Rev.\  {\bf D61} (2000)  077502
  [hep-ph/9911358].

\bibitem{Cattaruzza:2005nu}
  E.~Cattaruzza, A.~Del Fabbro, D.~Treleani,
  Phys.\ Rev.\  {\bf D72} (2005)  034022
  [hep-ph/0507052].

\bibitem{Maina:2009vx}
  E.~Maina,
  JHEP {\bf 0904} (2009)  098
  [arXiv:0904.2682].

\bibitem{Maina:2009sj}
  E.~Maina,
  JHEP {\bf 0909} (2009)  081
  [arXiv:0909.1586].

\bibitem{Maina:2010vh}
  E.~Maina,
  JHEP {\bf 1101} (2011)  061
  [arXiv:1010.5674].

\bibitem{Kulesza:1999zh}
  A.~Kulesza, W.~J.~Stirling,
  Phys.\ Lett.\  {\bf B475} (2000)  168
  [hep-ph/9912232].

\bibitem{Gaunt:2010pi}
  J.~R.~Gaunt, C.-H.~Kom, A.~Kulesza, W.~J.~Stirling,
  Eur.\ Phys.\ J.\  {\bf C69} (2010)  53
  [arXiv:1003.3953].

\bibitem{Berger:2011ep}
  E.~L.~Berger, C.~B.~Jackson, S.~Quackenbush, G.~Shaughnessy,
  Phys.\ Rev.\ D {\bf 84} (2011) 074021
  [arXiv:1107.3150].



\bibitem{Akesson:1986iv}
  T.~Akesson {\it et al.} [Axial Field Spectrometer Collaboration],
  Z.\ Phys.\  {\bf C34 } (1987)  163.

\bibitem{Alitti:1991rd}
  J.~Alitti {\it et al.} [UA2 Collaboration],
  Phys.\ Lett.\  {\bf B268 } (1991)  145.

\bibitem{Abe:1993rv}
  F.~Abe {\it et al.} [CDF Collaboration],
  Phys.\ Rev.\  {\bf D47} (1993)  4857.

\bibitem{Abe:1997bp}
  F.~Abe {\it et al.}  [CDF Collaboration],
  Phys.\ Rev.\ Lett.\  {\bf 79} (1997) 584.

\bibitem{Abe:1997xk}
  F.~Abe {\it et al.}  [CDF Collaboration],
  Phys.\ Rev.\  {\bf D56} (1997) 3811.

\bibitem{Abazov:2009gc}
  V.~M.~Abazov {\it et al.} [D0 Collaboration],
  Phys.\ Rev.\  {\bf D81} (2010)  052012
  [arXiv:0912.5104].

\bibitem{Abazov:2011rd}
  V.~M.~Abazov {\it et al.} [D0 Collaboration],
  Phys.\ Rev.\  {\bf D83} (2011)  052008
  [arXiv:1101.1509].


\bibitem{Sjostrand:2004pf}
  T.~Sj\"ostrand, P.~Z.~Skands,
  JHEP {\bf 0403} (2004) 053
  [hep-ph/0402078].

\bibitem{Buckley:2011ms}
  A.~Buckley {\it et al.},
  Phys.\ Rept.\  {\bf 504} (2011)  145
  [arXiv:1101.2599].


\bibitem{Alekhin:2005dx}
  S.~Alekhin {\it et al.},
  Proceedings on the Workshop on HERA and the LHC,
  DESY and CERN, 2004--2005, parts A and B
  [hep-ph/0601012] and [hep-ph/0601013].

\bibitem{Jung:2009eq}
  H.~Jung {\it et al.},
  Proceedings of the Workshop on HERA and the LHC,
  DESY and CERN, 2006--2008
  [arXiv:0903.3861].

\bibitem{Bartalini:2010su}
  P.~Bartalini {\it et al.},
  Proceedings of MPI 08, Perugia, Italy, October 27--31, 2008
  [arXiv:1003.4220].

\bibitem{Bartalini:2011jp}
  P.~Bartalini {\it et al.},
  [arXiv:1111.0469].

\bibitem{Atlas:2011co}
  ATLAS Collaboration,
  Note ATLAS-CONF-2011-160,
  \texttt{http://cdsweb.cern.ch/record/1404953}.

\bibitem{Bartalini:2011xj}
  P.~Bartalini, L.~Fan\`o,
  [arXiv:1103.6201].



\bibitem{Accardi:2001ih}
  A.~Accardi, D.~Treleani,
  Phys.\ Rev.\  D {\bf 64} (2001) 116004
  [hep-ph/0106306].

\bibitem{Strikman:2001gz}
  M.~Strikman, D.~Treleani,
  Phys.\ Rev.\ Lett.\  {\bf 88} (2002) 031801
  [hep-ph/0111468].

\bibitem{Cattaruzza:2004qb}
  E.~Cattaruzza, A.~Del Fabbro, D.~Treleani,
  Phys.\ Rev.\  D {\bf 70} (2004) 034022
  [hep-ph/0404177].

\bibitem{Calucci:2010wg}
  G.~Calucci, D.~Treleani,
  Phys.\ Rev.\  {\bf D83} (2011)  016012
  [arXiv:1009.5881].

\bibitem{Strikman:2010bg}
  M.~Strikman, W.~Vogelsang,
  Phys.\ Rev.\  {\bf D83} (2011)  034029
  [arXiv:1009.6123].

\bibitem{Diehl:2011tt}
  M.~Diehl, A.~Sch\"afer,
  Phys.\ Lett.\  {\bf B698 } (2011)  389
  [arXiv:1102.3081].


\bibitem{Collins:1981uk}
  J.~C.~Collins, D.~E.~Soper,
  Nucl.\ Phys.\  {\bf B193} (1981) 381.

\bibitem{Collins:2007ph}
  J.~C.~Collins, T.~C.~Rogers, A.~M.~Stasto,
  Phys.\ Rev.\  {\bf D77} (2008) 085009
  [arXiv:0708.2833].

\bibitem{Collins:2011}
  J.~C.~Collins, 
  \emph{The Foundations of Perturbative QCD},
  Cambridge University Press, Cambridge 2011.

\bibitem{Ji:2004wu}
  X.-D.~Ji, J.-P.~Ma, F.~Yuan,
  Phys.\ Rev.\  {\bf D71} (2005) 034005
  [hep-ph/0404183].



\bibitem{Paver:1982yp}
  N.~Paver, D.~Treleani,
  Nuovo Cim.\ {\bf A70} (1982) 215.

\bibitem{Mekhfi:1983az}
  M.~Mekhfi,
  Phys.\ Rev.\ {\bf D32} (1985) 2371.

\bibitem{Landshoff:1971xb}
  P.~V.~Landshoff, J.~C.~Polkinghorne,
  Phys.\ Rept.\ {\bf 5} (1972) 1.

\bibitem{Diehl:1998sm}
  M.~Diehl, T.~Gousset,
  Phys.\ Lett.\ B {\bf 428} (1998) 359
  [hep-ph/9801233].

\bibitem{Jaffe:1983hp} 
  R.~L.~Jaffe,
  Nucl.\ Phys.\ B {\bf 229} (1983) 205.

\bibitem{Hillery:1984}
  M.~Hillery, R.~F.~O'Connell, M.~O.~Scully, E.~P.~Wigner,
  Phys.\ Rept.\ {\bf 106} (1984) 121.

\bibitem{Belitsky:2003nz}
  A.~V.~Belitsky, X.-D.~Ji, F.~Yuan,
  Phys.\ Rev.\  {\bf D69} (2004) 074014
  [hep-ph/0307383].

\bibitem{Blok:2010ge}
  B.~Blok, Yu.~Dokshitzer, L.~Frankfurt, M.~Strikman,
  Phys.\ Rev.\  {\bf D83} (2011)  071501
  [arXiv:1009.2714].

\bibitem{Brodsky:1989}
  S.~J.~Brodsky and G.~P.~Lepage, in: \textit{Perturbative Quantum
    Chromodynamics}, edited by A.~H.~Mueller (World Scientific, Singapore
  1989).

\bibitem{Diehl:2000xz}
  M.~Diehl, T.~Feldmann, R.~Jakob, P.~Kroll,
  Nucl.\ Phys.\  {\bf B596} (2001) 33
  [hep-ph/0009255].

\bibitem{Diehl:2002he}
  M.~Diehl,
  Eur.\ Phys.\ J.\  C {\bf 25} (2002) 223,
  Erratum ibid.\  C {\bf 31} (2003) 277
  [hep-ph/0205208].

\bibitem{Politzer:1980me}
  H.~D.~Politzer,
  Nucl.\ Phys.\ {\bf B172} (1980) 349.
  
\bibitem{Soper:1976jc}
  D.~E.~Soper,
  Phys.\ Rev.\ {\bf D15} (1977) 1141.

\bibitem{Burkardt:2002hr}
  M.~Burkardt,
  Int.\ J.\ Mod.\ Phys.\ {\bf A18} (2003) 173
  [hep-ph/0207047].

\bibitem{Calucci:2009ea}
  G.~Calucci, D.~Treleani,
  Phys.\ Rev.\ D {\bf 80} (2009) 054025
  [arXiv:0907.4772].

\bibitem{Sjostrand:1986ep}
  T.~Sj\"ostrand, M.~van Zijl,
  Phys.\ Lett.\ {\bf B188} (1987) 149.

\bibitem{Sjostrand:1987su}
  T.~Sj\"ostrand, M.~van Zijl,
  Phys.\ Rev.\ {\bf D36} (1987) 2019.

\bibitem{Durand:1987yv}
  L.~Durand, H.~Pi,
  Phys.\ Rev.\ Lett.\ {\bf 58} (1987) 303.

\bibitem{Durand:1988ax}
  L.~Durand, H.~Pi,
  Phys.\ Rev.\ {\bf D40} (1989)  1436.

\bibitem{Ametller:1987ru}
  L.~Ametller, D.~Treleani,
  Int.\ J.\ Mod.\ Phys.\  {\bf A3} (1988)  521.

\bibitem{Frankfurt:2003td}
  L.~Frankfurt, M.~Strikman, C.~Weiss,
  Phys.\ Rev.\ {\bf D69} (2004) 114010
  [hep-ph/0311231].

\bibitem{Calucci:1997ii}
  G.~Calucci, D.~Treleani,
  Phys.\ Rev.\ {\bf D57} (1998) 503
  [hep-ph/9707389].

\bibitem{Calucci:1999yz}
  G.~Calucci, D.~Treleani,
  Phys.\ Rev.\  {\bf D60} (1999) 054023
  [hep-ph/9902479].

\bibitem{DelFabbro:2000ds}
  A.~Del Fabbro, D.~Treleani,
  Phys.\ Rev.\  D {\bf 63} (2001) 057901
  [hep-ph/0005273].

\bibitem{Rogers:2009ke}
  T.~C.~Rogers, M.~Strikman,
  Phys.\ Rev.\ {\bf D81} (2010)  016013
  [arXiv:0908.0251].

\bibitem{Flensburg:2011kj}
  C.~Flensburg, G.~Gustafson, L.~L\"onnblad, A.~Ster,
  JHEP {\bf 1106} (2011)  066
  [arXiv:1103.4320].

\bibitem{Ralston:1979ys}
  J.~P.~Ralston, D.~E.~Soper,
  Nucl.\ Phys.\  B {\bf 152} (1979) 109.

\bibitem{Tangerman:1994eh}
  R.~D.~Tangerman, P.~J.~Mulders,
  Phys.\ Rev.\  D {\bf 51} (1995) 3357
  [hep-ph/9403227].

\bibitem{Jaffe:1989xy}
  R.~L.~Jaffe, A.~Manohar,
  Phys.\ Lett.\  B {\bf 223} (1989) 218.

\bibitem{Belitsky:2000jk}
  A.~V.~Belitsky, D.~M\"uller,
  Phys.\ Lett.\  B {\bf 486} (2000) 369
  [hep-ph/0005028].

\bibitem{Mulders:2000sh}
  P.~J.~Mulders, J.~Rodrigues,
  Phys.\ Rev.\  {\bf D63} (2001) 094021
  [hep-ph/0009343].

\bibitem{Nadolsky:2007ba}
  P.~M.~Nadolsky, C.~Balazs, E.~L.~Berger, C.-P.~Yuan,
  Phys.\ Rev.\  {\bf D76} (2007)  013008
  [hep-ph/0702003].

\bibitem{Catani:2010pd}
  S.~Catani, M.~Grazzini,
  Nucl.\ Phys.\  {\bf B845} (2011)  297
  [arXiv:1011.3918].

\bibitem{Boer:2010zf}
  D.~Boer, S.~J.~Brodsky, P.~J.~Mulders, C.~Pisano,
  Phys.\ Rev.\ Lett.\  {\bf 106} (2011) 132001
  [arXiv:1011.4225].

\bibitem{Qiu:2011ai}
  J.~-W.~Qiu, M.~Schlegel, W.~Vogelsang,
  Phys.\ Rev.\ Lett.\  {\bf 107} (2011)  062001
  [arXiv:1103.3861].

\bibitem{Boer:2011kf}
  D.~Boer, W.~J.~den~Dunnen, C.~Pisano, M.~Schlegel, W.~Vogelsang,
  Phys.\ Rev.\ Lett.\  {\bf 108} (2012) 032002
  [arXiv:1109.1444].

\bibitem{Collins:2008sg}
  J.~C.~Collins, T.~C.~Rogers,
  Phys.\ Rev.\  {\bf D78 } (2008)  054012
  [arXiv:0805.1752].

\bibitem{Mekhfi:1985dv}
  M.~Mekhfi,
  Phys.\ Rev.\  D {\bf 32} (1985) 2380.

\bibitem{MacFarlane:1968vc}
  A.~J.~MacFarlane, A.~Sudbery, P.~H.~Weisz,
  Commun.\ Math.\ Phys.\  {\bf 11} (1968) 77.

\bibitem{Ragazzon:1995cb}
  R.~Ragazzon, D.~Treleani,
  Phys.\ Rev.\  D {\bf 53} (1996) 55
  [hep-ph/9508286].

\bibitem{Braun:2000ua}
  M.~Braun, D.~Treleani,
  Eur.\ Phys.\ J.\  C {\bf 18} (2001) 511
  [hep-ph/0005078].

\bibitem{Bartels:2005wa}
  J.~Bartels, M.~Salvadore, G.~P.~Vacca,
  Eur.\ Phys.\ J.\  {\bf C42} (2005)  53
  [hep-ph/0503049].

\bibitem{Levin:2008jf}
  E.~Levin, J.~Miller,
  Eur.\ Phys.\ J.\  {\bf C61 } (2009)  1
  [arXiv:0803.0646].

\bibitem{Bartels:2011qi}
  J.~Bartels, M.~G.~Ryskin,
  [arXiv:1105.1638].

\bibitem{Bartels:1999yt}
  J.~Bartels, L.~N.~Lipatov, G.~P.~Vacca,
  Phys.\ Lett.\  {\bf B477} (2000)  178
  [hep-ph/9912423].

\bibitem{Janik:1998xj}
  R.~A.~Janik, J.~Wosiek,
  Phys.\ Rev.\ Lett.\  {\bf 82} (1999)  1092
  [hep-th/9802100].

\bibitem{Blok:2011bu}
  B.~Blok, Yu.~Dokshitser, L.~Frankfurt, M.~Strikman,
  [arXiv:1106.5533v2].

\bibitem{Hagler:2009ni}
  Ph.~H\"agler,
  Phys.\ Rept.\  {\bf 490} (2010)  49
  [arXiv:0912.5483].

\bibitem{Aktas:2005xu}
  A.~Aktas {\it et al.} [H1 Collaboration],
  Eur.\ Phys.\ J.\  {\bf C46} (2006)  585
  [hep-ex/0510016].

\bibitem{Chekanov:2002xi}
  S.~Chekanov {\it et al.} [ZEUS Collaboration],
  Eur.\ Phys.\ J.\  {\bf C24 } (2002)  345
  [hep-ex/0201043].

\bibitem{Diehl:2007zu}
  M.~Diehl, W.~Kugler,
  Phys.\ Lett.\  {\bf B660 } (2008)  202
  [arXiv:0711.2184].

\bibitem{Frankfurt:2010ea}
  L.~Frankfurt, M.~Strikman, C.~Weiss,
  Phys.\ Rev.\  {\bf D83 } (2011)  054012
  [arXiv:1009.2559].

\bibitem{Corke:2011yy}
  R.~Corke, T.~Sj\"ostrand,
  JHEP {\bf 1105} (2011)  009
  [arXiv:1101.5953].

\bibitem{Frankfurt:2004kn}
  L.~Frankfurt, M.~Strikman, C.~Weiss,
  Annalen Phys.\  {\bf 13} (2004)  665
  [hep-ph/0410307].


\bibitem{Sterman:1978bi}
  G.~F.~Sterman,
  Phys.\ Rev.\  {\bf D17} (1978)  2773.

\bibitem{Libby:1978bx}
  S.~B.~Libby, G.~F.~Sterman,
  Phys.\ Rev.\  {\bf D18} (1978)  4737.

\bibitem{Coleman:1965xm}
  S.~Coleman, R.~E.~Norton,
  Nuovo Cim.\  {\bf 38} (1965)  438.

\bibitem{Paver:1984ux}
  N.~Paver, D.~Treleani,
  Z.\ Phys.\  {\bf C28} (1985)  187.
  
\bibitem{Corke:2009tk}
  R.~Corke, T.~Sj\"ostrand,
  JHEP {\bf 1001} (2010) 035
  [arXiv:0911.1909].


\bibitem{Rogers:2010dm}
  T.~C.~Rogers, P.~J.~Mulders,
  Phys.\ Rev.\ {\bf D81} (2010) 094006
  [arXiv:1001.2977].

\bibitem{Collins:2004nx}
  J.~C.~Collins, A.~Metz,
  Phys.\ Rev.\ Lett.\ {\bf 93} (2004) 252001
  [hep-ph/0408249].

\bibitem{Belitsky:2002sm}
  A.~V.~Belitsky, X.-D.~Ji, F.~Yuan,
  Nucl.\ Phys.\  {\bf B656} (2003) 165
  [hep-ph/0208038].

\bibitem{Boer:2003cm}
  D.~Boer, P.~J.~Mulders, F.~Pijlman,
  Nucl.\ Phys.\  {\bf B667} (2003) 201
  [hep-ph/0303034].

\bibitem{Aybat:2011zv}
  S.~M.~Aybat, T.~C.~Rogers,
  Phys.\ Rev.\  {\bf D83} (2011) 114042
  [arXiv:1101.5057].

\bibitem{Collins:2003fm}
  J.~C.~Collins,
  Acta Phys.\ Polon.\ {\bf B34} (2003) 3103
  [hep-ph/0304122].

\bibitem{Bacchetta:2008xw}
  A.~Bacchetta, D.~Boer, M.~Diehl, P.~J.~Mulders,
  JHEP {\bf 0808} (2008) 023
  [arXiv:0803.0227].

\bibitem{Collins:1984kg}
  J.~C.~Collins, D.~E.~Soper, G.~F.~Sterman,
  Nucl.\ Phys.\  {\bf B250} (1985) 199.

\bibitem{Idilbi:2004vb}
  A.~Idilbi, X.-D.~Ji, J.-P.~Ma, F.~Yuan,
  Phys.\ Rev.\  {\bf D70} (2004) 074021
  [hep-ph/0406302].

\bibitem{DescotesGenon:2001hm}
  S.~Descotes-Genon, C.~T.~Sachrajda,
  Nucl.\ Phys.\  {\bf B625} (2002) 239
  [hep-ph/0109260].

\bibitem{Mekhfi:1988kj}
  M.~Mekhfi, X.~Artru,
  Phys.\ Rev.\  {\bf D37} (1988) 2618.


\bibitem{Collins:2002kn}
  J.~C.~Collins,
  Phys.\ Lett.\  {\bf B536} (2002)  43
  [hep-ph/0204004].

\bibitem{Hagler:2009mb}
  Ph.~H\"agler, B.~U.~Musch, J.~W.~Negele, A.~Sch\"a\-fer,
  Europhys.\ Lett.\  {\bf 88} (2009) 61001
  [arXiv:0908.1283].

\bibitem{Musch:2010ka}
  B.~U.~Musch, Ph.~H\"agler, J.~W.~Negele, A.~Sch\"a\-fer,
  Phys.\ Rev.\  {\bf D83} (2011)  094507
  [arXiv:1011.1213].

\bibitem{Alexandrou:2008ru}
  C.~Alexandrou, G.~Koutsou,
  Phys.\ Rev.\  {\bf D78} (2008)  094506
  [arXiv:0809.2056].

\bibitem{Mueller:1998fv}
  D.~M\"uller, D.~Robaschik, B.~Geyer, F.~M.~Dittes, J.~Ho\v{r}ej\v{s}i,
  Fortsch.\ Phys.\  {\bf 42} (1994) 101
  [hep-ph/9812448].

\bibitem{Ji:1996ek}
  X.~D.~Ji,
  Phys.\ Rev.\ Lett.\  {\bf 78} (1997) 610
  [hep-ph/9603249].

\bibitem{Radyushkin:1997ki}
  A.~V.~Radyushkin,
  Phys.\ Rev.\  {\bf D56} (1997) 5524
  [hep-ph/9704207].

\bibitem{Goeke:2001tz}
  K.~Goeke, M.~V.~Polyakov, M.~Vanderhaeghen,
  Prog.\ Part.\ Nucl.\ Phys.\  {\bf 47} (2001)  401
  [hep-ph/0106012].

\bibitem{Diehl:2003ny}
  M.~Diehl,
  Phys.\ Rept.\  {\bf 388} (2003) 41
  [hep-ph/0307382].

\bibitem{Belitsky:2005qn}
  A.~V.~Belitsky, A.~V.~Radyushkin,
  Phys.\ Rept.\  {\bf 418} (2005)  1
  [hep-ph/0504030].

\bibitem{Meissner:2009ww}
  S.~Meissner, A.~Metz, M.~Schlegel,
  JHEP {\bf 0908} (2009) 056
  [arXiv:0906.5323].

\bibitem{Soper:1972xc}
  D.~E.~Soper,
  Phys.\ Rev.\  {\bf D5} (1972) 1956.



\bibitem{Ji:2006ub}
  X.-D.~Ji, J.-W.~Qiu, W.~Vogelsang, F.~Yuan,
  Phys.\ Rev.\ Lett.\  {\bf 97} (2006) 082002
  [hep-ph/0602239].

\bibitem{Ji:2006vf}
  X.-D.~Ji, J.-W.~Qiu, W.~Vogelsang, F.~Yuan,
  Phys.\ Rev.\  {\bf D73} (2006) 094017
  [hep-ph/0604023].

\bibitem{Ji:2006br}
  X.-D.~Ji, J.-W.~Qiu, W.~Vogelsang, F.~Yuan,
  Phys.\ Lett.\  {\bf B638} (2006) 178
  [hep-ph/0604128].

\bibitem{Koike:2007dg}
  Y.~Koike, W.~Vogelsang, F.~Yuan,
  Phys.\ Lett.\  {\bf B659} (2008) 878
  [arXiv:0711.0636].

\bibitem{Bukhvostov:1985rn}
  A.~P.~Bukhvostov, G.~V.~Frolov, L.~N.~Lipatov, E.~A.~Kuraev,
  Nucl.\ Phys.\  {\bf B258} (1985) 601.

\bibitem{Lipatov:2012pc}
  L.~N.~Lipatov,
  private communication (2012).

\bibitem{Bartels:1993ih}
  J.~Bartels,
  Z.\ Phys.\ C {\bf 60} (1993) 471.

\bibitem{Cacciari:2009dp}
  M.~Cacciari, G.~P.~Salam, S.~Sapeta,
  JHEP {\bf 1004} (2010) 065
  [arXiv:0912.4926].

\bibitem{Gaunt:2011xd}
  J.~R.~Gaunt, W.~J.~Stirling,
  JHEP {\bf 1106} (2011) 048
  [arXiv:1103.1888].

\bibitem{Ryskin:2011kk}
  M.~G.~Ryskin, A.~M.~Snigirev,
  Phys.\ Rev.\  {\bf D83} (2011) 114047
  [arXiv:1103.3495].

\bibitem{Kirschner:1979im}
  R.~Kirschner,
  Phys.\ Lett.\  {\bf B84} (1979) 266.

\bibitem{Shelest:1982dg}
  V.~P.~Shelest, A.~M.~Snigirev, G.~M.~Zinovev,
  Phys.\ Lett.\  {\bf B113} (1982) 325.
  
\bibitem{Snigirev:2003cq}
  A.~M.~Snigirev,
  Phys.\ Rev.\  {\bf D68} (2003) 114012
  [hep-ph/0304172].

\bibitem{Gaunt:2009re}
  J.~R.~Gaunt, W.~J.~Stirling,
  JHEP {\bf 1003} (2010) 005
  [arXiv:0910.4347].

\bibitem{Bozzi:2005wk}
  G.~Bozzi, S.~Catani, D.~de Florian, M.~Grazzini,
  Nucl.\ Phys.\  {\bf B737} (2006) 73
  [hep-ph/0508068].




\bibitem{Vermaseren:2000nd}
  J.~A.~M.~Vermaseren,
  [math-ph/0010025].

\bibitem{Binosi:2003yf}
  D.~Binosi, L.~Theussl,
  Comput.\ Phys.\ Commun.\  {\bf 161} (2004)  76
  [hep-ph/0309015].

\end{thebibliography}
\end{document}